\documentclass[a4paper,12pt]{article}
\pdfoutput=1
\usepackage{epsfig,graphicx,xcolor,amsbsy,amssymb,latexsym,amsfonts,amsmath,,tcolorbox,setspace}
\usepackage{pstricks}
\usepackage{color}
\usepackage{soul}
\usepackage[numbers,square,comma, compress]{natbib}
\usepackage{placeins}
\usepackage{wrapfig}

\usepackage{eurosym}
\usepackage[a4paper]{geometry}
\usepackage{graphicx}
\usepackage{tikz}
\usepackage{xcolor}
\usepackage{amsmath,amssymb,amsfonts,pstricks,setspace}
\usepackage[enableskew]{youngtab}

\numberwithin{equation}{section}

\newcommand{\bea}{\begin{eqnarray}\displaystyle}
\newcommand{\eea}{\end{eqnarray}}

\newcommand{\figref}[1]{Fig.~\protect\ref{#1}}

\title{
\begin{flushright}{\vspace{-2.5cm}\small LYCEN 2018-10\\}\end{flushright}
\vspace{2.3cm}
\bf{Dihedral Symmetries of Gauge Theories from Dual Calabi-Yau Threefolds}\\[45pt]}
\author{\large \textsc{Brice Bastian\footnote{\tt b.bastian@ipnl.in2p3.fr}}\,\,\, and\, \textsc{Stefan~Hohenegger\footnote{\tt s.hohenegger@ipnl.in2p3.fr}}}
\date{}

\begin{document}

\maketitle

\begin{center}
\renewcommand{\thefootnote}{\fnsymbol{footnote}}\vspace{-0.5cm}
${}^{\footnotemark[1]\footnotemark[2]}$ Universit\'e de Lyon\\
UMR 5822, CNRS/IN2P3, Institut de Physique Nucl\'eaire de Lyon\\ 4 rue Enrico Fermi, 69622 Villeurbanne Cedex, \rm FRANCE\\[2.5cm]
\end{center}

\begin{abstract}
Recent studies (arXiv:1610.07916, arXiv:1711.07921, arXiv:1807.00186) of six-di\-men\-sion\-al supersymmetric gauge theories that are engineered by a class of toric Calabi-Yau threefolds $X_{N,M}$, have uncovered a vast web of dualities. In this paper we analyse consequences of these dualities from the perspective of the partition functions $\mathcal{Z}_{N,M}$ (or the free energy $\mathcal{F}_{N,M}$) of these theories. Focusing on the case $M=1$, we find that the latter is invariant under the group $\mathbb{G}(N)\times S_N$: here $S_N$ corresponds to the Weyl group of the largest gauge group that can be engineered from $X_{N,1}$ and $\mathbb{G}(N)$ is a dihedral group, which acts in an intrinsically non-perturbative fashion and which is of infinite order for $N\geq 4$. We give an explicit representation of $\mathbb{G}(N)$ as a matrix group that is freely generated by two elements which act naturally on a specific basis of the K\"ahler moduli space of $X_{N,1}$. While we show the invariance of $\mathcal{Z}_{N,1}$ under $\mathbb{G}(N)\times S_N$ in full generality, we provide explicit checks by series expansions of $\mathcal{F}_{N,1}$ for a large number of examples. We also comment on the relation of $\mathbb{G}(N)$ to the modular group that arises due to the geometry of $X_{N,1}$ as a double elliptic fibration, as well as T-duality of Little String Theories that are constructed from $X_{N,1}$. 
\end{abstract}

\newpage

\tableofcontents

\onehalfspacing

\vskip1cm

%%%%%%%%%%%%%%%%%%%%%%%%%%%%%%%%%%%%%%
%%%%%%%%%%%%%%%%%%%%%%%%%%%%%%%%%%%%%%%%%%%%%%%%
\section{Introduction}
The engineering of supersymmetric gauge theories \cite{Katz:1996fh,Intriligator:1997pq} in dimensions $\leq 6$ through string- and M-theory constructions has been an active and fruitful field of study throughout the years. Indeed, the numerous dual approaches and formulations that are available on the string theory side provide us with a large range of tools (both computationally as well as conceptually) to explore hidden symmetries, dualities and even more sophisticated structures on the gauge theory side that would be very difficult to study otherwise. An important feature of this approach is that in many cases string theory methods give us access to non-perturbative aspects of the gauge theories and allow us to study them in an efficient manner \cite{Haghighat:2013gba,Haghighat:2013tka,Hohenegger:2013ala}. One very rich subclass of theories which has attracted a lot of attention recently \cite{Hohenegger:2015cba,Haghighat:2015coa,Hohenegger:2015btj,Bastian:2017jje} are supersymmetric, $U(M)$ circular quiver gauge theories on $\mathbb{R}^5\times S^1$, which can (among other methods) be approached through F-theory compactifications on a class of toric Calabi-Yau threefolds $X_{N,M}$\footnote{The numbers $N,M\in\mathbb{N}$ refer to the fact that $X_{N,M}$ has the structure of a double elliptic fibration, where the two fibrations have Kodaira singularities of the type $I_{N-1}$ and $I_{M-1}$ respectively \cite{Kanazawa:2016tnt}.}. The latter give rise to a quiver theory comprised of $N$ nodes of type $U(M)$ (which we shall denote as $[U(M)]^N$ in the following). A particularity of these theories is the fact that their UV-completion in general does not only contain point-like particles, but also stringy degrees of freedom, although gravity remains decoupled. Such theories are called Little String Theories (LSTs), which have originally been introduced over a decade ago \cite{Seiberg:1996vs,Berkooz:1997cq,Blum:1997mm,Seiberg:1997zk,Losev:1997hx,Intriligator:1997dh,Aharony:1998ub,Aharony:1999ks,Kutasov:2001uf} and recently have received a lot of renewed interest \cite{Bhardwaj:2015oru,Hohenegger:2015btj,Hohenegger:2016eqy,Hohenegger:2016yuv,Ahmed:2017hfr,Bastian:2017ary,Bastian:2017ing,Bastian:2018dfu,BHIR}. The fully refined, non-perturbative partition function $\mathcal{Z}_{N,M}$ of this theory is captured by the (refined) topological string partition function on $X_{N,M}$ and can very efficiently be computed \cite{Haghighat:2013gba,Haghighat:2013tka,Hohenegger:2013ala,Hohenegger:2015cba,Hohenegger:2015btj,Bastian:2017ing,Shabbir:2015oxa} with the help of the (refined) topological vertex \cite{Aganagic:2003db,Hollowood:2003cv,Iqbal:2007ii} (see \cite{Huang:2015sta,Klemm:2012sx} for a general discussion of the topological string partition function on elliptic Calabi-Yau threefolds). Since the latter (for technical reasons) requires a choice of preferred direction in the web diagram of $X_{N,M}$, this method provides different, but completely equivalent expansions of $\mathcal{Z}_{N,M}$, which can be interpreted as instanton expansions of different but dual gauge theories. While it is straightforward to see \cite{Haghighat:2013tka,Hohenegger:2013ala,Hohenegger:2015btj} that in this fashion the theory $[U(M)]^N$ is dual to $[U(N)]^M$, it was argued in \cite{Bastian:2017ary} that it is also dual to $[U(\tfrac{NM}{k})]^k$, where $k=\text{gcd}(N,M)$, thus leading to a \emph{triality} of gauge theories that are engineered by $X_{N,M}$. 

The Calabi-Yau manifolds $X_{N,M}$ depend on $NM+2$ independent K\"ahler parameters and the corresponding moduli space takes the form of a cone. The faces of the latter (which we shall call walls in the following) are (among others) comprised of singular loci where the area of one or more of the curves in the web diagram of $X_{N,M}$ vanish. From the perspective of the geometry of $X_{N,M}$, crossing such a wall (\emph{i.e.} continuing to negative area) gives rise to a new Calabi-Yau manifold, which corresponds to a different (but dual) resolution of the singularity. With the help of such \emph{flop transitions} \cite{TianYau,Kollar}, the K\"ahler moduli space of $X_{N,M}$ can be extended to include further regions that allow the engineering of yet new gauge theories. Indeed, it was argued in \cite{Hohenegger:2016yuv} that the Calabi-Yau manifolds $X_{N,M}$ and $X_{N',M'}$ can be related through a series of flop transformations if $NM=N'M'$ and $\text{gcd}(N,M)=\text{gcd}(N',M')$. Furthermore, non-trivial checks were presented in \cite{Hohenegger:2016yuv} that the topological string partition functions associated with $X_{N,M}$ and $X_{N',M'}$ are the same upon taking into account the non-trivial duality map. This was shown explicitly in \cite{Bastian:2017ing} for the cases $\text{gcd}(N,M)=1$ and a suitable basis of independent K\"ahler parameters was presented which is adapted to the invariance under a series of flop transformations that is instrumental in the duality $X_{N,M}\sim X_{N',M'}$.\footnote{This transformation is explained in detail in appendix~\ref{App:FlopTrafo} and the basis is reviewed in the following section.} Combining this invariance of $\mathcal{Z}_{N,M}$ with the triality of gauge theories proposed in \cite{Bastian:2017ary}, it was argued in \cite{Bastian:2018dfu} that the theory $[U(M)]^N$ is in fact dual to all theories of the form $[U(M')]^{N'}$ for any $N',M'$ with $NM=N'M'$ and $\text{gcd}(N,M)=\text{gcd}(N',M')$. It was furthermore argued in \cite{BHIR} that the extended moduli space~\cite{Reid,CT1,CT2,CT3,CT4} of $X_{N,M}$ contains different decompactification regions, which engineer different five-dimensional gauge theories with various gauge structures and matter content.

While previous works have focused on interpreting different expansions of $\mathcal{Z}_{N,M}$ as instanton partition functions of different gauge theories, thereby establishing a large network of dual theories, in this paper we discuss the consequences of these dualities from the perspective of symmetries of $\mathcal{Z}_{N,M}$. Focusing on the cases $M=1$, rather than switching between different expansions of the partition function $\mathcal{Z}_{N,1}$ (or more concretely the free energy $\mathcal{F}_{N,1}$), we shall focus on one particular expansion (as a power series in a suitable basis of K\"ahler parameters of $X_{N,1}$) and recast the results of \cite{Hohenegger:2016yuv,Bastian:2017ary,Bastian:2018dfu} in the form of highly non-trivial identities among the expansion coefficients of $\mathcal{F}_{N,1}$. From the perspective of any of the gauge theories of the type $[U(M')]^{N'}$, where $(N',M')$ are relative primes and $N'M'=N$, these correspond to generically non-perturbative symmetries that act in a highly non-trivial fashion on the spectrum of BPS states of the theory. Furthermore, since the combination of any two of these symmetries itself has to be another symmetry, they have the structure of a group $\widehat{\mathbb{G}}(N)$ which acts naturally on the vector space spanned by the independent K\"ahler parameters of $X_{N,1}$. 

We shall analyse $\widehat{\mathbb{G}}(N)$ first with the help of the explicit examples $N=1,2,3,4$, where we can study it (or its subgroups) explicitly as a matrix group. Based on these examples, we find a pattern, which allows us to prove for generic $N$ that $\widehat{\mathbb{G}}(N)$ has a subgroup of the form  
\begin{align}
&\widetilde{\mathbb{G}}(N)\cong \mathbb{G}(N)\,\times \,\widetilde{S}_N&&\text{with} &&\widetilde{\mathbb{G}}(N)\subset \widehat{\mathbb{G}}(N)\,,\label{DualityGeneric}
\end{align}
where $\widetilde{S}_N\subset S_N$ is a subgroup of the Weyl group of the largest simple gauge group that can be engineered from $X_{N,1}$ (\emph{i.e.} $U(N)$) and $\mathbb{G}(N)$ is isomorphic to a dihedral group\footnote{For $n\in\mathbb{N}$ the dihedral group $\text{Dih}_n$ is freely generated by two elements $a,b$ of order $2$ that satisfy a certain braid relation: $\text{Dih}_n=\left\langle \{ a,b| a^2=b^2=1\!\!1\text{ and }(ab)^n=1\!\!1\}\right\rangle$. The group $\text{Dih}_\infty$ corresponds to the limit $n\to \infty$ and is of infinite order.}, namely
\begin{align}
&\mathbb{G}(N)\cong\left\{ \begin{array}{lcl}\text{Dih}_3 & \text{if} & N=1\,, \\ \text{Dih}_2 & \text{if} & N=2\,, \\\text{Dih}_3 & \text{if} & N=3\,, \\ \text{Dih}_\infty & \text{if} & N\geq 4\,.\end{array}\right.\label{FormDihedralGen}
\end{align}
Here $\text{Dih}_\infty$ is a finitely generated group of infinite order (while $\text{ord}(\text{Dih}_n)=2n$ for finite $2\leq n\in\mathbb{N}$). 

In particular the group $\mathbb{G}(N)$ in (\ref{DualityGeneric}) combines non-trivially with other known symmetries and dualities of $X_{N,1}$:
\begin{itemize}
\item modularity: Owing to the fact that $X_{N,1}$ has the structure of a double elliptic fibration, the partition function transforms as a Jacobi form under two copies of the modular group $SL(2,\mathbb{Z})_\tau$ and $SL(2,\mathbb{Z})_\rho$.\footnote{Our notation follows the naming convention of the modular parameters as \emph{e.g.} in \cite{Hohenegger:2015btj}.} Since $\widetilde{\mathbb{G}}(N)$ acts non-trivially on the modular parameters $(\tau,\rho)$ the combined symmetry group is in general larger than simply $\widetilde{\mathbb{G}}(N)\times SL(2,\mathbb{Z})_\tau\times SL(2,\mathbb{Z})_\rho$. In the simplest case $N=1$, which we shall discuss in section~\ref{Sect:Example11}, we are in fact able to analyse explicitly the resulting group and we can show that it is isomorphic to $Sp(4,\mathbb{Z})$, which is the automorphism group of the genus-two curve that is the geometric mirror of the Calabi-Yau manifold $X_{1,1}$ (see \cite{Hollowood:2003cv,Kanazawa:2016tnt}). For $N>1$, the symmetry is more difficult to analyse, and we are only able to make statements about a specific region in the moduli space.

\item T-duality: As mentioned above, the UV completion of the gauge theory $[U(1)]^N$ is an LST with 8 supercharges, which was called type IIb little string in \cite{Hohenegger:2015btj}. The latter is T-dual to type IIa little string theory, whose low energy behaviour is described by the dual gauge theory $[U(N)]^1$ (see \cite{Bhardwaj:2015oru,Kim:2015gha,Hohenegger:2015btj,Hohenegger:2016eqy} for the discussion of T-duality of LSTs engineered from double elliptic Calabi-Yau threefolds). Denoting the partition functions of these little string theories by $Z_{\text{IIb}}$ and $Z_{\text{IIa}}$ respectively, it was proposed in \cite{Hohenegger:2015btj} that the partition functions of these two little string theories are captured by $\mathcal{Z}_{N,1}$
\begin{align}
&Z_{\text{IIa}}(\tau,\rho,\mathbf{K})=\mathcal{Z}_{N,1}(\tau,\rho,\mathbf{K})\,,&&\text{and} &&Z_{\text{IIb}}(\tau,\rho,\mathbf{K}')=\mathcal{Z}_{N,1}(\rho,\tau,\mathbf{K}')\,,
\end{align}
where for simplicity we have only explicitly displayed the dependence on the modular parameters $(\tau,\rho)$ and only schematically indicated the dependence on the remaining K\"ahler parameters through $\mathbf{K}$ and $\mathbf{K}'$ respectively. Furthermore, in \cite{Hohenegger:2015btj} it was proposed that T-duality of the IIa and IIb LSTs simply amounts to 
\begin{align}
Z_{\text{IIa}}(\tau,\rho,\mathbf{K})=Z_{\text{IIb}}(\rho,\tau,\mathbf{K}')\,,\label{Tduality}
\end{align}
which, from the perspective of the Calabi-Yau manifold $X_{N,1}$, corresponds to an exchange of the two elliptic curves, one in the fiber and one in the base (with a duality map relating $\mathbf{K}$ and $\mathbf{K}'$). Since the group $\widetilde{\mathbb{G}}(N)$ in (\ref{DualityGeneric}) acts non-trivially on the modular parameters $(\tau,\rho)$ (and in general mixes them in a non-trivial fashion), it extends the incarnation (\ref{Tduality}) of T-duality to a non-trivial group acting on the full spectrum of the LSTs.
\end{itemize}

\noindent
This paper is organised as follows: In section \ref{Sect:RSSummary} we first review important aspects of the computation of the partition function $\mathcal{Z}_{N,1}$, in particular the choice of basis of the independent K\"ahler parameters. Furthermore, we discuss in more detail our strategy in finding the group $\widetilde{\mathbb{G}}(N)$ in (\ref{DualityGeneric}). Finally, for the sake of readability of this paper, we also give a summary of the results obtained in the subsequent sections. In sections~\ref{Sect:Example11} -- \ref{Sect:Case41} we discuss in detail the examples $N=1,2,3,4$ respectively. For each of these cases we construct $\widetilde{\mathbb{G}}(N)$ and provide non-trivial evidence that it is a symmetry of the $\mathcal{F}_{N,1}$ by computing the leading orders in the expansion of the former as a power series of the K\"ahler parameters. In section~\ref{Sect:GeneralCase} we generalise a pattern that emerges from the previous examples and which allows us to prove (\ref{FormDihedralGen}) for generic $N\in\mathbb{N}$. Finally, section~\ref{Sect:Conclusions} contains our conclusions and directions for future research. Furthermore, this paper is accompanied by two appendices, which review a particular duality transformation for the web diagrams of $X_{N,1}$ and a finite presentation of the group $Sp(4,\mathbb{Z})$ respectively. These technical details are relevant for the computations performed in the main body of this work.

%%%%%%%%%%%%%%%%%%%%%%%%%%%%%%%%%%%%%%%%%%%%%%%%
%%%%%%%%%%%%%%%%%%%%%%%%%%%%%%%%%%%%%%%%%%%%%%%%
\section{Review, General Strategy and Summary of Results}\label{Sect:RSSummary}
\subsection{Review:  Partition Function and Free Energy}\label{Sect:Review}
\begin{wrapfigure}{l}{0.60\textwidth}
\begin{center}
\vspace{-0.7cm}
\scalebox{0.68}{\parbox{15cm}{\begin{tikzpicture}[scale = 1.50]
\draw[ultra thick] (-1,0) -- (0,0) -- (0,-1) -- (1,-1) -- (1,-1.5);
\node[rotate=315] at (1.5,-2) {\Huge $\cdots$};
\draw[ultra thick] (2,-2.5) -- (2,-3) -- (3,-3) -- (3,-4) -- (4,-4) -- (4,-4.5);
\node[rotate=315] at (4.5,-5) {\Huge $\cdots$};
\draw[ultra thick] (5,-5.5) -- (5,-6) -- (6,-6) -- (6,-7) -- (7,-7);
%diagonals
\draw[ultra thick] (0,0) -- (0.7,0.7);
\draw[ultra thick] (1,-1) -- (1.7,-0.3);
\draw[ultra thick] (3,-3) -- (3.7,-2.3);
\draw[ultra thick] (0,-1) -- (-0.7,-1.7);
\draw[ultra thick] (2,-3) -- (1.3,-3.7);
\draw[ultra thick] (3,-4) -- (2.3,-4.7);
\draw[ultra thick] (4,-4) -- (4.7,-3.3);
\draw[ultra thick] (5,-6) -- (4.3,-6.7);
\draw[ultra thick] (6,-6) -- (6.7,-5.3);
\draw[ultra thick] (6,-7) -- (5.3,-7.7);
%ends
\node at (-1.2,0) {\large {\bf $\mathbf a$}};
\node at (7.2,-7) {\large {\bf $\mathbf a$}};
\node at (0.85,0.85) {\large {$\mathbf 1$}};
\node at (1.85,-0.15) {\large {$\mathbf 2$}};
\node at (3.8,-2) {\large {$\mathbf{\delta}$}};
\node at (4.85,-3.15) {\large {$\mathbf{\delta+1}$}};
\node at (6.8,-5) {\large {$\mathbf{N}$}};
\node at (-0.85,-2.1) {\large {$\mathbf{N-\delta+1}$}};
\node at (1.15,-3.85) {\large {$\mathbf N$}};
\node at (2.25,-4.35) {\large {$\mathbf 1$}};
\node at (3.8,-6.2) {\large {$\mathbf N-\delta-1$}};
\node at (5.15,-7.9) {\large {$\mathbf N-\delta$}};
%lables hotizontal
\node at (-0.5,0.25) {\large  {\bf $h_1$}};
\node at (0.5,-0.775) {\large  {\bf $h_2$}};
\node at (2.5,-2.775) {\large  {\bf $h_{\delta}$}};
\node at (3.5,-3.775) {\large  {\bf $h_{\delta+1}$}};
\node at (5.5,-5.775) {\large  {\bf $h_N$}};
\node at (6.5,-6.775) {\large  {\bf $h_1$}};
%lables vertical
\node at (-0.2,-0.5) {\large  {\bf $v_1$}};
\node at (2.75,-3.5) {\large  {\bf $v_\delta$}};
\node at (5.75,-6.5) {\large  {\bf $v_N$}};
%lables diagonal
\node at (0.6,0.25) {\large  {\bf $m_1$}};
\node at (1.6,-0.75) {\large  {\bf $m_2$}};
\node at (3.65,-2.8) {\large  {\bf $m_\delta$}};
\node at (4.75,-3.8) {\large  {\bf $m_{\delta+1}$}};
\node at (6.75,-5.75) {\large  {\bf $m_N$}};
\node at (0.25,-1.45) {\large  {\bf $m_{N-\delta+1}$}};
\node at (1.25,-3.3) {\large  {\bf $m_{N}$}};
\node at (2.8,-4.65) {\large  {\bf $m_1$}};
\node at (5.1,-6.65) {\large  {\bf $m_{N-\delta-1}$}};
\node at (6.1,-7.45) {\large  {\bf $m_{N-\delta}$}};
%hexagons
\node[red] at (-0.2,0.65) {\large  {\bf $S_N$}};
\node[red] at (0.75,-0.3) {\large  {\bf $S_1$}};
%\node[red] at (1.75,-1.3) {\large  {\bf $S_2$}};
\node[red] at (2.75,-2.3) {\large  {\bf $S_{\delta-1}$}};
\node[red] at (3.75,-3.3) {\large  {\bf $S_\delta$}};
\node[red] at (5.75,-5.3) {\large  {\bf $S_{N-1}$}};
\node[red] at (6.75,-6.3) {\large  {\bf $S_N$}};
\node[red] at (-0.75,-0.8) {\large  {\bf $S_{N-\delta}$}};
\node[red] at (0.3,-1.8) {\large  {\bf $S_{N-\delta+1}$}};
%\node[red] at (1.25,-2.75) {\large  {\bf $S_2$}};
\node[red] at (2.2,-3.35) {\large  {\bf $S_{N}$}};
\node[red] at (3.3,-4.8) {\large  {\bf $S_1$}};
\node[red] at (5,-7.3) {\large  {\bf $S_{N-\delta-1}$}};
\node[red] at (6.6,-7.8) {\large  {\bf $S_{N-\delta}$}};
%roots
\draw[ultra thick,blue,<->] (1.05,0.95) -- (1.95,0.05);
\node[blue,rotate=315] at (1.75,0.65) {{\large {\bf {$\widehat{a}_1$}}}};
\draw[ultra thick,blue,<->] (2.05,-0.05) -- (2.95,-0.95);
\node[blue,rotate=315] at (2.75,-0.35) {{\large {\bf {$\widehat{a}_2$}}}};
\node[blue,rotate=315] at (3.5,-1.5) {\Huge $\cdots$};
%\draw[ultra thick,blue,<->] (3.05,-1.05) -- (3.95,-1.95);
%\node[blue,rotate=315] at (3.75,-1.35) {{\large {\bf {$\widehat{a}_3$}}}};
%
\draw[ultra thick,blue,<->] (4.05,-2.05) -- (4.95,-2.95);
\node[blue,rotate=315] at (4.75,-2.35) {{\large {\bf {$\widehat{a}_{\delta+1}$}}}};
\node[blue,rotate=315] at (6,-4) {\Huge $\cdots$};
\draw[ultra thick,blue,<->] (7.05,-5.05) -- (7.95,-5.95);
\node[blue,rotate=315] at (7.75,-5.35) {{\large {\bf {$\widehat{a}_N$}}}};
%definition S
\draw[dashed] (7,-7) -- (5.35,-8.65);
\draw[dashed] (2.3,-4.7) -- (1.8,-5.2);
\draw[ultra thick,blue,<->] (1.8,-5.2) -- (5.35,-8.65);
\node[blue] at (3.4,-7.3) {{\large {\bf {$S$}}}};
%parameter R
\draw[dashed] (2.3,-4.7) -- (1,-4.7);
\draw[dashed] (6,-7) -- (1,-7);
\draw[ultra thick,blue,<->] (1,-6.95) -- (1,-4.75);
\node[blue,rotate=90] at (0.75,-5.8) {{\large {\bf{$R-(N-\delta)S$}}}};
\end{tikzpicture}}}
\caption{\sl Web diagram of $X_{N,1}^{(\delta)}$. }
\label{Fig:N1web0}
\end{center}
${}$\\[-3cm]
\end{wrapfigure}
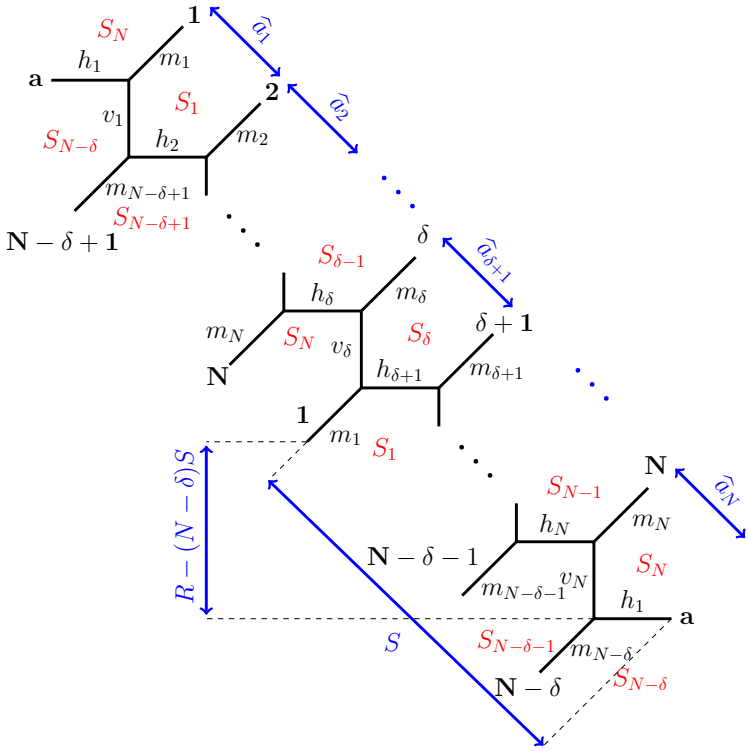

The web diagram for a general $X_{N,1}$ is shown in \figref{Fig:N1web0}. Each line is labelled by the area of the 

\noindent
curve they are representing: horizontal lines are labelled by $h_{1,\ldots,N}$, vertical lines by $v_{1,\ldots,N}$, and diagonal lines by $m_{1,\ldots,N}$. Not all of these areas are independent of one another, but they are subject to $2N$ consistency conditions (for $i=1,\ldots,N$), related to the $N$ hexagons $S_i$ of the web diagram
\begin{align}
S_i:\,&h_i+m_i=h_i+m_{i+1}\,,\nonumber\\
&v_i+m_i=v_{i+\delta}+m_{i+1},\label{ConsistGeneral}
\end{align} 
where $m_{i+N}=m_i$ and $v_{i+N}=v_i$. A general solution of these conditions is given by $v_i=v_{i+1}$ and $m_i=m_{i+1}$ for $i=1,\ldots,N-1$. Another solution, which is more adapted to the computations in the remainder of this work, is provided by the blue parameters in \figref{Fig:N1web0}, which equally represent an independent set of K\"ahler parameters of the Calabi-Yau manifold $X_{N,1}$. Physically, from the perspective of (one particular) gauge theory engineered by $X_{N,1}$, the parameters $\widehat{a}_{1,\ldots,N}$ correspond to the (affine) roots of the gauge group $U(N)$ (\emph{i.e.} the vacuum expectation values of the vector multiplet scalars), while the parameter $R$ is related to the coupling constant and $S$ to the mass parameter of the matter sector. As shown in \cite{Bastian:2017ary}, however, this assignment is not unique and the Calabi-Yau manifold $X_{N,1}$ in fact engineers several different gauge theories with different gauge groups\footnote{The non-affine part of the gauge groups, however, is in general a subgroup of $U(N)$.} and possibly different matter content. In the following we will therefore not be too much concerned with the physical interpretation of the parameters $(\widehat{a}_{1,\ldots,N},S)$. Instead, we shall treat the dependence of the partition function $\mathcal{Z}_{N,1}(\widehat{a}_{1,\ldots,N},S,R;\epsilon_{1,2})$ (associated with $X_{N,1}$) on all of these parameters on equal footing. The former can be computed from the web diagram in \figref{Fig:N1web0} with the help of the refined topological string. Here the constants $\epsilon_{1,2}\in\mathbb{R}$ represent the refinement and can be thought of as a means of regularising the partition function, which would otherwise be ill defined. 

An efficient method of computing $\mathcal{Z}_{N,1}$ from \figref{Fig:N1web0} (for arbitrary $\delta$) was given in \cite{Bastian:2017ing} (see also \cite{Haghighat:2013gba,Hohenegger:2013ala}) by computing a general building block $W^{\alpha_1,\ldots,\alpha_N}_{\beta_1,\ldots,\beta_N}$ that depends on the K\"ahler parameters $(\widehat{a}_{1,\ldots,N},S)$ and is labelled by $2N$ integer partitions $\alpha_{1,\ldots,N}$ and $\beta_{1,\ldots,N}$, which encode how the legs of the (various) building block(s) are glued together (for details see \cite{Bastian:2017ing}). While the formalism

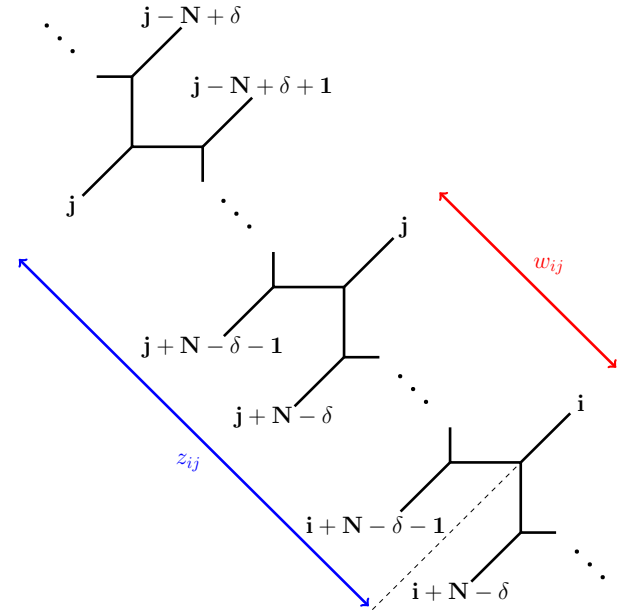
\begin{wrapfigure}{r}{0.45\textwidth}
\begin{center}
\scalebox{0.62}{\parbox{12.8cm}{\begin{tikzpicture}[scale = 1.50]
\node[rotate=315] at (-1,0.5) {\Huge $\cdots$};
\draw[ultra thick] (-0.5,0) -- (0,0) -- (0,-1) -- (1,-1) -- (1,-1.5);
\node[rotate=315] at (1.5,-2) {\Huge $\cdots$};
\draw[ultra thick] (2,-2.5) -- (2,-3) -- (3,-3) -- (3,-4) -- (3.5,-4);
\node[rotate=315] at (4,-4.5) {\Huge $\cdots$};
\draw[ultra thick] (4.5,-5) -- (4.5,-5.5) -- (5.5,-5.5) -- (5.5,-6.5) -- (6,-6.5);
\node[rotate=315] at (6.5,-7) {\Huge $\cdots$};
%diagonals top
\draw[ultra thick] (0,0) -- (0.7,0.7);
\draw[ultra thick] (1,-1) -- (1.7,-0.3);
\draw[ultra thick] (3,-3) -- (3.7,-2.3);
\draw[ultra thick] (5.5,-5.5) -- (6.2,-4.8); 
%diagonals bottom
\draw[ultra thick] (0,-1) -- (-0.7,-1.7);
\draw[ultra thick] (2,-3) -- (1.3,-3.7);
\draw[ultra thick] (3,-4) -- (2.3,-4.7);
\draw[ultra thick] (4.5,-5.5) -- (3.8,-6.2);
\draw[ultra thick] (5.5,-6.5) -- (4.8,-7.2);
%labels top
\node at (0.85,0.85) {\large {$\mathbf{j-N+\delta}$}};
\node at (1.85,-0.15) {\large {$\mathbf{j-N+\delta+1}$}};
\node at (3.85,-2.15) {\large {$\mathbf{j}$}};
\node at (6.35,-4.65) {\large {$\mathbf{i}$}};
%labels top
\node at (-0.85,-1.85) {\large {$\mathbf{j}$}};
\node at (1.15,-3.85) {\large {$\mathbf{j+N-\delta-1}$}};
\node at (2.15,-4.85) {\large {$\mathbf{j+N-\delta}$}};
\node at (3.45,-6.4) {\large {$\mathbf{i+N-\delta-1}$}};
\node at (4.65,-7.35) {\large {$\mathbf{i+N-\delta}$}};
%argument top
\draw[ultra thick,red,<->] (4.35,-1.65) -- (6.85,-4.15);
\node[red] at (5.9,-2.7) {\large $w_{ij}$};
%argument bottom 7.5
%\draw[ultra thick,red,<->] (-1.35,-2.35) -- (4.15,-7.85) ;
\draw[ultra thick,blue,<->] (-1.6,-2.6) -- (3.35,-7.55);
\draw[dashed] (3.4,-7.6) -- (5.5,-5.5);
\node[blue] at (0.8,-5.5) {\large $z_{ij}$};
\end{tikzpicture}}}
\caption{\sl Definition of the arguments of the $\vartheta$-functions appearing in $\mathcal{Z}_{N,1}$.}
\label{Fig:DefArgumentsPartFct}
\end{center}
${}$\\[-5cm]
\end{wrapfigure}

\noindent
developed in \cite{Bastian:2017ing} is more general and allows the computation of a much larger class of partition functions, in the present case we have
\begin{align}
&\mathcal{Z}_{N,1}(\widehat{a}_{1,\ldots,N},S,R;\epsilon_{1,2})\nonumber\\
&=\sum_{\{\alpha\}}\left(\prod_{i=1}^N Q_{m_i}^{|\alpha_i|}\right)\,W^{\alpha_1,\ldots,\alpha_N}_{\alpha_{N-\delta+1},\ldots,\alpha_{N-\delta}}(\widehat{a}_{1,\ldots,N},S;\epsilon_{1,2})\,,
\end{align}
with (our conventions for the normalisation of $W^{\alpha_1,\ldots,\alpha_N}_{\beta_1,\ldots,\beta_N}$ are adapted to \figref{Fig:N1web0}.)
\begin{align}
&W^{\alpha_1,\ldots,\alpha_N}_{\alpha_{N-\delta+1},\ldots,\alpha_{N-\delta}}(\widehat{a}_{1,\ldots,N},S;\epsilon_{1,2})\nonumber\\
&\hspace{0.2cm}=W_\emptyset^N(\widehat{a}_{1,\ldots,N})\,\left[\frac{\left(t/q\right)^{\frac{N-1}{2}}}{Q_\rho^{N-\delta-1}}\right]^{|\alpha_1|+\ldots+|\alpha_N|}\nonumber\\
&\hspace{0.7cm}\times \prod_{i,j=1}^N\frac{\vartheta_{\alpha_i\alpha_j}(\widehat{Q}_{i,j};\rho)}{\vartheta_{\alpha_i,\alpha_j}(\bar{Q}_{i,j}\,\sqrt{q/t};\rho)}\,.\nonumber
\end{align}
Here we have used the following notation:
\begin{align}
&Q_{m_i}=e^{-m_i}\,,&&\rho=\frac{i}{2\pi}\sum_{k=1}^N\widehat{a}_k\,,&&Q_\rho=e^{-\sum_{k=1}^N\widehat{a}_k}\,,&&q=e^{2\pi i \epsilon_1}\,,&&t=e^{-2\pi i\epsilon_2}\,,\nonumber
\end{align}
where $m_{i=1,\ldots,N}$ refer to the area of the diagonal lines in \figref{Fig:N1web0} expressed as functions of $(\widehat{\alpha}_{1,\ldots,N},S,R)$ with the help of the consistency conditions (\ref{ConsistGeneral}). Furthermore, $W_\emptyset^N$ is a normalisation factor (which from a physical perspective in particular encodes the perturbative contribution to the partition function) and $\vartheta_{\mu\nu}$ is a class of theta-functions that is labelled by two integer partitions $\mu$ and $\nu$ 
{\allowdisplaybreaks
\begin{align}
\vartheta_{\mu\nu}(x;\rho)&=\prod_{(i,j)\in \mu}\vartheta\left(x^{-1}q^{-\nu_j^t+i-\frac{1}{2}}\,t^{-\mu_i+j-\frac{1}{2}};\rho\right) \prod_{(i,j)\in\nu}\vartheta\left(x^{-1}q^{\mu_j^t-i+\frac{1}{2}}\,t^{\nu_i-j+\frac{1}{2}};\rho\right)\,,\nonumber
\end{align}}
with the further definition
\begin{align}
\vartheta(x;\rho)&=(x^{\frac{1}{2}}-x^{-\frac{1}{2}})\prod_{k=1}^\infty(1-x\, Q_\rho^k)(1-x^{-1}\,Q_\rho^k)\,.
\end{align}
Finally, the arguments of the $\vartheta$-functions can be defined as $\widehat{Q}_{i,j}=e^{-z_{ij}}$ and $\bar{Q}_{i,j}=e^{-w_{ij}}$ where $z_{ij}$ and $w_{ij}$ are implicitly defined in \figref{Fig:DefArgumentsPartFct} with respect to (part of) the web diagram (the labels on the diagonal and horizontal lines in \figref{Fig:DefArgumentsPartFct} (and \figref{Fig:N1web0}) indicate how they are glued together). 

With the partition function $\mathcal{Z}_{N,1}$, we can define the free energy as the plethystic logarithm
\begin{align}
\mathcal{F}_{N,1}(\widehat{a}_{1,\ldots,N},S,R;\epsilon_{1,2})=\text{PLog}\,\mathcal{Z}_{N,1}%(\widehat{a}_{1,\ldots,N},S,R;\epsilon_{1,2})
=\sum_{k=1}^\infty\frac{\mu(k)}{k}\,\ln\mathcal{Z}_{N,1}(k\,\widehat{a}_{1,\ldots,N},k\,S,k\,R;k\,\epsilon_{1,2})\,,\label{DefFreeEnergyGen}
\end{align}
where $\mu(k)$ is the M\"obius function. We can expand the free energy in the following fashion% in the various K\"ahler parameters of $X_{N,1}$
\begin{align}
\mathcal{F}_{N,1}(\widehat{a}_{1,\ldots,N},S,R;\epsilon_1,\epsilon_2)=\sum_{n=0}^\infty \sum_{i_1,\ldots,i_N=0}^\infty \sum_{k\in\mathbb{Z}}f_{i_1,\ldots,i_N,k,n}(\epsilon_1,\epsilon_2)\,\widehat{Q}_1^{i_1}\ldots\widehat{Q}_N^{i_N}\,Q_S^{k}\,Q_{R}^n\,,\label{FreeEnergyExpansionCoefficients}
\end{align}
with $\widehat{Q}_i=e^{-\widehat{a}_i}$ (for $i=1,\ldots,N$), $Q_S=e^{-S}$ and $Q_R=e^{-R}$. Apart from a first order pole, $\mathcal{F}_{N,1}$ has a power series expansion in $\epsilon_{1,2}$, which allows to compute the Nekrasov-Shatashvili-limit \cite{Nekrasov:2009rc,Mironov:2009uv} and the unrefined limit. For later convenience we therefore also introduce the expansion of the leading term in both parameters (which we simply denote NS)
\begin{align}
\lim_{\epsilon_{1,2}\to 0} \epsilon_1\,\epsilon_2\,\mathcal{F}_{N,1}(\widehat{a}_{1,\ldots,N},S,R;\epsilon_1,\epsilon_2)=\sum_{n=0}^\infty \sum_{i_1,\ldots,i_N=0}^\infty \sum_{k\in\mathbb{Z}}f^{\text{NS}}_{i_1,\ldots,i_N,k,n}\,\widehat{Q}_1^{i_1}\ldots\widehat{Q}_N^{i_N}\,Q_S^{k}\,Q_{R}^n\,,
\end{align}
where $f^{\text{NS}}_{i_1,\ldots,i_N,k,n}\in\mathbb{Z}$.
%%%%%%%%%%%%%%%%%%%%%%%%%%
\subsection{Symmetries Transformations: Strategy and Summary of Results}\label{Sect:GeneralStrategy}

\begin{wrapfigure}{l}{0.60\textwidth}
\begin{center}
\vspace{-0.7cm}
\scalebox{0.68}{\parbox{15cm}{\begin{tikzpicture}[scale = 1.50]
\draw[ultra thick] (-1,0) -- (0,0) -- (0,-1) -- (1,-1) -- (1,-1.5);
\node[rotate=315] at (1.5,-2) {\Huge $\cdots$};
\draw[ultra thick] (2,-2.5) -- (2,-3) -- (3,-3) -- (3,-4) -- (4,-4) -- (4,-4.5);
\node[rotate=315] at (4.5,-5) {\Huge $\cdots$};
\draw[ultra thick] (5,-5.5) -- (5,-6) -- (6,-6) -- (6,-7) -- (7,-7);
%diagonals
\draw[ultra thick] (0,0) -- (0.7,0.7);
\draw[ultra thick] (1,-1) -- (1.7,-0.3);
\draw[ultra thick] (3,-3) -- (3.7,-2.3);
\draw[ultra thick] (0,-1) -- (-0.7,-1.7);
\draw[ultra thick] (2,-3) -- (1.3,-3.7);
\draw[ultra thick] (3,-4) -- (2.3,-4.7);
\draw[ultra thick] (4,-4) -- (4.7,-3.3);
\draw[ultra thick] (5,-6) -- (4.3,-6.7);
\draw[ultra thick] (6,-6) -- (6.7,-5.3);
\draw[ultra thick] (6,-7) -- (5.3,-7.7);
%ends
\node at (-1.2,0) {\large {\bf $\mathbf a$}};
\node at (7.2,-7) {\large {\bf $\mathbf a$}};
\node at (0.85,0.85) {\large {$\mathbf 1$}};
\node at (1.85,-0.15) {\large {$\mathbf 2$}};
\node at (3.8,-2) {\large {$\mathbf{\delta'}$}};
\node at (4.85,-3.15) {\large {$\mathbf{\delta'+1}$}};
\node at (6.8,-5) {\large {$\mathbf{N}$}};
\node at (-0.85,-2.1) {\large {$\mathbf{N-\delta'+1}$}};
\node at (1.15,-3.85) {\large {$\mathbf N$}};
\node at (2.25,-4.35) {\large {$\mathbf 1$}};
\node at (3.8,-6.2) {\large {$\mathbf N-\delta'-1$}};
\node at (5.15,-7.9) {\large {$\mathbf N-\delta'$}};
%lables hotizontal
\node at (-0.5,0.25) {\large  {\bf $h'_1$}};
\node at (0.5,-0.775) {\large  {\bf $h'_2$}};
\node at (2.5,-2.775) {\large  {\bf $h'_{\delta'}$}};
\node at (3.5,-3.775) {\large  {\bf $h'_{\delta'+1}$}};
\node at (5.5,-5.775) {\large  {\bf $h'_N$}};
\node at (6.5,-6.775) {\large  {\bf $h'_1$}};
%lables vertical
\node at (-0.2,-0.5) {\large  {\bf $v'_1$}};
\node at (2.75,-3.5) {\large  {\bf $v'_{\delta'}$}};
\node at (5.75,-6.5) {\large  {\bf $v'_N$}};
%lables diagonal
\node at (0.6,0.25) {\large  {\bf $m'_1$}};
\node at (1.6,-0.75) {\large  {\bf $m'_2$}};
\node at (3.65,-2.8) {\large  {\bf $m'_{\delta'}$}};
\node at (4.75,-3.8) {\large  {\bf $m'_{\delta'+1}$}};
\node at (6.75,-5.75) {\large  {\bf $m'_N$}};
\node at (0.25,-1.45) {\large  {\bf $m'_{N-\delta'+1}$}};
\node at (1.25,-3.3) {\large  {\bf $m'_{N}$}};
\node at (2.8,-4.65) {\large  {\bf $m'_1$}};
\node at (5.05,-6.65) {\large  {\bf $m'_{N-\delta'-1}$}};
\node at (6.15,-7.4) {\large  {\bf $m'_{N-\delta'}$}};
%hexagons
\node[red] at (-0.2,0.65) {\large  {\bf $S'_N$}};
\node[red] at (0.75,-0.3) {\large  {\bf $S'_1$}};
%\node[red] at (1.75,-1.3) {\large  {\bf $S_2$}};
\node[red] at (2.75,-2.3) {\large  {\bf $S'_{\delta'-1}$}};
\node[red] at (3.75,-3.3) {\large  {\bf $S'_{\delta'}$}};
\node[red] at (5.75,-5.3) {\large  {\bf $S'_{N-1}$}};
\node[red] at (6.75,-6.3) {\large  {\bf $S'_N$}};
\node[red] at (-0.75,-0.8) {\large  {\bf $S'_{N-\delta'}$}};
\node[red] at (0.35,-1.85) {\large  {\bf $S'_{N-\delta'+1}$}};
%\node[red] at (1.25,-2.75) {\large  {\bf $S_2$}};
\node[red] at (2.2,-3.35) {\large  {\bf $S'_{N}$}};
\node[red] at (3.3,-4.8) {\large  {\bf $S'_1$}};
\node[red] at (5,-7.3) {\large  {\bf $S'_{N-\delta'-1}$}};
\node[red] at (6.6,-7.8) {\large  {\bf $S'_{N-\delta'}$}};
%roots
\draw[ultra thick,blue,<->] (1.05,0.95) -- (1.95,0.05);
\node[blue,rotate=315] at (1.75,0.65) {{\large {\bf {$\widehat{a}'_1$}}}};
\draw[ultra thick,blue,<->] (2.05,-0.05) -- (2.95,-0.95);
\node[blue,rotate=315] at (2.75,-0.35) {{\large {\bf {$\widehat{a}'_2$}}}};
\node[blue,rotate=315] at (3.5,-1.5) {\Huge $\cdots$};
%\draw[ultra thick,blue,<->] (3.05,-1.05) -- (3.95,-1.95);
%\node[blue,rotate=315] at (3.75,-1.35) {{\large {\bf {$\widehat{a}_3$}}}};
%
\draw[ultra thick,blue,<->] (4.05,-2.05) -- (4.95,-2.95);
\node[blue,rotate=315] at (4.75,-2.35) {{\large {\bf {$\widehat{a}'_{\delta'+1}$}}}};
\node[blue,rotate=315] at (6,-4) {\Huge $\cdots$};
\draw[ultra thick,blue,<->] (7.05,-5.05) -- (7.95,-5.95);
\node[blue,rotate=315] at (7.75,-5.35) {{\large {\bf {$\widehat{a}'_N$}}}};
%definition S
\draw[dashed] (7,-7) -- (5.35,-8.65);
\draw[dashed] (2.3,-4.7) -- (1.8,-5.2);
\draw[ultra thick,blue,<->] (1.8,-5.2) -- (5.35,-8.65);
\node[blue] at (3.9,-7.7) {{\large {\bf {$S'$}}}};
%parameter R
\draw[dashed] (2.3,-4.7) -- (1,-4.7);
\draw[dashed] (6,-7) -- (1,-7);
\draw[ultra thick,blue,<->] (1,-6.95) -- (1,-4.75);
\node[blue,rotate=90] at (0.75,-5.8) {{\large {\bf{$R'-(N-\delta')S'$}}}};
\end{tikzpicture}}}
\caption{\sl Web diagram of $X_{N,1}^{(\delta')}$ after a duality transformation of \figref{Fig:N1web0}.}
\label{Fig:N1web0trafo}
\end{center}
${}$\\[-2.5cm]
\end{wrapfigure}
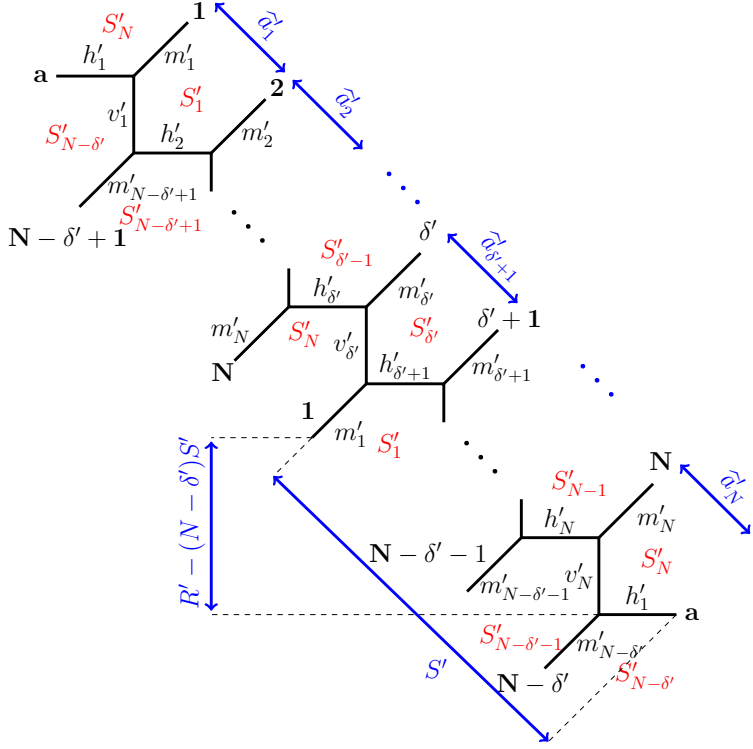 

In \cite{Hohenegger:2016yuv,Bastian:2017ing,Bastian:2017ary} different duality transformations have been discussed, which involve flop transformations \cite{TianYau,Kollar} of various curves of $X_{N,1}$, $SL(2,\mathbb{Z})$ transformations as well as cutting and re-gluing of the web diagram. While these duality transformations were shown in \cite{Bastian:2017ing} to leave $\mathcal{Z}_{N,1}$ (and thus also $\mathcal{F}_{N,1}$) invariant, they generically act in a rather non-trivial fashion on the web diagram \figref{Fig:N1web0}. Indeed, a particular example of such a transformation is reviewed in appendix~\ref{App:FlopTrafo}, which shifts $\delta\to \delta+1$ and transforms the areas of all curves $\{h_{1,\ldots,N},v_{1,\ldots,N},m_{1,\ldots,N}\}$ in a non-trivial fashion. In general, the web diagram \figref{Fig:N1web0} is transformed to a similar 'staircase' diagram as shown in \figref{Fig:N1web0trafo} (possibly with $\delta'\neq \delta$) where the areas of the new curves can be re-written as functions of the old areas
\begin{align}
&\{h'_{1,\ldots,N},v'_{1,\ldots,N},m'_{1,\ldots,N}\}\nonumber\\
&=\{h'_{1,\ldots,N}(h_{1,\ldots,N},v_{1,\ldots,N},m_{1,\ldots,N}),v'_{1,\ldots,N}(h_{1,\ldots,N},v_{1,\ldots,N},m_{1,\ldots,N}),m'_{1,\ldots,N}(h_{1,\ldots,N},v_{1,\ldots,N},m_{1,\ldots,N})\}\,.\label{RelAreaKaehler}
\end{align}
Furthermore, since both $(\widehat{a}_{1,\ldots,N},S,R)$ (as defined in \figref{Fig:N1web0}) and $(\widehat{a}'_{1,\ldots,N},S',R')$ (as defined in \figref{Fig:N1web0trafo}) are a maximal set of independent K\"ahler parameters, the areas $\{h_{1,\ldots,N},v_{1,\ldots,N},m_{1,\ldots,N}\}$ can be expressed as linear combinations of both of these bases. Therefore, (\ref{RelAreaKaehler}) gives a set of linear equations which have a unique solution of the form
\begin{align}
\left(\widehat{a}_1\,, \ldots \,, \widehat{a}_N \,, S \,, R\right)^T=G\cdot \left(\widehat{a}'_1\,, \ldots \,, \widehat{a}'_N \,, S' \,, R'\right)^T\,,\label{RelGenMatIdent}
\end{align}
where $G$ is an invertible $(N+2)\times (N+2)$ matrix with integer entries. Finally, using the result \cite{Bastian:2017ing} that the partition function $\mathcal{Z}_{N,1}$ is invariant under the duality transformation, \emph{i.e.} $\mathcal{Z}_{N,1}(\widehat{a}_{1,\ldots,N},S,R)=\mathcal{Z}_{N,1}(\widehat{a}'_{1,\ldots,N},S',R')$, the matrix $G$ in (\ref{RelGenMatIdent}) is a symmetry of the partition function. More concretely, at the level of the free energy, we have the following relations for the expansion coefficients appearing in (\ref{FreeEnergyExpansionCoefficients})
\begin{align}
&f_{i_1,\ldots,i_N,k,n}(\epsilon_1,\epsilon_2)=f_{i'_1,\ldots,i'_N,k',n'}(\epsilon_1,\epsilon_2)&&\text{for} && (i'_1,\ldots,i'_N,k', n')^T=G^T\cdot (i_1,\ldots,i_N,k, n)^T\,.\label{SymTrafoFreeEnergy}
\end{align}
The transposition of $G$ in this relation is due to the fact, that the transformation (\ref{RelGenMatIdent}) is a passive one from the perspective of the coefficients $f_{i_1,\ldots,i_N,k,n}$.

For given $X_{N,1}$ there are in general numerous different transformations $G$ of the type described above. Since the concatenation of two such transformations defines a new transformation, the latter form a group. In the following sections we shall determine at least a subgroup of this group for the simplest examples $N=1,2,3,4$, which in section~\ref{Sect:GeneralCase} can be generalised to generic $N\in\mathbb{N}$. However, before doing so and for ease of readability, we summarise our results: For generic $N\in\mathbb{N}$, we identify a finitely generated group of symmetry transformations of the type (\ref{RelGenMatIdent}), which can be written as\\[-10pt]

\begin{tcolorbox}
${}$\\[-35pt]
\begin{align}
&\widetilde{\mathbb{G}}(N)\cong \mathbb{G}(N)\,\times \,\widetilde{S}_N&&\text{with} &&\mathbb{G}(N)\cong\left\{ \begin{array}{lcl}\text{Dih}_3 & \text{if} & N=1\,, \\ \text{Dih}_2 & \text{if} & N=2\,, \\\text{Dih}_3 & \text{if} & N=3\,, \\ \text{Dih}_\infty & \text{if} & N\geq 4\,.\end{array}\right.
\end{align}
${}$\\[-35pt]
\end{tcolorbox}

\noindent 
The group $\widetilde{S}_N\subset S_N$ is generated by simple relabellings of the web diagram of $X_{N,1}$ and physically corresponds to a subgroup of the Weyl group of $U(N)$, which is the largest gauge group that can be engineered by $X_{N,1}$. For generic $N$, the group $\mathbb{G}(N)$ is freely generated by two $(N+2)\times (N+2)$ matrices of order 2, which satisfy a specific braid relation\footnote{In the following $\langle \mathcal{E}\rangle$ denotes the group freely generated by the ensemble $\mathcal{E}$.}\\[-10pt]
\begin{tcolorbox}
${}$\\[-35pt]
\begin{align}
&\mathbb{G}(N)\cong \left\langle\{\mathcal{G}_2(N),\mathcal{G}'_2(N)\big|(\mathcal{G}_2(N))^2=(\mathcal{G}'_2(N))^2=(\mathcal{G}_2(N)\cdot \mathcal{G}'_2(N))^n=1\!\!1\}\right\rangle\,,
\end{align}
${}$\\[-44pt]
\end{tcolorbox}

\noindent
where $n=3$ for $N=1,3$ and $n=2$ for $N=2$ but for $N\geq 4$ we find $n\to \infty$, which means that there is no braid relation in these cases. Explicitly, the generators are given by the following lower- and upper triangular matrices 
\begin{align}
&\mathcal{G}_2(N)=\left(\begin{array}{ccccc} & & & 0 & 0 \\ & 1\!\!1_{N\times N} & & \vdots & \vdots \\ & & & 0 & 0 \\ 1 & \cdots & 1 & -1 & 0 \\ N & \cdots & N & -2N & 1 \end{array}\right)\,,&&\text{and}&&\mathcal{G}'_2(N)=\left(\begin{array}{ccccc} & & & -2 & 1 \\ & 1\!\!1_{N\times N} & & \vdots & \vdots \\ & & & -2 & 1 \\ 0 & \cdots & 0 & -1 & 1 \\ 0 & \cdots & 0 & 0 & 1 \end{array}\right)\,.
\end{align}
These matrices are symmetry transformations of the partition function $\mathcal{Z}_{N,1}$ and the free energy $\mathcal{F}_{N,1}$ in the sense of (\ref{SymTrafoFreeEnergy}), which can be checked in explicit examples. In the case $N=1$, combining the group $\widetilde{\mathbb{G}}(N)$ with the modular group $SL(2,\mathbb{Z})$ acting on one of the modular parameters of $X_{1,1}$ generates the group $Sp(4,\mathbb{Z})$. For the cases $N>1$, the combination with the modular group is more difficult to analyse at a general point in the moduli space of $X_{N,1}$. However, in the region in moduli space where $\widehat{a}_{1,\ldots,N}=\widehat{a}$ in \figref{Fig:N1web0}, this analysis is simpler to perform and we can prove that the combination of $\mathbb{G}(N)$ with the modular group is a subgroup of $Sp(4,\mathbb{Z})$. This is in line with the checks performed in \cite{Hohenegger:2016yuv} to provide evidence for the duality $X_{N,M}\sim X_{N',M'}$ (for $NM=N'M'$ and $\text{gcd}(N,M)=\text{gcd}(N',M')$) of Calabi-Yau threefolds.

%%%%%%%%%%%%%%%%%%%%%%%%%%%%%%%%%%%%%%%%%%%%%%%%
\section{Example: $(N,M)=(1,1)$}\label{Sect:Example11}
\subsection{Dualities and $\text{Dih}_3$ Group Action}
\begin{figure}[h]
\begin{center}
\scalebox{0.80}{\parbox{21.2cm}{\begin{tikzpicture}[scale = 1.50]
%%%%%%%%%%%%%%%%%%%%%%%%%%%%%%%%%%%%%%%%%%%%%%%%
%%%%%%%%%%%%%%%%%%%%%%%%%%%%%%%%%%%%%%%%%%%%%%%%
\draw[ultra thick] (-1,0) -- (0,0) -- (0,-1) -- (1,-1);
%diagonals
\draw[ultra thick] (0,0) -- (0.7,0.7);
\draw[ultra thick] (0,-1) -- (-0.7,-1.7);
%ends
\node at (-1.2,0) {\large {\bf $\mathbf a$}};
\node at (1.2,-1) {\large {\bf $\mathbf a$}};
\node at (0.85,0.85) {\large {$\mathbf 1$}};
\node at (-0.85,-1.5) {\large {$\mathbf 1$}};
%lables hotizontal
\node at (-0.5,0.25) {\large  {\bf $h$}};
\node at (0.5,-0.75) {\large  {\bf $h$}};
%lables vertical
\node at (-0.2,-0.5) {\large  {\bf $v$}};
%lables diagonal
\node at (0.6,0.25) {\large  {\bf $m$}};
\node at (-0.25,-1.6) {\large  {\bf $m$}};
%roots
\draw[ultra thick,blue,<->] (1.1,1) -- (2,0.1);
\node[blue,rotate=315] at (1.8,0.7) {{\large {\bf {$\widehat{a}$}}}};
%definition S
\draw[dashed] (1,-1) -- (-0.7,-2.7);
\draw[dashed] (-0.7,-1.7) -- (-1.2,-2.2);
\draw[ultra thick,blue,<->] (-1.2,-2.2) -- (-0.7,-2.7);
\node[blue] at (-1.175,-2.625) {{\large {\bf {$S$}}}};
%parameter R
\draw[dashed] (-0.7,-1.7) -- (-1.5,-1.7);
\draw[dashed] (0,-1) -- (-1.5,-1);
\draw[ultra thick,blue,<->] (-1.5,-1) -- (-1.5,-1.7);
\node[blue,rotate=90] at (-1.8,-1.3) {{\large {\bf{$R-S$}}}};
%stamp
\node at (0,-3.25) {\Large {\bf $\mathbf (a)$}};
%%%%%%%%%%%%%%%%%%
%%%%%%%%%%%%%%%%%%%%%%%%%%%%%%%%%%%%%%%%%%%%%%%%
%%%%%%%%%%%%%%%%%%%%%%%%%%%%%%%%%%%%%%%%%%%%%%%%
\begin{scope}[xshift=5cm]
\draw[ultra thick] (-1,0) -- (0,0) -- (0,-1) -- (1,-1);
%diagonals
\draw[ultra thick] (0,0) -- (0.7,0.7);
\draw[ultra thick] (0,-1) -- (-0.7,-1.7);
%ends
\node at (-1.2,0) {\large {\bf $\mathbf a$}};
\node at (1.2,-1) {\large {\bf $\mathbf a$}};
\node at (0.85,0.85) {\large {$\mathbf 1$}};
\node at (-0.85,-1.5) {\large {$\mathbf 1$}};
%lables hotizontal
\node at (-0.5,0.25) {\large  {\bf $m$}};
\node at (0.5,-0.75) {\large  {\bf $m$}};
%lables vertical
\node at (-0.2,-0.5) {\large  {\bf $h$}};
%lables diagonal
\node at (0.6,0.25) {\large  {\bf $v$}};
\node at (-0.25,-1.6) {\large  {\bf $v$}};
%roots
\draw[ultra thick,blue,<->] (1.1,1) -- (2,0.1);
\node[blue,rotate=315] at (1.8,0.7) {{\large {\bf {$\widehat{a}'$}}}};
%definition S
\draw[dashed] (1,-1) -- (-0.7,-2.7);
\draw[dashed] (-0.7,-1.7) -- (-1.2,-2.2);
\draw[ultra thick,blue,<->] (-1.2,-2.2) -- (-0.7,-2.7);
\node[blue] at (-1.2,-2.65) {{\large {\bf {$S'$}}}};
%parameter R
\draw[dashed] (-0.7,-1.7) -- (-1.5,-1.7);
\draw[dashed] (0,-1) -- (-1.5,-1);
\draw[ultra thick,blue,<->] (-1.5,-1) -- (-1.5,-1.7);
\node[blue,rotate=90] at (-1.8,-1.275) {{\large {\bf{$R'-S'$}}}};
%stamp
\node at (0,-3.25) {\Large {\bf $\mathbf (b)$}};
\end{scope}
%%%%%%%%%%%%%%%%%%%%%%%%%%%%%%%%%%%%%%%%%%%%%%%%
%%%%%%%%%%%%%%%%%%%%%%%%%%%%%%%%%%%%%%%%%%%%%%%%
\begin{scope}[xshift=10cm]
\draw[ultra thick] (-1,0) -- (0,0) -- (0,-1) -- (1,-1);
%diagonals
\draw[ultra thick] (0,0) -- (0.7,0.7);
\draw[ultra thick] (0,-1) -- (-0.7,-1.7);
%ends
\node at (-1.2,0) {\large {\bf $\mathbf a$}};
\node at (1.2,-1) {\large {\bf $\mathbf a$}};
\node at (0.85,0.85) {\large {$\mathbf 1$}};
\node at (-0.85,-1.5) {\large {$\mathbf 1$}};
%lables hotizontal
\node at (-0.5,0.25) {\large  {\bf $h$}};
\node at (0.5,-0.75) {\large  {\bf $h$}};
%lables vertical
\node at (-0.225,-0.5) {\large  {\bf $m$}};
%lables diagonal
\node at (0.6,0.25) {\large  {\bf $v$}};
\node at (-0.25,-1.6) {\large  {\bf $v$}};
%roots
\draw[ultra thick,blue,<->] (1.1,1) -- (2,0.1);
\node[blue,rotate=315] at (1.8,0.7) {{\large {\bf {$\widehat{a}''$}}}};
%definition S
\draw[dashed] (1,-1) -- (-0.7,-2.7);
\draw[dashed] (-0.7,-1.7) -- (-1.2,-2.2);
\draw[ultra thick,blue,<->] (-1.2,-2.2) -- (-0.7,-2.7);
\node[blue] at (-1.2,-2.65) {{\large {\bf {$S''$}}}};
%parameter R
\draw[dashed] (-0.7,-1.7) -- (-1.5,-1.7);
\draw[dashed] (0,-1) -- (-1.5,-1);
\draw[ultra thick,blue,<->] (-1.5,-1) -- (-1.5,-1.7);
\node[blue,rotate=90] at (-1.8,-1.25) {{\large {\bf{$R''-S''$}}}};
%stamp
\node at (0,-3.25) {\Large {\bf $\mathbf (c)$}};
\end{scope}
%%%%%%%%%%%%%%%%%%%%%%%%%%%%%%%%%%%%%%
%%%%%%%%%%%%%%%%%%%%%%%%%%%%%%%%%%%%%%
%\begin{scope}[xshift=10cm]
%\draw[ultra thick] (-1,0) -- (0,0) -- (0,-1) -- (1,-1);
%diagonals
%\draw[ultra thick] (0,0) -- (0.7,0.7);
%\draw[ultra thick] (0,-1) -- (-0.7,-1.7);
%ends
%\node at (-1.2,0) {\large {\bf $\mathbf a$}};
%\node at (1.2,-1) {\large {\bf $\mathbf a$}};
%\node at (0.85,0.85) {\large {$\mathbf 1$}};
%\node at (-0.85,-1.5) {\large {$\mathbf 1$}};
%lables hotizontal
%\node at (-0.5,0.25) {\large  {\bf $v$}};
%\node at (0.5,-0.75) {\large  {\bf $v$}};
%lables vertical
%\node at (-0.2,-0.5) {\large  {\bf $m$}};
%lables diagonal
%\node at (0.6,0.25) {\large  {\bf $h$}};
%\node at (-0.25,-1.6) {\large  {\bf $h$}};
%roots
%\draw[ultra thick,blue,<->] (1.1,1) -- (2,0.1);
%\node[blue,rotate=315] at (1.8,0.7) {{\large {\bf {$\widehat{a}'''$}}}};
%definition S
%\draw[dashed] (1,-1) -- (-0.7,-2.7);
%\draw[dashed] (-0.7,-1.7) -- (-1.2,-2.2);
%\draw[ultra thick,blue,<->] (-1.2,-2.2) -- (-0.7,-2.7);
%\node[blue] at (-1.175,-2.625) {{\large {\bf {$S'''$}}}};
%parameter R
%\draw[dashed] (-0.7,-1.7) -- (-1.5,-1.7);
%\draw[dashed] (0,-1) -- (-1.5,-1);
%\draw[ultra thick,blue,<->] (-1.5,-1) -- (-1.5,-1.7);
%\node[blue,rotate=90] at (-1.8,-1.3) {{\large {\bf{$R'''-S''''$}}}};
%stamp
%\node at (0,-3.25) {\Large {\bf $\mathbf (c)$}};
%\end{scope}
%%%%%%%%%%%%%%%%%%
\end{tikzpicture}}}
\caption{\sl Three different presentations of the web diagram of $X_{1,1}$ with a parametrisation of the areas of all curves. The parameters $(h,v,m)$ are independent of each other and the blue parameters represent an alternative parametrisation in line with \figref{Fig:N1web0}.}
\label{Fig:11webs}
\end{center}
\end{figure}
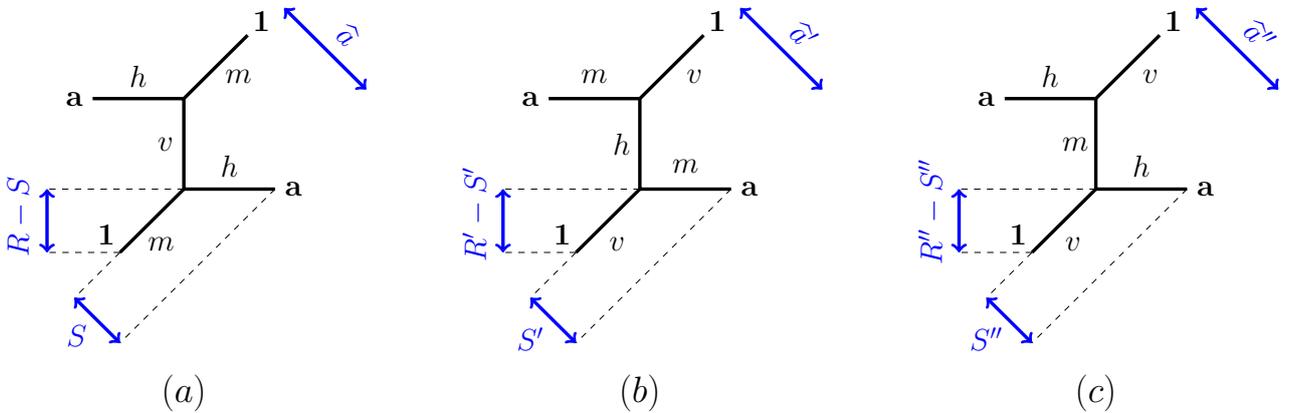

The simplest (albeit somewhat trivial) example to illustrate the idea explained in Section~\ref{Sect:GeneralStrategy} is the configuration $(N,M)=(1,1)$. The corresponding web diagram is shown in \figref{Fig:11webs}(a). Through simple $SL(2,\mathbb{Z})$ transformations (as well as cutting and re-gluing) the former can also be presented (among other ways) in the form of \figref{Fig:11webs}(b) and \figref{Fig:11webs}(c).

\noindent
Each diagram can be parametrised in terms of the parameters $(h,v,m)$ or respectively $(\widehat{a},S,R)$, $(\widehat{a}',S',R')$ or $(\widehat{a}'',S'',R'')$. The latter can be expressed in terms of $(h,v,m)$ as
{\allowdisplaybreaks
\begin{align}
&\widehat{a}=h+v\,,&&S=h\,,&&R-S=m\,,\nonumber\\
&\widehat{a}'=h+m\,,&&S'=m\,,&&R'-S'=v\,,\nonumber\\
&\widehat{a}''=h+m\,,&&S''=h\,,&&R''-S''=v\,.
%&\widehat{a}'''=m+v\,,&&S'''=v\,,&&R'''-S'''=v\,.
\end{align}}
Inverting these relations, $(h,v,m)$ can be expressed as linear combinations of $(\widehat{a},S,R)$, $(\widehat{a}',S',R')$ or $(\widehat{a}'',S'',R'')$ respectively
{\allowdisplaybreaks
\begin{align}
&h=S=\widehat{a}'-S'=S''\,,&&v=\widehat{a}-S=R'-S'=R''-S''\,,&&m=R-S=S'=\widehat{a}''-S''\,.
\end{align}}
These equations also furnish linear transformations between $(\widehat{a},S,R)$, $(\widehat{a}',S',R')$ or $(\widehat{a}'',S'',R'')$
\begin{align}
&\left(\begin{array}{c}\widehat{a} \\ S \\ R\end{array}\right)=G_1\cdot \left(\begin{array}{c}\widehat{a}' \\ S' \\ R'\end{array}\right)=G_2\cdot \left(\begin{array}{c}\widehat{a}'' \\ S'' \\ R''\end{array}\right)\,,&&\text{with} &&G_1=\left(\begin{array}{ccc} 1 & -2 & 1 \\ 1 & -1 & 0 \\ 1 & 0 & 0 \end{array}\right)\,,&&G_2=\left(\begin{array}{ccc} 0 & 0 & 1 \\ 0 & 1 & 0 \\ 1 & 0 & 0 \end{array}\right)\,.\label{Def11G1}%=G_3\cdot \left(\begin{array}{c}\widehat{a}''' \\ S''' \\ R'''\end{array}\right)\,,
\end{align}
The matrix $G_1$ is of order $3$ (\emph{i.e.} $G_1\cdot G_1\cdot G_1=1\!\!1_{3\times 3}$) while $G_2$ is of order $2$ (\emph{i.e.} $G_2\cdot G_2=1\!\!1_{3\times 3}$). Thus, introducing also the matrices\footnote{In the same manner as $G_1$ and $G_2$, these matrices can also be read off from web diagrams as in \figref{Fig:11webs} with a suitable exchange of $(h,v,m)$, which, however, we do not show explicitly.}
\begin{align}
&E=1\!\!1_{3\times 3}\,,&&G_3=G_1\cdot G_1\,,&&G_4=G_1\cdot G_2\,,&&G_5=G_2\cdot G_1\,,
\end{align}
the ensemble $\mathbb{G}(1)=\{E,G_1,G_2,G_3,G_4,G_5\}$ forms a finite group, whose multiplication table is
\begin{align}
\begin{array}{c|cccccc}
 & E & G_1 & G_2 & G_3 & G_4 & G_5\\\hline
E & E & G_1 & G_2 & G_3 & G_4 & G_5 \\
G_1 & G_1 & G_3 & G_4 & E & G_5 & G_2 \\
G_2 & G_2 & G_5 & E & G_4 & G_3 & G_1 \\
G_3 & G_3 & E & G_5 & G_1 & G_2 & G_4 \\
G_4 & G_4 & G_2 & G_1 & G_5 & E & G_3 \\
G_5 & G_5 & G_4 & G_3 & G_2 & G_1 & E \\
\end{array}
\end{align}
from which we can read off $\mathbb{G}(1)=\{E,G_1,G_2,G_3,G_4,G_5\}\cong \text{Dih}_3\cong S_3$. The latter can be formulated more elegantly as the free group generated by the elements 
\begin{align}
&a=G_4=G_1\cdot G_2=\left(\begin{array}{ccc}1 & -2 & 1 \\ 0 & - 1 & 1 \\ 0 & 0 & 1\end{array}\right)\,,&&\text{and} && b=G_5=G_2\cdot G_1=\left(\begin{array}{ccc}1 & 0 & 0 \\ 1 & - 1 & 0 \\ 1 & -2 & 1\end{array}\right)\,,\label{Def11ab}
\end{align}
furnishing the following presentation
\begin{align}
\mathbb{G}(1)\cong\text{Dih}_3\cong \left\langle \{a,b| a^2=b^2=1\!\!1_{3\times 3}, (ab)^3=1\!\!1_{3\times 3}\}\right\rangle\,.\label{Dih3Con11}
\end{align}
%%%%%%%%%%%%%%%%%%%%%%%%%%%
\subsection{Invariance of the Non-perturbative Free Energy}
As a check of the fact that $G_{1,2}$ defined in (\ref{Def11G1}) are indeed symmetry transformations of $\mathcal{Z}_{1,1}$, we can consider the coefficients in the expansion of the associated free energy $\mathcal{F}_{1,1}$. Indeed, for $N=1$, the expansion (\ref{FreeEnergyExpansionCoefficients}) can be written as
\begin{align}
\mathcal{F}_{1,1}(\widehat{a},S,R;\epsilon_1,\epsilon_2)=\sum_{n,i=0}^\infty \sum_{k\in\mathbb{Z}}f_{i,k,n}(\epsilon_1,\epsilon_2)\,\widehat{Q}^{i}\,Q_S^{k}\,Q_{R}^n\,,
\end{align}
with $\widehat{Q}=e^{-\widehat{a}}$. As explained in section~\ref{Sect:GeneralStrategy}, in order to be a symmetry, the coefficients $f_{i,k,n}(\epsilon_1,\epsilon_2)$ (which are functions of $\epsilon_{1,2}$ with a first order pole) need to satisfy
\begin{align}
&f_{i,k,n}(\epsilon_1,\epsilon_2)=f_{i',k',n'}(\epsilon_1,\epsilon_2)&&\text{for} && (i',k', n')^T=G^T_\ell\cdot (i,k, n)^T && \forall \ell=1,2\,.
\end{align}
Below we tabulate examples of coefficients $f_{i,k,n}$ with $i\leq 8$ for $n=1$, $i\leq 4$ for $n=2$ and $i\leq 2$ for $n=3$ that are related by $G_{1,2}$: Table~\ref{Tab:SymTrafo11G1} shows the relations for $G_1$ and Table~\ref{Tab:SymTrafo11G2} for $G_2$.
\begin{table}[htbp]
\begin{center}
\begin{tabular}{|c|c|c|}\hline
&&\\[-10pt]
$(i,k,n)$ & $(i',k',n')$ & $f_{i,k,n}(\epsilon_{1,2})=f_{i',k',n'}(\epsilon_{1,2})$\\[4pt]\hline\hline
&&\\[-11pt]
$(1,0,1)$ & $(2,-2,1)$ & $\frac{(q  t +1) \left(q ^2 t +q  (t +1)^2+t \right)}{(q -1) q  (t -1) t }$\\[5pt]\hline
&&\\[-11pt]
$(1,1,1)$ & $(3,-3,1)$ & $-\frac{(q +1) (t +1) (q  t +1)}{(q -1) \sqrt{q } (t -1) \sqrt{t }}$\\[5pt]\hline
&&\\[-11pt]
$(1,2,1)$ & $(4,-4,1)$ & $\frac{q  t +1}{(q -1) (t -1)}$\\[5pt]\hline
&&\\[-11pt]
$(2,-2,1)$ & $(1,-2,2)$ & $\frac{q ^3 t ^2+q ^2 t  \left(t ^2+2 t +2\right)+q  \left(2 t ^2+2 t +1\right)+t }{(q -1) q  (t -1) t }$\\[5pt]\hline
&&\\[-11pt]
$(2,1,1)$ & $(4,-5,2)$ & $\frac{q ^4 \left(-t ^2\right) (t +1)-q ^3 t  \left(t ^3+3 t ^2+4 t +1\right)-q ^2 \left(t ^4+4 t ^3+7 t ^2+4 t +1\right)-q  \left(t ^3+4 t ^2+3 t +1\right)-t  (t +1)}{(q -1) q ^{3/2} (t -1) t ^{3/2}}$\\[5pt]\hline
&&\\[-11pt]
$(3,-3,1)$ & $(1,-3,3)$ & $-\frac{(q +1) (t +1) (q  t +1)}{(q -1) \sqrt{q } (t -1) \sqrt{t }}$\\[5pt]\hline
&&\\[-11pt]
$(1,-1,2)$ & $(2,-1,1)$ & $\frac{q ^4 \left(-t ^2\right) (t +1)-q ^3 t  \left(t ^3+3 t ^2+4 t +1\right)-q ^2 \left(t ^4+4 t ^3+7 t ^2+4 t +1\right)-q  \left(t ^3+4 t ^2+3 t +1\right)-t  (t +1)}{(q -1) q ^{3/2} (t -1) t ^{3/2}}$\\[5pt]\hline
&&\\[-11pt]
$(1,1,2)$ & $(4,-3,1)$ & $\frac{q ^4 \left(-t ^2\right) (t +1)-q ^3 t  \left(t ^3+3 t ^2+4 t +1\right)-q ^2 \left(t ^4+4 t ^3+7 t ^2+4 t +1\right)-q  \left(t ^3+4 t ^2+3 t +1\right)-t  (t +1)}{(q -1) q ^{3/2} (t -1) t ^{3/2}}$\\[5pt]\hline
&&\\[-11pt]
$(1,3,2)$ & $(6,-5,1)$ & $-\frac{\sqrt{qt}}{(q -1) (t -1)}$\\[5pt]\hline
&&\\[-11pt]
$(2,-3,2)$ & $(1,-1,2)$ & $\frac{q ^4 \left(-t ^2\right) (t +1)-q ^3 t  \left(t ^3+3 t ^2+4 t +1\right)-q ^2 \left(t ^4+4 t ^3+7 t ^2+4 t +1\right)-q  \left(t ^3+4 t ^2+3 t +1\right)-t  (t +1)}{(q -1) q ^{3/2} (t -1) t ^{3/2}}$\\[5pt]\hline
&&\\[-11pt]
$(1,-2,3)$ & $(2,0,1)$ & $\frac{q ^5 t ^3+q ^4 t ^2 \left(2 t ^2+3 t +2\right)+q ^3 t  \left(t ^4+3 t ^3+8 t ^2+6 t +2\right)+q ^2 \left(2 t ^4+6 t ^3+8 t ^2+3 t +1\right)+q  t  \left(2 t ^2+3 t +2\right)+t ^2}{(q -1) q ^2 (t -1) t ^2}$\\[5pt]\hline
&&\\[-11pt]
$(1,1,3)$ & $(5,-3,1)$ & $-\frac{(q +1) (t +1) \left(q ^5 t ^3+q ^4 t ^2 (t +1)^2+q ^3 t  \left(t ^4+2 t ^3+6 t ^2+4 t +1\right)+q ^2 \left(t ^4+4 t ^3+6 t ^2+2 t +1\right)+q  t  (t +1)^2+t ^2\right)}{(q -1) q ^{5/2} (t -1) t ^{5/2}}$\\[5pt]\hline
&&\\[-11pt]
$(1,2,3)$ & $(6,-4,1)$ & $\frac{q ^5 t ^3+q ^4 t ^2 \left(2 t ^2+3 t +2\right)+q ^3 t  \left(t ^4+3 t ^3+8 t ^2+6 t +2\right)+q ^2 \left(2 t ^4+6 t ^3+8 t ^2+3 t +1\right)+q  t  \left(2 t ^2+3 t +2\right)+t ^2}{(q -1) q ^2 (t -1) t ^2}$\\[5pt]\hline
&&\\[-11pt]
$(1,3,3)$ & $(7,-5,1)$ & $-\frac{(q +1) (t +1) (q  t +1)}{(q -1) \sqrt{q } (t -1) \sqrt{t }}$\\[5pt]\hline
\end{tabular}
\end{center}
\caption{\sl Action of $G_1$: the indices are related by $(i'_1,i'_2,k',n')^T=G_1^T\cdot (i_1,i_2,k,n)^T$.}
\label{Tab:SymTrafo11G1}
\end{table}

\begin{table}[htbp]
\begin{center}
\begin{tabular}{|c|c|c|}\hline
&&\\[-10pt]
$(i,k,n)$ & $(i',k',n')$ & $f_{i,k,n}(\epsilon_{1,2})=f_{i',k',n'}(\epsilon_{1,2})$\\[4pt]\hline\hline
&&\\[-11pt]
$(2,-3,1)$ & $(1,-3,2)$ & $-\frac{\sqrt{qt}}{(q -1) (t -1)}$\\[5pt]\hline
&&\\[-11pt]
$(2,-2,1)$ & $(1,-2,2)$ & $\frac{q ^3 t ^2+q ^2 t  \left(t ^2+2 t +2\right)+q  \left(2 t ^2+2 t +1\right)+t }{(q -1) q  (t -1) t }$\\[5pt]\hline
&&\\[-11pt]
$(2,-1,1)$ & $(1,-1,2)$ & $\frac{q ^4 \left(-t ^2\right) (t +1)-q ^3 t  \left(t ^3+3 t ^2+4 t +1\right)-q ^2 \left(t ^4+4 t ^3+7 t ^2+4 t +1\right)-q  \left(t ^3+4 t ^2+3 t +1\right)-t  (t +1)}{(q -1) q ^{3/2} (t -1) t ^{3/2}}$\\[5pt]\hline
&&\\[-11pt]
$(2,0,1)$ & $(1,0,2)$ & $\frac{q ^5 t ^3+q ^4 t ^2 \left(2 t ^2+3 t +2\right)+q ^3 t  \left(t ^4+3 t ^3+8 t ^2+6 t +2\right)+q ^2 \left(2 t ^4+6 t ^3+8 t ^2+3 t +1\right)+q  t  \left(2 t ^2+3 t +2\right)+t ^2}{(q -1) q ^2 (t -1) t ^2}$\\[5pt]\hline
&&\\[-11pt]
$(2,1,1)$ & $(1,1,2)$ & $\frac{q ^4 \left(-t ^2\right) (t +1)-q ^3 t  \left(t ^3+3 t ^2+4 t +1\right)-q ^2 \left(t ^4+4 t ^3+7 t ^2+4 t +1\right)-q  \left(t ^3+4 t ^2+3 t +1\right)-t  (t +1)}{(q -1) q ^{3/2} (t -1) t ^{3/2}}$\\[5pt]\hline
&&\\[-11pt]
$(2,2,1)$ & $(1,2,2)$ & $\frac{q ^3 t ^2+q ^2 t  \left(t ^2+2 t +2\right)+q  \left(2 t ^2+2 t +1\right)+t }{(q -1) q  (t -1) t }$\\[5pt]\hline
&&\\[-11pt]
$(2,3,1)$ & $(1,3,2)$ & $-\frac{\sqrt{qt}}{(q -1) (t -1)}$\\[5pt]\hline
&&\\[-11pt]
$(3,-3,1)$ & $(1,-3,3)$ & $-\frac{(q +1) (t +1) (q  t +1)}{(q -1) \sqrt{q } (t -1) \sqrt{t }}$\\[5pt]\hline
&&\\[-11pt]
$(3,-2,1)$ & $(1,-2,3)$ & $\frac{q ^5 t ^3+q ^4 t ^2 \left(2 t ^2+3 t +2\right)+q ^3 t  \left(t ^4+3 t ^3+8 t ^2+6 t +2\right)+q ^2 \left(2 t ^4+6 t ^3+8 t ^2+3 t +1\right)+q  t  \left(2 t ^2+3 t +2\right)+t ^2}{(q -1) q ^2 (t -1) t ^2}$\\[5pt]\hline
&&\\[-11pt]
$(3,-1,1)$ & $(1,-1,3)$ & $-\frac{(q +1) (t +1) \left(q ^5 t ^3+q ^4 t ^2 (t +1)^2+q ^3 t  \left(t ^4+2 t ^3+6 t ^2+4 t +1\right)+q ^2 \left(t ^4+4 t ^3+6 t ^2+2 t +1\right)+q  t  (t +1)^2+t ^2\right)}{(q -1) q ^{5/2} (t -1) t ^{5/2}}$\\[5pt]\hline
&&\\[-11pt]
$(3,1,1)$ & $(1,1,3)$ & $-\frac{(q +1) (t +1) \left(q ^5 t ^3+q ^4 t ^2 (t +1)^2+q ^3 t  \left(t ^4+2 t ^3+6 t ^2+4 t +1\right)+q ^2 \left(t ^4+4 t ^3+6 t ^2+2 t +1\right)+q  t  (t +1)^2+t ^2\right)}{(q -1) q ^{5/2} (t -1) t ^{5/2}}$\\[5pt]\hline
&&\\[-11pt]
$(3,2,1)$ & $(1,2,3)$ & $\frac{q ^5 t ^3+q ^4 t ^2 \left(2 t ^2+3 t +2\right)+q ^3 t  \left(t ^4+3 t ^3+8 t ^2+6 t +2\right)+q ^2 \left(2 t ^4+6 t ^3+8 t ^2+3 t +1\right)+q  t  \left(2 t ^2+3 t +2\right)+t ^2}{(q -1) q ^2 (t -1) t ^2}$\\[5pt]\hline
&&\\[-11pt]
$(3,3,1)$ & $(1,3,3)$ & $-\frac{(q +1) (t +1) (q  t +1)}{(q -1) \sqrt{q } (t -1) \sqrt{t }}$\\[5pt]\hline
&&\\[-11pt]
$(1,-3,2)$ & $(2,-3,1)$ & $-\frac{\sqrt{qt}}{(q -1) (t -1)}$\\[5pt]\hline
&&\\[-11pt]
$(1,-2,2)$ & $(2,-2,1)$ & $\frac{(q  t +1) \left(q ^2 t +q  (t +1)^2+t \right)}{(q -1) q  (t -1) t }$\\[5pt]\hline
&&\\[-11pt]
$(1,2,2)$ & $(2,2,1)$ & $\frac{(q  t +1) \left(q ^2 t +q  (t +1)^2+t \right)}{(q -1) q  (t -1) t }$\\[5pt]\hline
&&\\[-11pt]
$(1,3,2)$ & $(2,3,1)$ & $-\frac{\sqrt{qt}}{(q -1) (t -1)}$\\[5pt]\hline
&&\\[-11pt]
$(1,-3,3)$ & $(3,-3,1)$ & $-\frac{(q +1) (t +1) (q  t +1)}{(q -1) \sqrt{q } (t -1) \sqrt{t }}$\\[5pt]\hline
&&\\[-11pt]
$(1,3,3)$ & $(3,3,1)$ & $-\frac{(q +1) (t +1) (q  t +1)}{(q -1) \sqrt{q } (t -1) \sqrt{t }}$\\[5pt]\hline
\end{tabular}
\end{center}
\caption{\sl Action of $G_2$: the indices are related by $(i'_1,i'_2,k',n')^T=G_2^T\cdot (i_1,i_2,k,n)^T$.}
\label{Tab:SymTrafo11G2}
\end{table}

%%%%%%%%%%%%%%%%%%%%%%%%%%%
\subsection{Modularity and $Sp(4,\mathbb{Z})$ Symmetry}
The action of $\mathbb{G}(1)$ as presented in (\ref{Dih3Con11}) combines with $SL(2,\mathbb{Z})\times SL(2,\mathbb{Z})$ into $Sp(4,\mathbb{Z})$, which is (a subgroup of) the automorphism group of $X_{1,1}$. To see this, instead of considering the action of $\mathbb{G}(1)$ on the vector space spanned by $(\widehat{a},S,R)$, we consider the vector space spanned by $(\tau=h+v,\rho=m+v,v)$. Arranging the latter in the period matrix
\begin{align}
\Omega=\left(\begin{array}{cc}\tau & v \\ v & \rho\end{array}\right)\,,
\end{align}
there is a natural action of $Sp(4,\mathbb{Z})$, as reviewed in appendix~\ref{App:Sp4}. The action of $G_{1,2}$ on $\Omega$ is
\begin{align}
&G_1:\hspace{0.25cm}\Omega\to \left(
\begin{array}{cc}
 -2 v+\rho +\tau  & \tau -v \\
 \tau -v & \tau  \\
\end{array}
\right)\,,&&G_2:\hspace{0.25cm} \Omega\to \left(
\begin{array}{cc}
 \tau  & \tau -v \\
 \tau -v & -2 v+\rho +\tau  \\
\end{array}
\right)
\end{align} 
Based on this action, we can equivalently represent the action of $\mathbb{G}(1)$ by $G'_{1,2}\in Sp(4,\mathbb{Z})$
\begin{align}
&G'_1=HK=\left(
\begin{array}{cccc}
 1 & -1 & 0 & 0 \\
 1 & 0 & 0 & 0 \\
 0 & 0 & 0 & -1 \\
 0 & 0 & 1 & 1 \\
\end{array}
\right)\,,&&\text{and} &&G'_2=K=\left(
\begin{array}{cccc}
 1 & 0 & 0 & 0 \\
 1 & -1 & 0 & 0 \\
 0 & 0 & 1 & 1 \\
 0 & 0 & 0 & -1 \\
\end{array}
\right)\,,\label{DefsGp1Gp2}
\end{align}
where $K$ and $H$ are defined as in appendix~\ref{App:Sp4}. This implies that $\mathbb{G}(1)\subset Sp(4,\mathbb{Z})$. Moreover, combining $\mathbb{G}(1)$ with the $SL(2,\mathbb{Z})_\rho$ symmetry\footnote{Notice that the symmetry group is isomorphic to $SL(2,\mathbb{Z})$ rather than $PSL(2,\mathbb{Z})$, since $S_\rho^2\neq 1\!\!1$, as can be seen from the action of 
$S_\rho^2$ on the period matrix $\Omega\to \left(\begin{array}{cc}\tau & -v \\ -v & \rho\end{array}\right)$.} acting on the modular parameter\footnote{We could also choose the modular group $SL(2,\mathbb{Z})_\tau$ which acts in a similar fashion on the modular parameter $\tau$. More precisely, $SL(2,\mathbb{Z})_\tau$ is generated by $S_\tau=H S_\rho H$ and $T_\tau=H T_\rho H$.} $\rho$ as
\begin{align}
&S_\rho:\hspace{0.25cm}(\tau,\rho,v)\longmapsto \left(\tau-\tfrac{v^2}{\rho},-\tfrac{1}{\rho},\tfrac{v}{\rho}\right)\,,&&T_\rho:\hspace{0.25cm}(\tau,\rho,v)\longmapsto(\tau,\rho+1,v)\,,\label{Sl2rho}
\end{align}
generates the complete action of $Sp(4,\mathbb{Z})$: the generators $(S_\rho,T_\rho)$ can be expressed as $S_\rho=L^3$ and $T_\rho=L^9 H L^{10} H=X_2$. Furthermore, we have $G'_2 G'_1=L^5 K L^7$ such that we can write
{\allowdisplaybreaks
\begin{align}
&X_1=G'_2 G'_1 S_\rho^2\,,&& X_2=T_\rho\,,&&X_3=S_\rho G'_1 G'_1 S_\rho\,,\nonumber\\
&X_4=G'_1G'_2T_\rho G'_1G'_2\,,&&X_5=G'_1G'_2 S_\rho^2\,,&&X_6=S_\rho^3 G'_1 G'_2 S_\rho^2 G'_1 G'_2\,,
\end{align}}
with $X_{1,2,3,4,5,6}$ defined in (\ref{DefXGeneratorsBehr}). This indicates that
\begin{align}
\langle G'_1\,,G'_2\,,S_\rho\,,T_\rho \rangle\supset \langle X_1\,,X_2\,,X_3\,,X_4\,,X_5\,,X_6\rangle\cong Sp(4,\mathbb{Z})\,,
\end{align}
where the last relation was shown in \cite{Behr}. From (\ref{DefsGp1Gp2}), using the presentation of $Sp(4,\mathbb{Z})$ given in \cite{Bender}, it follows that 
\begin{align}
\langle G'_1\,,G'_2\,,S_\rho\,,T_\rho \rangle\subset \langle K,L\rangle\cong Sp(4,\mathbb{Z})\,,
\end{align}
which implies $\langle G'_1\,,G'_2\,,S_\rho\,,T_\rho \rangle\cong Sp(4,\mathbb{Z})$.

%%%%%%%%%%%%%%%%%%%%%%%%%%%%%%%%%%%%%%%%%%%%%%%%
\section{Example: $(N,M)=(2,1)$}
\subsection{Dualities and $\text{Dih}_2$ Group Action}
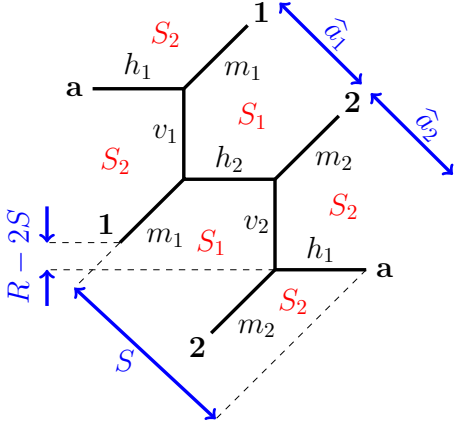
\begin{wrapfigure}{l}{0.38\textwidth}
\begin{center}
${}$\\[-0.5cm]
\scalebox{0.80}{\parbox{7.5cm}{\begin{tikzpicture}[scale = 1.50]
\draw[ultra thick] (-1,0) -- (0,0) -- (0,-1) -- (1,-1) -- (1,-2) -- (2,-2);
%diagonals
\draw[ultra thick] (0,0) -- (0.7,0.7);
\draw[ultra thick] (1,-1) -- (1.7,-0.3);
\draw[ultra thick] (0,-1) -- (-0.7,-1.7);
\draw[ultra thick] (1,-2) -- (0.3,-2.7);
%ends
\node at (-1.2,0) {\large {\bf $\mathbf a$}};
\node at (2.2,-2) {\large {\bf $\mathbf a$}};
\node at (0.85,0.85) {\large {$\mathbf 1$}};
\node at (1.85,-0.15) {\large {$\mathbf 2$}};
\node at (-0.85,-1.5) {\large {$\mathbf 1$}};
\node at (0.15,-2.85) {\large {$\mathbf 2$}};
%lables hotizontal
\node at (-0.5,0.25) {\large  {\bf $h_1$}};
\node at (0.5,-0.775) {\large  {\bf $h_2$}};
\node at (1.5,-1.775) {\large  {\bf $h_1$}};
%lables vertical
\node at (-0.2,-0.5) {\large  {\bf $v_1$}};
\node at (0.8,-1.5) {\large  {\bf $v_2$}};
%lables diagonal
\node at (0.65,0.2) {\large  {\bf $m_1$}};
\node at (1.65,-0.8) {\large  {\bf $m_2$}};
\node at (-0.2,-1.65) {\large  {\bf $m_1$}};
\node at (0.8,-2.65) {\large  {\bf $m_2$}};
%hexagons
\node[red] at (-0.2,0.6) {\large  {\bf $S_2$}};
\node[red] at (0.75,-0.3) {\large  {\bf $S_1$}};
\node[red] at (1.75,-1.3) {\large  {\bf $S_2$}};
\node[red] at (-0.75,-0.8) {\large  {\bf $S_2$}};
\node[red] at (0.3,-1.7) {\large  {\bf $S_1$}};
\node[red] at (1.2,-2.35) {\large  {\bf $S_2$}};
%roots
\draw[ultra thick,blue,<->] (1.05,0.95) -- (1.95,0.05);
\node[blue,rotate=315] at (1.7,0.6) {{\large {\bf {$\widehat{a}_1$}}}};
\draw[ultra thick,blue,<->] (2.05,-0.05) -- (2.95,-0.95);
\node[blue,rotate=315] at (2.7,-0.4) {{\large {\bf {$\widehat{a}_2$}}}};
%definition S
\draw[dashed] (2,-2) -- (0.35,-3.65);
\draw[dashed] (-0.7,-1.7) -- (-1.2,-2.2);
\draw[ultra thick,blue,<->] (-1.2,-2.2) -- (0.35,-3.65);
\node[blue] at (-0.65,-3) {{\large {\bf {$S$}}}};
%parameter R
\draw[dashed] (-0.7,-1.7) -- (-1.5,-1.7);
\draw[dashed] (1,-2) -- (-1.5,-2);
\draw[ultra thick,blue,->] (-1.5,-1.3) -- (-1.5,-1.7);
\draw[ultra thick,blue,->] (-1.5,-2.4) -- (-1.5,-2);
\node[blue,rotate=90] at (-1.8,-1.85) {{\large {\bf{$R-2S$}}}};
\end{tikzpicture}}}
\caption{\sl Web diagram of $X_{2,1}$ with a parametrisation of the areas of all curves. The blue parameters represent an independent set of K\"ahler parameters.}
\label{Fig:21web0}
\end{center}
${}$\\[-2.5cm]
\end{wrapfigure}

In this section we generalise the analysis of the previous section and, using the simplest non-trivial example (namely $(N,M)=(2,1)$), explain how the duality transformations advocated in \cite{Bastian:2017ary,Bastian:2018dfu} lead to non-trivial symmetries at the level of the set of independent K\"ahler parameters of $X_{2,1}$. In the following subsection we give further evidence for this symmetry at the level of the partition function $\mathcal{Z}_{2,1}$. The starting point is the web diagram shown in \figref{Fig:21web0} along with a parametrisation of the areas of all curves involved. The latter are not all independent of one another, but for each of the two hexagons $S_{1,2}$, they have to satisfy the following consistency 

\begin{figure}
\begin{center}
\scalebox{0.75}{\parbox{20.5cm}{\begin{tikzpicture}[scale = 1.50]
\draw[ultra thick] (-1,0) -- (0,0) -- (1,1) -- (2,1) -- (3,2) -- (4,2);
%verticals
\draw[ultra thick] (1,1) -- (1,2);
\draw[ultra thick] (3,2) -- (3,3);
\draw[ultra thick] (0,0) -- (0,-1);
\draw[ultra thick] (2,1) -- (2,0);
%ends
\node at (-1.2,0) {\large {\bf $\mathbf a$}};
\node at (4.2,2) {\large {\bf $\mathbf a$}};
\node at (0,-1.25) {\large {$\mathbf 2$}};
\node at (2,-0.25) {\large {$\mathbf 1$}};
\node at (1,2.25) {\large {$\mathbf 2$}};
\node at (3,3.25) {\large {$\mathbf 1$}};
%lables hotizontal
\node at (-0.5,0.25) {\large  {\bf $h_1$}};
\node at (1.5,1.25) {\large  {\bf $h_2$}};
\node at (3.5,2.25) {\large  {\bf $h_1$}};
%lables diagonal
\node at (0.7,0.3) {\large  {\bf $v_2$}};
\node at (2.7,1.3) {\large  {\bf $v_1$}};
%labels vertical
\node at (0.7,1.5) {\large  {\bf $m_2$}};
\node at (2.7,2.5) {\large  {\bf $m_1$}};
\node at (-0.3,-0.5) {\large  {\bf $m_2$}};
\node at (1.7,0.5) {\large  {\bf $m_1$}};
%stamp
\node at (1.5,-2) {\Large {\bf $\mathbf (a)$}};
%%%%%%%%%%%%%%%%%%%%%%%
%%%%%%%%%%%%%%%%%%%%%%%
\begin{scope}[xshift=9cm,yshift=3cm]
\draw[ultra thick] (-1,0) -- (0,0) -- (0,-1) -- (1,-1) -- (1,-2) -- (2,-2);
%diagonals
\draw[ultra thick] (0,0) -- (0.7,0.7);
\draw[ultra thick] (1,-1) -- (1.7,-0.3);
\draw[ultra thick] (0,-1) -- (-0.7,-1.7);
\draw[ultra thick] (1,-2) -- (0.3,-2.7);
%ends
\node at (-1.2,0) {\large {\bf $\mathbf a$}};
\node at (2.2,-2) {\large {\bf $\mathbf a$}};
\node at (0.85,0.85) {\large {$\mathbf I$}};
\node at (1.9,-0.1) {\large {$\mathbf{II}$}};
\node at (-0.85,-1.5) {\large {$\mathbf I$}};
\node at (0.1,-2.9) {\large {$\mathbf{II}$}};
%lables hotizontal
\node at (-0.5,0.25) {\large  {\bf $h_1$}};
\node at (0.5,-0.775) {\large  {\bf $h_2$}};
\node at (1.5,-1.775) {\large  {\bf $h_1$}};
%lables vertical
\node at (-0.25,-0.5) {\large  {\bf $m_2$}};
\node at (0.75,-1.5) {\large  {\bf $m_1$}};
%lables diagonal
\node at (0.6,0.25) {\large  {\bf $v_2$}};
\node at (1.6,-0.75) {\large  {\bf $v_1$}};
\node at (-0.2,-1.6) {\large  {\bf $v_2$}};
\node at (0.75,-2.6) {\large  {\bf $v_1$}};
%hexagons
\node[red] at (-0.2,0.6) {\large  {\bf $S'_2$}};
\node[red] at (0.75,-0.3) {\large  {\bf $S'_1$}};
\node[red] at (1.75,-1.3) {\large  {\bf $S'_2$}};
\node[red] at (-0.75,-0.8) {\large  {\bf $S'_2$}};
\node[red] at (0.3,-1.7) {\large  {\bf $S'_1$}};
\node[red] at (1.2,-2.35) {\large  {\bf $S'_2$}};
%roots
\draw[ultra thick,blue,<->] (1.1,1) -- (2,0.1);
\node[blue,rotate=315] at (1.8,0.7) {{\large {\bf {$\widehat{a}'_1$}}}};
\draw[ultra thick,blue,<->] (2.1,0) -- (3,-0.9);
\node[blue,rotate=315] at (2.8,-0.3) {{\large {\bf {$\widehat{a}'_2$}}}};
%definition S
\draw[dashed] (2,-2) -- (0.35,-3.65);
\draw[dashed] (-0.7,-1.7) -- (-1.2,-2.2);
\draw[ultra thick,blue,<->] (-1.2,-2.2) -- (0.35,-3.65);
\node[blue] at (-0.7,-3.05) {{\large {\bf {$S'$}}}};
%parameter R
\draw[dashed] (-0.7,-1.7) -- (-1.5,-1.7);
\draw[dashed] (1,-2) -- (-1.5,-2);
\draw[ultra thick,blue,->] (-1.5,-1.3) -- (-1.5,-1.7);
\draw[ultra thick,blue,->] (-1.5,-2.4) -- (-1.5,-2);
\node[blue,rotate=90] at (-1.8,-1.85) {{\large {\bf{$R'-2S'$}}}};
\end{scope}
%stamp
\node at (9.5,-2) {\Large {\bf $\mathbf (b)$}};
\end{tikzpicture}}}
\caption{\sl (a) web diagram of \figref{Fig:21web0} after mirroring and an $SL(2,\mathbb{Z})$ transformation. (b) same web diagram after cutting along the lines $v_{1,2}$ and re-gluing along the lines $m_{1,2}$.}
\label{Fig:21web1}
\end{center}
\end{figure}
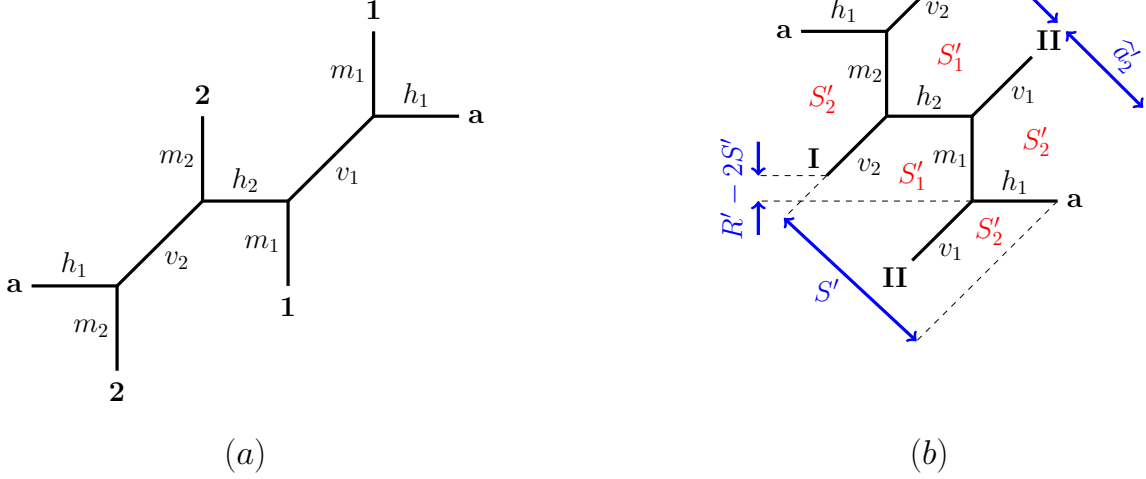

\noindent
conditions:
\begin{align}
&S_1:\,\,&h_2+m_2=m_1+h_2\,,&&v_1+m_1=m_2+v_2\,,\nonumber\\
&S_2:\,\,&h_1+m_1=m_2+h_1\,,&&m_1+v_1=m_2+v_2\,.\label{ConsistencyWeb21}
\end{align}
A solution for these conditions was provided in \cite{Bastian:2017ing} in the form of the parameters $(\widehat{a}_{1,2},S,R)$ as indicated in \figref{Fig:21web0}
\begin{align}
&\widehat{a}_1=v_1+h_2\,,&&\widehat{a}_2=v_2+h_1\,,\nonumber\\
&S=h_2+v_2+h_1\,,&&R-2S=m_1-v_2\,.
\end{align}
Indeed, all of the areas $(h_{1,2},v_{1,2},m_{1,2})$ can be expressed as a linear combination of $(\widehat{a}_1,\widehat{a}_2,S,R)$:
\begin{align}
&h_1=S-\widehat{a}_1\,,\hspace{0.5cm}h_2=S-\widehat{a}_2\,,\hspace{0.5cm}v_1=v_2=\widehat{a}_1+\widehat{a}_2\,,\nonumber\\
&m_1=m_2=\widehat{a}_1+\widehat{a}_2+R-3S\,.\label{Base1web21}
\end{align}
Mirroring the diagram and performing an $SL(2,\mathbb{Z})$ transformation, \figref{Fig:21web0} can also be presented in the form of \figref{Fig:21web1}(a). After cutting the latter along the curve labelled $v_{1,2}$ and re-gluing along the curves labelled $m_{1,2}$ leads to the presentation in \figref{Fig:21web1}(b). The consistency conditions of this web are the same as (\ref{ConsistencyWeb21}). Furthermore, the web diagram \figref{Fig:21web1}(b) is of the same form as \figref{Fig:21web0} and thus allows for a solution of (\ref{ConsistencyWeb21}) in terms of the parameters $(\widehat{a}'_1,\widehat{a}'_2,S',R')$: 
\begin{align}
&\widehat{a}'_1=m_1+h_1\,,&&\widehat{a}'_2=m_2+h_2\,,&&S'=h_2+m_1+h_1\,,&&R'-2S'=v_2-m_1\,.
\end{align}
Indeed, we can express the areas $(h_{1,2},v_{1,2},m_{1,2})$ in terms of the latter
\begin{align}
&h_1=S'-\widehat{a}'_2\,,&&h_2=S'-\widehat{a}'_2\,,&&v_1=v_2=\widehat{a}'_1+\widehat{a}'_2-3S+R'\,,&&m_1=m_2=\widehat{a}'_1+\widehat{a}'_2-S'\,.\label{Base2web21}
\end{align}
Comparing (\ref{Base1web21}) with (\ref{Base2web21}) gives rise to a linear relation between $(\widehat{a}_1,\widehat{a}_2,S,R)$ to $(\widehat{a}'_1,\widehat{a}'_2,S',R')$:
\begin{align}
&\left(\begin{array}{c}\widehat{a}_1\\ \widehat{a}_2 \\ S \\ R\end{array}\right)=G_1\cdot \left(\begin{array}{c}\widehat{a}'_1\\ \widehat{a}'_2 \\ S' \\ R'\end{array}\right)\,,&&\text{where} &&G_1=\left(
\begin{array}{cccc}
 1 & 0 & -2 & 1 \\
 0 & 1 & -2 & 1 \\
 0 & 0 & -1 & 1 \\
 0 & 0 & 0 & 1 \\
\end{array}
\right) && \text{with} &&\begin{array}{l}\text{det}\, G_1=1\,, \\ G_1\cdot G_1=1\!\!1_{4\times 4}\,.\end{array}\label{Gen21G1}
\end{align}

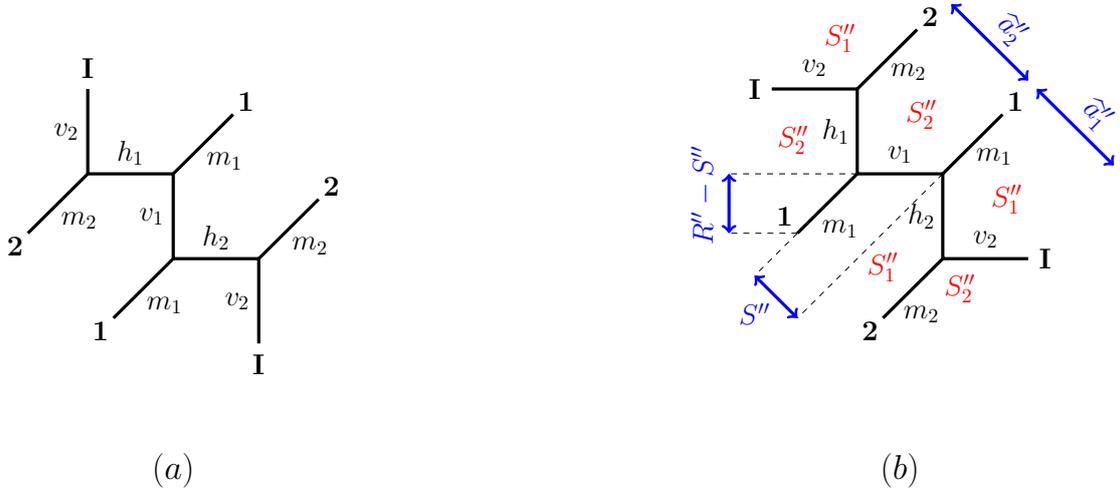
\begin{figure}
\begin{center}
\scalebox{0.75}{\parbox{19.75cm}{\begin{tikzpicture}[scale = 1.50]
\draw[ultra thick] (0,1) -- (0,0) -- (1,0) -- (1,-1) -- (2,-1) -- (2,-2);
\draw[ultra thick] (1,0) -- (1.7,0.7);
\draw[ultra thick] (2,-1) -- (2.7,-0.3);
\draw[ultra thick] (0,0) -- (-0.7,-0.7);
\draw[ultra thick] (1,-1) -- (0.3,-1.7);
%ends
\node at (1.85,0.85) {\large {$\mathbf 1$}};
\node at (2.85,-0.15) {\large {$\mathbf 2$}};
\node at (-0.85,-0.85) {\large {$\mathbf 2$}};
\node at (0.15,-1.85) {\large {$\mathbf 1$}};
\node at (0,1.25) {\large {$\mathbf I$}};
\node at (2,-2.25) {\large {$\mathbf I$}};
%lables horizontal
\node at (0.5,0.25) {\large  {\bf $h_1$}};
\node at (1.5,-0.75) {\large  {\bf $h_2$}};
%lables vertical
\node at (-0.25,0.5) {\large  {\bf $v_2$}};
\node at (0.75,-0.5) {\large  {\bf $v_1$}};
\node at (1.75,-1.5) {\large  {\bf $v_2$}};
%lables diagonal
\node at (-0.1,-0.55) {\large  {\bf $m_2$}};
\node at (0.9,-1.55) {\large  {\bf $m_1$}};
\node at (1.6,0.15) {\large  {\bf $m_1$}};
\node at (2.6,-0.85) {\large  {\bf $m_2$}};
%stamp
\node at (1,-3.5) {\Large {\bf $\mathbf (a)$}};
%%%%%%%%%%%%%%%%%%%%%%%
%%%%%%%%%%%%%%%%%%%%%%%
\begin{scope}[xshift=9cm,yshift=1cm]
\draw[ultra thick] (-1,0) -- (0,0) -- (0,-1) -- (1,-1) -- (1,-2) -- (2,-2);
%diagonals
\draw[ultra thick] (0,0) -- (0.7,0.7);
\draw[ultra thick] (1,-1) -- (1.7,-0.3);
\draw[ultra thick] (0,-1) -- (-0.7,-1.7);
\draw[ultra thick] (1,-2) -- (0.3,-2.7);
%ends
\node at (-1.2,0) {\large {\bf $\mathbf I$}};
\node at (2.2,-2) {\large {\bf $\mathbf{I}$}};
\node at (0.85,0.85) {\large {$\mathbf 2$}};
\node at (1.85,-0.15) {\large {$\mathbf{1}$}};
\node at (-0.85,-1.5) {\large {$\mathbf 1$}};
\node at (0.15,-2.85) {\large {$\mathbf 2$}};
%lables hotizontal
\node at (-0.5,0.25) {\large  {\bf $v_2$}};
\node at (0.5,-0.775) {\large  {\bf $v_1$}};
\node at (1.5,-1.775) {\large  {\bf $v_2$}};
%lables vertical
\node at (-0.25,-0.5) {\large  {\bf $h_1$}};
\node at (0.75,-1.5) {\large  {\bf $h_2$}};
%lables diagonal
\node at (0.6,0.2) {\large  {\bf $m_2$}};
\node at (1.6,-0.8) {\large  {\bf $m_1$}};
\node at (-0.2,-1.65) {\large  {\bf $m_1$}};
\node at (0.75,-2.65) {\large  {\bf $m_2$}};
%hexagons
\node[red] at (-0.2,0.6) {\large  {\bf $S''_1$}};
\node[red] at (0.75,-0.3) {\large  {\bf $S''_2$}};
\node[red] at (1.75,-1.3) {\large  {\bf $S''_1$}};
\node[red] at (-0.75,-0.6) {\large  {\bf $S''_2$}};
\node[red] at (0.3,-2.1) {\large  {\bf $S''_1$}};
\node[red] at (1.2,-2.35) {\large  {\bf $S''_2$}};
%roots
\draw[ultra thick,blue,<->] (1.1,1) -- (2,0.1);
\node[blue,rotate=315] at (1.8,0.7) {{\large {\bf {$\widehat{a}''_2$}}}};
\draw[ultra thick,blue,<->] (2.1,0) -- (3,-0.9);
\node[blue,rotate=315] at (2.8,-0.3) {{\large {\bf {$\widehat{a}''_1$}}}};
%definition S
\draw[dashed] (1,-1) -- (-0.7,-2.7);
\draw[dashed] (-0.7,-1.7) -- (-1.2,-2.2);
\draw[ultra thick,blue,<->] (-1.2,-2.2) -- (-0.7,-2.7);
\node[blue] at (-1.2,-2.65) {{\large {\bf {$S''$}}}};
%parameter R
\draw[dashed] (-0.7,-1.7) -- (-1.5,-1.7);
\draw[dashed] (0,-1) -- (-1.5,-1);
\draw[ultra thick,blue,<->] (-1.5,-1) -- (-1.5,-1.7);
\node[blue,rotate=90] at (-1.8,-1.25) {{\large {\bf{$R''-S''$}}}};
\end{scope}
%stamp
\node at (9.5,-3.5) {\Large {\bf $\mathbf (b)$}};
\end{tikzpicture}}}
\caption{\sl (a) web diagram obtained from \figref{Fig:21web0} after cutting along the line labelled $v_2$ and re-gluing along $h_1$. (b) alternative presentation of the same diagram.}
\label{Fig:21web2}
\end{center}
\end{figure}

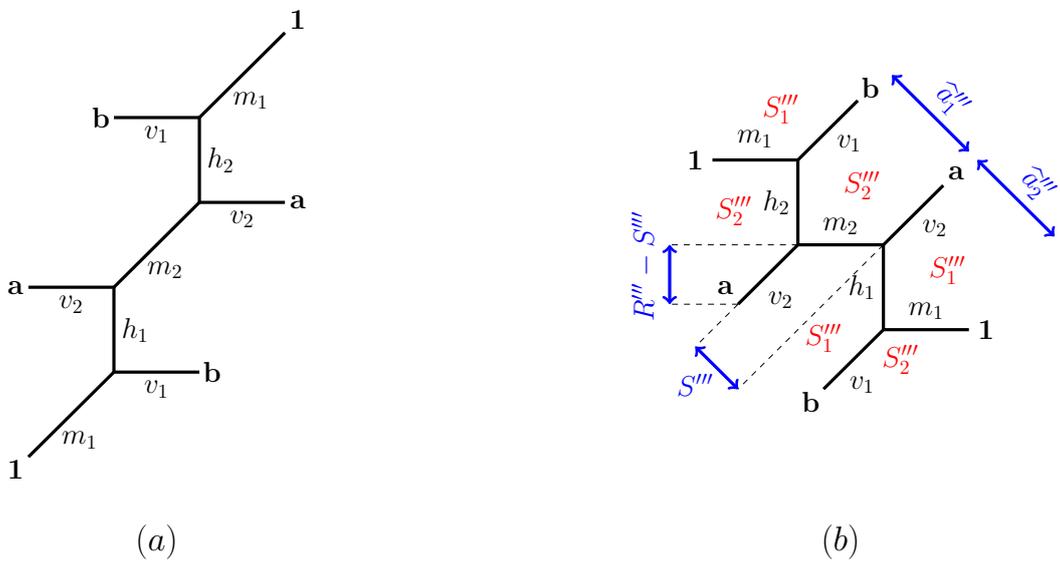
\begin{figure}
\begin{center}
\scalebox{0.75}{\parbox{18.75cm}{\begin{tikzpicture}[scale = 1.50]
\draw[ultra thick] (-1,-1) -- (0,0) -- (0,1) -- (1,2) -- (1,3) -- (2,4);
%horizontals
\draw[ultra thick] (0,0) -- (1,0);
\draw[ultra thick] (1,2) -- (2,2);
\draw[ultra thick] (0,1) -- (-1,1);
\draw[ultra thick] (1,3) -- (0,3);
%ends
\node at (-1.15,-1.15) {\large {$\mathbf{1}$}};
\node at (2.15,4.15) {\large {$\mathbf{1}$}};
\node at (1.15,0) {\large {\bf $\mathbf b$}};
\node at (2.15,2) {\large {\bf $\mathbf a$}};
\node at (-1.15,1) {\large {\bf $\mathbf a$}};
\node at (-0.15,3) {\large {\bf $\mathbf b$}};
%stamp
\node at (0.5,-2) {\Large {\bf $\mathbf (a)$}};
%diagonals
\node at (-0.4,-0.8) {\large  {\bf $m_1$}};
\node at (0.6,1.2) {\large  {\bf $m_2$}};
\node at (1.6,3.2) {\large  {\bf $m_1$}};
%horizontals
\node at (0.5,-0.2) {\large  {\bf $v_1$}};
\node at (1.5,1.8) {\large  {\bf $v_2$}};
\node at (-0.5,0.8) {\large  {\bf $v_2$}};
\node at (0.5,2.8) {\large  {\bf $v_1$}};
%verticals
\node at (0.25,0.5) {\large  {\bf $h_1$}};
\node at (1.25,2.5) {\large  {\bf $h_2$}};
%%%%%%%%%%%%%%%%%%%%%%%
%%%%%%%%%%%%%%%%%%%%%%%
\begin{scope}[xshift=8cm,yshift=2.5cm]
\draw[ultra thick] (-1,0) -- (0,0) -- (0,-1) -- (1,-1) -- (1,-2) -- (2,-2);
%diagonals
\draw[ultra thick] (0,0) -- (0.7,0.7);
\draw[ultra thick] (1,-1) -- (1.7,-0.3);
\draw[ultra thick] (0,-1) -- (-0.7,-1.7);
\draw[ultra thick] (1,-2) -- (0.3,-2.7);
%ends
\node at (-1.2,0) {\large {\bf $\mathbf 1$}};
\node at (2.2,-2) {\large {\bf $\mathbf 1$}};
\node at (0.85,0.85) {\large {$\mathbf b$}};
\node at (1.85,-0.15) {\large {$\mathbf a$}};
\node at (-0.85,-1.5) {\large {$\mathbf a$}};
\node at (0.15,-2.85) {\large {$\mathbf b$}};
%lables hotizontal
\node at (-0.5,0.25) {\large  {\bf $m_1$}};
\node at (0.5,-0.775) {\large  {\bf $m_2$}};
\node at (1.5,-1.775) {\large  {\bf $m_1$}};
%lables vertical
\node at (-0.25,-0.5) {\large  {\bf $h_2$}};
\node at (0.75,-1.5) {\large  {\bf $h_1$}};
%lables diagonal
\node at (0.6,0.2) {\large  {\bf $v_1$}};
\node at (1.6,-0.8) {\large  {\bf $v_2$}};
\node at (-0.2,-1.65) {\large  {\bf $v_2$}};
\node at (0.75,-2.65) {\large  {\bf $v_1$}};
%hexagons
\node[red] at (-0.2,0.6) {\large  {\bf $S^{\prime\prime\prime}_1$}};
\node[red] at (0.75,-0.3) {\large  {\bf $S^{\prime\prime\prime}_2$}};
\node[red] at (1.75,-1.3) {\large  {\bf $S^{\prime\prime\prime}_1$}};
\node[red] at (-0.75,-0.6) {\large  {\bf $S^{\prime\prime\prime}_2$}};
\node[red] at (0.3,-2.1) {\large  {\bf $S^{\prime\prime\prime}_1$}};
\node[red] at (1.2,-2.35) {\large  {\bf $S^{\prime\prime\prime}_2$}};
%roots
\draw[ultra thick,blue,<->] (1.1,1) -- (2,0.1);
\node[blue,rotate=315] at (1.8,0.7) {{\large {\bf {$\widehat{a}^{\prime\prime\prime}_1$}}}};
\draw[ultra thick,blue,<->] (2.1,0) -- (3,-0.9);
\node[blue,rotate=315] at (2.8,-0.3) {{\large {\bf {$\widehat{a}^{\prime\prime\prime}_2$}}}};
%definition S
\draw[dashed] (1,-1) -- (-0.7,-2.7);
\draw[dashed] (-0.7,-1.7) -- (-1.2,-2.2);
\draw[ultra thick,blue,<->] (-1.2,-2.2) -- (-0.7,-2.7);
\node[blue] at (-1.2,-2.65) {{\large {\bf {$S^{\prime\prime\prime}$}}}};
%parameter R
\draw[dashed] (-0.7,-1.7) -- (-1.5,-1.7);
\draw[dashed] (0,-1) -- (-1.5,-1);
\draw[ultra thick,blue,<->] (-1.5,-1) -- (-1.5,-1.7);
\node[blue,rotate=90] at (-1.8,-1.25) {{\large {\bf{$R^{\prime\prime\prime}-S^{\prime\prime\prime}$}}}};
\end{scope}
%stamp
\node at (8.5,-2) {\Large {\bf $\mathbf (b)$}};
\end{tikzpicture}}}
\caption{\sl (a) web diagram obtained from \figref{Fig:21web2}(b) by cutting the curve labelled $v_1$ and re-gluing along $m_2$. (b) alternative presentation of the same diagram.}
\label{Fig:21web3}
\end{center}
\end{figure}

\begin{figure}
\begin{center}
\scalebox{0.53}{\parbox{28.3cm}{\begin{tikzpicture}[scale = 1.50]
%%%%%%%%%%%%%%%%%%%%%%%%%%%%%
%central graph
%%%%%%%%%%%%%%%%%%%%%%%%%%%%%
\draw (-2.05,1.15) -- (3.1,1.15) -- (3.1,-4.5) -- (-2.05,-4.5) -- (-2.05,1.15);
\draw[ultra thick] (-1,0) -- (0,0) -- (0,-1) -- (1,-1) -- (1,-2) -- (2,-2);
%diagonals
\draw[ultra thick] (0,0) -- (0.7,0.7);
\draw[ultra thick] (1,-1) -- (1.7,-0.3);
\draw[ultra thick] (0,-1) -- (-0.7,-1.7);
\draw[ultra thick] (1,-2) -- (0.3,-2.7);
%ends
\node at (-1.2,0) {\large {\bf $\mathbf a$}};
\node at (2.2,-2) {\large {\bf $\mathbf a$}};
\node at (0.85,0.85) {\large {$\mathbf 1$}};
\node at (1.85,-0.15) {\large {$\mathbf 2$}};
\node at (-0.85,-1.5) {\large {$\mathbf 1$}};
\node at (0.15,-2.85) {\large {$\mathbf 2$}};
%lables hotizontal
\node at (-0.5,0.25) {\large  {\bf $h_1$}};
\node at (0.5,-0.775) {\large  {\bf $h_2$}};
\node at (1.5,-1.775) {\large  {\bf $h_1$}};
%lables vertical
\node at (-0.2,-0.5) {\large  {\bf $v_1$}};
\node at (0.8,-1.5) {\large  {\bf $v_2$}};
%lables diagonal
\node at (0.65,0.2) {\large  {\bf $m_1$}};
\node at (1.65,-0.8) {\large  {\bf $m_2$}};
\node at (-0.2,-1.65) {\large  {\bf $m_1$}};
\node at (0.8,-2.65) {\large  {\bf $m_2$}};
%hexagons
%\node[red] at (-0.2,0.6) {\large  {\bf $S_2$}};
%\node[red] at (0.75,-0.3) {\large  {\bf $S_1$}};
%\node[red] at (1.75,-1.3) {\large  {\bf $S_2$}};
%\node[red] at (-0.75,-0.8) {\large  {\bf $S_2$}};
%\node[red] at (0.3,-1.7) {\large  {\bf $S_1$}};
%\node[red] at (1.2,-2.35) {\large  {\bf $S_2$}};
%roots
\draw[ultra thick,blue,<->] (1.05,0.95) -- (1.95,0.05);
\node[blue,rotate=315] at (1.7,0.6) {{\large {\bf {$\widehat{a}_1$}}}};
\draw[ultra thick,blue,<->] (2.05,-0.05) -- (2.95,-0.95);
\node[blue,rotate=315] at (2.7,-0.4) {{\large {\bf {$\widehat{a}_2$}}}};
%definition S
\draw[dashed] (2,-2) -- (0.35,-3.65);
\draw[dashed] (-0.7,-1.7) -- (-1.2,-2.2);
\draw[ultra thick,blue,<->] (-1.2,-2.2) -- (0.35,-3.65);
\node[blue] at (-0.65,-3) {{\large {\bf {$S$}}}};
%parameter R
\draw[dashed] (-0.7,-1.7) -- (-1.5,-1.7);
\draw[dashed] (1,-2) -- (-1.5,-2);
\draw[ultra thick,blue,->] (-1.5,-1.3) -- (-1.5,-1.7);
\draw[ultra thick,blue,->] (-1.5,-2.4) -- (-1.5,-2);
\node[blue,rotate=90] at (-1.8,-1.85) {{\large {\bf{$R-2S$}}}};
%label
\node at (0.5,-4.15) {{\large web diagram in \figref{Fig:21web0}}};
%center
%\draw[dashed] (-2.05,1.15) -- (3.1,-4.5);
%\draw[dashed] (3.1,1.15) -- (-2.05,-4.5);
%\node[red] at (0.5,-1.65) {\Huge $\bullet$};
%%%%%%%%%%%%%%%%%%%%%%%%%%%%%
%top graph
%%%%%%%%%%%%%%%%%%%%%%%%%%%%%
\begin{scope}[yshift=8cm]
\draw (-2.05,1.15) -- (3.1,1.15) -- (3.1,-4.5) -- (-2.05,-4.5) -- (-2.05,1.15);
\draw[ultra thick] (-1,0) -- (0,0) -- (0,-1) -- (1,-1) -- (1,-2) -- (2,-2);
%diagonals
\draw[ultra thick] (0,0) -- (0.7,0.7);
\draw[ultra thick] (1,-1) -- (1.7,-0.3);
\draw[ultra thick] (0,-1) -- (-0.7,-1.7);
\draw[ultra thick] (1,-2) -- (0.3,-2.7);
%ends
\node at (-1.2,0) {\large {\bf $\mathbf a$}};
\node at (2.2,-2) {\large {\bf $\mathbf a$}};
\node at (0.85,0.85) {\large {$\mathbf 1$}};
\node at (1.85,-0.15) {\large {$\mathbf 2$}};
\node at (-0.85,-1.5) {\large {$\mathbf 1$}};
\node at (0.15,-2.85) {\large {$\mathbf 2$}};
%lables hotizontal
\node at (-0.5,0.25) {\large  {\bf $h_1$}};
\node at (0.5,-0.775) {\large  {\bf $h_2$}};
\node at (1.5,-1.775) {\large  {\bf $h_1$}};
%lables vertical
\node at (-0.25,-0.5) {\large  {\bf $m_2$}};
\node at (0.75,-1.5) {\large  {\bf $m_1$}};
%lables diagonal
\node at (0.6,0.25) {\large  {\bf $v_2$}};
\node at (1.6,-0.75) {\large  {\bf $v_1$}};
\node at (-0.2,-1.6) {\large  {\bf $v_2$}};
\node at (0.75,-2.6) {\large  {\bf $v_1$}};
%hexagons
%\node[red] at (-0.2,0.6) {\large  {\bf $S'_2$}};
%\node[red] at (0.75,-0.3) {\large  {\bf $S'_1$}};
%\node[red] at (1.75,-1.3) {\large  {\bf $S'_2$}};
%\node[red] at (-0.75,-0.8) {\large  {\bf $S'_2$}};
%\node[red] at (0.3,-1.7) {\large  {\bf $S'_1$}};
%\node[red] at (1.2,-2.35) {\large  {\bf $S'_2$}};
%roots
\draw[ultra thick,blue,<->] (1.1,1) -- (2,0.1);
\node[blue,rotate=315] at (1.8,0.7) {{\large {\bf {$\widehat{a}'_1$}}}};
\draw[ultra thick,blue,<->] (2.1,0) -- (3,-0.9);
\node[blue,rotate=315] at (2.8,-0.3) {{\large {\bf {$\widehat{a}'_2$}}}};
%definition S
\draw[dashed] (2,-2) -- (0.35,-3.65);
\draw[dashed] (-0.7,-1.7) -- (-1.2,-2.2);
\draw[ultra thick,blue,<->] (-1.2,-2.2) -- (0.35,-3.65);
\node[blue] at (-0.7,-3.05) {{\large {\bf {$S'$}}}};
%parameter R
\draw[dashed] (-0.7,-1.7) -- (-1.5,-1.7);
\draw[dashed] (1,-2) -- (-1.5,-2);
\draw[ultra thick,blue,->] (-1.5,-1.3) -- (-1.5,-1.7);
\draw[ultra thick,blue,->] (-1.5,-2.4) -- (-1.5,-2);
\node[blue,rotate=90] at (-1.8,-1.85) {{\large {\bf{$R'-2S'$}}}};
%label
\node at (0.5,-4.15) {{\large web diagram in \figref{Fig:21web1}(b)}};
%center
%\draw[dashed] (-2.05,1.15) -- (3.1,-4.5);
%\draw[dashed] (3.1,1.15) -- (-2.05,-4.5);
%\node[red] at (0.5,-1.65) {\Huge $\bullet$};
\end{scope}
%%%%%%%%%%%%%%%%%%%%%%%%%%%%%
%graph left bottom
%%%%%%%%%%%%%%%%%%%%%%%%%%%%%
\begin{scope}[xshift=-6.85cm,yshift=-6cm]
\draw (-2.05,1.15) -- (3.1,1.15) -- (3.1,-3.85) -- (-2.05,-3.85) -- (-2.05,1.15);
\draw[ultra thick] (-1,0) -- (0,0) -- (0,-1) -- (1,-1) -- (1,-2) -- (2,-2);
%diagonals
\draw[ultra thick] (0,0) -- (0.7,0.7);
\draw[ultra thick] (1,-1) -- (1.7,-0.3);
\draw[ultra thick] (0,-1) -- (-0.7,-1.7);
\draw[ultra thick] (1,-2) -- (0.3,-2.7);
%ends
\node at (-1.2,0) {\large {\bf $\mathbf a$}};
\node at (2.2,-2) {\large {\bf $\mathbf a$}};
\node at (0.85,0.85) {\large {$\mathbf 1$}};
\node at (1.85,-0.15) {\large {$\mathbf{2}$}};
\node at (-0.85,-1.5) {\large {$\mathbf 2$}};
\node at (0.15,-2.85) {\large {$\mathbf 1$}};
%lables hotizontal
\node at (-0.5,0.25) {\large  {\bf $v_2$}};
\node at (0.5,-0.775) {\large  {\bf $v_1$}};
\node at (1.5,-1.775) {\large  {\bf $v_2$}};
%lables vertical
\node at (-0.25,-0.5) {\large  {\bf $h_1$}};
\node at (0.75,-1.5) {\large  {\bf $h_2$}};
%lables diagonal
\node at (0.6,0.2) {\large  {\bf $m_2$}};
\node at (1.6,-0.8) {\large  {\bf $m_1$}};
\node at (-0.2,-1.65) {\large  {\bf $m_1$}};
\node at (0.75,-2.65) {\large  {\bf $m_2$}};
%hexagons
%\node[red] at (-0.2,0.6) {\large  {\bf $S''_2$}};
%\node[red] at (0.75,-0.3) {\large  {\bf $S''_1$}};
%\node[red] at (1.75,-1.3) {\large  {\bf $S''_2$}};
%\node[red] at (-0.75,-0.6) {\large  {\bf $S''_2$}};
%\node[red] at (0.3,-2.1) {\large  {\bf $S''_1$}};
%\node[red] at (1.2,-2.35) {\large  {\bf $S''_2$}};
%roots
\draw[ultra thick,blue,<->] (1.1,1) -- (2,0.1);
\node[blue,rotate=315] at (1.8,0.7) {{\large {\bf {$\widehat{a}''_2$}}}};
\draw[ultra thick,blue,<->] (2.1,0) -- (3,-0.9);
\node[blue,rotate=315] at (2.8,-0.3) {{\large {\bf {$\widehat{a}''_1$}}}};
%definition S
\draw[dashed] (1,-1) -- (-0.7,-2.7);
\draw[dashed] (-0.7,-1.7) -- (-1.2,-2.2);
\draw[ultra thick,blue,<->] (-1.2,-2.2) -- (-0.7,-2.7);
\node[blue] at (-1.2,-2.65) {{\large {\bf {$S''$}}}};
%parameter R
\draw[dashed] (-0.7,-1.7) -- (-1.5,-1.7);
\draw[dashed] (0,-1) -- (-1.5,-1);
\draw[ultra thick,blue,<->] (-1.5,-1) -- (-1.5,-1.7);
\node[blue,rotate=90] at (-1.8,-1.25) {{\large {\bf{$R''-S''$}}}};
%label
\node at (0.5,-3.4) {{\large web diagram in \figref{Fig:21web2}(b)}};
%center
%\draw[dashed] (-2.05,1.15) -- (3.1,-4.5);
%\draw[dashed] (3.1,1.15) -- (-2.05,-4.5);
%\node[red] at (0.5,-1.65) {\Huge $\bullet$};
\end{scope}
%%%%%%%%%%%%%%%%%%%%%%%%%%%%%
%graph right bottom
%%%%%%%%%%%%%%%%%%%%%%%%%%%%%
\begin{scope}[xshift=6.85cm,yshift=-6cm]
\draw (-2.05,1.15) -- (3.1,1.15) -- (3.1,-3.85) -- (-2.05,-3.85) -- (-2.05,1.15);
\draw[ultra thick] (-1,0) -- (0,0) -- (0,-1) -- (1,-1) -- (1,-2) -- (2,-2);
%diagonals
\draw[ultra thick] (0,0) -- (0.7,0.7);
\draw[ultra thick] (1,-1) -- (1.7,-0.3);
\draw[ultra thick] (0,-1) -- (-0.7,-1.7);
\draw[ultra thick] (1,-2) -- (0.3,-2.7);
%ends
\node at (-1.2,0) {\large {\bf $\mathbf a$}};
\node at (2.2,-2) {\large {\bf $\mathbf a$}};
\node at (0.85,0.85) {\large {$\mathbf 1$}};
\node at (1.85,-0.15) {\large {$\mathbf 2$}};
\node at (-0.85,-1.5) {\large {$\mathbf 2$}};
\node at (0.15,-2.85) {\large {$\mathbf 1$}};
%lables hotizontal
\node at (-0.5,0.25) {\large  {\bf $m_1$}};
\node at (0.5,-0.775) {\large  {\bf $m_2$}};
\node at (1.5,-1.775) {\large  {\bf $m_1$}};
%lables vertical
\node at (-0.25,-0.5) {\large  {\bf $h_2$}};
\node at (0.75,-1.5) {\large  {\bf $h_1$}};
%lables diagonal
\node at (0.6,0.2) {\large  {\bf $v_1$}};
\node at (1.6,-0.8) {\large  {\bf $v_2$}};
\node at (-0.2,-1.65) {\large  {\bf $v_2$}};
\node at (0.75,-2.65) {\large  {\bf $v_1$}};
%hexagons
%\node[red] at (-0.2,0.6) {\large  {\bf $S^{\prime\prime\prime}_2$}};
%\node[red] at (0.75,-0.3) {\large  {\bf $S^{\prime\prime\prime}_1$}};
%\node[red] at (1.75,-1.3) {\large  {\bf $S^{\prime\prime\prime}_2$}};
%\node[red] at (-0.75,-0.6) {\large  {\bf $S^{\prime\prime\prime}_2$}};
%\node[red] at (0.3,-2.1) {\large  {\bf $S^{\prime\prime\prime}_1$}};
%\node[red] at (1.2,-2.35) {\large  {\bf $S^{\prime\prime\prime}_2$}};
%roots
\draw[ultra thick,blue,<->] (1.1,1) -- (2,0.1);
\node[blue,rotate=315] at (1.8,0.7) {{\large {\bf {$\widehat{a}^{\prime\prime\prime}_1$}}}};
\draw[ultra thick,blue,<->] (2.1,0) -- (3,-0.9);
\node[blue,rotate=315] at (2.8,-0.3) {{\large {\bf {$\widehat{a}^{\prime\prime\prime}_2$}}}};
%definition S
\draw[dashed] (1,-1) -- (-0.7,-2.7);
\draw[dashed] (-0.7,-1.7) -- (-1.2,-2.2);
\draw[ultra thick,blue,<->] (-1.2,-2.2) -- (-0.7,-2.7);
\node[blue] at (-1.2,-2.65) {{\large {\bf {$S^{\prime\prime\prime}$}}}};
%parameter R
\draw[dashed] (-0.7,-1.7) -- (-1.5,-1.7);
\draw[dashed] (0,-1) -- (-1.5,-1);
\draw[ultra thick,blue,<->] (-1.5,-1) -- (-1.5,-1.7);
\node[blue,rotate=90] at (-1.8,-1.25) {{\large {\bf{$R^{\prime\prime\prime}-S^{\prime\prime\prime}$}}}};
%label
\node at (0.5,-3.4) {{\large web diagram in \figref{Fig:21web3}(b)}};
%center
%\draw[dashed] (-2.05,1.15) -- (3.1,-4.5);
%\draw[dashed] (3.1,1.15) -- (-2.05,-4.5);
%\node[red] at (0.5,-1.65) {\Huge $\bullet$};
\end{scope}
%%%%%%%%%%%%%%%%%%%%%%%%%%%%%
%arrows and elements
%\draw[ultra thick,blue,->] (0.5,-1.65) -- (0.5,6.35);
%\draw[ultra thick,blue,->] (0.5,-1.65) -- (-6.4282, -5.65);
%\draw[ultra thick,blue,->] (0.5,-1.65) -- (7.4282, -5.65);
\draw [line width=1.5mm, red,->] (0.7,1.25) -- (0.7,3.45);
\draw [line width=1.5mm, red,->] (0.3,3.45) -- (0.3,1.25);
\node[red] at (1.2,2.25) {\LARGE $G_1$}; 
\node[red] at (-0.15,2.25) {\LARGE $G_1$}; 
\draw [line width=1.5mm, red,->] (-2.2,-1.75) -- (-5,-4.7);
\draw [line width=1.5mm, red,->] (-5.6,-4.7) -- (-2.2,-1.25);
\node[red] at (-3.25,-3.5) {\LARGE $G_2$}; 
\node[red] at (-4.4,-2.7) {\LARGE $G_2$}; 
\draw [line width=1.5mm, red,->] (3.2,-1.75) -- (6,-4.7);
\draw [line width=1.5mm, red,->] (6.6,-4.7) -- (3.2,-1.25);
\node[red] at (4.25,-3.5) {\LARGE $G_3$}; 
\node[red] at (5.4,-2.7) {\LARGE $G_3$}; 
%
%\draw [line width=1.5mm, red,->,dashed] (-3.5,-7.1) -- (4.6,-7.1);
%\draw [line width=1.5mm, red,->,dashed] (4.6,-7.5) -- (-3.5,-7.5);
%\node[red] at (0.5,-6.7) {\LARGE $G_1$}; 
%\node[red] at (0.5,-7.95) {\LARGE $G_1$}; 
%
%\draw [line width=1.5mm, red,->,dashed] (-7,-4.7) to [out=90,in=255] (-6.3,1) to [out=75,in=180] (-2.15,6);
%\draw [line width=1.5mm, red,<-,dashed] (-7.4,-4.7) to [out=90,in=255] (-6.6,1.3) to [out=75,in=180] (-2.15,6.4);
%\node[red] at (-5.8,1) {\LARGE $G_3$}; 
%\node[red] at (-7.2,1) {\LARGE $G_3$}; 
%
%\draw [line width=1.5mm, red,->,dashed] (8,-4.7) to [out=90,in=285] (7.3,1) to [out=105,in=0] (3.15,6);
%\draw [line width=1.5mm, red,<-,dashed] (8.4,-4.7) to [out=90,in=285] (7.6,1.3) to [out=105,in=0] (3.15,6.4);
%\node[red] at (6.8,1) {\LARGE $G_2$}; 
%\node[red] at (8.2,1) {\LARGE $G_2$}; 
%%%%%%%%%%%%%%%%%%%%%%%%%%%%%
\end{tikzpicture}}}
\caption{\sl Presentations of web diagrams related to $X_{2,1}$. The transformations $G_{1,2,3}$ act on the basis of independent K\"ahler parameters $(\widehat{a}_1,\widehat{a}_2,S,R)$. The organisation of web diagrams and transformations is reminiscent of the cycle graph of $\text{Dih}_2$.}
\label{Fig:Overview21Webs}
\end{center}
\end{figure}
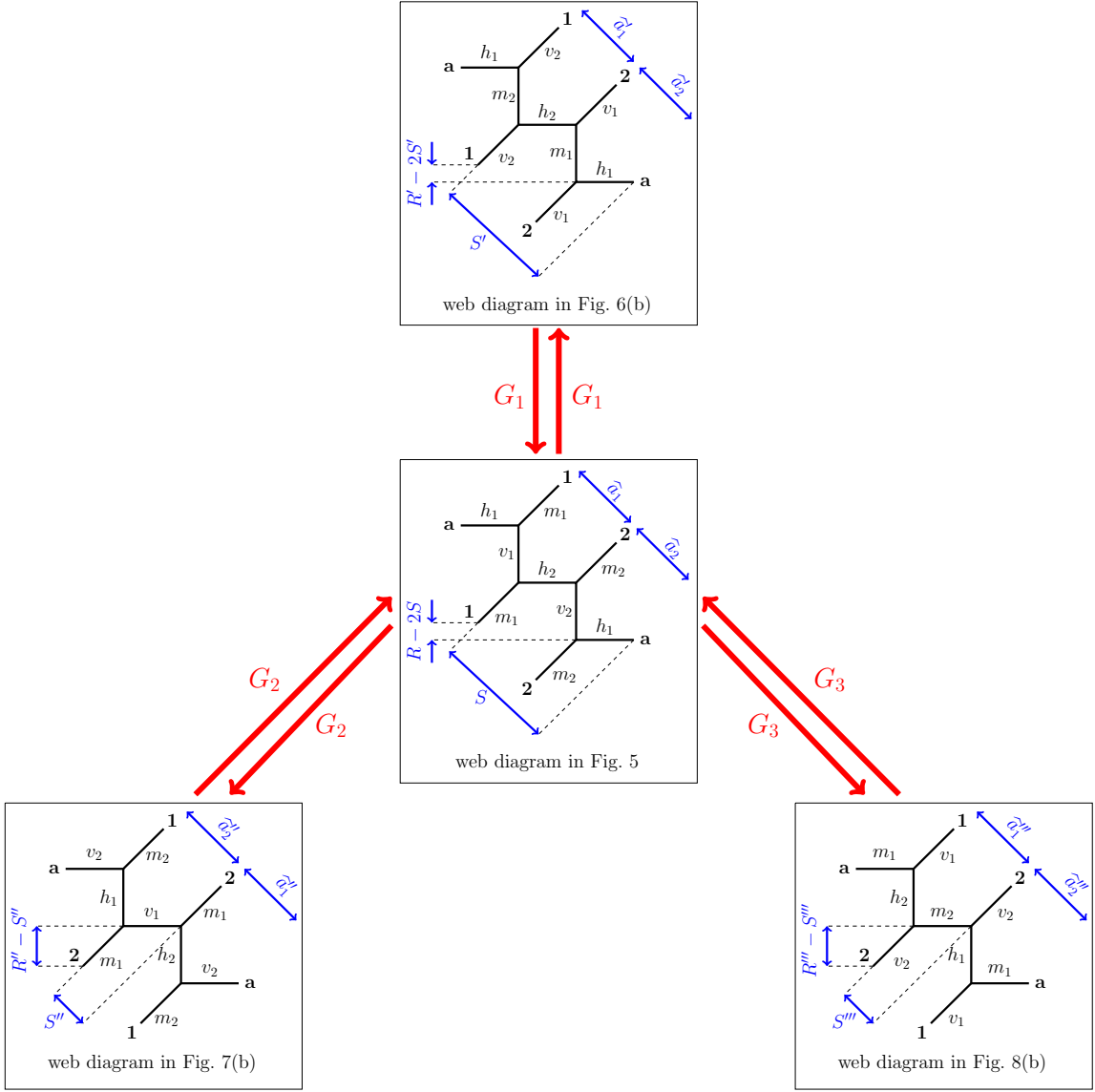

\noindent
We can obtain another symmetry transformation by cutting the diagram \figref{Fig:21web0} along the line labelled $v_2$ and re-gluing it along the line $h_1$ to obtain \figref{Fig:21web2}(a). Mirroring the latter, it can also be presented in the form of \figref{Fig:21web2}(b) which takes the form of a web with the shift $\delta=1$.
The latter can be parametrised by $(\widehat{a}''_1,\widehat{a}''_2,S'',R'')$
\begin{align}
&\widehat{a}''_1=h_2+v_2\,,&&\widehat{a}''_2=h_1+v_1\,,&&S''=v_1\,,&&R''-S''=m_1\,,
\end{align}
which allows to uniquely express all areas $(h_{1,2},v_{1,2},m_{1,2})$
\begin{align}
&h_1=\widehat{a}''_2-S''\,,&&h_2=\widehat{a}''_1=S''\,,&&v_1=v_2=S''\,,&&m_1=m_2=R''-S''\,.\label{Base1web22}
\end{align}
Comparing (\ref{Base1web22}) with (\ref{Base2web21}) gives rise to a transformation between $(\widehat{a}_1,\widehat{a}_2,S,R)$ and $(\widehat{a}''_1,\widehat{a}''_2,S'',R'')$
\begin{align}
&\left(\begin{array}{c}\widehat{a}_1\\ \widehat{a}_2 \\ S \\ R\end{array}\right)=G_2\cdot \left(\begin{array}{c}\widehat{a}''_1\\ \widehat{a}''_2 \\ S'' \\ R''\end{array}\right)\,,&&\text{where} &&G_2=\left(
\begin{array}{cccc}
 1 & 0 & 0 & 0 \\
 0 & 1 & 0 & 0 \\
 1 & 1 & -1 & 0 \\
 2 & 2 & -4 & 1 \\
\end{array}
\right)\,,&& \text{with} &&\begin{array}{l}\text{det}\, G_2=-1\,, \\ G_2\cdot G_2=1\!\!1_{4\times 4}\,.\end{array}\label{Gen21G2}
\end{align}
Finally, cutting the diagram \figref{Fig:21web2}(b) along the curve labelled $v_1$ and re-gluing it along the line $m_2$ yields the diagram \figref{Fig:21web3}(a), which (after mirroring and performing an $SL(2,\mathbb{Z})$-transformation) can also be presented in the form \figref{Fig:21web3}(b).
This diagram is parametrised by $(\widehat{a}^{\prime\prime\prime}_1,\widehat{a}^{\prime\prime\prime}_2,S^{\prime\prime\prime},R^{\prime\prime\prime})$
\begin{align}
&\widehat{a}^{\prime\prime\prime}_1=h_1+m_1\,,&&\widehat{a}^{\prime\prime\prime}_2=h_2+m_2\,,&&S^{\prime\prime\prime}=m_2\,,&&R^{\prime\prime\prime}-S^{\prime\prime\prime}=v_2\,,
\end{align}
which provide a parametrisation of all the areas
\begin{align}
&h_1=\widehat{a}^{\prime\prime\prime}_1-S^{\prime\prime\prime}\,,&&h_2=\widehat{a}^{\prime\prime\prime}_2-S^{\prime\prime\prime}\,,&&v_1=v_2=R^{\prime\prime\prime}-S^{\prime\prime\prime}\,,&&m_1=m_2=S^{\prime\prime\prime}\,.\label{Base1web23}
\end{align}
Comparing (\ref{Base1web23}) with (\ref{Base1web21}) provides a linear transformation between the parameters $(\widehat{a}_1,\widehat{a}_2,S,R)$ and $(\widehat{a}^{\prime\prime\prime}_1,\widehat{a}^{\prime\prime\prime}_2,S^{\prime\prime\prime},R^{\prime\prime\prime})$
\begin{align}
&\left(\begin{array}{c}\widehat{a}_1\\ \widehat{a}_2 \\ S \\ R\end{array}\right)=G_3\cdot \left(\begin{array}{c}\widehat{a}^{\prime\prime\prime}_1\\ \widehat{a}^{\prime\prime\prime}_2 \\ S^{\prime\prime\prime} \\ R^{\prime\prime\prime}\end{array}\right)\,,&&\text{with} &&G_3=\left(
\begin{array}{cccc}
 1 & 0 & -2 & 1 \\
 0 & 1 & -2 & 1 \\
 1 & 1 & -3 & 1 \\
 2 & 2 & -4 & 1 \\
\end{array}
\right)\,,&& \text{and} &&\begin{array}{l}\text{det}\, G_3=-1\,, \\ G_3\cdot G_3=1\!\!1_{4\times 4}\,.\end{array}\label{Gen21G3}
\end{align}
The matrices $G_{1,2,3}$ together with the identity matrix $E=1\!\!1_{4\times 4}$ form a discrete group of order 4, whose multiplication table is given by
\begin{align}
\begin{array}{c|cccc}
 & E & G_1 & G_2 & G_3 \\\hline
E & E & G_1 & G_2 & G_3 \\
G_1 & G_1 & E & G_3 & G_2 \\
G_2 & G_2 & G_3 & E & G_1 \\
G_3 & G_3 & G_2 & G_1 & E \\
\end{array}
\end{align}
The latter is identical to the multiplication table of $\text{Dih}_2$, \emph{i.e.} the dihedral group of order $4$ (which is isomorphic to the Klein four-group). We therefore have\footnote{For further reference, we remark that $\mathbb{G}(2)$ can also be presented as the group freely generated by $G_{1,2}$, \emph{i.e.} $\mathbb{G}(2)\cong\langle \{G_1,G_2\}\rangle$, where $G_1^2=1\!\!1_{4\times 4}=G_2^2$ and $(G_1\cdot G_2)^2=1\!\!1_{4\times 4}$.}
\begin{align}
\mathbb{G}(2)\cong\{E,G_1,G_2,G_3\}\cong \text{Dih}_2\,.\label{DefDih2}
\end{align}
An overview over $G_{1,2,3}$ and their relation to different presentations of the web diagram \figref{Fig:21web0} is given in \figref{Fig:Overview21Webs} (which corresponds to the cycle graph of $\text{Dih}_2$). We remark that all other presentations of the web (including webs related by a transformation $\mathcal{F}$ (appendix~\ref{App:FlopTrafo})) only give rise to coordinate transformations that differ from $\{E,G_1,G_2,G_3\}$ by the action of 
\begin{align}
R=\left(
\begin{array}{cccc}
 0 & 1 & 0 & 0 \\
 1 & 0 & 0 & 0 \\
 0 & 0 & 1 & 0 \\
 0 & 0 & 0 & 1 \\
\end{array}
\right)\,,\label{DefS2R}
\end{align}
which exchanges $\widehat{a}_1\longleftrightarrow \widehat{a}_2$ and commutes with $G_{1,2,3}$. Since $R$ generates the group $S_2$, we can define $\widetilde{\mathbb{G}}(2)=\mathbb{G}(2)\times S_2$ as a non-trivial symmetry group of $\mathcal{F}_{2,1}$.

%%%%%%%%%%%%%%%%%%%%%%%%%%%%%%%%%%%
\subsection{Invariance of the Non-perturbative Free Energy}
It was shown in \cite{Bastian:2017ing} that the web diagrams \figref{Fig:21web0}, \figref{Fig:21web1}(b), \figref{Fig:21web2}(b) and \figref{Fig:21web3}(b) give rise to the same partition function, the linear transformations $G_{1,2,3}$ in eqs.~(\ref{Gen21G1}), (\ref{Gen21G2}) and (\ref{Gen21G3}) correspond to symmetries of the free energy $\mathcal{F}_{2,1}(\widehat{a}_{1,2},S,R;\epsilon_1,\epsilon_2)$, as defined in (\ref{DefFreeEnergyGen}). In this section we provide evidence for this symmetry by considering the expansion 
\begin{align}
\mathcal{F}_{2,1}(\widehat{a}_1,\widehat{a}_2,S,R;\epsilon_1,\epsilon_2)=\sum_{n=0}^\infty \sum_{i_1,i_2=0}^\infty \sum_{k\in\mathbb{Z}}f_{i_1,i_2,k,n}(\epsilon_1,\epsilon_2)\,\widehat{Q}_1^{i_1}\,\widehat{Q}_2^{i_2}\,Q_S^{k}\,Q_{R}^n\,,
\end{align}
with $\widehat{Q}_i=e^{-\widehat{a}_i}$ (for $i=1,2$), $Q_S=e^{-S}$ and $Q_R=e^{-R}$. As explained in section~\ref{Sect:GeneralStrategy}, we have
\begin{align}
&f_{i_1,i_2,k,n}(\epsilon_1,\epsilon_2)=f_{i'_1,i'_2,k',n'}(\epsilon_1,\epsilon_2)&&\text{for} && (i'_1,i'_2,k', n')^T=G^T_\ell\cdot (i_1,i_2,k, n)^T && \forall \ell=1,2,3\,.
%\left(\begin{array}{c}i'_1 \\ i'_2 \\ k' \\ n'\end{array}\right)=G^T_\ell\cdot \left(\begin{array}{c}i_1 \\ i_2 \\ k \\ n\end{array}\right)&&\forall \ell=1,2,3\,.
\end{align}
Below we tabulate coefficients $f_{i_1,i_2,k,n}$ with $i_1+i_2\leq 3$ for $n=1$ and $i_1+i_2\leq 2$ for $n=2$ that are related by $G_{1,2,3}$: Table~\ref{Tab:SymTrafo21G1} shows relations for $G_1$, Table~\ref{Tab:SymTrafo21G2} those for $G_2$ and Table~\ref{Tab:SymTrafo21G3} for $G_3$.
\begin{table}[hbp]
\begin{center}
\begin{tabular}{|c|c|c|}\hline
&&\\[-10pt]
$(i_1,i_2,k,n)$ & $(i'_1,i'_2,k',n')$ & $f_{i_1,i_2,k,n}(\epsilon_{1,2})=f_{i'_1,i'_2,k',n'}(\epsilon_{1,2})$\\[4pt]\hline\hline
&&\\[-11pt]
$(0,1,0,1)$ & $(0,1,-2,2)$ & $\frac{(q+t) (q (1+t (q+t+2))+t)}{(q-1) q (t-1) t}$\\[5pt]\hline
&&\\[-11pt]
$(0,2,-1,1)$ & $(0,2,-3,2)$ & $-\frac{(q+1) (t+1) (q+t) \left(q^2+t^2\right)}{(q-1) q^{3/2} (t-1) t^{3/2}}$\\[5pt]\hline
&&\\[-11pt]
$(1,0,0,1)$ & $(1,0,-2,2)$ & $\frac{(q+t) (q (t (q+t+2)+1)+t)}{(q-1) q (t-1) t}$\\[5pt]\hline
&&\\[-11pt]
$(1,1,-1,1)$ & $(1,1,-3,2)$ & $-\frac{2 \left(q^2 (t (t+3)+1)+q (t (3 t+7)+3)+t (t+3)+1\right)}{(q-1) \sqrt{q} (t-1) \sqrt{t}}$\\[5pt]\hline
&&\\[-11pt]
$(2,0,-1,1)$ & $(2,0,-3,2)$ & $-\frac{(q+1) (t+1) (q+t) \left(q^2+t^2\right)}{(q-1) q^{3/2} (t-1) t^{3/2}}$\\[5pt]\hline
&&\\[-11pt]
$(0,1,-2,2)$ & $(0,1,0,1)$ & $\frac{(q+t) (q (t (q+t+2)+1)+t)}{(q-1) q (t-1) t}$\\[5pt]\hline
&&\\[-11pt]
$(0,2,-3,2)$ & $(0,2,-1,1)$ & $-\frac{(q+1) (t+1) (q+t) \left(q^2+t^2\right)}{(q-1) q^{3/2} (t-1) t^{3/2}}$\\[5pt]\hline
&&\\[-11pt]
$(1,0,-2,2)$ & $(1,0,0,1)$ & $\frac{(q+t) (q (t (q+t+2)+1)+t)}{(q-1) q (t-1) t}$\\[5pt]\hline
&&\\[-11pt]
$(1,1,-3,2)$ & $(1,1,-1,1)$ & $-\frac{2 \left(q^2 (t (t+3)+1)+q (t (3 t+7)+3)+t (t+3)+1\right)}{(q-1) \sqrt{q} (t-1) \sqrt{t}}$\\[5pt]\hline
&&\\[-11pt]
$(2,0,-3,2)$ & $(2,0,-1,1)$ & $-\frac{(q+1) (t+1) (q+t) \left(q^2+t^2\right)}{(q-1) q^{3/2} (t-1) t^{3/2}}$\\[5pt]\hline
\end{tabular}
\end{center}
\caption{\sl Action of $G_1$: the indices are related by $(i'_1,i'_2,k',n')^T=G_1^T\cdot (i_1,i_2,k,n)^T$.}
\label{Tab:SymTrafo21G1}
\end{table}

\begin{table}[hp]
\begin{center}
\begin{tabular}{|c|c|c|}\hline
&&\\[-10pt]
$(i_1,i_2,k,n)$ & $(i'_1,i'_2,k',n')$ & $f_{i_1,i_2,k,n}(\epsilon_{1,2})=f_{i'_1,i'_2,k',n'}(\epsilon_{1,2})$\\[4pt]\hline\hline
&&\\[-11pt]
$(0,0,-1,1)$ & $(1,1,-3,1)$ & $-\frac{2 \sqrt{q} \sqrt{t}}{(q-1) (t-1)}$\\[5pt]\hline
&&\\[-11pt]
$(1,2,-3,1)$ & $(0,1,-1,1)$ & $-\frac{(q+1) (t+1) (q+t)}{(q-1) \sqrt{q} (t-1) \sqrt{t}}$\\[5pt]\hline
&&\\[-11pt]
$(2,1,-3,1)$ & $(1,0,-1,1)$ & $-\frac{(q+1) (t+1) (q+t)}{(q-1) \sqrt{q} (t-1) \sqrt{t}}$\\[5pt]\hline
\end{tabular}
\end{center}
\caption{\sl Action of $G_2$: the indices are related by $(i'_1,i'_2,k',n')^T=G_2\cdot (i_1,i_2,k,n)^T$.}
\label{Tab:SymTrafo21G2}
\end{table}

\begin{table}[hpt]
\begin{center}
\begin{tabular}{|c|c|c|}\hline
&&\\[-10pt]
$(i_1,i_2,k,n)$ & $(i'_1,i'_2,k',n')$ & $f_{i_1,i_2,k,n}(\epsilon_{1,2})=f_{i'_1,i'_2,k',n'}(\epsilon_{1,2})$\\[4pt]\hline\hline
&&\\[-11pt]
$(0,2,-2,1)$ & $(0,2,-2,1)$ & $\frac{(q+t) \left(q^2+t^2\right)}{(q-1) q (t-1) t}$\\[5pt]\hline
&&\\[-11pt]
$(1,1,-2,1)$ & $(1,1,-2,1)$ & $\frac{4 (q+1) (t+1)}{(q-1) (t-1)}$\\[5pt]\hline
&&\\[-11pt]
$(1,2,-3,1)$ & $(0,1,-1,1)$ & $-\frac{(q+1) (t+1) (q+t)}{(q-1) \sqrt{q} (t-1) \sqrt{t}}$\\[5pt]\hline
&&\\[-11pt]
$(2,0,-2,1)$ & $(2,0,-2,1)$ & $\frac{(q+t) \left(q^2+t^2\right)}{(q-1) q (t-1) t}$\\[5pt]\hline
%&&\\[-10pt]
%$(2,1,-3,1)$ & $(0,1,-1,1)$ & $-\frac{(q+1) (t+1) (q+t)}{(q-1) \sqrt{q} (t-1) \sqrt{t}}$\\[6pt]\hline
\end{tabular}
\end{center}
\caption{\sl Action of $G_3$: the indices are related by $(i'_1,i'_2,k',n')^T=G_3\cdot (i_1,i_2,k,n)^T$.}
\label{Tab:SymTrafo21G3}
\end{table}
%%%%%%%%%%%%%%%%%%%%%%%%%%%%%%%%%%%%%%%%%%%%%%%%%
%%%%%%%%%%%%%%%%%%%%%%%%%%%%%%%%%%%%%%%%%%%%%%%%%
\subsection{Modularity at a Particular Point of the Moduli Space}
For the case $N=1$, we showed that the combination of $\mathbb{G}(1)\cong \text{Dih}_3$ with the modular group acting as in (\ref{Sl2rho}) generates the group $Sp(4,\mathbb{Z})$. The case $N=2$ is more complicated. However, in the following we shall show in a particular region of the moduli space that $\mathbb{G}(2)\cong\text{Dih}_2$ in (\ref{DefDih2}) can be understood as a subgroup of $Sp(4,\mathbb{Z})$. This region is characterised by imposing $\widehat{a}_1^{(0)}=\widehat{a}_2^{(0)}=\widehat{a}$,\footnote{This is the same region in the moduli space which was used in \cite{Hohenegger:2016yuv} for a non-trivial check that $\mathcal{Z}_{N,M}=\mathcal{Z}_{N',M'}$ for $NM=N'M'$ and $\text{gcd}(N,M)=\text{gcd}(N',M')$.} which implies $h_1=h_2=h$ (while the consistency conditions (\ref{ConsistencyWeb21}) already impose $v_1=v_2=v$ and $m_1=m_2=m$). This region is also a fixed point of $S_2$ generated by $R$ in (\ref{DefS2R}). The remaining independent parameters can be organised in the period matrix
\begin{align}
&\Omega=\left(\begin{array}{cc}\tau & v \\ v & \rho\end{array}\right)\,,&&\text{with} &&\begin{array}{l}\tau=m+v\,,\\ \rho=h+m\,.\end{array}\label{Def2PeriodMatrix}
\end{align} 
Furthermore, the symmetry transformations $G_1$ in (\ref{Gen21G1}) and $G_2$ in (\ref{Gen21G2}) can be reduced to act on the subspace $(\widehat{a},S,R)$
\begin{align}
&G_1^{(\text{red})}=\left(\begin{array}{ccc}1 & -2 & 1 \\ 0 & -1 & 1 \\ 0 & 0 & 1\end{array}\right)\,,&&\text{and} &&G_2^{(\text{red})}=\left(\begin{array}{ccc}1 & 0 & 0 \\ 2 & -1 & 0 \\ 4 & -4 & 1\end{array}\right)\,,
\end{align}
or on the space $(\tau,\rho,v)$
\begin{align}
&\widetilde{G}_1^{(\text{red})}=D_2^{-1}\cdot G_1^{(\text{red})}\cdot D_2=\left(\begin{array}{ccc}1 & 0 & 0 \\ 1 & 1 & -2 \\ 1 & 0 & -1\end{array}\right)\,,&&\text{with} && D_2=\left(\begin{array}{ccc} 0 & 1 & 0 \\ 0 & 2 & -1 \\ 1 & 4 & -4\end{array}\right)\,, \nonumber\\
&\widetilde{G}_2^{(\text{red})}=D_2^{-1}\cdot G_2^{(\text{red})}\cdot D_2=\left(\begin{array}{ccc}1 & 4 & -4 \\ 0 & 1 & 0 \\ 0 & 2 & -1\end{array}\right)\,.\nonumber
\end{align}
Rewriting the latter as elements of $Sp(4,\mathbb{Z})$ that act like in (\ref{ActionSp4Period}) on the period matrix $\Omega$ in (\ref{Def2PeriodMatrix}), they take the form
\begin{align}
&\widetilde{G}_1^{(\text{red,Sp})}=K\,,&&\text{and} &&\widetilde{G}_2^{(\text{red,Sp})}=HK L^6KH\,,
\end{align}
where $K$, $L$ and $H$ are defined in appendix~\ref{App:Sp4}. This implies that the restriction of $\mathbb{G}(2)$ to the particular region of the K\"ahler moduli space explained above is a subgroup of $Sp(4,\mathbb{Z})$. However, unlike the case $N=1$, we cannot conclude that the group freely generated as $\langle \widetilde{G}_1^{(\text{red,Sp})},\widetilde{G}_2^{(\text{red,Sp})}, S_\rho,T_\rho,S_\tau,T_\tau \rangle$ is isomorphic to $Sp(4,\mathbb{Z})$.
%%%%%%%%%%%%%%%%%%%%%%%%%%%%%%%%%%%
%%%%%%%%%%%%%%%%%%%%%%%%%%%%%%%%%%%
%%%%%%%%%%%%%%%%%%%%%%%%%%%%%%%%%%%
%%%%%%%%%%%%%%%%%%%%%%%%%%%%%%%%%%%
%%%%%%%%%%%%%%%%%%%%%%%%%%%%%%%%%%%
%%%%%%%%%%%%%%%%%%%%%%%%%%%%%%%%%%%%%%%%%%%%%%%%
\section{Example: $(N,M)=(3,1)$}
\subsection{Dualities and $\text{Dih}_3$ Group Action}
Following the previous example of $X_{2,1}^{(\delta=0)}$, we can also analyse $X_{3,1}^{(\delta=0)}$ in a similar fashion. 
\begin{wrapfigure}{l}{0.42\textwidth}
\scalebox{0.75}{\parbox{9cm}{\begin{tikzpicture}[scale = 1.50]
\draw[ultra thick] (-1,0) -- (0,0) -- (0,-1) -- (1,-1) -- (1,-2) -- (2,-2) -- (2,-3) -- (3,-3);
%diagonals
\draw[ultra thick] (0,0) -- (0.7,0.7);
\draw[ultra thick] (1,-1) -- (1.7,-0.3);
\draw[ultra thick] (2,-2) -- (2.7,-1.3);
\draw[ultra thick] (0,-1) -- (-0.7,-1.7);
\draw[ultra thick] (1,-2) -- (0.3,-2.7);
\draw[ultra thick] (2,-3) -- (1.3,-3.7);
%ends
\node at (-1.2,0) {\large {\bf $\mathbf a$}};
\node at (3.2,-3) {\large {\bf $\mathbf a$}};
\node at (0.85,0.85) {\large {$\mathbf 1$}};
\node at (1.85,-0.15) {\large {$\mathbf 2$}};
\node at (2.85,-1.15) {\large {$\mathbf 3$}};
\node at (-0.85,-1.5) {\large {$\mathbf 1$}};
\node at (0.15,-2.85) {\large {$\mathbf 2$}};
\node at (1.15,-3.85) {\large {$\mathbf 3$}};
%lables hotizontal
\node at (-0.5,0.25) {\large  {\bf $h_1$}};
\node at (0.5,-0.775) {\large  {\bf $h_2$}};
\node at (1.5,-1.775) {\large  {\bf $h_3$}};
\node at (2.5,-2.775) {\large  {\bf $h_1$}};
%lables vertical
\node at (-0.2,-0.5) {\large  {\bf $v_1$}};
\node at (0.8,-1.5) {\large  {\bf $v_2$}};
\node at (1.8,-2.5) {\large  {\bf $v_3$}};
%lables diagonal
\node at (0.65,0.2) {\large  {\bf $m_1$}};
\node at (1.65,-0.8) {\large  {\bf $m_2$}};
\node at (2.65,-1.8) {\large  {\bf $m_3$}};
\node at (-0.2,-1.65) {\large  {\bf $m_1$}};
\node at (0.8,-2.65) {\large  {\bf $m_2$}};
\node at (1.8,-3.65) {\large  {\bf $m_3$}};
%hexagons
\node[red] at (-0.2,0.6) {\large  {\bf $S^{(0)}_3$}};
\node[red] at (0.75,-0.3) {\large  {\bf $S^{(0)}_1$}};
\node[red] at (1.75,-1.3) {\large  {\bf $S^{(0)}_2$}};
\node[red] at (2.75,-2.3) {\large  {\bf $S^{(0)}_3$}};
\node[red] at (-0.75,-0.8) {\large  {\bf $S^{(0)}_3$}};
\node[red] at (0.3,-1.7) {\large  {\bf $S^{(0)}_1$}};
\node[red] at (1.3,-2.7) {\large  {\bf $S^{(0)}_2$}};
\node[red] at (2.2,-3.35) {\large  {\bf $S^{(0)}_3$}};
%roots
\draw[ultra thick,blue,<->] (1.05,0.95) -- (1.95,0.05);
\node[blue,rotate=315] at (1.75,0.65) {{\large {\bf {$\widehat{a}^{(0)}_1$}}}};
\draw[ultra thick,blue,<->] (2.05,-0.05) -- (2.95,-0.95);
\node[blue,rotate=315] at (2.75,-0.35) {{\large {\bf {$\widehat{a}^{(0)}_2$}}}};
\draw[ultra thick,blue,<->] (3.05,-1.05) -- (3.95,-1.95);
\node[blue,rotate=315] at (3.75,-1.35) {{\large {\bf {$\widehat{a}^{(0)}_3$}}}};
%definition S
\draw[dashed] (3,-3) -- (1.35,-4.65);
\draw[dashed] (-0.7,-1.7) -- (-1.2,-2.2);
\draw[ultra thick,blue,<->] (-1.2,-2.2) -- (1.35,-4.65);
\node[blue] at (-0.1,-3.7) {{\large {\bf {$S^{(0)}$}}}};
%parameter R
\draw[dashed] (-0.7,-1.7) -- (-1.5,-1.7);
\draw[dashed] (2,-3) -- (-1.5,-3);
\draw[ultra thick,blue,<->] (-1.5,-2.95) -- (-1.5,-1.75);
\node[blue,rotate=90] at (-1.75,-2.3) {{\large {\bf{$R^{(0)}-3S^{(0)}$}}}};
\end{tikzpicture}}}
\caption{\sl Web diagram of $X_{3,1}$ with a parametrisation of the areas of all curves. The blue parameters represent an independent set of K\"ahler parameters, as explained in eq.~(\ref{IndepKaehler310}).}
\label{Fig:31web0}
${}$\\[-4cm]
\end{wrapfigure} 
The starting point is the web diagram shown in \figref{Fig:31web0}, which includes labels for the areas of all curves. The consistency conditions associated with the three hexagons $S^{(0)}_{1,2,3}$ take the form
\begin{align}
&S^{(0)}_1:&&h_2+m_2=m_1+h_2\,,&&v_1+m_1=m_2+v_2\,,\nonumber\\
&S^{(0)}_2:&&h_3+m_3=m_2+h_3\,,&&v_2+m_2=m_3+v_3\,,\nonumber\\
&S^{(0)}_3:&&h_1+m_1=m_3+h_1\,,&&m_1+v_1=v_3+m_3\,.
\end{align}

\noindent
A solution of these conditions is provided by the parameters $(\widehat{a}^{(0)}_{1,2,3},S^{(0)},R^{(0)})$
\begin{align}
&\widehat{a}^{(0)}_1=v_1+h_2\,,\hspace{0.5cm}\widehat{a}^{(0)}_2=v_2+h_3\,,\hspace{0.5cm}\widehat{a}^{(0)}_3=v_3+h_1\,,\nonumber\\
&S^{(0)}=h_2+v_2+h_3+v_3+h_1\,,\nonumber\\
&R^{(0)}-3S^{(0)}=m_1-v_2-v_3\,,\label{IndepKaehler310}
\end{align}
such that the areas $(h_{1,2,3},v_{1,2,3},m_{1,2,3})$ can be expressed as the linear combinations
\begin{align}
&h_1=S^{(0)}-\widehat{a}^{(0)}_1-\widehat{a}^{(0)}_2\,,\hspace{1cm}h_2=S^{(0)}-\widehat{a}^{(0)}_2-\widehat{a}^{(0)}_3\,,\hspace{1cm}h_3=S^{(0)}-\widehat{a}^{(0)}_1-\widehat{a}^{(0)}_3\,,\nonumber\\
&m_1=m_2=m_3=2(\widehat{a}^{(0)}_1+\widehat{a}^{(0)}_2+\widehat{a}^{(0)}_3)+R^{(0)}-5S^{(0)}\,,\hspace{0.5cm}v_1=v_2=v_3=\widehat{a}^{(0)}_1+\widehat{a}^{(0)}_2+\widehat{a}^{(0)}_3-S^{(0)}\,.\label{SolutionWeb310}
\end{align}

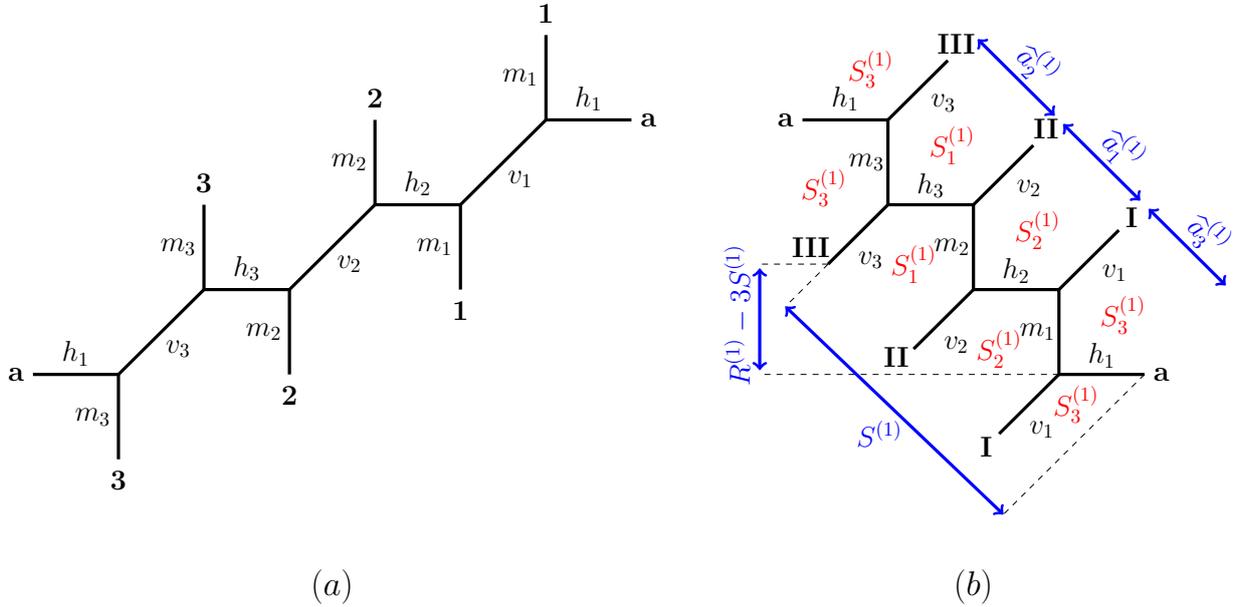
\begin{figure}
\begin{center}
\scalebox{0.75}{\parbox{21.75cm}{\begin{tikzpicture}[scale = 1.50]
\draw[ultra thick] (-1,0) -- (0,0) -- (1,1) -- (2,1) -- (3,2) -- (4,2) -- (5,3) -- (6,3);
%verticals
\draw[ultra thick] (1,1) -- (1,2);
\draw[ultra thick] (3,2) -- (3,3);
\draw[ultra thick] (5,3) -- (5,4);
\draw[ultra thick] (0,0) -- (0,-1);
\draw[ultra thick] (2,1) -- (2,0);
\draw[ultra thick] (4,2) -- (4,1);
%ends
\node at (-1.2,0) {\large {\bf $\mathbf a$}};
\node at (6.2,3) {\large {\bf $\mathbf a$}};
\node at (0,-1.25) {\large {$\mathbf 3$}};
\node at (2,-0.25) {\large {$\mathbf 2$}};
\node at (4,0.75) {\large {$\mathbf 1$}};
\node at (1,2.25) {\large {$\mathbf 3$}};
\node at (3,3.25) {\large {$\mathbf 2$}};
\node at (5,4.25) {\large {$\mathbf 1$}};
%lables hotizontal
\node at (-0.5,0.25) {\large  {\bf $h_1$}};
\node at (1.5,1.25) {\large  {\bf $h_3$}};
\node at (3.5,2.25) {\large  {\bf $h_2$}};
\node at (5.5,3.25) {\large  {\bf $h_1$}};
%lables diagonal
\node at (0.7,0.3) {\large  {\bf $v_3$}};
\node at (2.7,1.3) {\large  {\bf $v_2$}};
\node at (4.7,2.3) {\large  {\bf $v_1$}};
%labels vertical
\node at (0.7,1.5) {\large  {\bf $m_3$}};
\node at (2.7,2.5) {\large  {\bf $m_2$}};
\node at (4.7,3.5) {\large  {\bf $m_1$}};
\node at (-0.3,-0.5) {\large  {\bf $m_3$}};
\node at (1.7,0.5) {\large  {\bf $m_2$}};
\node at (3.7,1.5) {\large  {\bf $m_1$}};
%stamp
\node at (2.5,-2.5) {\Large {\bf $\mathbf (a)$}};
%%%%%%%%%%%%%%%%%%%%%%%
%%%%%%%%%%%%%%%%%%%%%%%
\begin{scope}[xshift=9cm,yshift=3cm]
\draw[ultra thick] (-1,0) -- (0,0) -- (0,-1) -- (1,-1) -- (1,-2) -- (2,-2) -- (2,-3) -- (3,-3);
%diagonals
\draw[ultra thick] (0,0) -- (0.7,0.7);
\draw[ultra thick] (1,-1) -- (1.7,-0.3);
\draw[ultra thick] (2,-2) -- (2.7,-1.3);
\draw[ultra thick] (0,-1) -- (-0.7,-1.7);
\draw[ultra thick] (1,-2) -- (0.3,-2.7);
\draw[ultra thick] (2,-3) -- (1.3,-3.7);
%ends
\node at (-1.2,0) {\large {\bf $\mathbf a$}};
\node at (3.2,-3) {\large {\bf $\mathbf a$}};
\node at (0.8,0.9) {\large {$\mathbf{III}$}};
\node at (1.85,-0.1) {\large {$\mathbf{II}$}};
\node at (2.85,-1.15) {\large {$\mathbf{I}$}};
\node at (-0.9,-1.5) {\large {$\mathbf{III}$}};
\node at (0.1,-2.8) {\large {$\mathbf{II}$}};
\node at (1.15,-3.85) {\large {$\mathbf{I}$}};
%lables hotizontal
\node at (-0.5,0.25) {\large  {\bf $h_1$}};
\node at (0.5,-0.775) {\large  {\bf $h_3$}};
\node at (1.5,-1.775) {\large  {\bf $h_2$}};
\node at (2.5,-2.775) {\large  {\bf $h_1$}};
%lables vertical
\node at (-0.25,-0.5) {\large  {\bf $m_3$}};
\node at (0.75,-1.5) {\large  {\bf $m_2$}};
\node at (1.75,-2.5) {\large  {\bf $m_1$}};
%lables diagonal
\node at (0.65,0.2) {\large  {\bf $v_3$}};
\node at (1.65,-0.8) {\large  {\bf $v_2$}};
\node at (2.65,-1.8) {\large  {\bf $v_1$}};
\node at (-0.2,-1.65) {\large  {\bf $v_3$}};
\node at (0.8,-2.65) {\large  {\bf $v_2$}};
\node at (1.8,-3.65) {\large  {\bf $v_1$}};
%hexagons
\node[red] at (-0.2,0.6) {\large  {\bf $S^{(1)}_3$}};
\node[red] at (0.75,-0.3) {\large  {\bf $S^{(1)}_1$}};
\node[red] at (1.75,-1.3) {\large  {\bf $S^{(1)}_2$}};
\node[red] at (2.75,-2.3) {\large  {\bf $S^{(1)}_3$}};
\node[red] at (-0.75,-0.8) {\large  {\bf $S^{(1)}_3$}};
\node[red] at (0.3,-1.7) {\large  {\bf $S^{(1)}_1$}};
\node[red] at (1.3,-2.7) {\large  {\bf $S^{(1)}_2$}};
\node[red] at (2.2,-3.35) {\large  {\bf $S^{(1)}_3$}};
%roots
\draw[ultra thick,blue,<->] (1.05,0.95) -- (1.95,0.05);
\node[blue,rotate=315] at (1.75,0.65) {{\large {\bf {$\widehat{a}^{(1)}_2$}}}};
\draw[ultra thick,blue,<->] (2.05,-0.05) -- (2.95,-0.95);
\node[blue,rotate=315] at (2.75,-0.35) {{\large {\bf {$\widehat{a}^{(1)}_1$}}}};
\draw[ultra thick,blue,<->] (3.05,-1.05) -- (3.95,-1.95);
\node[blue,rotate=315] at (3.75,-1.35) {{\large {\bf {$\widehat{a}^{(1)}_3$}}}};
%definition S
\draw[dashed] (3,-3) -- (1.35,-4.65);
\draw[dashed] (-0.7,-1.7) -- (-1.2,-2.2);
\draw[ultra thick,blue,<->] (-1.2,-2.2) -- (1.35,-4.65);
\node[blue] at (-0.1,-3.7) {{\large {\bf {$S^{(1)}$}}}};
%parameter R
\draw[dashed] (-0.7,-1.7) -- (-1.5,-1.7);
\draw[dashed] (2,-3) -- (-1.5,-3);
\draw[ultra thick,blue,<->] (-1.5,-2.95) -- (-1.5,-1.75);
\node[blue,rotate=90] at (-1.75,-2.3) {{\large {\bf{$R^{(1)}-3S^{(1)}$}}}};
\end{scope}
%stamp
\node at (10,-2.5) {\Large {\bf $\mathbf (b)$}};
\end{tikzpicture}}}
\caption{\sl (a) alternative presentation of the web diagram in \figref{Fig:31web0}. (b) cutting the web diagram (a) along the lines labelled $v_{1,2,3}$ and re-gluing them along the curves labelled $m_{1,2,3}$.}
\label{Fig:31web1}
\end{center}
\end{figure}

\noindent
The web diagram of $X_{3,1}^{(\delta=0)}$ allows various other presentations:  mirroring the diagram and performing an $SL(2,\mathbb{Z})$ transformation, the web can be drawn in the form of \figref{Fig:31web1}(a). Furthermore, cutting the diagram along the lines labelled $v_{1,2,3}$ and re-gluing them along the lines labelled $m_{1,2,3}$ one obtains \figref{Fig:31web1}(b).
The latter is again a web diagram with $\delta=0$, which can thus be parametrised by $(\widehat{a}^{(1)}_{1,2,3},S^{(1)},R^{(1)})$, as indicated in \figref{Fig:31web1}(b)
\begin{align}
&\widehat{a}_1^{(1)}=m_3+h_3\,,&&\widehat{a}_2^{(1)}=m_2+h_2\,,&&\widehat{a}_3^{(1)}=m_1+h_1\,,&&S^{(1)}=h_3+m_2+h_2+m_1+h_1\,,\nonumber\\
&R^{(1)}=v_3-m_2-m_1\,,
\end{align}
such that the areas can be expressed in the following manner
{\allowdisplaybreaks
\begin{align}
&h_1=S^{(1)}-\widehat{a}_1^{(1)}-\widehat{a}_2^{(1)}\,,\hspace{1cm}h_2=S^{(1)}-\widehat{a}_1^{(1)}-\widehat{a}_3^{(1)}\,,\hspace{1cm}h_1=S^{(1)}-\widehat{a}_2^{(1)}-\widehat{a}_3^{(1)}\,,\nonumber\\
&v_1=v_2=v_3=R^{(1)}+2\left(\widehat{a}_1^{(1)}+\widehat{a}_2^{(1)}+\widehat{a}_3^{(1)}\right)-5S^{(1)}\,,\hspace{0.5cm}m_1=m_2=m_3=\widehat{a}_1^{(1)}+\widehat{a}_2^{(1)}+\widehat{a}_3^{(1)}-S^{(1)}\,.\label{SolutionWeb311}
\end{align}}
Moreover, as explained in section~\ref{Sect:GeneralStrategy}, comparing (\ref{SolutionWeb311}) with (\ref{SolutionWeb310}) gives rise to a symmetry of the partition function as a linear transformation relating $(\widehat{a}^{(1)}_{1,2,3},S^{(1)},R^{(1)})$ to $(\widehat{a}^{(0)}_{1,2,3},S^{(0)},R^{(0)})$
\begin{align}
&\left(\begin{array}{c}\widehat{a}^{(0)}_1\\ \widehat{a}^{(0)}_2  \\ \widehat{a}^{(0)}_3  \\ S^{(0)} \\ R^{(0)}\end{array}\right)=G_1\cdot \left(\begin{array}{c}\widehat{a}^{(1)}_1\\ \widehat{a}^{(1)}_2  \\ \widehat{a}^{(1)}_3  \\ S^{(1)} \\ R^{(1)}\end{array}\right)\,,&&\text{where} &&G_1=\left(
\begin{array}{ccccc}
 2 & 1 & 1 & -4 & 1 \\
 1 & 2 & 1 & -4 & 1 \\
 1 & 1 & 2 & -4 & 1 \\
 2 & 2 & 2 & -7 & 2 \\
 3 & 3 & 3 & -12 & 4 \\
\end{array}
\right) && \text{with} &&\begin{array}{l}\text{det }G_1=1\,, \\ G_1\cdot G_1=1\!\!1_{5\times 5}\,.\end{array}\label{Gen31G1}
\end{align}

\noindent
In order to obtain another symmetry generator we first perform a transformation $\mathcal{F}$ as explained in appendix~\ref{App:FlopTrafo}. The corresponding geometry is of the type $X_{3,1}^{(\delta=1)}$ and a parametrisation of the various curves through an independent set of K\"ahler parameters is shown in~\figref{Fig:31webA}.
\begin{wrapfigure}{l}{0.39\textwidth}
\scalebox{0.75}{\parbox{8.7cm}{\begin{tikzpicture}[scale = 1.50]
\draw[ultra thick] (-1,0) -- (0,0) -- (0,-1) -- (1,-1) -- (1,-2) -- (2,-2) -- (2,-3) -- (3,-3);
%diagonals
\draw[ultra thick] (0,0) -- (0.7,0.7);
\draw[ultra thick] (1,-1) -- (1.7,-0.3);
\draw[ultra thick] (2,-2) -- (2.7,-1.3);
\draw[ultra thick] (0,-1) -- (-0.7,-1.7);
\draw[ultra thick] (1,-2) -- (0.3,-2.7);
\draw[ultra thick] (2,-3) -- (1.3,-3.7);
%ends
\node at (-1.2,0) {\large {\bf $\mathbf a$}};
\node at (3.2,-3) {\large {\bf $\mathbf a$}};
\node at (0.85,0.85) {\large {$\mathbf 1$}};
\node at (1.85,-0.15) {\large {$\mathbf 2$}};
\node at (2.85,-1.15) {\large {$\mathbf 3$}};
\node at (-0.85,-1.85) {\large {$\mathbf 3$}};
\node at (0.15,-2.45) {\large {$\mathbf 1$}};
\node at (1.15,-3.85) {\large {$\mathbf 2$}};
%lables hotizontal
\node at (-0.5,0.25) {\large  {\bf $v'_3$}};
\node at (0.5,-0.775) {\large  {\bf $v'_1$}};
\node at (1.5,-1.775) {\large  {\bf $v'_2$}};
\node at (2.5,-2.775) {\large  {\bf $v'_3$}};
%lables vertical
\node at (-0.35,-0.4) {\large  {\bf $-h_1$}};
\node at (0.65,-1.4) {\large  {\bf $-h_2$}};
\node at (1.65,-2.4) {\large  {\bf $-h_3$}};
%lables diagonal
\node at (0.65,0.2) {\large  {\bf $m'_1$}};
\node at (1.65,-0.8) {\large  {\bf $m'_2$}};
\node at (2.65,-1.8) {\large  {\bf $m'_3$}};
\node at (-0.2,-1.65) {\large  {\bf $m'_3$}};
\node at (0.8,-2.65) {\large  {\bf $m'_1$}};
\node at (1.8,-3.65) {\large  {\bf $m'_2$}};
%hexagons
\node[red] at (-0.1,0.7) {\large  {\bf $S'_3$}};
\node[red] at (0.75,-0.3) {\large  {\bf $S'_1$}};
\node[red] at (1.75,-1.3) {\large  {\bf $S'_2$}};
\node[red] at (2.75,-2.3) {\large  {\bf $S'_3$}};
\node[red] at (-0.75,-0.8) {\large  {\bf $S'_2$}};
\node[red] at (0.4,-1.8) {\large  {\bf $S'_3$}};
\node[red] at (1.3,-2.7) {\large  {\bf $S'_1$}};
\node[red] at (2.2,-3.35) {\large  {\bf $S'_2$}};
%roots
\draw[ultra thick,blue,<->] (1.05,0.95) -- (1.95,0.05);
\node[blue,rotate=315] at (1.75,0.65) {{\large {\bf {$\widehat{a}^{(0)}_1$}}}};
\draw[ultra thick,blue,<->] (2.05,-0.05) -- (2.95,-0.95);
\node[blue,rotate=315] at (2.75,-0.35) {{\large {\bf {$\widehat{a}^{(0)}_2$}}}};
\draw[ultra thick,blue,<->] (3.05,-1.05) -- (3.95,-1.95);
\node[blue,rotate=315] at (3.75,-1.35) {{\large {\bf {$\widehat{a}^{(0)}_3$}}}};
%definition S
\draw[dashed] (3,-3) -- (1.35,-4.65);
\draw[dashed] (0.3,-2.7) -- (-0.25,-3.25);
\draw[ultra thick,blue,<->] (-0.15,-3.25) -- (1.35,-4.65);
\node[blue] at (0.3,-4.2) {{\large {\bf {$S^{(0)}$}}}};
%parameter R
\draw[dashed] (0.3,-2.7) -- (-1.2,-2.7);
\draw[dashed] (2,-3) -- (-1.2,-3);
\draw[ultra thick,blue,->] (-1.2,-3.5) -- (-1.2,-3.05) ;
\draw[ultra thick,blue,<-] (-1.2,-2.65) -- (-1.2,-2.1) ;
\node[blue,rotate=90] at (-1.45,-2.7) {{\large {\bf{$R^{(0)}-2S^{(0)}$}}}};
\end{tikzpicture}}}
\caption{\sl Web diagram after a transformation $\mathcal{F}$ of \figref{Fig:31web0}. The blue parameters are the same as defined in eq.~(\ref{IndepKaehler310}).}
\label{Fig:31webA}
${}$\\[-1.5cm]
\end{wrapfigure} 
The duality map of $\mathcal{F}$ is explicitly given by
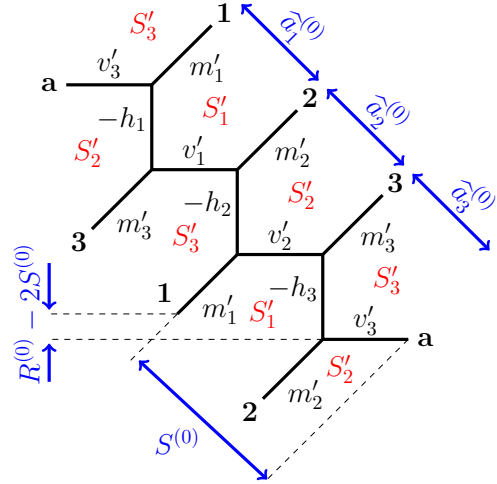
\begin{align}
&v'_1=v_1+h_1+h_2\,,&&v'_2=v_2+h_2+h_3\,,\nonumber\\
&v'_3=v_3+h_1+h_3\,,&&m'_1=m_1+h_1+h_2\,,\nonumber\\
&m'_2=m_2+h_2+h_3\,,&&m'_3=m_3+h_1+h_3\,.\label{construct31}
\end{align}
As was shown in \cite{Bastian:2017ing} for generic $X_{N,1}^{(\delta)}$, the independent parameters $(\widehat{a}_{1,2,3}^{(0)},S^{(0)},R^{(0)})$ are invariants of $\mathcal{F}$ in the sense that the parameters appearing in \figref{Fig:31webA} are the same as the ones defined in (\ref{IndepKaehler310}).\footnote{The only $\delta$-dependence (and thus dependence on $\mathcal{F}$) appears in the coefficient of $S^{(0)}$ in the defining equation of $R^{(0)}$ (see the generic parametrisation of $X_{N,1}^{(\delta)}$ in \figref{Fig:N1web0}).} While the transformation $\mathcal{F}$ itself therefore does not generate a new non-trivial symmetry transformation, one can consider different presentations of \figref{Fig:31webA}. Indeed, mirroring the latter and performing an $SL(2,\mathbb{Z})$ transformation, on obtains \figref{Fig:31web2}(a). Cutting the latter along the lines labelled $-h_{1,2,3}$ and re-gluing them along the lines labelled 

\begin{figure}[h]
\begin{center}
\scalebox{0.75}{\parbox{21.75cm}{\begin{tikzpicture}[scale = 1.50]
\draw[ultra thick] (-1,0) -- (0,0) -- (1,1) -- (2,1) -- (3,2) -- (4,2) -- (5,3) -- (6,3);
%verticals
\draw[ultra thick] (1,1) -- (1,2);
\draw[ultra thick] (3,2) -- (3,3);
\draw[ultra thick] (5,3) -- (5,4);
\draw[ultra thick] (0,0) -- (0,-1);
\draw[ultra thick] (2,1) -- (2,0);
\draw[ultra thick] (4,2) -- (4,1);
%ends
\node at (-1.2,0) {\large {\bf $\mathbf a$}};
\node at (6.2,3) {\large {\bf $\mathbf a$}};
\node at (0,-1.25) {\large {$\mathbf 3$}};
\node at (2,-0.25) {\large {$\mathbf 2$}};
\node at (4,0.75) {\large {$\mathbf 1$}};
\node at (1,2.25) {\large {$\mathbf 2$}};
\node at (3,3.25) {\large {$\mathbf 1$}};
\node at (5,4.25) {\large {$\mathbf 3$}};
%lables hotizontal
\node at (-0.5,0.25) {\large  {\bf $v'_3$}};
\node at (1.5,1.25) {\large  {\bf $v'_2$}};
\node at (3.5,2.25) {\large  {\bf $v'_1$}};
\node at (5.5,3.25) {\large  {\bf $v'_3$}};
%lables diagonal
\node at (0.8,0.3) {\large  {\bf $-h_3$}};
\node at (2.8,1.3) {\large  {\bf $-h_2$}};
\node at (4.8,2.3) {\large  {\bf $-h_1$}};
%labels vertical
\node at (0.7,1.5) {\large  {\bf $m'_3$}};
\node at (2.7,2.5) {\large  {\bf $m'_2$}};
\node at (4.7,3.5) {\large  {\bf $m'_1$}};
\node at (-0.3,-0.5) {\large  {\bf $m'_2$}};
\node at (1.7,0.5) {\large  {\bf $m'_1$}};
\node at (3.7,1.5) {\large  {\bf $m'_3$}};
%stamp
\node at (2.5,-2.5) {\Large {\bf $\mathbf (a)$}};
%%%%%%%%%%%%%%%%%%%%%%%
%%%%%%%%%%%%%%%%%%%%%%%
\begin{scope}[xshift=9cm,yshift=3cm]
\draw[ultra thick] (-1,0) -- (0,0) -- (0,-1) -- (1,-1) -- (1,-2) -- (2,-2) -- (2,-3) -- (3,-3);
%diagonals
\draw[ultra thick] (0,0) -- (0.7,0.7);
\draw[ultra thick] (1,-1) -- (1.7,-0.3);
\draw[ultra thick] (2,-2) -- (2.7,-1.3);
\draw[ultra thick] (0,-1) -- (-0.7,-1.7);
\draw[ultra thick] (1,-2) -- (0.3,-2.7);
\draw[ultra thick] (2,-3) -- (1.3,-3.7);
%ends
\node at (-1.2,0) {\large {\bf $\mathbf a$}};
\node at (3.2,-3) {\large {\bf $\mathbf a$}};
\node at (0.85,0.85) {\large {$\mathbf I$}};
\node at (1.8,-0.1) {\large {$\mathbf{III}$}};
\node at (2.85,-1.15) {\large {$\mathbf{II}$}};
\node at (-0.85,-1.85) {\large {$\mathbf{II}$}};
\node at (0.15,-2.45) {\large {$\mathbf{I}$}};
\node at (1.15,-3.9) {\large {$\mathbf{III}$}};
%lables hotizontal
\node at (-0.5,0.3) {\large  {\bf $v'_3$}};
\node at (0.5,-0.75) {\large  {\bf $v'_1$}};
\node at (1.5,-1.75) {\large  {\bf $v'_2$}};
\node at (2.5,-2.75) {\large  {\bf $v'_3$}};
%lables vertical
\node at (-0.25,-0.4) {\large  {\bf $m'_2$}};
\node at (0.75,-1.4) {\large  {\bf $m'_3$}};
\node at (1.75,-2.4) {\large  {\bf $m'_1$}};
%lables diagonal
\node at (0.65,0.2) {\large  {\bf $-h_3$}};
\node at (1.65,-0.8) {\large  {\bf $-h_1$}};
\node at (2.65,-1.8) {\large  {\bf $-h_2$}};
\node at (-0.2,-1.65) {\large  {\bf $-h_2$}};
\node at (0.7,-2.75) {\large  {\bf $-h_3$}};
\node at (1.8,-3.65) {\large  {\bf $-h_1$}};
%hexagons
\node[red] at (-0.1,0.7) {\large  {\bf $S^{(2)}_3$}};
\node[red] at (0.75,-0.3) {\large  {\bf $S^{(2)}_1$}};
\node[red] at (1.75,-1.3) {\large  {\bf $S^{(2)}_2$}};
\node[red] at (2.75,-2.3) {\large  {\bf $S^{(2)}_3$}};
\node[red] at (-0.75,-0.8) {\large  {\bf $S^{(2)}_2$}};
\node[red] at (0.4,-1.8) {\large  {\bf $S^{(2)}_3$}};
\node[red] at (1.3,-2.7) {\large  {\bf $S^{(2)}_1$}};
\node[red] at (2.2,-3.35) {\large  {\bf $S^{(2)}_2$}};
%roots
\draw[ultra thick,blue,<->] (1.05,0.95) -- (1.95,0.05);
\node[blue,rotate=315] at (1.75,0.65) {{\large {\bf {$\widehat{a}^{(2)}_1$}}}};
\draw[ultra thick,blue,<->] (2.05,-0.05) -- (2.95,-0.95);
\node[blue,rotate=315] at (2.75,-0.35) {{\large {\bf {$\widehat{a}^{(2)}_2$}}}};
\draw[ultra thick,blue,<->] (3.05,-1.05) -- (3.95,-1.95);
\node[blue,rotate=315] at (3.75,-1.35) {{\large {\bf {$\widehat{a}^{(2)}_3$}}}};
%definition S
\draw[dashed] (3,-3) -- (1.35,-4.65);
\draw[dashed] (0.3,-2.7) -- (-0.25,-3.25);
\draw[ultra thick,blue,<->] (-0.15,-3.25) -- (1.35,-4.65);
\node[blue] at (0.3,-4.2) {{\large {\bf {$S^{(2)}$}}}};
%parameter R
\draw[dashed] (0.3,-2.7) -- (-1.2,-2.7);
\draw[dashed] (2,-3) -- (-1.2,-3);
\draw[ultra thick,blue,->] (-1.2,-3.5) -- (-1.2,-3.05) ;
\draw[ultra thick,blue,<-] (-1.2,-2.65) -- (-1.2,-2.1) ;
\node[blue,rotate=90] at (-1.45,-2.7) {{\large {\bf{$R^{(2)}-2S^{(2)}$}}}};
\end{scope}
%stamp
\node at (10,-2.5) {\Large {\bf $\mathbf (b)$}};
\end{tikzpicture}}}
\caption{\sl (a) alternative presentation of the web diagram \figref{Fig:31webA}. (b) web diagram obtained by cutting the lines labelled $-h_{1,2,3}$ and re-gluing along the lines $m_{1,2,3}$.}
\label{Fig:31web2}
\end{center}
\end{figure}
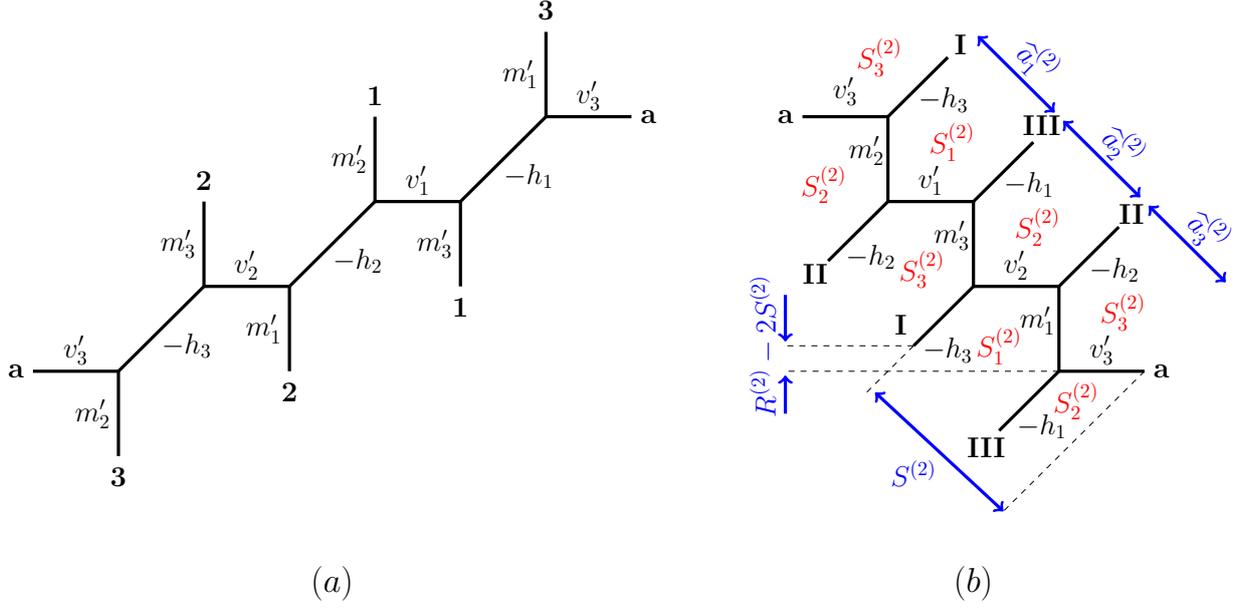 

\noindent
$m_{1,2,3}$ yields the presentation \figref{Fig:31web2}(b). The set of independent parameters $(\widehat{a}^{(2)}_{1,2,3},S^{(2)},R^{(2)})$ 
\begin{align}
&\widehat{a}^{(2)}_1=v'_2+m'_3\,,&&\widehat{a}^{(2)}_2=v'_1+m'_2\,,&&\widehat{a}^{(2)}_3=v'_3+m'_1\,,\nonumber\\
&S^{(2)}=v'_2+m'_1+v'_3\,,&&R^{(2)}-2S^{(2)}=-h_3-m'_1\,,
\end{align}
gives rise to a new parametrisation of all curves of the original diagram \figref{Fig:31web2}
\begin{align}
&h_1=-\widehat{a}^{(2)}_1-\widehat{a}^{(2)}_2-R^{(2)}+3S^{(2)}\,,\hspace{1cm}h_2=-\widehat{a}^{(2)}_1-\widehat{a}^{(2)}_3-R^{(2)}+3S^{(2)}\,,\nonumber\\
&h_3=-\widehat{a}^{(2)}_2-\widehat{a}^{(2)}_3-R^{(2)}+3S^{(2)}\,,\hspace{1cm}v_1=v_2=v_3=\widehat{a}^{(2)}_1+\widehat{a}^{(2)}_2+\widehat{a}^{(2)}_3+2R^{(2)}-5S^{(2)}\,,\nonumber\\
&m_1=m_2=m_3=2\left(\widehat{a}^{(2)}_1+\widehat{a}^{(2)}_2+\widehat{a}^{(2)}_3\right)+2R^{(2)}-7S^{(2)}\,.\label{SolutionWeb312}
\end{align}
Comparing (\ref{SolutionWeb312}) with (\ref{SolutionWeb310}) gives rise to a symmetry of the partition function as a linear transformation relating $(\widehat{a}^{(2)}_{1,2,3},S^{(2)},R^{(2)})$ to $(\widehat{a}^{(0)}_{1,2,3},S^{(0)},R^{(0)})$
\begin{align}
&\left(\begin{array}{c}\widehat{a}^{(0)}_1\\ \widehat{a}^{(0)}_2  \\ \widehat{a}^{(0)}_3  \\ S^{(0)} \\ R^{(0)}\end{array}\right)=G_2\cdot \left(\begin{array}{c}\widehat{a}^{(2)}_1\\ \widehat{a}^{(2)}_2  \\ \widehat{a}^{(2)}_3  \\ S^{(2)} \\ R^{(2)}\end{array}\right)\,,&&\text{where} &&G_2=\left(
\begin{array}{ccccc}
 1 & 0 & 0 & -2 & 1 \\
 0 & 1 & 0 & -2 & 1 \\
 0 & 0 & 1 & -2 & 1 \\
 0 & 0 & 0 & -1 & 1 \\
 0 & 0 & 0 & 0 & 1 \\
\end{array}
\right) && \text{with} &&\begin{array}{l}\text{det }G_2=1\,, \\ G_2\cdot G_2=1\!\!1_{5\times 5}\,.\end{array}\label{Gen31G2}
\end{align}

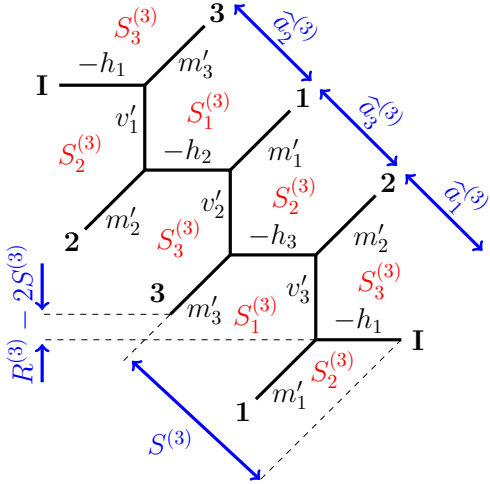
\begin{wrapfigure}{l}{0.39\textwidth}
\scalebox{0.75}{\parbox{8.7cm}{\begin{tikzpicture}[scale = 1.50]
\draw[ultra thick] (-1,0) -- (0,0) -- (0,-1) -- (1,-1) -- (1,-2) -- (2,-2) -- (2,-3) -- (3,-3);
%diagonals
\draw[ultra thick] (0,0) -- (0.7,0.7);
\draw[ultra thick] (1,-1) -- (1.7,-0.3);
\draw[ultra thick] (2,-2) -- (2.7,-1.3);
\draw[ultra thick] (0,-1) -- (-0.7,-1.7);
\draw[ultra thick] (1,-2) -- (0.3,-2.7);
\draw[ultra thick] (2,-3) -- (1.3,-3.7);
%ends
\node at (-1.2,0) {\large {\bf $\mathbf I$}};
\node at (3.2,-3) {\large {\bf $\mathbf I$}};
\node at (0.85,0.85) {\large {$\mathbf 3$}};
\node at (1.85,-0.15) {\large {$\mathbf 1$}};
\node at (2.85,-1.15) {\large {$\mathbf 2$}};
\node at (-0.85,-1.85) {\large {$\mathbf 2$}};
\node at (0.15,-2.45) {\large {$\mathbf 3$}};
\node at (1.15,-3.85) {\large {$\mathbf 1$}};
%lables hotizontal
\node at (-0.5,0.25) {\large  {\bf $-h_1$}};
\node at (0.5,-0.775) {\large  {\bf $-h_2$}};
\node at (1.5,-1.775) {\large  {\bf $-h_3$}};
\node at (2.5,-2.775) {\large  {\bf $-h_1$}};
%lables vertical
\node at (-0.2,-0.4) {\large  {\bf $v'_1$}};
\node at (0.8,-1.4) {\large  {\bf $v'_2$}};
\node at (1.8,-2.4) {\large  {\bf $v'_3$}};
%lables diagonal
\node at (0.6,0.25) {\large  {\bf $m'_3$}};
\node at (1.65,-0.8) {\large  {\bf $m'_1$}};
\node at (2.65,-1.8) {\large  {\bf $m'_2$}};
\node at (-0.25,-1.6) {\large  {\bf $m'_2$}};
\node at (0.7,-2.65) {\large  {\bf $m'_3$}};
\node at (1.7,-3.65) {\large  {\bf $m'_1$}};
%hexagons
\node[red] at (-0.1,0.7) {\large  {\bf $S^{(3)}_3$}};
\node[red] at (0.75,-0.3) {\large  {\bf $S^{(3)}_1$}};
\node[red] at (1.75,-1.3) {\large  {\bf $S^{(3)}_2$}};
\node[red] at (2.75,-2.3) {\large  {\bf $S^{(3)}_3$}};
\node[red] at (-0.75,-0.8) {\large  {\bf $S^{(3)}_2$}};
\node[red] at (0.4,-1.8) {\large  {\bf $S^{(3)}_3$}};
\node[red] at (1.3,-2.7) {\large  {\bf $S^{(3)}_1$}};
\node[red] at (2.2,-3.35) {\large  {\bf $S^{(3)}_2$}};
%roots
\draw[ultra thick,blue,<->] (1.05,0.95) -- (1.95,0.05);
\node[blue,rotate=315] at (1.75,0.65) {{\large {\bf {$\widehat{a}^{(3)}_2$}}}};
\draw[ultra thick,blue,<->] (2.05,-0.05) -- (2.95,-0.95);
\node[blue,rotate=315] at (2.75,-0.35) {{\large {\bf {$\widehat{a}^{(3)}_3$}}}};
\draw[ultra thick,blue,<->] (3.05,-1.05) -- (3.95,-1.95);
\node[blue,rotate=315] at (3.75,-1.35) {{\large {\bf {$\widehat{a}^{(3)}_1$}}}};
%definition S
\draw[dashed] (3,-3) -- (1.35,-4.65);
\draw[dashed] (0.3,-2.7) -- (-0.25,-3.25);
\draw[ultra thick,blue,<->] (-0.15,-3.25) -- (1.35,-4.65);
\node[blue] at (0.3,-4.2) {{\large {\bf {$S^{(3)}$}}}};
%parameter R
\draw[dashed] (0.3,-2.7) -- (-1.2,-2.7);
\draw[dashed] (2,-3) -- (-1.2,-3);
\draw[ultra thick,blue,->] (-1.2,-3.5) -- (-1.2,-3.05) ;
\draw[ultra thick,blue,<-] (-1.2,-2.65) -- (-1.2,-2.1) ;
\node[blue,rotate=90] at (-1.45,-2.7) {{\large {\bf{$R^{(3)}-2S^{(3)}$}}}};
\end{tikzpicture}}}
\caption{\sl Presentation of the web diagram obtained by cutting \figref{Fig:31webA} along the line $-h_1$ and gluing along the line $v'_3$.}
\label{Fig:31web3}
${}$\\[-6cm]
\end{wrapfigure}

One can find another symmetry transformation by cutting the diagram \figref{Fig:31webA} along the line labelled $-h_1$ and re-gluing it along the line labelled $v'_{3}$. After mirroring the diagram, it can also be presented in the form of \figref{Fig:31web3}, which corresponds to a web diagram of the form $X_{3,1}^{(\delta=1)}$. The latter can thus be parametrised by $(\widehat{a}^{(3)}_{1,2,3},S^{(3)},R^{(3)})$, as shown in \figref{Fig:21web3}:
\begin{align}
&\widehat{a}^{(3)}_1=v'_3-h_1\,,&&\widehat{a}^{(3)}_2=v'_2-h_3\,,&&\widehat{a}^{(3)}_3=v'_1-h_2\,,&&S^{(2)}=v'_2-h_2-h_3\,,&&R^{(2)}-2S^{(2)}=m'_2-v'_2\,.
\end{align}
Indeed, the areas $(h_{1,2,3},v_{1,2,3},m_{1,2,3})$ can be expressed in terms of $(\widehat{a}^{(3)}_{1,2,3},S^{(3)},R^{(3)})$
{\allowdisplaybreaks
\begin{align}
&h_1=\widehat{a}^{(3)}_3-S^{(3)}\,,\hspace{2.7cm}h_2=\widehat{a}^{(3)}_2-S^{(3)}\,,\hspace{2.7cm}h_3=\widehat{a}^{(3)}_1-S^{(3)}\,,\nonumber\\
&v_1=v_2=v_3=S^{(3)}\,,\hspace{2cm}m_1=m_2=m_3=R^{(3)}-S^{(3)}\,.\label{SolutionWeb312}
\end{align}}
Since the partition functions computed from \figref{Fig:31web3} and \figref{Fig:31web0} are the same \cite{Bastian:2017ing}, comparing eq.~(\ref{SolutionWeb312}) to eq.~(\ref{SolutionWeb310}) gives rise to a linear transformation that is a symmetry of $\mathcal{Z}_{3,1}$.

\noindent
Explicitly, one finds
%\begin{align}
%&(\widehat{a}^{(0)}_1\,,\widehat{a}^{(0)}_2\,,\widehat{a}^{(0)}_3\,, S^{(0)}\,,R^{(0)})^T=G_3\cdot (\widehat{a}^{(3)}_1\,, \widehat{a}^{(3)}_2  \,, \widehat{a}^{(3)}_3  \,, S^{(3)} \,, R^{(3)})^T\,,
%\end{align}
\begin{align}
&\left(\begin{array}{c}\widehat{a}^{(0)}_1\\ \widehat{a}^{(0)}_2  \\ \widehat{a}^{(0)}_3  \\ S^{(0)} \\ R^{(0)}\end{array}\right)=G_3\cdot \left(\begin{array}{c}\widehat{a}^{(3)}_1\\ \widehat{a}^{(3)}_2  \\ \widehat{a}^{(3)}_3  \\ S^{(3)} \\ R^{(3)}\end{array}\right)\,,
\end{align}
where the $5\times 5$ matrix $G_3$ is given by
\begin{align}
&G_3=\left(
\begin{array}{ccccc}
 1 & 0 & 0 & 0 & 0 \\
 0 & 1 & 0 & 0 & 0 \\
 0 & 0 & 1 & 0 & 0 \\
 1 & 1 & 1 & -1 & 0 \\
 3 & 3 & 3 & -6 & 1 \\
\end{array}
\right) && \text{with} &&\begin{array}{l}\text{det }G_3=1\,, \\ G_3\cdot G_3=1\!\!1_{5\times 5}\,.\end{array}\label{Gen31G3}
\end{align}

\begin{figure}[htbp]
\begin{center}
\scalebox{0.73}{\parbox{18.75cm}{\begin{tikzpicture}[scale = 1.50]
\draw[ultra thick] (-1,-1) -- (0,0) -- (0,1) -- (1,2) -- (1,3) -- (2,4) -- (2,5) -- (3,6);
%horizontals
\draw[ultra thick] (0,0) -- (1,0);
\draw[ultra thick] (1,2) -- (2,2);
\draw[ultra thick] (2,4) -- (3,4);
\draw[ultra thick] (0,1) -- (-1,1);
\draw[ultra thick] (1,3) -- (0,3);
\draw[ultra thick] (2,5) -- (1,5);
%ends
\node at (-1.15,-1.15) {\large {$\mathbf{3}$}};
\node at (3.15,6.15) {\large {$\mathbf{3}$}};
\node at (1.15,0) {\large {\bf $\mathbf b$}};
\node at (2.15,2) {\large {\bf $\mathbf c$}};
\node at (3.15,4) {\large {\bf $\mathbf a$}};
\node at (-1.15,1) {\large {\bf $\mathbf a$}};
\node at (-0.15,3) {\large {\bf $\mathbf b$}};
\node at (0.85,5) {\large {\bf $\mathbf c$}};
%stamp
\node at (1,-2) {\Large {\bf $\mathbf (a)$}};
%diagonals
\node at (-0.4,-0.8) {\large  {\bf $m'_3$}};
\node at (0.6,1.2) {\large  {\bf $m'_1$}};
\node at (1.6,3.2) {\large  {\bf $m'_2$}};
\node at (2.6,5.2) {\large  {\bf $m'_3$}};
%horizontals
\node at (0.5,-0.25) {\large  {\bf $v'_1$}};
\node at (1.5,1.75) {\large  {\bf $v'_2$}};
\node at (2.5,3.75) {\large  {\bf $v'_3$}};
\node at (-0.5,0.75) {\large  {\bf $v'_3$}};
\node at (0.5,2.75) {\large  {\bf $v'_1$}};
\node at (1.5,4.75) {\large  {\bf $v'_2$}};
%verticals
\node at (0.4,0.5) {\large  {\bf $-h_1$}};
\node at (1.4,2.5) {\large  {\bf $-h_2$}};
\node at (2.4,4.5) {\large  {\bf $-h_3$}};
%%%%%%%%%%%%%%%%%%%%%%%
%%%%%%%%%%%%%%%%%%%%%%%
\begin{scope}[xshift=7.5cm,yshift=4cm]
\draw[ultra thick] (-1,0) -- (0,0) -- (0,-1) -- (1,-1) -- (1,-2) -- (2,-2) -- (2,-3) -- (3,-3);
%diagonals
\draw[ultra thick] (0,0) -- (0.7,0.7);
\draw[ultra thick] (1,-1) -- (1.7,-0.3);
\draw[ultra thick] (2,-2) -- (2.7,-1.3);
\draw[ultra thick] (0,-1) -- (-0.7,-1.7);
\draw[ultra thick] (1,-2) -- (0.3,-2.7);
\draw[ultra thick] (2,-3) -- (1.3,-3.7);
%ends
\node at (-1.2,0) {\large {\bf $\mathbf I$}};
\node at (3.2,-3) {\large {\bf $\mathbf I$}};
\node at (0.85,0.85) {\large {$\mathbf a$}};
\node at (1.85,-0.15) {\large {$\mathbf b$}};
\node at (2.85,-1.15) {\large {$\mathbf c$}};
\node at (-0.85,-1.85) {\large {$\mathbf c$}};
\node at (0.15,-2.45) {\large {$\mathbf a$}};
\node at (1.15,-3.85) {\large {$\mathbf b$}};
%lables hotizontal
\node at (-0.5,0.25) {\large  {\bf $-h_1$}};
\node at (0.5,-0.775) {\large  {\bf $-h_2$}};
\node at (1.5,-1.775) {\large  {\bf $-h_3$}};
\node at (2.5,-2.775) {\large  {\bf $-h_1$}};
%lables vertical
\node at (-0.25,-0.4) {\large  {\bf $m'_1$}};
\node at (0.75,-1.4) {\large  {\bf $m'_2$}};
\node at (1.75,-2.4) {\large  {\bf $m'_3$}};
%lables diagonal
\node at (0.6,0.25) {\large  {\bf $v'_3$}};
\node at (1.65,-0.8) {\large  {\bf $v'_1$}};
\node at (2.65,-1.8) {\large  {\bf $v'_2$}};
\node at (-0.25,-1.6) {\large  {\bf $v'_2$}};
\node at (0.7,-2.65) {\large  {\bf $v'_3$}};
\node at (1.7,-3.65) {\large  {\bf $v'_1$}};
%hexagons
\node[red] at (-0.1,0.7) {\large  {\bf $S^{(4)}_3$}};
\node[red] at (0.75,-0.3) {\large  {\bf $S^{(4)}_1$}};
\node[red] at (1.75,-1.3) {\large  {\bf $S^{(4)}_2$}};
\node[red] at (2.75,-2.3) {\large  {\bf $S^{(4)}_3$}};
\node[red] at (-0.75,-0.8) {\large  {\bf $S^{(4)}_2$}};
\node[red] at (0.4,-1.8) {\large  {\bf $S^{(4)}_3$}};
\node[red] at (1.3,-2.7) {\large  {\bf $S^{(4)}_1$}};
\node[red] at (2.2,-3.35) {\large  {\bf $S^{(4)}_2$}};
%roots
\draw[ultra thick,blue,<->] (1.05,0.95) -- (1.95,0.05);
\node[blue,rotate=315] at (1.75,0.65) {{\large {\bf {$\widehat{a}^{(4)}_3$}}}};
\draw[ultra thick,blue,<->] (2.05,-0.05) -- (2.95,-0.95);
\node[blue,rotate=315] at (2.75,-0.35) {{\large {\bf {$\widehat{a}^{(4)}_1$}}}};
\draw[ultra thick,blue,<->] (3.05,-1.05) -- (3.95,-1.95);
\node[blue,rotate=315] at (3.75,-1.35) {{\large {\bf {$\widehat{a}^{(4)}_2$}}}};
%definition S
\draw[dashed] (3,-3) -- (1.35,-4.65);
\draw[dashed] (0.3,-2.7) -- (-0.25,-3.25);
\draw[ultra thick,blue,<->] (-0.15,-3.25) -- (1.35,-4.65);
\node[blue] at (0.3,-4.2) {{\large {\bf {$S^{(4)}$}}}};
%parameter R
\draw[dashed] (0.3,-2.7) -- (-1.2,-2.7);
\draw[dashed] (2,-3) -- (-1.2,-3);
\draw[ultra thick,blue,->] (-1.2,-3.5) -- (-1.2,-3.05) ;
\draw[ultra thick,blue,<-] (-1.2,-2.65) -- (-1.2,-2.1) ;
\node[blue,rotate=90] at (-1.45,-2.7) {{\large {\bf{$R^{(4)}-2S^{(4)}$}}}};
\end{scope}
%stamp
\node at (8.5,-2) {\Large {\bf $\mathbf (b)$}};
\end{tikzpicture}}}
\caption{\sl (a) web diagram obtained by cutting the lines labelled $v'_{1,2,3}$ in \figref{Fig:31webA} and re-gluing along the lines $m'_{1,2,3}$. (b) alternative presentation of the same web diagram after cutting the line $-h_1$ and re-gluing along the line $m'_3$.}
\label{Fig:31web4}
\end{center}
\end{figure}
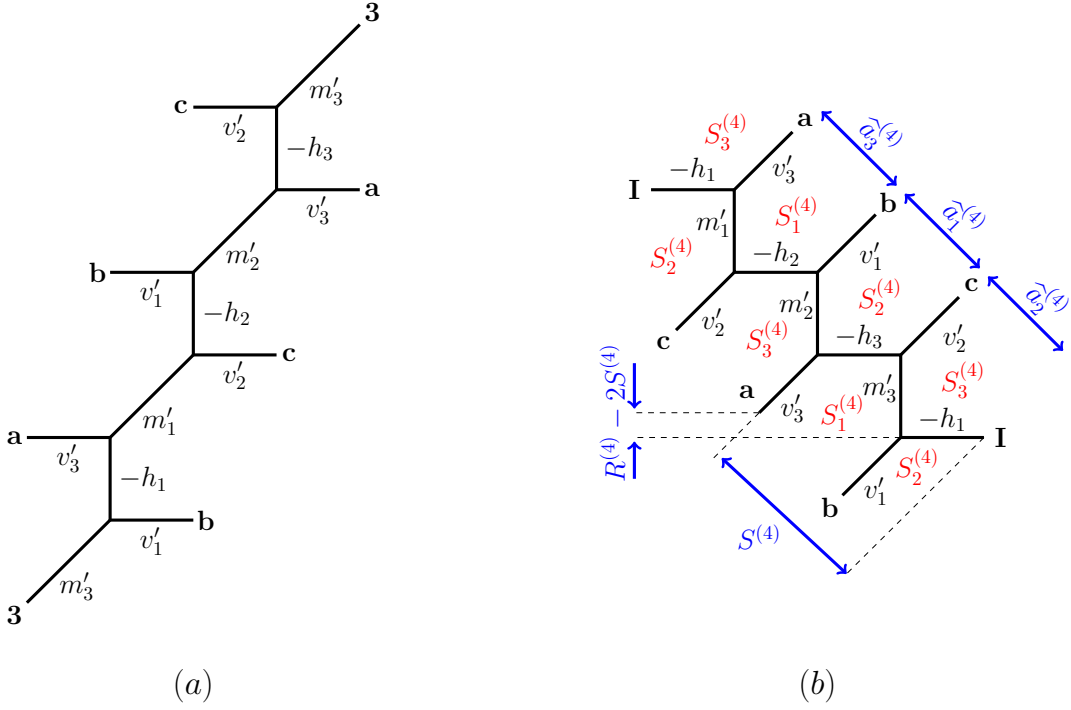

\noindent
From \figref{Fig:31webA} one can extract yet another symmetry generator. Indeed, cutting the diagram along the curves $v'_{1,2,3}$ and re-gluing it along the lines $m'_{1,2,3}$ one obtains \figref{Fig:31web4}(a). Cutting furthermore along the line labelled $-h_1$ and re-gluing along the line $m'_3$ one obtains \figref{Fig:31web4}(b) after performing an $SL(2,\mathbb{Z})$ transformation.

An independent set of parameters is given by
\begin{align}
&\widehat{a}^{(4)}_1=m'_2-h_3\,,\hspace{1.5cm}\widehat{a}^{(4)}_2=m'_3-h_1\,,\hspace{1.5cm}\widehat{a}^{(4)}_3=m'_1-h_2\,,\hspace{1.5cm}S^{(4)}=m'_2-h_2-h_3\,,\nonumber\\
&R^{(4)}-2S^{(2)}=v'_2-m'_2\,,
\end{align}
which allows to express $(h_{1,2,3},v_{1,2,3},m_{1,2,3})$ in the following fashion
\begin{align}
&h_1=\widehat{a}^{(4)}_3-S^{(4)}\,,\hspace{1cm}h_2=\widehat{a}^{(4)}_1-S^{(4)}\,,\hspace{1cm}h_3=\widehat{a}^{(4)}_2-S^{(4)}\,,\hspace{1cm}v_1=v_2=v_3=R^{(4)}-S^{(4)}\,,\nonumber\\
&m_1=m_2=m_3=S^{(4)}\,.\label{SolutionWeb313}
\end{align}

\noindent
Comparing eq.~(\ref{SolutionWeb313}) to eq.~(\ref{SolutionWeb310}) gives rise to the following linear transformation
\begin{align}
&\left(\begin{array}{c}\widehat{a}^{(0)}_1\\ \widehat{a}^{(0)}_2  \\ \widehat{a}^{(0)}_3  \\ S^{(0)} \\ R^{(0)}\end{array}\right)=G_4\cdot \left(\begin{array}{c}\widehat{a}^{(4)}_1\\ \widehat{a}^{(4)}_2  \\ \widehat{a}^{(4)}_3  \\ S^{(4)} \\ R^{(4)}\end{array}\right)\,,&&\text{where} &&G_4=\left(
\begin{array}{ccccc}
 1 & 0 & 0 & -2 & 1 \\
 0 & 1 & 0 & -2 & 1 \\
 0 & 0 & 1 & -2 & 1 \\
 1 & 1 & 1 & -5 & 2 \\
 3 & 3 & 3 & -12 & 4 \\
\end{array}
\right) && \text{with} &&\begin{array}{l}\text{det }G_4=1\,, \\ G_4\cdot G_4\cdot G_4=1\!\!1_{5\times 5}\,.\end{array}\label{Gen31G4}
\end{align}

\noindent
The matrix $G_4$ is of order 3, which means that $G_5=G_4\cdot G_4$ is a new symmetry element. It can also be associated to a particular presentation of the web diagram of $X_{3,1}$. To see this, we first perform a transformation $\mathcal{F}$ on the web diagram in \figref{Fig:31webA} to obtain \figref{Fig:31webB}.

\begin{wrapfigure}{l}{0.42\textwidth}
\scalebox{0.79}{\parbox{8.7cm}{\begin{tikzpicture}[scale = 1.50]
\draw[ultra thick] (-1,0) -- (0,0) -- (0,-1) -- (1,-1) -- (1,-2) -- (2,-2) -- (2,-3) -- (3,-3);
%diagonals
\draw[ultra thick] (0,0) -- (0.7,0.7);
\draw[ultra thick] (1,-1) -- (1.7,-0.3);
\draw[ultra thick] (2,-2) -- (2.7,-1.3);
\draw[ultra thick] (0,-1) -- (-0.7,-1.7);
\draw[ultra thick] (1,-2) -- (0.3,-2.7);
\draw[ultra thick] (2,-3) -- (1.3,-3.7);
%ends
\node at (-1.2,0) {\large {\bf $\mathbf I$}};
\node at (3.2,-3) {\large {\bf $\mathbf I$}};
\node at (0.85,0.85) {\large {$\mathbf 1$}};
\node at (1.85,-0.15) {\large {$\mathbf 2$}};
\node at (2.85,-1.15) {\large {$\mathbf 3$}};
\node at (-0.85,-1.8) {\large {$\mathbf 2$}};
\node at (0.15,-2.45) {\large {$\mathbf 3$}};
\node at (1.15,-3.85) {\large {$\mathbf 1$}};
%lables hotizontal
\node at (-0.5,0.25) {\large  {\bf $v''_3$}};
\node at (0.5,-0.775) {\large  {\bf $v''_1$}};
\node at (1.5,-1.775) {\large  {\bf $v''_2$}};
\node at (2.75,-3.225) {\large  {\bf $v''_3$}};
%lables vertical
\node at (-0.35,-0.4) {\large  {\bf $-v'_3$}};
\node at (0.65,-1.4) {\large  {\bf $-v'_1$}};
\node at (2.35,-2.3) {\large  {\bf $-v'_2$}};
%lables diagonal
\node at (0.65,0.2) {\large  {\bf $m''_1$}};
\node at (1.65,-0.8) {\large  {\bf $m''_2$}};
\node at (2.65,-1.8) {\large  {\bf $m''_3$}};
\node at (-0.2,-1.65) {\large  {\bf $m''_2$}};
\node at (0.8,-2.65) {\large  {\bf $m''_3$}};
\node at (1.8,-3.65) {\large  {\bf $m''_1$}};
%hexagons
\node[red] at (-0.1,0.7) {\large  {\bf $S''_3$}};
\node[red] at (0.75,-0.3) {\large  {\bf $S''_1$}};
\node[red] at (1.75,-1.3) {\large  {\bf $S''_2$}};
\node[red] at (2.95,-2.3) {\large  {\bf $S''_3$}};
\node[red] at (-0.75,-0.8) {\large  {\bf $S''_1$}};
\node[red] at (0.4,-1.8) {\large  {\bf $S''_2$}};
\node[red] at (1.4,-3.1) {\large  {\bf $S''_3$}};
\node[red] at (2.25,-3.35) {\large  {\bf $S''_1$}};
%roots
\draw[ultra thick,blue,<->] (1.05,0.95) -- (1.95,0.05);
\node[blue,rotate=315] at (1.75,0.65) {{\large {\bf {$\widehat{a}^{(0)}_1$}}}};
\draw[ultra thick,blue,<->] (2.05,-0.05) -- (2.95,-0.95);
\node[blue,rotate=315] at (2.75,-0.35) {{\large {\bf {$\widehat{a}^{(0)}_2$}}}};
\draw[ultra thick,blue,<->] (3.05,-1.05) -- (3.95,-1.95);
\node[blue,rotate=315] at (3.75,-1.35) {{\large {\bf {$\widehat{a}^{(0)}_3$}}}};
%definition S
\draw[dashed] (2,-2) -- (0.35,-3.65);
\draw[dashed] (0.3,-2.7) -- (-0.15,-3.15);
\draw[ultra thick,blue,<->] (-0.15,-3.15) -- (0.35,-3.65);
\node[blue] at (-0.2,-3.6) {{\large {\bf {$S^{(0)}$}}}};
%parameter R
\draw[dashed] (0.3,-2.7) -- (-1.2,-2.7);
\draw[dashed] (1,-2) -- (-1.2,-2);
%\draw[ultra thick,blue,->] (-1.2,-3.5) -- (-1.2,-3.05) ;
\draw[ultra thick,blue,<->] (-1.2,-2.65) -- (-1.2,-2.1) ;
\node[blue,rotate=90] at (-1.45,-2.3) {{\large {\bf{$R^{(0)}-S^{(0)}$}}}};
\end{tikzpicture}}}
\caption{\sl Web diagram after a transformation $\mathcal{F}$ of \figref{Fig:31webA}. The blue parameters are the same as defined in eq.~(\ref{IndepKaehler310}).}
\label{Fig:31webB}
${}$\\[-2.8cm]
\end{wrapfigure} 

\noindent
Since $\mathcal{F}$ leaves the partition function invariant, the parameters $(\widehat{a}^{(0)}_{1,2,3},S^{(0)},R^{(0)})$ are the same as introduced in eq.~(\ref{IndepKaehler310}). Furthermore, we have introduced the areas
\begin{align}
&v''_1=-h_1+v'_1+v'_3\,,&&v''_2=-h_2+v'_1+v'_2\,,\nonumber\\
&v''_3=-h_3+v'_2+'_3\,,&&m''_1=m'_1+v'_2+v'_3\,,\nonumber\\
&m''_2=m'_2+v'_1+v'_3\,,&&m''_3=m'_3+v'_1+v'_2\,,
\end{align}
where we have used the definitions (\ref{construct31}). Next, we cut the diagram \figref{Fig:31webB} along the lines labelled $v''_{1,2,3}$ and re-glue it along the lines labelled $m''_{1,2,3}$ to obtain \figref{Fig:31web5}(a). Cutting the diagram again along the line $-v'_3$, it can also be presented in the form of \figref{Fig:31web5}(b), which is a diagram with shift $\delta=0$. It can be parametrised by $(\widehat{a}^{(5)}_{1,2,3},S^{(5)},R^{(5)})$

\begin{figure}[h]
\begin{center}
\scalebox{0.75}{\parbox{18.75cm}{\begin{tikzpicture}[scale = 1.50]
\draw[ultra thick] (-1,-1) -- (0,0) -- (0,1) -- (1,2) -- (1,3) -- (2,4) -- (2,5) -- (3,6);
%horizontals
\draw[ultra thick] (0,0) -- (1,0);
\draw[ultra thick] (1,2) -- (2,2);
\draw[ultra thick] (2,4) -- (3,4);
\draw[ultra thick] (0,1) -- (-1,1);
\draw[ultra thick] (1,3) -- (0,3);
\draw[ultra thick] (2,5) -- (1,5);
%ends
\node at (-1.15,-1.15) {\large {$\mathbf{3}$}};
\node at (3.15,6.15) {\large {$\mathbf{3}$}};
\node at (1.15,0) {\large {\bf $\mathbf a$}};
\node at (2.15,2) {\large {\bf $\mathbf c$}};
\node at (3.15,4) {\large {\bf $\mathbf b$}};
\node at (-1.15,1) {\large {\bf $\mathbf c$}};
\node at (-0.15,3) {\large {\bf $\mathbf b$}};
\node at (0.85,5) {\large {\bf $\mathbf a$}};
%stamp
\node at (1,-2) {\Large {\bf $\mathbf (a)$}};
%diagonals
\node at (-0.4,-0.8) {\large  {\bf $m''_2$}};
\node at (0.6,1.2) {\large  {\bf $m''_1$}};
\node at (1.6,3.2) {\large  {\bf $m''_3$}};
\node at (2.6,5.2) {\large  {\bf $m''_2$}};
%horizontals
\node at (0.5,-0.25) {\large  {\bf $v''_1$}};
\node at (1.5,1.75) {\large  {\bf $v''_3$}};
\node at (2.5,3.75) {\large  {\bf $v''_2$}};
\node at (-0.5,0.75) {\large  {\bf $v''_3$}};
\node at (0.5,2.75) {\large  {\bf $v''_2$}};
\node at (1.5,4.75) {\large  {\bf $v''_1$}};
%verticals
\node at (0.4,0.5) {\large  {\bf $-v'_3$}};
\node at (1.4,2.5) {\large  {\bf $-v'_2$}};
\node at (2.4,4.5) {\large  {\bf $-v'_1$}};
%%%%%%%%%%%%%%%%%%%%%%%
%%%%%%%%%%%%%%%%%%%%%%%
\begin{scope}[xshift=7.5cm,yshift=4cm]
\draw[ultra thick] (-1,0) -- (0,0) -- (0,-1) -- (1,-1) -- (1,-2) -- (2,-2) -- (2,-3) -- (3,-3);
%diagonals
\draw[ultra thick] (0,0) -- (0.7,0.7);
\draw[ultra thick] (1,-1) -- (1.7,-0.3);
\draw[ultra thick] (2,-2) -- (2.7,-1.3);
\draw[ultra thick] (0,-1) -- (-0.7,-1.7);
\draw[ultra thick] (1,-2) -- (0.3,-2.7);
\draw[ultra thick] (2,-3) -- (1.3,-3.7);
%ends
\node at (-1.2,0) {\large {\bf $\mathbf I$}};
\node at (3.2,-3) {\large {\bf $\mathbf I$}};
\node at (0.85,0.85) {\large {$\mathbf c$}};
\node at (1.85,-0.15) {\large {$\mathbf b$}};
\node at (2.85,-1.15) {\large {$\mathbf a$}};
\node at (-0.85,-1.5) {\large {$\mathbf c$}};
\node at (0.15,-2.75) {\large {$\mathbf b$}};
\node at (1.15,-3.85) {\large {$\mathbf a$}};
%lables hotizontal
\node at (-0.5,0.25) {\large  {\bf $-v'_3$}};
\node at (0.5,-0.775) {\large  {\bf $-v'_2$}};
\node at (1.5,-1.775) {\large  {\bf $-v'_1$}};
\node at (2.5,-2.775) {\large  {\bf $-v'_3$}};
%lables vertical
\node at (-0.25,-0.5) {\large  {\bf $m''_1$}};
\node at (0.75,-1.5) {\large  {\bf $m''_3$}};
\node at (1.75,-2.5) {\large  {\bf $m''_2$}};
%lables diagonal
\node at (0.6,0.3) {\large  {\bf $v''_3$}};
\node at (1.6,-0.7) {\large  {\bf $v''_2$}};
\node at (2.6,-1.7) {\large  {\bf $v''_1$}};
\node at (-0.25,-1.55) {\large  {\bf $v''_3$}};
\node at (0.8,-2.55) {\large  {\bf $v''_2$}};
\node at (1.7,-3.6) {\large  {\bf $v''_1$}};
%hexagons
\node[red] at (-0.2,0.7) {\large  {\bf $S^{(5)}_3$}};
\node[red] at (0.75,-0.3) {\large  {\bf $S^{(5)}_1$}};
\node[red] at (1.75,-1.3) {\large  {\bf $S^{(5)}_2$}};
\node[red] at (2.75,-2.3) {\large  {\bf $S^{(5)}_3$}};
\node[red] at (-0.75,-0.8) {\large  {\bf $S^{(5)}_3$}};
\node[red] at (0.25,-1.7) {\large  {\bf $S^{(5)}_1$}};
\node[red] at (1.3,-2.7) {\large  {\bf $S^{(5)}_2$}};
\node[red] at (2.2,-3.35) {\large  {\bf $S^{(5)}_3$}};
%roots
\draw[ultra thick,blue,<->] (1.05,0.95) -- (1.95,0.05);
\node[blue,rotate=315] at (1.75,0.65) {{\large {\bf {$\widehat{a}^{(5)}_3$}}}};
\draw[ultra thick,blue,<->] (2.05,-0.05) -- (2.95,-0.95);
\node[blue,rotate=315] at (2.75,-0.35) {{\large {\bf {$\widehat{a}^{(5)}_2$}}}};
\draw[ultra thick,blue,<->] (3.05,-1.05) -- (3.95,-1.95);
\node[blue,rotate=315] at (3.75,-1.35) {{\large {\bf {$\widehat{a}^{(5)}_1$}}}};
%definition S
\draw[dashed] (3,-3) -- (1.35,-4.65);
\draw[dashed] (-0.7,-1.7) -- (-1.2,-2.2);
\draw[ultra thick,blue,<->] (-1.2,-2.2) -- (1.35,-4.65);
\node[blue] at (-0.1,-3.7) {{\large {\bf {$S^{(5)}$}}}};
%parameter R
\draw[dashed] (-0.7,-1.7) -- (-1.5,-1.7);
\draw[dashed] (2,-3) -- (-1.5,-3);
\draw[ultra thick,blue,<->] (-1.5,-2.95) -- (-1.5,-1.75);
\node[blue,rotate=90] at (-1.75,-2.3) {{\large {\bf{$R^{(5)}-3S^{(5)}$}}}};
\end{scope}
%stamp
\node at (8.5,-2) {\Large {\bf $\mathbf (b)$}};
\end{tikzpicture}}}
\caption{\sl (a) web diagram \figref{Fig:31webB} after cutting the lines $v''_{1,2,3}$ and re-gluing along $m''_{1,2,3}$. (b) presentation of the web diagram after cutting along the line $-v'_3$ and gluing along $m''_2$.  }
\label{Fig:31web5}
\end{center}
\end{figure}

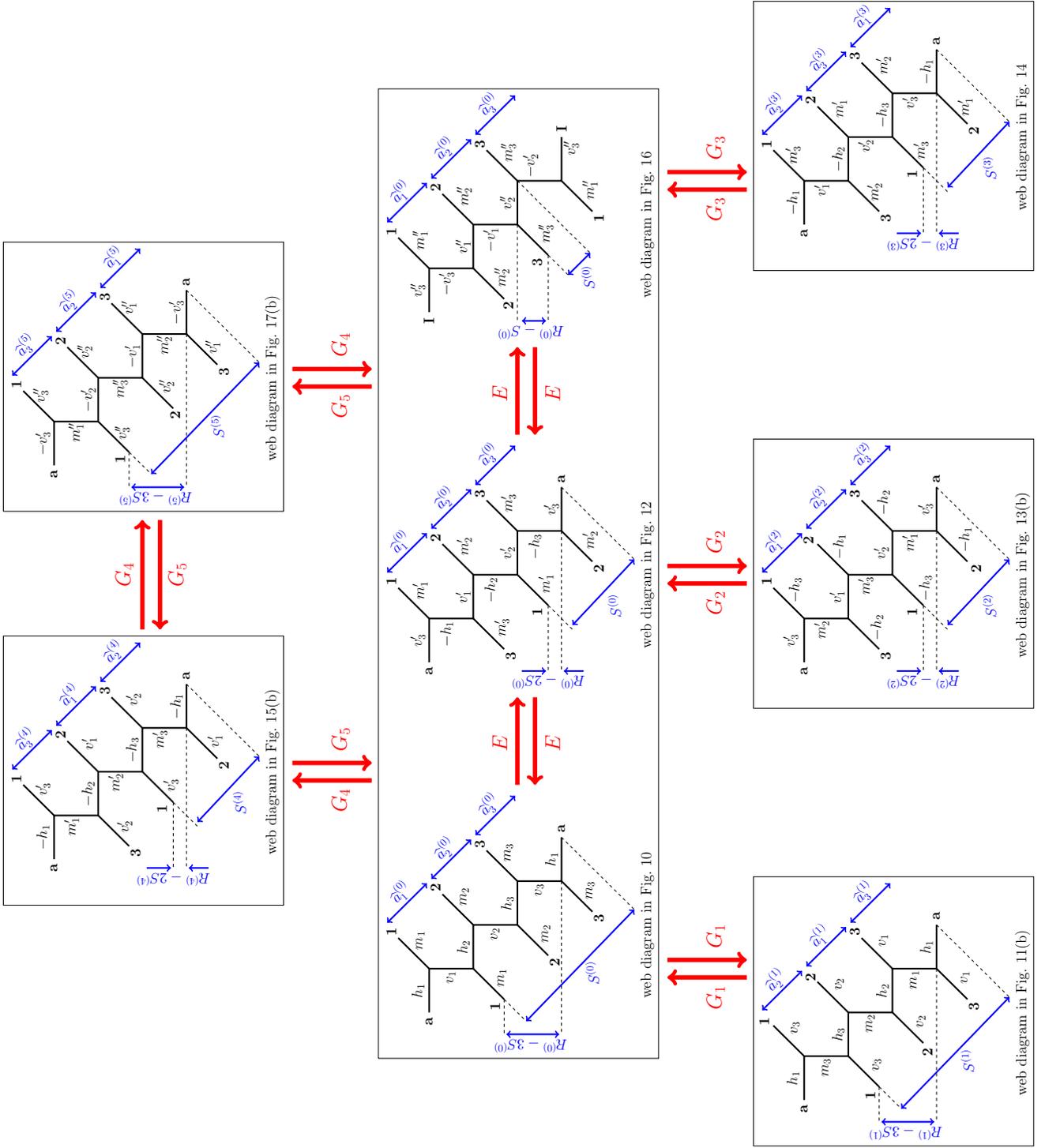
\begin{figure}
\begin{center}
\rotatebox{90}{\scalebox{0.49}{\parbox{42.5cm}{\begin{tikzpicture}[scale = 1.50]
%%%%%%%%%%%%%%%%%%%%%%%%%%%%%
%central graph 1
%%%%%%%%%%%%%%%%%%%%%%%%%%%%%
\draw (-2.05,1.15) -- (20.1,1.15) -- (20.1,-5.2) -- (-2.05,-5.2) -- (-2.05,1.15);
\draw[ultra thick] (-1,0) -- (0,0) -- (0,-1) -- (1,-1) -- (1,-2) -- (2,-2) -- (2,-3) -- (3,-3);
%diagonals
\draw[ultra thick] (0,0) -- (0.7,0.7);
\draw[ultra thick] (1,-1) -- (1.7,-0.3);
\draw[ultra thick] (2,-2) -- (2.7,-1.3);
\draw[ultra thick] (0,-1) -- (-0.7,-1.7);
\draw[ultra thick] (1,-2) -- (0.3,-2.7);
\draw[ultra thick] (2,-3) -- (1.3,-3.7);
%ends
\node at (-1.2,0) {\large {\bf $\mathbf a$}};
\node at (3.2,-3) {\large {\bf $\mathbf a$}};
\node at (0.85,0.85) {\large {$\mathbf 1$}};
\node at (1.85,-0.15) {\large {$\mathbf 2$}};
\node at (2.85,-1.15) {\large {$\mathbf 3$}};
\node at (-0.85,-1.5) {\large {$\mathbf 1$}};
\node at (0.15,-2.85) {\large {$\mathbf 2$}};
\node at (1.15,-3.85) {\large {$\mathbf 3$}};
%lables hotizontal
\node at (-0.5,0.25) {\large  {\bf $h_1$}};
\node at (0.5,-0.775) {\large  {\bf $h_2$}};
\node at (1.5,-1.775) {\large  {\bf $h_3$}};
\node at (2.5,-2.775) {\large  {\bf $h_1$}};
%lables vertical
\node at (-0.2,-0.5) {\large  {\bf $v_1$}};
\node at (0.8,-1.5) {\large  {\bf $v_2$}};
\node at (1.8,-2.5) {\large  {\bf $v_3$}};
%lables diagonal
\node at (0.65,0.2) {\large  {\bf $m_1$}};
\node at (1.65,-0.8) {\large  {\bf $m_2$}};
\node at (2.65,-1.8) {\large  {\bf $m_3$}};
\node at (-0.2,-1.65) {\large  {\bf $m_1$}};
\node at (0.8,-2.65) {\large  {\bf $m_2$}};
\node at (1.8,-3.65) {\large  {\bf $m_3$}};
%hexagons
%\node[red] at (-0.2,0.6) {\large  {\bf $S^{(0)}_3$}};
%\node[red] at (0.75,-0.3) {\large  {\bf $S^{(0)}_1$}};
%\node[red] at (1.75,-1.3) {\large  {\bf $S^{(0)}_2$}};
%\node[red] at (2.75,-2.3) {\large  {\bf $S^{(0)}_3$}};
%\node[red] at (-0.75,-0.8) {\large  {\bf $S^{(0)}_3$}};
%\node[red] at (0.3,-1.7) {\large  {\bf $S^{(0)}_1$}};
%\node[red] at (1.3,-2.7) {\large  {\bf $S^{(0)}_2$}};
%\node[red] at (2.2,-3.35) {\large  {\bf $S^{(0)}_3$}};
%roots
\draw[ultra thick,blue,<->] (1.05,0.95) -- (1.95,0.05);
\node[blue,rotate=315] at (1.75,0.65) {{\large {\bf {$\widehat{a}^{(0)}_1$}}}};
\draw[ultra thick,blue,<->] (2.05,-0.05) -- (2.95,-0.95);
\node[blue,rotate=315] at (2.75,-0.35) {{\large {\bf {$\widehat{a}^{(0)}_2$}}}};
\draw[ultra thick,blue,<->] (3.05,-1.05) -- (3.95,-1.95);
\node[blue,rotate=315] at (3.75,-1.35) {{\large {\bf {$\widehat{a}^{(0)}_3$}}}};
%definition S
\draw[dashed] (3,-3) -- (1.35,-4.65);
\draw[dashed] (-0.7,-1.7) -- (-1.2,-2.2);
\draw[ultra thick,blue,<->] (-1.2,-2.2) -- (1.35,-4.65);
\node[blue] at (-0.1,-3.7) {{\large {\bf {$S^{(0)}$}}}};
%parameter R
\draw[dashed] (-0.7,-1.7) -- (-1.5,-1.7);
\draw[dashed] (2,-3) -- (-1.5,-3);
\draw[ultra thick,blue,<->] (-1.5,-2.95) -- (-1.5,-1.75);
\node[blue,rotate=90] at (-1.75,-2.3) {{\large {\bf{$R^{(0)}-3S^{(0)}$}}}};
%label
\node at (1,-4.95) {{\large web diagram in \figref{Fig:31web0}}};
%%%%%%%%%%%%%%%%%%%%%%%%%%%%%
\node[red] at (5.15,-1.6) {\LARGE $E$};
\draw[line width=1.5mm,red,->] (4.2,-2) -- (6.2,-2);
\draw[line width=1.5mm,red,<-] (4.2,-2.4) -- (6.2,-2.4);
\node[red] at (5.15,-2.8) {\LARGE $E$};
%%%%%%%%%%%%%%%%%%%%%%%%%%%%%
%central graph 1
%%%%%%%%%%%%%%%%%%%%%%%%%%%%%
\begin{scope}[xshift=8cm]
\draw[ultra thick] (-1,0) -- (0,0) -- (0,-1) -- (1,-1) -- (1,-2) -- (2,-2) -- (2,-3) -- (3,-3);
%diagonals
\draw[ultra thick] (0,0) -- (0.7,0.7);
\draw[ultra thick] (1,-1) -- (1.7,-0.3);
\draw[ultra thick] (2,-2) -- (2.7,-1.3);
\draw[ultra thick] (0,-1) -- (-0.7,-1.7);
\draw[ultra thick] (1,-2) -- (0.3,-2.7);
\draw[ultra thick] (2,-3) -- (1.3,-3.7);
%ends
\node at (-1.2,0) {\large {\bf $\mathbf a$}};
\node at (3.2,-3) {\large {\bf $\mathbf a$}};
\node at (0.85,0.85) {\large {$\mathbf 1$}};
\node at (1.85,-0.15) {\large {$\mathbf 2$}};
\node at (2.85,-1.15) {\large {$\mathbf 3$}};
\node at (-0.85,-1.85) {\large {$\mathbf 3$}};
\node at (0.15,-2.45) {\large {$\mathbf 1$}};
\node at (1.15,-3.85) {\large {$\mathbf 2$}};
%lables hotizontal
\node at (-0.5,0.25) {\large  {\bf $v'_3$}};
\node at (0.5,-0.775) {\large  {\bf $v'_1$}};
\node at (1.5,-1.775) {\large  {\bf $v'_2$}};
\node at (2.5,-2.775) {\large  {\bf $v'_3$}};
%lables vertical
\node at (-0.35,-0.4) {\large  {\bf $-h_1$}};
\node at (0.65,-1.4) {\large  {\bf $-h_2$}};
\node at (1.65,-2.4) {\large  {\bf $-h_3$}};
%lables diagonal
\node at (0.65,0.2) {\large  {\bf $m'_1$}};
\node at (1.65,-0.8) {\large  {\bf $m'_2$}};
\node at (2.65,-1.8) {\large  {\bf $m'_3$}};
\node at (-0.2,-1.65) {\large  {\bf $m'_3$}};
\node at (0.8,-2.65) {\large  {\bf $m'_1$}};
\node at (1.8,-3.65) {\large  {\bf $m'_2$}};
%hexagons
%\node[red] at (-0.1,0.7) {\large  {\bf $S'_3$}};
%\node[red] at (0.75,-0.3) {\large  {\bf $S'_1$}};
%\node[red] at (1.75,-1.3) {\large  {\bf $S'_2$}};
%\node[red] at (2.75,-2.3) {\large  {\bf $S'_3$}};
%\node[red] at (-0.75,-0.8) {\large  {\bf $S'_3$}};
%\node[red] at (0.4,-1.8) {\large  {\bf $S'_1$}};
%\node[red] at (1.3,-2.7) {\large  {\bf $S'_2$}};
%\node[red] at (2.2,-3.35) {\large  {\bf $S'_3$}};
%roots
\draw[ultra thick,blue,<->] (1.05,0.95) -- (1.95,0.05);
\node[blue,rotate=315] at (1.75,0.65) {{\large {\bf {$\widehat{a}^{(0)}_1$}}}};
\draw[ultra thick,blue,<->] (2.05,-0.05) -- (2.95,-0.95);
\node[blue,rotate=315] at (2.75,-0.35) {{\large {\bf {$\widehat{a}^{(0)}_2$}}}};
\draw[ultra thick,blue,<->] (3.05,-1.05) -- (3.95,-1.95);
\node[blue,rotate=315] at (3.75,-1.35) {{\large {\bf {$\widehat{a}^{(0)}_3$}}}};
%definition S
\draw[dashed] (3,-3) -- (1.35,-4.65);
\draw[dashed] (0.3,-2.7) -- (-0.25,-3.25);
\draw[ultra thick,blue,<->] (-0.15,-3.25) -- (1.35,-4.65);
\node[blue] at (0.3,-4.2) {{\large {\bf {$S^{(0)}$}}}};
%parameter R
\draw[dashed] (0.3,-2.7) -- (-1.2,-2.7);
\draw[dashed] (2,-3) -- (-1.2,-3);
\draw[ultra thick,blue,->] (-1.2,-3.5) -- (-1.2,-3.05) ;
\draw[ultra thick,blue,<-] (-1.2,-2.65) -- (-1.2,-2.1) ;
\node[blue,rotate=90] at (-1.45,-2.7) {{\large {\bf{$R^{(0)}-2S^{(0)}$}}}};
%label
\node at (1,-4.95) {{\large web diagram in \figref{Fig:31webA}}};
\end{scope}
%%%%%%%%%%%%%%%%%%%%%%%%%%%%%
\begin{scope}[xshift=8cm]
\node[red] at (5.15,-1.6) {\LARGE $E$};
\draw[line width=1.5mm,red,->] (4.2,-2) -- (6.2,-2);
\draw[line width=1.5mm,red,<-] (4.2,-2.4) -- (6.2,-2.4);
\node[red] at (5.15,-2.8) {\LARGE $E$};
\end{scope}
%%%%%%%%%%%%%%%%%%%%%%%%%%%%%
%%%%%%%%%%%%%%%%%%%%%%%%%%%%%
%central graph 2
%%%%%%%%%%%%%%%%%%%%%%%%%%%%%
\begin{scope}[xshift=16cm]
\draw[ultra thick] (-1,0) -- (0,0) -- (0,-1) -- (1,-1) -- (1,-2) -- (2,-2) -- (2,-3) -- (3,-3);
%diagonals
\draw[ultra thick] (0,0) -- (0.7,0.7);
\draw[ultra thick] (1,-1) -- (1.7,-0.3);
\draw[ultra thick] (2,-2) -- (2.7,-1.3);
\draw[ultra thick] (0,-1) -- (-0.7,-1.7);
\draw[ultra thick] (1,-2) -- (0.3,-2.7);
\draw[ultra thick] (2,-3) -- (1.3,-3.7);
%ends
\node at (-1.2,0) {\large {\bf $\mathbf I$}};
\node at (3.2,-3) {\large {\bf $\mathbf I$}};
\node at (0.85,0.85) {\large {$\mathbf 1$}};
\node at (1.85,-0.15) {\large {$\mathbf 2$}};
\node at (2.85,-1.15) {\large {$\mathbf 3$}};
\node at (-0.85,-1.8) {\large {$\mathbf 2$}};
\node at (0.15,-2.45) {\large {$\mathbf 3$}};
\node at (1.15,-3.85) {\large {$\mathbf 1$}};
%lables hotizontal
\node at (-0.5,0.25) {\large  {\bf $v''_3$}};
\node at (0.5,-0.775) {\large  {\bf $v''_1$}};
\node at (1.5,-1.775) {\large  {\bf $v''_2$}};
\node at (2.75,-3.225) {\large  {\bf $v''_3$}};
%lables vertical
\node at (-0.35,-0.4) {\large  {\bf $-v'_3$}};
\node at (0.65,-1.4) {\large  {\bf $-v'_1$}};
\node at (2.35,-2.3) {\large  {\bf $-v'_2$}};
%lables diagonal
\node at (0.65,0.2) {\large  {\bf $m''_1$}};
\node at (1.65,-0.8) {\large  {\bf $m''_2$}};
\node at (2.65,-1.8) {\large  {\bf $m''_3$}};
\node at (-0.2,-1.65) {\large  {\bf $m''_2$}};
\node at (0.8,-2.65) {\large  {\bf $m''_3$}};
\node at (1.8,-3.65) {\large  {\bf $m''_1$}};
%hexagons
%\node[red] at (-0.1,0.7) {\large  {\bf $S''_3$}};
%\node[red] at (0.75,-0.3) {\large  {\bf $S''_1$}};
%\node[red] at (1.75,-1.3) {\large  {\bf $S''_2$}};
%\node[red] at (2.95,-2.3) {\large  {\bf $S''_3$}};
%\node[red] at (-0.75,-0.8) {\large  {\bf $S''_1$}};
%\node[red] at (0.4,-1.8) {\large  {\bf $S''_2$}};
%\node[red] at (1.4,-3.1) {\large  {\bf $S''_3$}};
%\node[red] at (2.25,-3.35) {\large  {\bf $S''_1$}};
%roots
\draw[ultra thick,blue,<->] (1.05,0.95) -- (1.95,0.05);
\node[blue,rotate=315] at (1.75,0.65) {{\large {\bf {$\widehat{a}^{(0)}_1$}}}};
\draw[ultra thick,blue,<->] (2.05,-0.05) -- (2.95,-0.95);
\node[blue,rotate=315] at (2.75,-0.35) {{\large {\bf {$\widehat{a}^{(0)}_2$}}}};
\draw[ultra thick,blue,<->] (3.05,-1.05) -- (3.95,-1.95);
\node[blue,rotate=315] at (3.75,-1.35) {{\large {\bf {$\widehat{a}^{(0)}_3$}}}};
%definition S
\draw[dashed] (2,-2) -- (0.35,-3.65);
\draw[dashed] (0.3,-2.7) -- (-0.15,-3.15);
\draw[ultra thick,blue,<->] (-0.15,-3.15) -- (0.35,-3.65);
\node[blue] at (-0.2,-3.6) {{\large {\bf {$S^{(0)}$}}}};
%parameter R
\draw[dashed] (0.3,-2.7) -- (-1.2,-2.7);
\draw[dashed] (1,-2) -- (-1.2,-2);
%\draw[ultra thick,blue,->] (-1.2,-3.5) -- (-1.2,-3.05) ;
\draw[ultra thick,blue,<->] (-1.2,-2.65) -- (-1.2,-2.1) ;
\node[blue,rotate=90] at (-1.45,-2.3) {{\large {\bf{$R^{(0)}-S^{(0)}$}}}};
%label
\node at (1,-4.95) {{\large web diagram in \figref{Fig:31webB}}};
\end{scope}
%%%%%%%%%%%%%%%%%%%%%%%%%%%%%
\begin{scope}[xshift=3cm]
\node[red] at (0.85,2) {\LARGE $G_4$};
\draw[line width=1.5mm,red,->] (1.3,1.3) -- (1.3,3.1);
\draw[line width=1.5mm,red,<-] (1.7,1.3) -- (1.7,3.1);
\node[red] at (2.25,2) {\LARGE $G_5$};
\end{scope}
%%%%%%%%%%%%%%%%%%%%%%%%%%%%%
%%%%%%%%%%%%%%%%%%%%%%%%%%%%%
%graph top left
%%%%%%%%%%%%%%%%%%%%%%%%%%%%%
\begin{scope}[xshift=3.5cm,yshift=8.5cm]
\draw (-2.05,1.15) -- (4.1,1.15) -- (4.1,-5.2) -- (-2.05,-5.2) -- (-2.05,1.15);
\draw[ultra thick] (-1,0) -- (0,0) -- (0,-1) -- (1,-1) -- (1,-2) -- (2,-2) -- (2,-3) -- (3,-3);
%diagonals
\draw[ultra thick] (0,0) -- (0.7,0.7);
\draw[ultra thick] (1,-1) -- (1.7,-0.3);
\draw[ultra thick] (2,-2) -- (2.7,-1.3);
\draw[ultra thick] (0,-1) -- (-0.7,-1.7);
\draw[ultra thick] (1,-2) -- (0.3,-2.7);
\draw[ultra thick] (2,-3) -- (1.3,-3.7);
%ends
\node at (-1.2,0) {\large {\bf $\mathbf a$}};
\node at (3.2,-3) {\large {\bf $\mathbf a$}};
\node at (0.85,0.85) {\large {$\mathbf 1$}};
\node at (1.85,-0.15) {\large {$\mathbf 2$}};
\node at (2.85,-1.15) {\large {$\mathbf 3$}};
\node at (-0.85,-1.85) {\large {$\mathbf 3$}};
\node at (0.15,-2.45) {\large {$\mathbf 1$}};
\node at (1.15,-3.85) {\large {$\mathbf 2$}};
%lables hotizontal
\node at (-0.5,0.25) {\large  {\bf $-h_1$}};
\node at (0.5,-0.775) {\large  {\bf $-h_2$}};
\node at (1.5,-1.775) {\large  {\bf $-h_3$}};
\node at (2.5,-2.775) {\large  {\bf $-h_1$}};
%lables vertical
\node at (-0.25,-0.4) {\large  {\bf $m'_1$}};
\node at (0.75,-1.4) {\large  {\bf $m'_2$}};
\node at (1.75,-2.4) {\large  {\bf $m'_3$}};
%lables diagonal
\node at (0.6,0.25) {\large  {\bf $v'_3$}};
\node at (1.65,-0.8) {\large  {\bf $v'_1$}};
\node at (2.65,-1.8) {\large  {\bf $v'_2$}};
\node at (-0.25,-1.6) {\large  {\bf $v'_2$}};
\node at (0.7,-2.65) {\large  {\bf $v'_3$}};
\node at (1.7,-3.65) {\large  {\bf $v'_1$}};
%hexagons
%\node[red] at (-0.1,0.7) {\large  {\bf $S^{(4)}_3$}};
%\node[red] at (0.75,-0.3) {\large  {\bf $S^{(4)}_1$}};
%\node[red] at (1.75,-1.3) {\large  {\bf $S^{(4)}_2$}};
%\node[red] at (2.75,-2.3) {\large  {\bf $S^{(4)}_3$}};
%\node[red] at (-0.75,-0.8) {\large  {\bf $S^{(4)}_2$}};
%\node[red] at (0.4,-1.8) {\large  {\bf $S^{(4)}_3$}};
%\node[red] at (1.3,-2.7) {\large  {\bf $S^{(4)}_1$}};
%\node[red] at (2.2,-3.35) {\large  {\bf $S^{(4)}_2$}};
%roots
\draw[ultra thick,blue,<->] (1.05,0.95) -- (1.95,0.05);
\node[blue,rotate=315] at (1.75,0.65) {{\large {\bf {$\widehat{a}^{(4)}_3$}}}};
\draw[ultra thick,blue,<->] (2.05,-0.05) -- (2.95,-0.95);
\node[blue,rotate=315] at (2.75,-0.35) {{\large {\bf {$\widehat{a}^{(4)}_1$}}}};
\draw[ultra thick,blue,<->] (3.05,-1.05) -- (3.95,-1.95);
\node[blue,rotate=315] at (3.75,-1.35) {{\large {\bf {$\widehat{a}^{(4)}_2$}}}};
%definition S
\draw[dashed] (3,-3) -- (1.35,-4.65);
\draw[dashed] (0.3,-2.7) -- (-0.25,-3.25);
\draw[ultra thick,blue,<->] (-0.15,-3.25) -- (1.35,-4.65);
\node[blue] at (0.3,-4.2) {{\large {\bf {$S^{(4)}$}}}};
%parameter R
\draw[dashed] (0.3,-2.7) -- (-1.2,-2.7);
\draw[dashed] (2,-3) -- (-1.2,-3);
\draw[ultra thick,blue,->] (-1.2,-3.5) -- (-1.2,-3.05) ;
\draw[ultra thick,blue,<-] (-1.2,-2.65) -- (-1.2,-2.1) ;
\node[blue,rotate=90] at (-1.45,-2.7) {{\large {\bf{$R^{(4)}-2S^{(4)}$}}}};
%label
\node at (1,-4.95) {{\large web diagram in \figref{Fig:31web4}(b)}};
\end{scope}
%%%%%%%%%%%%%%%%%%%%%%%%%%%%%
\begin{scope}[xshift=12cm]
\node[red] at (0.85,2) {\LARGE $G_5$};
\draw[line width=1.5mm,red,->] (1.3,1.3) -- (1.3,3.1);
\draw[line width=1.5mm,red,<-] (1.7,1.3) -- (1.7,3.1);
\node[red] at (2.25,2) {\LARGE $G_4$};
\end{scope}
%%%%%%%%%%%%%%%%%%%%%%%%%%%%%
%%%%%%%%%%%%%%%%%%%%%%%%%%%%%
%graph top right
%%%%%%%%%%%%%%%%%%%%%%%%%%%%%
\begin{scope}[xshift=12.5cm,yshift=8.5cm]
\draw (-2.05,1.15) -- (4.1,1.15) -- (4.1,-5.2) -- (-2.05,-5.2) -- (-2.05,1.15);
\draw[ultra thick] (-1,0) -- (0,0) -- (0,-1) -- (1,-1) -- (1,-2) -- (2,-2) -- (2,-3) -- (3,-3);
%diagonals
\draw[ultra thick] (0,0) -- (0.7,0.7);
\draw[ultra thick] (1,-1) -- (1.7,-0.3);
\draw[ultra thick] (2,-2) -- (2.7,-1.3);
\draw[ultra thick] (0,-1) -- (-0.7,-1.7);
\draw[ultra thick] (1,-2) -- (0.3,-2.7);
\draw[ultra thick] (2,-3) -- (1.3,-3.7);
%ends
\node at (-1.2,0) {\large {\bf $\mathbf a$}};
\node at (3.2,-3) {\large {\bf $\mathbf a$}};
\node at (0.85,0.85) {\large {$\mathbf 1$}};
\node at (1.85,-0.15) {\large {$\mathbf 2$}};
\node at (2.85,-1.15) {\large {$\mathbf 3$}};
\node at (-0.85,-1.5) {\large {$\mathbf 1$}};
\node at (0.15,-2.75) {\large {$\mathbf 2$}};
\node at (1.15,-3.85) {\large {$\mathbf 3$}};
%lables hotizontal
\node at (-0.5,0.25) {\large  {\bf $-v'_3$}};
\node at (0.5,-0.775) {\large  {\bf $-v'_2$}};
\node at (1.5,-1.775) {\large  {\bf $-v'_1$}};
\node at (2.5,-2.775) {\large  {\bf $-v'_3$}};
%lables vertical
\node at (-0.25,-0.5) {\large  {\bf $m''_1$}};
\node at (0.75,-1.5) {\large  {\bf $m''_3$}};
\node at (1.75,-2.5) {\large  {\bf $m''_2$}};
%lables diagonal
\node at (0.6,0.3) {\large  {\bf $v''_3$}};
\node at (1.6,-0.7) {\large  {\bf $v''_2$}};
\node at (2.6,-1.7) {\large  {\bf $v''_1$}};
\node at (-0.25,-1.55) {\large  {\bf $v''_3$}};
\node at (0.8,-2.55) {\large  {\bf $v''_2$}};
\node at (1.7,-3.6) {\large  {\bf $v''_1$}};
%hexagons
%\node[red] at (-0.2,0.7) {\large  {\bf $S^{(5)}_3$}};
%\node[red] at (0.75,-0.3) {\large  {\bf $S^{(5)}_1$}};
%\node[red] at (1.75,-1.3) {\large  {\bf $S^{(5)}_2$}};
%\node[red] at (2.75,-2.3) {\large  {\bf $S^{(5)}_3$}};
%\node[red] at (-0.75,-0.8) {\large  {\bf $S^{(5)}_3$}};
%\node[red] at (0.25,-1.7) {\large  {\bf $S^{(5)}_1$}};
%\node[red] at (1.3,-2.7) {\large  {\bf $S^{(5)}_2$}};
%\node[red] at (2.2,-3.35) {\large  {\bf $S^{(5)}_3$}};
%roots
\draw[ultra thick,blue,<->] (1.05,0.95) -- (1.95,0.05);
\node[blue,rotate=315] at (1.75,0.65) {{\large {\bf {$\widehat{a}^{(5)}_3$}}}};
\draw[ultra thick,blue,<->] (2.05,-0.05) -- (2.95,-0.95);
\node[blue,rotate=315] at (2.75,-0.35) {{\large {\bf {$\widehat{a}^{(5)}_2$}}}};
\draw[ultra thick,blue,<->] (3.05,-1.05) -- (3.95,-1.95);
\node[blue,rotate=315] at (3.75,-1.35) {{\large {\bf {$\widehat{a}^{(5)}_1$}}}};
%definition S
\draw[dashed] (3,-3) -- (1.35,-4.65);
\draw[dashed] (-0.7,-1.7) -- (-1.2,-2.2);
\draw[ultra thick,blue,<->] (-1.2,-2.2) -- (1.35,-4.65);
\node[blue] at (-0.1,-3.7) {{\large {\bf {$S^{(5)}$}}}};
%parameter R
\draw[dashed] (-0.7,-1.7) -- (-1.5,-1.7);
\draw[dashed] (2,-3) -- (-1.5,-3);
\draw[ultra thick,blue,<->] (-1.5,-2.95) -- (-1.5,-1.75);
\node[blue,rotate=90] at (-1.75,-2.3) {{\large {\bf{$R^{(5)}-3S^{(5)}$}}}};
%label
\node at (1,-4.95) {{\large web diagram in \figref{Fig:31web5}(b)}};
\end{scope}
%%%%%%%%%%%%%%%%%%%%%%%%%%%%%
%graph bottom left
%%%%%%%%%%%%%%%%%%%%%%%%%%%%%
\begin{scope}[xshift=-2cm,yshift=-8.5cm]
\draw (-2.05,1.15) -- (4.1,1.15) -- (4.1,-5.2) -- (-2.05,-5.2) -- (-2.05,1.15);
\draw[ultra thick] (-1,0) -- (0,0) -- (0,-1) -- (1,-1) -- (1,-2) -- (2,-2) -- (2,-3) -- (3,-3);
%diagonals
\draw[ultra thick] (0,0) -- (0.7,0.7);
\draw[ultra thick] (1,-1) -- (1.7,-0.3);
\draw[ultra thick] (2,-2) -- (2.7,-1.3);
\draw[ultra thick] (0,-1) -- (-0.7,-1.7);
\draw[ultra thick] (1,-2) -- (0.3,-2.7);
\draw[ultra thick] (2,-3) -- (1.3,-3.7);
%ends
\node at (-1.2,0) {\large {\bf $\mathbf a$}};
\node at (3.2,-3) {\large {\bf $\mathbf a$}};
\node at (0.8,0.9) {\large {$\mathbf{1}$}};
\node at (1.8,-0.15) {\large {$\mathbf{2}$}};
\node at (2.85,-1.15) {\large {$\mathbf{3}$}};
\node at (-0.9,-1.5) {\large {$\mathbf{1}$}};
\node at (0.1,-2.8) {\large {$\mathbf{2}$}};
\node at (1.15,-3.85) {\large {$\mathbf{3}$}};
%lables hotizontal
\node at (-0.5,0.25) {\large  {\bf $h_1$}};
\node at (0.5,-0.775) {\large  {\bf $h_3$}};
\node at (1.5,-1.775) {\large  {\bf $h_2$}};
\node at (2.5,-2.775) {\large  {\bf $h_1$}};
%lables vertical
\node at (-0.25,-0.5) {\large  {\bf $m_3$}};
\node at (0.75,-1.5) {\large  {\bf $m_2$}};
\node at (1.75,-2.5) {\large  {\bf $m_1$}};
%lables diagonal
\node at (0.65,0.2) {\large  {\bf $v_3$}};
\node at (1.65,-0.8) {\large  {\bf $v_2$}};
\node at (2.65,-1.8) {\large  {\bf $v_1$}};
\node at (-0.2,-1.65) {\large  {\bf $v_3$}};
\node at (0.8,-2.65) {\large  {\bf $v_2$}};
\node at (1.8,-3.65) {\large  {\bf $v_1$}};
%hexagons
%\node[red] at (-0.2,0.6) {\large  {\bf $S^{(1)}_3$}};
%\node[red] at (0.75,-0.3) {\large  {\bf $S^{(1)}_1$}};
%\node[red] at (1.75,-1.3) {\large  {\bf $S^{(1)}_2$}};
%\node[red] at (2.75,-2.3) {\large  {\bf $S^{(1)}_3$}};
%\node[red] at (-0.75,-0.8) {\large  {\bf $S^{(1)}_3$}};
%\node[red] at (0.3,-1.7) {\large  {\bf $S^{(1)}_1$}};
%\node[red] at (1.3,-2.7) {\large  {\bf $S^{(1)}_2$}};
%\node[red] at (2.2,-3.35) {\large  {\bf $S^{(1)}_3$}};
%roots
\draw[ultra thick,blue,<->] (1.05,0.95) -- (1.95,0.05);
\node[blue,rotate=315] at (1.75,0.65) {{\large {\bf {$\widehat{a}^{(1)}_2$}}}};
\draw[ultra thick,blue,<->] (2.05,-0.05) -- (2.95,-0.95);
\node[blue,rotate=315] at (2.75,-0.35) {{\large {\bf {$\widehat{a}^{(1)}_1$}}}};
\draw[ultra thick,blue,<->] (3.05,-1.05) -- (3.95,-1.95);
\node[blue,rotate=315] at (3.75,-1.35) {{\large {\bf {$\widehat{a}^{(1)}_3$}}}};
%definition S
\draw[dashed] (3,-3) -- (1.35,-4.65);
\draw[dashed] (-0.7,-1.7) -- (-1.2,-2.2);
\draw[ultra thick,blue,<->] (-1.2,-2.2) -- (1.35,-4.65);
\node[blue] at (-0.1,-3.7) {{\large {\bf {$S^{(1)}$}}}};
%parameter R
\draw[dashed] (-0.7,-1.7) -- (-1.5,-1.7);
\draw[dashed] (2,-3) -- (-1.5,-3);
\draw[ultra thick,blue,<->] (-1.5,-2.95) -- (-1.5,-1.75);
\node[blue,rotate=90] at (-1.75,-2.3) {{\large {\bf{$R^{(1)}-3S^{(1)}$}}}};
%label
\node at (1,-4.95) {{\large web diagram in \figref{Fig:31web1}(b)}};
\end{scope}
%%%%%%%%%%%%%%%%%%%%%%%%%%%%%
%graph bottom center
%%%%%%%%%%%%%%%%%%%%%%%%%%%%%
\begin{scope}[xshift=8cm,yshift=-8.5cm]
\draw (-2.05,1.15) -- (4.1,1.15) -- (4.1,-5.2) -- (-2.05,-5.2) -- (-2.05,1.15);
\draw[ultra thick] (-1,0) -- (0,0) -- (0,-1) -- (1,-1) -- (1,-2) -- (2,-2) -- (2,-3) -- (3,-3);
%diagonals
\draw[ultra thick] (0,0) -- (0.7,0.7);
\draw[ultra thick] (1,-1) -- (1.7,-0.3);
\draw[ultra thick] (2,-2) -- (2.7,-1.3);
\draw[ultra thick] (0,-1) -- (-0.7,-1.7);
\draw[ultra thick] (1,-2) -- (0.3,-2.7);
\draw[ultra thick] (2,-3) -- (1.3,-3.7);
%ends
\node at (-1.2,0) {\large {\bf $\mathbf a$}};
\node at (3.2,-3) {\large {\bf $\mathbf a$}};
\node at (0.85,0.85) {\large {$\mathbf 1$}};
\node at (1.8,-0.1) {\large {$\mathbf{2}$}};
\node at (2.85,-1.15) {\large {$\mathbf{3}$}};
\node at (-0.85,-1.85) {\large {$\mathbf{3}$}};
\node at (0.15,-2.45) {\large {$\mathbf{1}$}};
\node at (1.15,-3.9) {\large {$\mathbf{2}$}};
%lables hotizontal
\node at (-0.5,0.3) {\large  {\bf $v'_3$}};
\node at (0.5,-0.75) {\large  {\bf $v'_1$}};
\node at (1.5,-1.75) {\large  {\bf $v'_2$}};
\node at (2.5,-2.75) {\large  {\bf $v'_3$}};
%lables vertical
\node at (-0.25,-0.4) {\large  {\bf $m'_2$}};
\node at (0.75,-1.4) {\large  {\bf $m'_3$}};
\node at (1.75,-2.4) {\large  {\bf $m'_1$}};
%lables diagonal
\node at (0.65,0.2) {\large  {\bf $-h_3$}};
\node at (1.65,-0.8) {\large  {\bf $-h_1$}};
\node at (2.65,-1.8) {\large  {\bf $-h_2$}};
\node at (-0.2,-1.65) {\large  {\bf $-h_2$}};
\node at (0.7,-2.75) {\large  {\bf $-h_3$}};
\node at (1.8,-3.65) {\large  {\bf $-h_1$}};
%hexagons
%\node[red] at (-0.1,0.7) {\large  {\bf $S^{(2)}_3$}};
%\node[red] at (0.75,-0.3) {\large  {\bf $S^{(2)}_1$}};
%\node[red] at (1.75,-1.3) {\large  {\bf $S^{(2)}_2$}};
%\node[red] at (2.75,-2.3) {\large  {\bf $S^{(2)}_3$}};
%\node[red] at (-0.75,-0.8) {\large  {\bf $S^{(2)}_2$}};
%\node[red] at (0.4,-1.8) {\large  {\bf $S^{(2)}_3$}};
%\node[red] at (1.3,-2.7) {\large  {\bf $S^{(2)}_1$}};
%\node[red] at (2.2,-3.35) {\large  {\bf $S^{(2)}_2$}};
%roots
\draw[ultra thick,blue,<->] (1.05,0.95) -- (1.95,0.05);
\node[blue,rotate=315] at (1.75,0.65) {{\large {\bf {$\widehat{a}^{(2)}_1$}}}};
\draw[ultra thick,blue,<->] (2.05,-0.05) -- (2.95,-0.95);
\node[blue,rotate=315] at (2.75,-0.35) {{\large {\bf {$\widehat{a}^{(2)}_2$}}}};
\draw[ultra thick,blue,<->] (3.05,-1.05) -- (3.95,-1.95);
\node[blue,rotate=315] at (3.75,-1.35) {{\large {\bf {$\widehat{a}^{(2)}_3$}}}};
%definition S
\draw[dashed] (3,-3) -- (1.35,-4.65);
\draw[dashed] (0.3,-2.7) -- (-0.25,-3.25);
\draw[ultra thick,blue,<->] (-0.15,-3.25) -- (1.35,-4.65);
\node[blue] at (0.3,-4.2) {{\large {\bf {$S^{(2)}$}}}};
%parameter R
\draw[dashed] (0.3,-2.7) -- (-1.2,-2.7);
\draw[dashed] (2,-3) -- (-1.2,-3);
\draw[ultra thick,blue,->] (-1.2,-3.5) -- (-1.2,-3.05) ;
\draw[ultra thick,blue,<-] (-1.2,-2.65) -- (-1.2,-2.1) ;
\node[blue,rotate=90] at (-1.45,-2.7) {{\large {\bf{$R^{(2)}-2S^{(2)}$}}}};
%label
\node at (1,-4.95) {{\large web diagram in \figref{Fig:31web2}(b)}};
\end{scope}
%%%%%%%%%%%%%%%%%%%%%%%%%%%%%
%graph bottom center
%%%%%%%%%%%%%%%%%%%%%%%%%%%%%
\begin{scope}[xshift=18cm,yshift=-8.5cm]
\draw (-2.05,1.15) -- (4.1,1.15) -- (4.1,-5.2) -- (-2.05,-5.2) -- (-2.05,1.15);
\draw[ultra thick] (-1,0) -- (0,0) -- (0,-1) -- (1,-1) -- (1,-2) -- (2,-2) -- (2,-3) -- (3,-3);
%diagonals
\draw[ultra thick] (0,0) -- (0.7,0.7);
\draw[ultra thick] (1,-1) -- (1.7,-0.3);
\draw[ultra thick] (2,-2) -- (2.7,-1.3);
\draw[ultra thick] (0,-1) -- (-0.7,-1.7);
\draw[ultra thick] (1,-2) -- (0.3,-2.7);
\draw[ultra thick] (2,-3) -- (1.3,-3.7);
%ends
\node at (-1.2,0) {\large {\bf $\mathbf a$}};
\node at (3.2,-3) {\large {\bf $\mathbf a$}};
\node at (0.85,0.85) {\large {$\mathbf 1$}};
\node at (1.85,-0.15) {\large {$\mathbf 2$}};
\node at (2.85,-1.15) {\large {$\mathbf 3$}};
\node at (-0.85,-1.85) {\large {$\mathbf 3$}};
\node at (0.15,-2.45) {\large {$\mathbf 1$}};
\node at (1.15,-3.85) {\large {$\mathbf 2$}};
%lables hotizontal
\node at (-0.5,0.25) {\large  {\bf $-h_1$}};
\node at (0.5,-0.775) {\large  {\bf $-h_2$}};
\node at (1.5,-1.775) {\large  {\bf $-h_3$}};
\node at (2.5,-2.775) {\large  {\bf $-h_1$}};
%lables vertical
\node at (-0.2,-0.4) {\large  {\bf $v'_1$}};
\node at (0.8,-1.4) {\large  {\bf $v'_2$}};
\node at (1.8,-2.4) {\large  {\bf $v'_3$}};
%lables diagonal
\node at (0.6,0.25) {\large  {\bf $m'_3$}};
\node at (1.65,-0.8) {\large  {\bf $m'_1$}};
\node at (2.65,-1.8) {\large  {\bf $m'_2$}};
\node at (-0.25,-1.6) {\large  {\bf $m'_2$}};
\node at (0.7,-2.65) {\large  {\bf $m'_3$}};
\node at (1.7,-3.65) {\large  {\bf $m'_1$}};
%hexagons
%\node[red] at (-0.1,0.7) {\large  {\bf $S^{(3)}_3$}};
%\node[red] at (0.75,-0.3) {\large  {\bf $S^{(3)}_1$}};
%\node[red] at (1.75,-1.3) {\large  {\bf $S^{(3)}_2$}};
%\node[red] at (2.75,-2.3) {\large  {\bf $S^{(3)}_3$}};
%\node[red] at (-0.75,-0.8) {\large  {\bf $S^{(3)}_2$}};
%\node[red] at (0.4,-1.8) {\large  {\bf $S^{(3)}_3$}};
%\node[red] at (1.3,-2.7) {\large  {\bf $S^{(3)}_1$}};
%\node[red] at (2.2,-3.35) {\large  {\bf $S^{(3)}_2$}};
%roots
\draw[ultra thick,blue,<->] (1.05,0.95) -- (1.95,0.05);
\node[blue,rotate=315] at (1.75,0.65) {{\large {\bf {$\widehat{a}^{(3)}_2$}}}};
\draw[ultra thick,blue,<->] (2.05,-0.05) -- (2.95,-0.95);
\node[blue,rotate=315] at (2.75,-0.35) {{\large {\bf {$\widehat{a}^{(3)}_3$}}}};
\draw[ultra thick,blue,<->] (3.05,-1.05) -- (3.95,-1.95);
\node[blue,rotate=315] at (3.75,-1.35) {{\large {\bf {$\widehat{a}^{(3)}_1$}}}};
%definition S
\draw[dashed] (3,-3) -- (1.35,-4.65);
\draw[dashed] (0.3,-2.7) -- (-0.25,-3.25);
\draw[ultra thick,blue,<->] (-0.15,-3.25) -- (1.35,-4.65);
\node[blue] at (0.3,-4.2) {{\large {\bf {$S^{(3)}$}}}};
%parameter R
\draw[dashed] (0.3,-2.7) -- (-1.2,-2.7);
\draw[dashed] (2,-3) -- (-1.2,-3);
\draw[ultra thick,blue,->] (-1.2,-3.5) -- (-1.2,-3.05) ;
\draw[ultra thick,blue,<-] (-1.2,-2.65) -- (-1.2,-2.1) ;
\node[blue,rotate=90] at (-1.45,-2.7) {{\large {\bf{$R^{(3)}-2S^{(3)}$}}}};
%label
\node at (1,-4.95) {{\large web diagram in \figref{Fig:31web3}}};
\end{scope}
%%%%%%%%%%%%%%%%%%%%%%%%%%%%%
\begin{scope}[xshift=4cm,yshift=8.5cm]
\node[red] at (5,-1.6) {\LARGE $G_4$};
\draw[line width=1.5mm,red,->] (3.75,-2) -- (6.25,-2);
\draw[line width=1.5mm,red,<-] (3.75,-2.4) -- (6.25,-2.4);
\node[red] at (5,-2.8) {\LARGE $G_5$};
\end{scope}
%%%%%%%%%%%%%%%%%%%%%%%%%%%%%
%%%%%%%%%%%%%%%%%%%%%%%%%%%%%
\begin{scope}[xshift=-1.5cm,yshift=-8.5cm]
\node[red] at (0.85,2) {\LARGE $G_1$};
\draw[line width=1.5mm,red,->] (1.3,1.3) -- (1.3,3.1);
\draw[line width=1.5mm,red,<-] (1.7,1.3) -- (1.7,3.1);
\node[red] at (2.25,2) {\LARGE $G_1$};
\end{scope}
%%%%%%%%%%%%%%%%%%%%%%%%%%%%%
%%%%%%%%%%%%%%%%%%%%%%%%%%%%%
\begin{scope}[xshift=7.5cm,yshift=-8.5cm]
\node[red] at (0.85,2) {\LARGE $G_2$};
\draw[line width=1.5mm,red,->] (1.3,1.3) -- (1.3,3.1);
\draw[line width=1.5mm,red,<-] (1.7,1.3) -- (1.7,3.1);
\node[red] at (2.25,2) {\LARGE $G_2$};
\end{scope}
%%%%%%%%%%%%%%%%%%%%%%%%%%%%%
%%%%%%%%%%%%%%%%%%%%%%%%%%%%%
\begin{scope}[xshift=16.5cm,yshift=-8.5cm]
\node[red] at (0.85,2) {\LARGE $G_3$};
\draw[line width=1.5mm,red,->] (1.3,1.3) -- (1.3,3.1);
\draw[line width=1.5mm,red,<-] (1.7,1.3) -- (1.7,3.1);
\node[red] at (2.25,2) {\LARGE $G_3$};
\end{scope}
%%%%%%%%%%%%%%%%%%%%%%%%%%%%%
\end{tikzpicture}}}}
\caption{\sl Presentations of web diagrams related to $X_{3,1}$. The transformations $G_{1,2,3,4,5}$ act on the basis of independent K\"ahler parameters $(\widehat{a}_1^{(0)},\widehat{a}_2^{(0)},\widehat{a}_3^{(0)},S^{(0)},R^{(0)})$. The organisation of web diagrams and transformations is reminiscent of the cycle graph of $\text{Dih}_3$.}
\label{Fig:Overview31Webs}
\end{center}
\end{figure}

\noindent
{\allowdisplaybreaks
\begin{align}
&h_1=3S^{(5)}-\widehat{a}^{(5)}_1-\widehat{a}^{(5)}_2-R^{(5)}\,,\hspace{0.25cm}h_2=3S^{(5)}-\widehat{a}^{(5)}_2-\widehat{a}^{(5)}_3-R^{(5)}\,,\hspace{0.25cm}h_3=3S^{(5)}-\widehat{a}^{(5)}_1-\widehat{a}^{(5)}_3-R^{(5)}\,,\nonumber\\
&v_1=v_2=v_3=2\left(\widehat{a}^{(5)}_1+\widehat{a}^{(5)}_2+\widehat{a}^{(5)}_3\right)-7S^{(5)}+2R^{(5)}\,,\nonumber\\
&m_1=m_2=m_3=\widehat{a}^{(5)}_1+\widehat{a}^{(5)}_2+\widehat{a}^{(5)}_3-5S^{(5)}+2R^{(5)}\,.\label{SolutionWeb315}
\end{align}}
Comparing (\ref{SolutionWeb315}) to (\ref{SolutionWeb310}) indeed gives rise to the following symmetry transformation
\begin{align}
&\left(\begin{array}{c}\widehat{a}^{(0)}_1\\ \widehat{a}^{(0)}_2  \\ \widehat{a}^{(0)}_3  \\ S^{(0)} \\ R^{(0)}\end{array}\right)=G_5\cdot \left(\begin{array}{c}\widehat{a}^{(5)}_1\\ \widehat{a}^{(5)}_2  \\ \widehat{a}^{(5)}_3  \\ S^{(5)} \\ R^{(5)}\end{array}\right)\,,&&\text{where} &&G_5=\left(
\begin{array}{ccccc}
 2 & 1 & 1 & -4 & 1 \\
 1 & 2 & 1 & -4 & 1 \\
 1 & 1 & 2 & -4 & 1 \\
 2 & 2 & 2 & -5 & 1 \\
 3 & 3 & 3 & -6 & 1 \\
\end{array}
\right) && \text{with} &&\begin{array}{l}\text{det }G_5=1\,, \\ G_5\cdot G_5\cdot G_5=1\!\!1_{5\times 5}\,.\end{array}\label{Gen31G5}
\end{align}
Other presentations of the web diagram of $X_{3,1}$ do not give rise to other symmetries than $G_{1,2,3,4,5}$, apart from a permutation of the parameters $\widehat{a}_{1,2,3}$. These latter symmetries form the group $S_3$, which, from the point of view of the gauge theory engineered by $X_{3,1}$, corresponds to the Weyl group of the gauge group $U(3)$. Factoring out this $S_3$,  the $5\times 5$ identity matrix $E=1\!\!1_{5\times 5}$ and the linear transformations $G_{1,2,3,4,5}$ form a finite group of order 6, which commute with $S_3$ and whose multiplication table is given by
\begin{align}
\begin{array}{c|cccccc}
& E & G_1 & G_2 & G_3 & G_4 & G_5 \\\hline
E & E  & G_1  & G_2  & G_3  & G_4  & G_5  \\
G_1 & G_1  & E  & G_5  & G_4  & G_3  & G_2  \\
G_2 & G_2  & G_4  & E  & G_5  & G_1  & G_3  \\
G_3 & G_3  & G_5  & G_4  & E  & G_2  & G_1  \\
G_4 & G_4  & G_2  & G_3  & G_1  & G_5  & E  \\
G_5 & G_5  & G_3  & G_1  & G_2  & E  & G_4  \\
\end{array}
\end{align} 
This table is the same as the one of the dihedral group $\text{Dih}_3$, such that we have
\begin{align}
\mathbb{G}(3)\cong\{E,G_1,G_2,G_3,G_4,G_5\}\cong \text{Dih}_3\,.\label{DefDih3}
\end{align}
An overview over the group elements $G_{1,2,3,4,5}$ and their relation to different presentations of the web diagram \figref{Fig:31web0} are shown in \figref{Fig:Overview31Webs}. 

For later use, we remark that the dihedral group (\ref{DefDih3}) can also be represented as the group that is freely generated by $G_2$ and $G_3$
\begin{align}
&\mathbb{G}(3)\cong\langle\{ G_2,G_3\}\rangle\,,&&\text{with} &&\begin{array}{l} G_2\cdot G_2=1\!\!1_{5\times 5}=G_3\cdot G_3\,,\\ (G_2\cdot G_3)^3=1\!\!1_{5\times 5}\,.\end{array}\label{Dih3FreeGeneration}
\end{align}
%%%%%%%%%%%%%%%%%%%%%%%%%%%%%%%%%%%
%%%%%%%%%%%%%%%%%%%%%%%%%%%%%%%%%%%
\subsection{Invariance of the Non-perturbative Free Energy}\label{Sect:Order3Checks}
As in the previous example, following the result of \cite{Bastian:2017ing}, the linear transformations $G_{1,2,3,4,5}$ in eqs.~(\ref{Gen31G1}), (\ref{Gen31G2}), (\ref{Gen31G3}), (\ref{Gen31G4}) and (\ref{Gen31G5}) correspond to symmetries of the free energy $\mathcal{F}_{3,1}(\widehat{a}_{1,2,3},S,R;\epsilon_1,\epsilon_2)$, as defined in (\ref{DefFreeEnergyGen}). In this section we provide evidence for this symmetry, however, for simplicity we limit ourselves to checking the leading limit in $\epsilon_{1,2}$ of the free energy. To this end, we introduce the following expansion 
\begin{align}
&\lim_{\epsilon_{1,2}\to 0} \epsilon_1\,\epsilon_2\,\mathcal{F}_{3,1}(\widehat{a}_{1,2,3},S,R;\epsilon_1,\epsilon_2)=\sum_{n=0}^\infty \sum_{i_1,i_2,i_3=0}^\infty \sum_{k\in\mathbb{Z}}f^{\text{NS}}_{i_1,i_2,i_3,k,n}\,\widehat{Q}_1^{i_1}\,\widehat{Q}_2^{i_2}\,\widehat{Q}_3^{i_3}\,Q_S^{k}\,Q_{R}^n\,,
\end{align}
where $f^{\text{NS}}_{i_1,i_2,i_3,k,n}\in\mathbb{Z}$ and $\widehat{Q}_i=e^{-\widehat{a}_i}$ (for $i=1,2,3$), $Q_S=e^{- S}$ and $Q_R=e^{-R}$. %{\color{red}{Draw triangle diagrams to show symmetry under $S_3$}}
As explained in section~\ref{Sect:GeneralStrategy}, the fact that the (shifted) web diagrams in \figref{Fig:Overview21Webs} all give rise to the same partition functions implies
\begin{align}
&f^{\text{NS}}_{i_1,i_2,i_3,k,n}=f^{\text{NS}}_{i'_1,i'_2,i'_3,k',n'}&&\text{for} && (i'_1,i'_2,i'_3,k', n')^T=G^T_\ell\cdot (i_1,i_2,i_3,k, n)^T&&\forall \ell=1,2,3,4,5\,.
%\left(\begin{array}{c}i'_1 \\ i'_2 \\ i'_3 \\ k' \\ n'\end{array}\right)=G^T_\ell\cdot \left(\begin{array}{c}i_1 \\ i_2 \\ i_3 \\ k \\ n\end{array}\right)&&\forall \ell=1,2,3,4,5\,.
\end{align}
In Tables~\ref{Tab:SymTrafo31G1}, \ref{Tab:SymTrafo31G3} and \ref{Tab:SymTrafo31G4} we tabulate coefficients $f^{\text{NS}}_{i_1,i_2,i_3,k,n}$ with $i_1+i_2+i_3\leq 7$ for $n=1$ and $n=2$ that are related by $G_{1,2,3,4,5}$.\footnote{We do not display symmetries between coefficients that also involve purely $S_3$ transformations.}
\begin{table}[hbp]
\begin{center}
\parbox{8cm}{\begin{tabular}{|c|c|c|}\hline
&&\\[-10pt]
$(i_1,i_2,i_3,k,n)$ & $(i'_1,i'_2,i'_3,k',n')$ & $f^{\text{NS}}_{i_1,i_2,i_3,k,n}$\\[4pt]\hline\hline
&&\\[-13pt]
$(0,0,0,-1,1) $ & $(1,1,1,-5,2)$ & $-3$\\\hline
&&\\[-13pt]
$(0,0,2,-2,1)$ & $(1,1,3,-6,2)$ & 4\\\hline
&&\\[-13pt]
$(0,1,1,-2,1)$ & $(1,2,2,-6,2)$ & 8\\\hline
&&\\[-13pt]
$(0,1,3,-3,1)$ & $(1,2,4,-7,2)$ & $-5$\\\hline
&&\\[-13pt]
%$(0,2,0,-2,1)$ & $(3,1,1,-6,2)$ & 4\\\hline
%&&\\[-13pt]
$(0,2,2,-3,1)$ & $(1,3,3,-7,2)$ & $-4$\\\hline
&&\\[-13pt]
%$(0,3,1,-3,1)$ & $(4,1,2,-7,2)$ & $-5$\\\hline
%&&\\[-13pt]
%$(1,0,1,-2,1)$ & $(1,2,2,-6,2)$ & 8\\\hline
$(1,1,2,-3,1)$ & $(2,2,3,-7,2)$ & $-25$\\\hline
\end{tabular}
\vspace{2.85cm}}
\hspace{0.1cm}
\begin{tabular}{|c|c|c|}\hline
&&\\[-10pt]
$(i_1,i_2,i_3,k,n)$ & $(i''_1,i''_2,i''_3,k'',n'')$ & $f^{\text{NS}}_{i_1,i_2,i_3,k,n}$\\[4pt]\hline\hline
&&\\[-13pt]
$(0,0,1,0,1)$ & $(0,0,1,-2,2)$ & 12\\\hline
&&\\[-13pt]
$(0,0,2,-1,1)$ & $(0,0,2,-3,2)$ & $-16$\\\hline
&&\\[-13pt]
$(0,0,3,-2,1)$ & $(0,0,3,-4,2)$ & 6\\\hline
&&\\[-13pt]
$(0,1,1,-1,1)$ & $(0,1,1,-3,2)$ & $-23$\\\hline
&&\\[-13pt]
$(0,1,2,-2,1)$ & $(0,1,2,-4,2)$ & 18\\\hline
&&\\[-13pt]
$(0,1,3,-3,1)$ & $(0,1,3,-5,2)$ & $-5$\\\hline
&&\\[-13pt]
$(0,2,2,-3,1)$ & $(0,2,2,-5,2)$ & $-4$\\\hline
&&\\[-13pt]
$(1,1,1,-2,1)$ & $(1,1,1,-4,2)$ & 42\\\hline
&&\\[-13pt]
$(1,1,2,-3,1)$ & $(1,1,2,-5,2)$ & $-25$\\\hline
&&\\[-13pt]
$(1,1,3,-4,1)$ & $(1,1,3,-6,2)$ & 4 \\\hline
&&\\[-13pt]
$(2,2,2,-5,1)$ & $(2,2,2,-7,2)$ & $-3$\\\hline
\end{tabular}
\end{center}
\caption{\sl Action of $G_1$ (left) and $G_2$ (right): the indices are related by $(i'_1,i'_2,i'_3,k',n')^T=G_1^T\cdot (i_1,i_2,i_3,k,n)^T$ and $(i''_1,i''_2,i''_3,k'',n')^T=G_2^T\cdot (i_1,i_2,i_3,k,n)^T$.}
\label{Tab:SymTrafo31G1}
\end{table}

\begin{table}[hbp]
\begin{center}
\begin{tabular}{|c|c|c|}\hline
&&\\[-10pt]
$(i_1,i_2,i_3,k,n)$ & $(i'_1,i'_2,i'_3,k',n')$ & $f^{\text{NS}}_{i_1,i_2,i_3,k,n}$\\[4pt]\hline\hline
&&\\[-13pt]
$(0,0,0,-1,1)$ & $(2,2,2,-5,1)$ & $-3$\\\hline
&&\\[-13pt]
$(0,0,1,-2,1)$ & $(1,1,2,-4,1)$ & $2$\\\hline
&&\\[-13pt]
$(0,0,1,-1,1)$ & $(2,2,3,-5,1)$ & $-8$\\\hline
&&\\[-13pt]
$(0,0,2,-2,1)$ & $(1,1,3,-4,1)$ & $4$\\\hline
&&\\[-13pt]
$(0,0,3,-2,1)$ & $(1,1,4,-4,1)$ & $6$\\\hline
&&\\[-13pt]
$(0,0,4,-2,1)$ & $(1,1,5,-4,1)$ & $8$\\\hline
&&\\[-13pt]
$(0,1,2,-2,1)$ & $(1,2,3,-4,1)$ & $18$\\\hline
&&\\[-13pt]
$(0,1,3,-2,1)$ & $(1,2,4,-4,1)$ & 30\\\hline
&&\\[-13pt]
$(0,2,2,-2,1)$ & $(1,3,3,-4,1)$ & 28\\\hline
&&\\[-13pt]
$(1,1,1,-2,1)$ & $(2,2,2,-4,1)$ & 42\\\hline
&&\\[-13pt]
$(1,1,2,-2,1)$ & $(2,2,3,-4,1)$ & 112\\\hline
&&\\[-13pt]
$(1,1,6,-4,1)$ & $(0,0,5,-2,1)$ & 10\\\hline
&&\\[-13pt]
$(1,1,7,-4,1)$ & $(0,0,6,-2,1)$ & 12\\\hline
&&\\[-13pt]
$(1,2,6,-4,1)$ & $(0,1,5,-2,1)$ & 54\\\hline
&&\\[-13pt]
$(1,3,4,-4,1)$ & $(0,2,3,-2,1)$ & 48\\\hline
\end{tabular}
\hspace{0.1cm}
\begin{tabular}{|c|c|c|}\hline
&&\\[-10pt]
$(i_1,i_2,i_3,k,n)$ & $(i'_1,i'_2,i'_3,k',n')$ & $f^{\text{NS}}_{i_1,i_2,i_3,k,n}$\\[4pt]\hline\hline
&&\\[-13pt]
$(1,3,5,-4,1)$ & $(0,2,4,-2,1)$ & 72\\\hline
&&\\[-13pt]
$(1,4,4,-4,1)$ & $(0,3,3,-2,1)$ & 60\\\hline
&&\\[-13pt]
$(2,2,4,-5,1)$ & $(0,0,2,-1,1)$ & $-16$\\\hline
&&\\[-13pt]
$(2,2,4,-4,1)$ & $(1,1,3,-2,1)$ & 208\\\hline
&&\\[-13pt]
$(2,2,5,-5,1)$ & $(0,0,3,-1,1)$ & $-24$\\\hline
&&\\[-13pt]
$(2,2,5,-4,1)$ & $(1,1,4,-2,1)$ & 312\\\hline
&&\\[-13pt]
$(2,3,3,-5,1)$ & $(0,1,1,-1,1)$ & $-23$\\\hline
&&\\[-13pt]
$(2,3,3,-4,1)$ & $(1,2,2,-2,1)$ & 286\\\hline
&&\\[-13pt]
$(2,3,4,-5,1)$ & $(0,1,2,-1,1)$ & $-45$\\\hline
&&\\[-13pt]
$(2,3,4,-4,1)$ & $(1,2,3,-2,1)$ & 540\\\hline
&&\\[-13pt]
$(3,3,3,-4,1)$ & $(2,2,2,-2,1)$ & 948\\\hline
&&\\[-13pt]
$(0,1,3,-5,2)$ & $(1,2,4,-7,2)$ & $-5$\\\hline
&&\\[-13pt]
$(0,2,2,-5,2)$ & $(1,3,3,-7,2)$ & $-4$\\\hline
&&\\[-13pt]
$(1,1,1,-5,2)$ & $(2,2,2,-7,2)$ & $-3$\\\hline
&&\\[-13pt]
$(1,1,2,-5,2)$ & $(2,2,3,-7,2)$ & $-25$\\\hline
\end{tabular}
\end{center}
\caption{\sl Action of $G_3$: the indices are related by $(i'_1,i'_2,i'_3,k',n')^T=G_3^T\cdot (i_1,i_2,i_3,k,n)^T$.}
\label{Tab:SymTrafo31G3}
\end{table}

\begin{table}[htbp]
\begin{center}
\begin{tabular}{|c|c|c|c|}\hline
&&&\\[-10pt]
$(i_1,i_2,i_3,k,n)$ & $(i'_1,i'_2,i'_3,k',n')$ & $(i''_1,i''_2,i''_3,k'',n'')$ & $f^{\text{NS}}_{i_1,i_2,i_3,k,n}$\\[4pt]\hline\hline
&&&\\[-13pt]
$(0,0,2,-2,1)$ & $(1,1,3,-6,2)$ & $(1,1,3,-4,1)$ & 4\\\hline
&&&\\[-13pt]
$(0,1,1,-2,1)$ & $(1,2,2,-6,2)$ & $(1,2,2,-4,1)$ & 8\\\hline
&&&\\[-13pt]
$(0,1,3,-3,1)$ & $(0,1,3,-5,2)$ & $(1,2,4,-7,2)$ & $-5$\\\hline
&&&\\[-13pt]
$(0,2,2,-3,1)$ & $(0,2,2,-5,2)$ & $(1,3,3,-7,2)$ & $-4$\\\hline
&&&\\[-13pt]
$(1,1,2,-3,1)$ & $(1,1,2,-5,2)$ & $(2,2,3,-7,2)$ & $-25$\\\hline
\end{tabular}
\end{center}
\caption{\sl Action of $G_4$ and $G_5$: the indices are related by $(i'_1,i'_2,i'_3,k',n')^T=G_4^T\cdot (i_1,i_2,i_3,k,n)^T$ and $(i''_1,i''_2,i''_3,k'',n'')^T=G_4^T\cdot G_4^T\cdot (i_1,i_2,i_3,k,n)^T$, as well as $(i_1,i_2,i_3,k,n)^T=G_5^T\cdot (i'_1,i'_2,i'_3,k',n')^T$ and $(i_1,i_2,i_3,k,n)^T=G_5^T\cdot G_5^T\cdot (i''_1,i''_2,i''_3,k'',n'')^T$.}
\label{Tab:SymTrafo31G4}
\end{table}
%%%%%%%%%%%%%%%%%%%%%%%%%%%%%%%%%%%%%%%%%%%%%%%%%
%%%%%%%%%%%%%%%%%%%%%%%%%%%%%%%%%%%%%%%%%%%%%%%%%
\subsection{Modularity at a Particular Point of the Moduli Space}
Similarly to the case $N=2$ above, we can analyse how the group $\mathbb{G}(3)$ is related to $Sp(4,\mathbb{Z})$ at the particular region in the moduli space, which is characterised by $\widehat{a}_1^{(0)}=\widehat{a}_2^{(0)}=\widehat{a}_3^{(0)}=\widehat{a}$, which implies $h_1=h_2=h_3=h$ (while the consistency conditions (\ref{ConsistencyWeb21}) already impose $v_1=v_2=v_3=v$ and $m_1=m_2=m_3=m$). As in the previous section, we can introduce the period matrix
\begin{align}
&\Omega=\left(\begin{array}{cc}\tau & v \\ v & \rho\end{array}\right)\,,&&\text{with} &&\begin{array}{l}\tau=m+v\,,\\ \rho=h+m\,.\end{array}\label{Def3PeriodMatrix}
\end{align} 
Using the parametrisation (\ref{Dih3FreeGeneration}) of $\mathbb{G}(3)$, it is sufficient to analyse the relation of the generators $G_2$ and $G_3$ to $Sp(4,\mathbb{Z})$. The restriction of these generators to the subspace $(\widehat{a},S,R)$ can be written in the form
\begin{align}
&G_2^{(\text{red})}=\left(\begin{array}{ccc}1 & -2 & 1 \\ 0 & -1 & 1 \\ 0 & 0 & 1\end{array}\right)\,,&&\text{and} &&G_3^{(\text{red})}=\left(\begin{array}{ccc}1 & 0 & 0 \\ 3 & -1 & 0 \\ 9 & -6 & 1\end{array}\right)\,,
\end{align}
Rewriting them furthermore to act as elements of $Sp(4,\mathbb{Z})$ in the form of (\ref{ActionSp4Period}) on the period matrix $\Omega$ in (\ref{Def3PeriodMatrix}), they take the form
\begin{align}
&\widetilde{G}_2^{(\text{red,Sp})}=HKL^6HKHL^6KH\,,&&\text{and} &&\widetilde{G}_3^{(\text{red,Sp})}=HKL^6KL^6KH\,,
\end{align}
where $K$, $L$ and $H$ are defined in appendix~\ref{App:Sp4}. As in the case of $N=2$, this implies that the restriction of $\mathbb{G}(3)$ to the particular region of the K\"ahler moduli space $(\widehat{a},S,R)$ is a subgroup of $Sp(4,\mathbb{Z})$. However, unlike the case $N=1$, we cannot conclude that the freely generated group $\langle \widetilde{G}_2^{(\text{red,Sp})},\widetilde{G}_3^{(\text{red,Sp})}, S_\rho,T_\rho,S_\tau,T_\tau \rangle$ is isomorphic to $Sp(4,\mathbb{Z})$.

%%%%%%%%%%%%%%%%%%%%%%%%%%%%%%%%%%%
%%%%%%%%%%%%%%%%%%%%%%%%%%%%%%%%%%%
%%%%%%%%%%%%%%%%%%%%%%%%%%%%%%%%%%%
%%%%%%%%%%%%%%%%%%%%%%%%%%%%%%%%%%%%%%%%%%%%%%%%
\section{Example: $(N,M)=(4,1)$}\label{Sect:Case41}
\subsection{Dualities and $\text{Dih}_\infty$ Group Action}
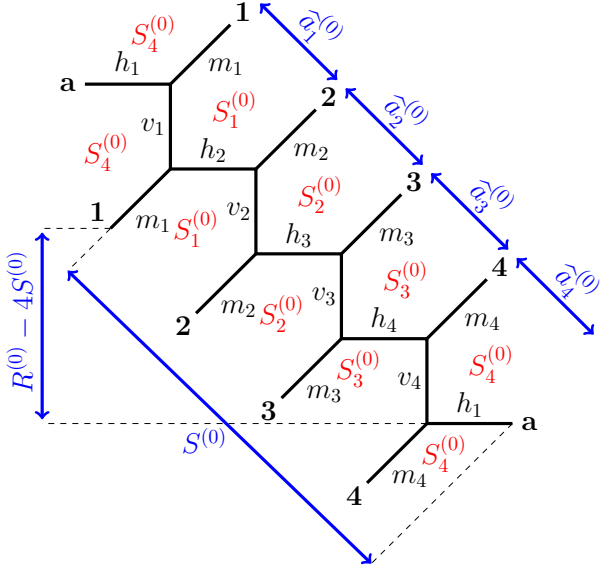
\begin{wrapfigure}{l}{0.49\textwidth}
\scalebox{0.75}{\parbox{10.5cm}{\begin{tikzpicture}[scale = 1.50]
\draw[ultra thick] (-1,0) -- (0,0) -- (0,-1) -- (1,-1) -- (1,-2) -- (2,-2) -- (2,-3) -- (3,-3) -- (3,-4) -- (4,-4);
%diagonals
\draw[ultra thick] (0,0) -- (0.7,0.7);
\draw[ultra thick] (1,-1) -- (1.7,-0.3);
\draw[ultra thick] (2,-2) -- (2.7,-1.3);
\draw[ultra thick] (3,-3) -- (3.7,-2.3);
\draw[ultra thick] (0,-1) -- (-0.7,-1.7);
\draw[ultra thick] (1,-2) -- (0.3,-2.7);
\draw[ultra thick] (2,-3) -- (1.3,-3.7);
\draw[ultra thick] (3,-4) -- (2.3,-4.7);
%ends
\node at (-1.2,0) {\large {\bf $\mathbf a$}};
\node at (4.2,-4) {\large {\bf $\mathbf a$}};
\node at (0.85,0.85) {\large {$\mathbf 1$}};
\node at (1.85,-0.15) {\large {$\mathbf 2$}};
\node at (2.85,-1.15) {\large {$\mathbf 3$}};
\node at (3.85,-2.15) {\large {$\mathbf 4$}};
\node at (-0.85,-1.5) {\large {$\mathbf 1$}};
\node at (0.15,-2.85) {\large {$\mathbf 2$}};
\node at (1.15,-3.85) {\large {$\mathbf 3$}};
\node at (2.15,-4.85) {\large {$\mathbf 4$}};
%lables hotizontal
\node at (-0.5,0.25) {\large  {\bf $h_1$}};
\node at (0.5,-0.775) {\large  {\bf $h_2$}};
\node at (1.5,-1.775) {\large  {\bf $h_3$}};
\node at (2.5,-2.775) {\large  {\bf $h_4$}};
\node at (3.5,-3.775) {\large  {\bf $h_1$}};
%lables vertical
\node at (-0.2,-0.5) {\large  {\bf $v_1$}};
\node at (0.8,-1.5) {\large  {\bf $v_2$}};
\node at (1.8,-2.5) {\large  {\bf $v_3$}};
\node at (2.8,-3.5) {\large  {\bf $v_4$}};
%lables diagonal
\node at (0.65,0.2) {\large  {\bf $m_1$}};
\node at (1.65,-0.8) {\large  {\bf $m_2$}};
\node at (2.65,-1.8) {\large  {\bf $m_3$}};
\node at (3.65,-2.8) {\large  {\bf $m_4$}};
\node at (-0.2,-1.65) {\large  {\bf $m_1$}};
\node at (0.8,-2.65) {\large  {\bf $m_2$}};
\node at (1.8,-3.65) {\large  {\bf $m_3$}};
\node at (2.8,-4.65) {\large  {\bf $m_4$}};
%hexagons
\node[red] at (-0.2,0.6) {\large  {\bf $S^{(0)}_4$}};
\node[red] at (0.75,-0.3) {\large  {\bf $S^{(0)}_1$}};
\node[red] at (1.75,-1.3) {\large  {\bf $S^{(0)}_2$}};
\node[red] at (2.75,-2.3) {\large  {\bf $S^{(0)}_3$}};
\node[red] at (3.75,-3.3) {\large  {\bf $S^{(0)}_4$}};
\node[red] at (-0.75,-0.8) {\large  {\bf $S^{(0)}_4$}};
\node[red] at (0.3,-1.7) {\large  {\bf $S^{(0)}_1$}};
\node[red] at (1.3,-2.7) {\large  {\bf $S^{(0)}_2$}};
\node[red] at (2.2,-3.35) {\large  {\bf $S^{(0)}_3$}};
\node[red] at (3.2,-4.35) {\large  {\bf $S^{(0)}_4$}};
%roots
\draw[ultra thick,blue,<->] (1.05,0.95) -- (1.95,0.05);
\node[blue,rotate=315] at (1.75,0.65) {{\large {\bf {$\widehat{a}^{(0)}_1$}}}};
\draw[ultra thick,blue,<->] (2.05,-0.05) -- (2.95,-0.95);
\node[blue,rotate=315] at (2.75,-0.35) {{\large {\bf {$\widehat{a}^{(0)}_2$}}}};
\draw[ultra thick,blue,<->] (3.05,-1.05) -- (3.95,-1.95);
\node[blue,rotate=315] at (3.75,-1.35) {{\large {\bf {$\widehat{a}^{(0)}_3$}}}};
\draw[ultra thick,blue,<->] (4.05,-2.05) -- (4.95,-2.95);
\node[blue,rotate=315] at (4.75,-2.35) {{\large {\bf {$\widehat{a}^{(0)}_4$}}}};
%definition S
\draw[dashed] (4,-4) -- (2.35,-5.65);
\draw[dashed] (-0.7,-1.7) -- (-1.2,-2.2);
\draw[ultra thick,blue,<->] (-1.2,-2.2) -- (2.35,-5.65);
\node[blue] at (0.4,-4.2) {{\large {\bf {$S^{(0)}$}}}};
%parameter R
\draw[dashed] (-0.7,-1.7) -- (-1.5,-1.7);
\draw[dashed] (3,-4) -- (-1.5,-4);
\draw[ultra thick,blue,<->] (-1.5,-3.95) -- (-1.5,-1.75);
\node[blue,rotate=90] at (-1.75,-2.8) {{\large {\bf{$R^{(0)}-4S^{(0)}$}}}};
\end{tikzpicture}}}
\caption{\sl Web diagram of $X_{4,1}$. An independent set of K\"ahler parameters is shown in blue.}
\label{Fig:41web0}
${}$\\[-4cm]
\end{wrapfigure}

Continuing the previous examples, we next consider $X_{4,1}^{(\delta=0)}$, whose web diagram is shown in \figref{Fig:41web0}. While the method we employ to study it is the same as in the previous cases, we shall encounter a novel twist. The consistency conditions stemming from the web diagram are
\begin{align}
&S^{(0)}_1:\,h_2+m_2=m_1+h_2\,,\hspace{0.1cm}v_1+m_1=m_2+v_2\,;&&S^{(0)}_2:\,h_3+m_3=m_2+h_3\,,\hspace{0.1cm}v_2+m_2=m_3+v_3\,;\nonumber\\
&S^{(0)}_3:\,h_4+m_4=m_3+h_4\,,\hspace{0.1cm}v_3+m_3=m_4+v_4\,;&&S^{(0)}_4:\,h_1+m_1=m_4+h_1\,,\hspace{0.1cm}m_1+v_1=m_4+v_4\,,\label{Consistency41}
\end{align}

\noindent
while a solution is provided by the parameters $(\widehat{a}^{(0)}_{1,2,3,4},S^{(0)},R^{(0)})$
\begin{align}
&\widehat{a}_1^{(0)}=v_1+h_2\,,\hspace{1cm}\widehat{a}_2^{(0)}=v_2+h_3\,,\nonumber\\
&\widehat{a}_3^{(0)}=v_3+h_4\,,\hspace{1cm}\widehat{a}_4^{(0)}=v_4+h_1\,,\nonumber\\
&S^{(0)}=h_2+v_2+h_3+v_3+h_4+v_4+h_1\,,\nonumber\\
&R^{(0)}-4S^{(0)}=m_1-v_2-v_3-v_4\,.\label{IndepKaehler410}
\end{align}
The dihedral groups found in the previous examples were generated by two transformations. The latter can in fact be obtained in a simple fashion by considering two diagrams that are obtained from \figref{Fig:41web0} through a rearrangement and a flop transformation respectively:
%%%%%%%%%%
\subsubsection*{\bf 1) rearrangement:} 
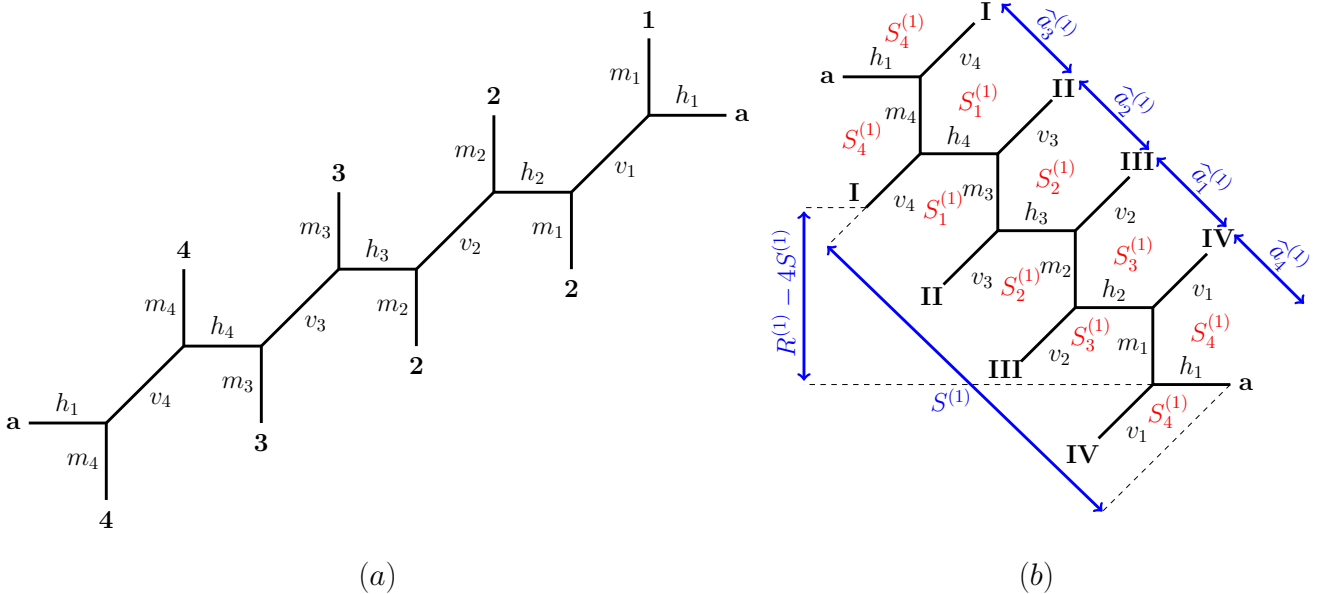
\begin{figure}[hbp]
\begin{center}
\scalebox{0.68}{\parbox{25.5cm}{\begin{tikzpicture}[scale = 1.50]
\draw[ultra thick] (-1,0) -- (0,0) -- (1,1) -- (2,1) -- (3,2) -- (4,2) -- (5,3) -- (6,3) -- (7,4) -- (8,4);
%verticals
\draw[ultra thick] (1,1) -- (1,2);
\draw[ultra thick] (3,2) -- (3,3);
\draw[ultra thick] (5,3) -- (5,4);
\draw[ultra thick] (7,4) -- (7,5);
\draw[ultra thick] (0,0) -- (0,-1);
\draw[ultra thick] (2,1) -- (2,0);
\draw[ultra thick] (4,2) -- (4,1);
\draw[ultra thick] (6,3) -- (6,2);
%ends
\node at (-1.2,0) {\large {\bf $\mathbf a$}};
\node at (8.2,4) {\large {\bf $\mathbf a$}};
\node at (0,-1.25) {\large {$\mathbf 4$}};
\node at (2,-0.25) {\large {$\mathbf 3$}};
\node at (4,0.75) {\large {$\mathbf 2$}};
\node at (6,1.75) {\large {$\mathbf 2$}};
\node at (1,2.25) {\large {$\mathbf 4$}};
\node at (3,3.25) {\large {$\mathbf 3$}};
\node at (5,4.25) {\large {$\mathbf 2$}};
\node at (7,5.25) {\large {$\mathbf 1$}};
%lables hotizontal
\node at (-0.5,0.25) {\large  {\bf $h_1$}};
\node at (1.5,1.25) {\large  {\bf $h_4$}};
\node at (3.5,2.25) {\large  {\bf $h_3$}};
\node at (5.5,3.25) {\large  {\bf $h_2$}};
\node at (7.5,4.25) {\large  {\bf $h_1$}};
%lables diagonal
\node at (0.7,0.3) {\large  {\bf $v_4$}};
\node at (2.7,1.3) {\large  {\bf $v_3$}};
\node at (4.7,2.3) {\large  {\bf $v_2$}};
\node at (6.7,3.3) {\large  {\bf $v_1$}};
%labels vertical
\node at (0.7,1.5) {\large  {\bf $m_4$}};
\node at (2.7,2.5) {\large  {\bf $m_3$}};
\node at (4.7,3.5) {\large  {\bf $m_2$}};
\node at (6.7,4.5) {\large  {\bf $m_1$}};
\node at (-0.3,-0.5) {\large  {\bf $m_4$}};
\node at (1.7,0.5) {\large  {\bf $m_3$}};
\node at (3.7,1.5) {\large  {\bf $m_2$}};
\node at (5.7,2.5) {\large  {\bf $m_1$}};
%stamp
\node at (3.5,-2) {\Large {\bf $\mathbf (a)$}};
%%%%%%%%%%%%%%%%%%%%%%%
%%%%%%%%%%%%%%%%%%%%%%%
\begin{scope}[xshift=10.5cm,yshift=4.5cm]
\draw[ultra thick] (-1,0) -- (0,0) -- (0,-1) -- (1,-1) -- (1,-2) -- (2,-2) -- (2,-3) -- (3,-3) -- (3,-4) -- (4,-4);
%diagonals
\draw[ultra thick] (0,0) -- (0.7,0.7);
\draw[ultra thick] (1,-1) -- (1.7,-0.3);
\draw[ultra thick] (2,-2) -- (2.7,-1.3);
\draw[ultra thick] (3,-3) -- (3.7,-2.3);
\draw[ultra thick] (0,-1) -- (-0.7,-1.7);
\draw[ultra thick] (1,-2) -- (0.3,-2.7);
\draw[ultra thick] (2,-3) -- (1.3,-3.7);
\draw[ultra thick] (3,-4) -- (2.3,-4.7);
%ends
\node at (-1.2,0) {\large {\bf $\mathbf a$}};
\node at (4.2,-4) {\large {\bf $\mathbf a$}};
\node at (0.85,0.85) {\large {$\mathbf{I}$}};
\node at (1.85,-0.15) {\large {$\mathbf{II}$}};
\node at (2.8,-1.1) {\large {$\mathbf{III}$}};
\node at (3.85,-2.1) {\large {$\mathbf{IV}$}};
\node at (-0.85,-1.5) {\large {$\mathbf{I}$}};
\node at (0.15,-2.85) {\large {$\mathbf{II}$}};
\node at (1.1,-3.8) {\large {$\mathbf{III}$}};
\node at (2.1,-4.9) {\large {$\mathbf{IV}$}};
%lables hotizontal
\node at (-0.5,0.25) {\large  {\bf $h_1$}};
\node at (0.5,-0.775) {\large  {\bf $h_4$}};
\node at (1.5,-1.775) {\large  {\bf $h_3$}};
\node at (2.5,-2.775) {\large  {\bf $h_2$}};
\node at (3.5,-3.775) {\large  {\bf $h_1$}};
%lables vertical
\node at (-0.25,-0.5) {\large  {\bf $m_4$}};
\node at (0.75,-1.5) {\large  {\bf $m_3$}};
\node at (1.75,-2.5) {\large  {\bf $m_2$}};
\node at (2.75,-3.5) {\large  {\bf $m_1$}};
%lables diagonal
\node at (0.65,0.2) {\large  {\bf $v_4$}};
\node at (1.65,-0.8) {\large  {\bf $v_3$}};
\node at (2.65,-1.8) {\large  {\bf $v_2$}};
\node at (3.65,-2.8) {\large  {\bf $v_1$}};
\node at (-0.2,-1.65) {\large  {\bf $v_4$}};
\node at (0.8,-2.65) {\large  {\bf $v_3$}};
\node at (1.8,-3.65) {\large  {\bf $v_2$}};
\node at (2.8,-4.65) {\large  {\bf $v_1$}};
%hexagons
\node[red] at (-0.2,0.6) {\large  {\bf $S^{(1)}_4$}};
\node[red] at (0.75,-0.3) {\large  {\bf $S^{(1)}_1$}};
\node[red] at (1.75,-1.3) {\large  {\bf $S^{(1)}_2$}};
\node[red] at (2.75,-2.3) {\large  {\bf $S^{(1)}_3$}};
\node[red] at (3.75,-3.3) {\large  {\bf $S^{(1)}_4$}};
\node[red] at (-0.75,-0.8) {\large  {\bf $S^{(1)}_4$}};
\node[red] at (0.3,-1.7) {\large  {\bf $S^{(1)}_1$}};
\node[red] at (1.3,-2.7) {\large  {\bf $S^{(1)}_2$}};
\node[red] at (2.2,-3.35) {\large  {\bf $S^{(1)}_3$}};
\node[red] at (3.2,-4.35) {\large  {\bf $S^{(1)}_4$}};
%roots
\draw[ultra thick,blue,<->] (1.05,0.95) -- (1.95,0.05);
\node[blue,rotate=315] at (1.75,0.65) {{\large {\bf {$\widehat{a}^{(1)}_3$}}}};
\draw[ultra thick,blue,<->] (2.05,-0.05) -- (2.95,-0.95);
\node[blue,rotate=315] at (2.75,-0.35) {{\large {\bf {$\widehat{a}^{(1)}_2$}}}};
\draw[ultra thick,blue,<->] (3.05,-1.05) -- (3.95,-1.95);
\node[blue,rotate=315] at (3.75,-1.35) {{\large {\bf {$\widehat{a}^{(1)}_1$}}}};
\draw[ultra thick,blue,<->] (4.05,-2.05) -- (4.95,-2.95);
\node[blue,rotate=315] at (4.75,-2.35) {{\large {\bf {$\widehat{a}^{(1)}_4$}}}};
%definition S
\draw[dashed] (4,-4) -- (2.35,-5.65);
\draw[dashed] (-0.7,-1.7) -- (-1.2,-2.2);
\draw[ultra thick,blue,<->] (-1.2,-2.2) -- (2.35,-5.65);
\node[blue] at (0.4,-4.2) {{\large {\bf {$S^{(1)}$}}}};
%parameter R
\draw[dashed] (-0.7,-1.7) -- (-1.5,-1.7);
\draw[dashed] (3,-4) -- (-1.5,-4);
\draw[ultra thick,blue,<->] (-1.5,-3.95) -- (-1.5,-1.75);
\node[blue,rotate=90] at (-1.75,-2.8) {{\large {\bf{$R^{(1)}-4S^{(1)}$}}}};
\end{scope}
%stamp
\node at (12,-2) {\Large {\bf $\mathbf (b)$}};
\end{tikzpicture}}}
\caption{\sl (a) mirrored web diagram \figref{Fig:41web0} after an $SL(2,\mathbb{Z})$ transformation. (b) Same diagram after cutting the lines $v_{1,2,3,4}$ and re-gluing the lines $m_{1,2,3,4}$ (and performing an $SL(2,\mathbb{Z})$ transformation).}
\label{Fig:41web1}
\end{center}
\end{figure}

A simple rearrangement of \figref{Fig:41web0} is shown in \figref{Fig:41web1}(a). The parametrisation in terms of $(\widehat{a}^{(1)}_{1,2,3,4},S^{(1)},R^{(1)})$ as indicated in the \figref{Fig:41web1}(b) is distinct to the one in \figref{Fig:41web0} by $(\widehat{a}^{(0)}_{1,2,3,4},S^{(0)},R^{(0)})$. Indeed, the two bases are related through a linear transformation given by

\begin{align}
&\left(\begin{array}{c}\widehat{a}^{(0)}_1\\ \widehat{a}^{(0)}_2  \\ \widehat{a}^{(0)}_3 \\ \widehat{a}^{(0)}_4  \\ S^{(0)} \\ R^{(0)}\end{array}\right)=G_1\cdot \left(\begin{array}{c}\widehat{a}^{(1)}_1\\ \widehat{a}^{(1)}_2  \\ \widehat{a}^{(1)}_3  \\ \widehat{a}^{(1)}_4  \\ S^{(1)} \\ R^{(1)}\end{array}\right)\,,&&\text{where} &&G_1=\left(
\begin{array}{cccccc}
 3 & 2 & 2 & 2 & -6 & 1 \\
 2 & 3 & 2 & 2 & -6 & 1 \\
 2 & 2 & 3 & 2 & -6 & 1 \\
 2 & 2 & 2 & 3 & -6 & 1 \\
 6 & 6 & 6 & 6 & -17 & 3 \\
 16 & 16 & 16 & 16 & -48 & 9 \\
\end{array}
\right)\,.\label{Gen41G1}
\end{align}
The matrix $G_1$ satisfies $\text{det }G_1=-1$ and $ G_1^2=1\!\!1_{6\times 6}$.
%%%%%%%%%%%%%%%%%%%%%%%%%%%%
\subsubsection*{\bf 2) transformation $\mathcal{F}$:} 
Another symmetry transformation can be obtained after performing a transformation $\mathcal{F}$ on \figref{Fig:41web0}, as shown in \figref{Fig:41webA}.

\begin{wrapfigure}{l}{0.49\textwidth}
\scalebox{0.75}{\parbox{8.7cm}{\begin{tikzpicture}[scale = 1.50]
\draw[ultra thick] (-1,0) -- (0,0) -- (0,-1) -- (1,-1) -- (1,-2) -- (2,-2) -- (2,-3) -- (3,-3) -- (3,-4) -- (4,-4);
%diagonals
\draw[ultra thick] (0,0) -- (0.7,0.7);
\draw[ultra thick] (1,-1) -- (1.7,-0.3);
\draw[ultra thick] (2,-2) -- (2.7,-1.3);
\draw[ultra thick] (3,-3) -- (3.7,-2.3);
\draw[ultra thick] (0,-1) -- (-0.7,-1.7);
\draw[ultra thick] (1,-2) -- (0.3,-2.7);
\draw[ultra thick] (2,-3) -- (1.3,-3.7);
\draw[ultra thick] (3,-4) -- (2.3,-4.7);
%ends
\node at (-1.2,0) {\large {\bf $\mathbf a$}};
\node at (4.2,-4) {\large {\bf $\mathbf a$}};
\node at (0.85,0.85) {\large {$\mathbf 1$}};
\node at (1.85,-0.15) {\large {$\mathbf 2$}};
\node at (2.85,-1.15) {\large {$\mathbf 3$}};
\node at (3.85,-2.15) {\large {$\mathbf 4$}};
\node at (-0.85,-1.5) {\large {$\mathbf 4$}};
\node at (0.15,-2.85) {\large {$\mathbf 1$}};
\node at (1.15,-3.85) {\large {$\mathbf 2$}};
\node at (2.15,-4.85) {\large {$\mathbf 3$}};
%lables hotizontal
\node at (-0.5,0.25) {\large  {\bf $v'_4$}};
\node at (0.5,-0.775) {\large  {\bf $v'_1$}};
\node at (1.5,-1.775) {\large  {\bf $v'_2$}};
\node at (2.5,-2.775) {\large  {\bf $v'_3$}};
\node at (3.5,-3.775) {\large  {\bf $v'_4$}};
%lables vertical
\node at (-0.35,-0.5) {\large  {\bf $-h_1$}};
\node at (0.65,-1.5) {\large  {\bf $-h_2$}};
\node at (1.65,-2.5) {\large  {\bf $-h_3$}};
\node at (2.65,-3.5) {\large  {\bf $-h_4$}};
%lables diagonal
\node at (0.65,0.2) {\large  {\bf $m'_1$}};
\node at (1.65,-0.8) {\large  {\bf $m'_2$}};
\node at (2.65,-1.8) {\large  {\bf $m'_3$}};
\node at (3.65,-2.8) {\large  {\bf $m'_4$}};
\node at (-0.2,-1.65) {\large  {\bf $m'_4$}};
\node at (0.8,-2.65) {\large  {\bf $m'_1$}};
\node at (1.8,-3.65) {\large  {\bf $m'_2$}};
\node at (2.8,-4.65) {\large  {\bf $m'_3$}};
%hexagons
\node[red] at (-0.2,0.65) {\large  {\bf $S^{(0)}_1$}};
\node[red] at (0.75,-0.3) {\large  {\bf $S^{(0)}_2$}};
\node[red] at (1.75,-1.3) {\large  {\bf $S^{(0)}_3$}};
\node[red] at (2.75,-2.3) {\large  {\bf $S^{(0)}_4$}};
\node[red] at (3.75,-3.3) {\large  {\bf $S^{(0)}_1$}};
\node[red] at (-0.75,-0.8) {\large  {\bf $S^{(0)}_4$}};
\node[red] at (0.3,-1.8) {\large  {\bf $S^{(0)}_1$}};
\node[red] at (1.25,-2.75) {\large  {\bf $S^{(0)}_2$}};
\node[red] at (2.2,-3.35) {\large  {\bf $S^{(0)}_3$}};
\node[red] at (3.2,-4.35) {\large  {\bf $S^{(0)}_4$}};
%roots
\draw[ultra thick,blue,<->] (1.05,0.95) -- (1.95,0.05);
\node[blue,rotate=315] at (1.75,0.65) {{\large {\bf {$\widehat{a}^{(0)}_1$}}}};
\draw[ultra thick,blue,<->] (2.05,-0.05) -- (2.95,-0.95);
\node[blue,rotate=315] at (2.75,-0.35) {{\large {\bf {$\widehat{a}^{(0)}_2$}}}};
\draw[ultra thick,blue,<->] (3.05,-1.05) -- (3.95,-1.95);
\node[blue,rotate=315] at (3.75,-1.35) {{\large {\bf {$\widehat{a}^{(0)}_3$}}}};
\draw[ultra thick,blue,<->] (4.05,-2.05) -- (4.95,-2.95);
\node[blue,rotate=315] at (4.75,-2.35) {{\large {\bf {$\widehat{a}^{(0)}_4$}}}};
%definition S
\draw[dashed] (3,-3) -- (1.35,-4.65);
\draw[dashed] (-0.7,-1.7) -- (-1.2,-2.2);
\draw[ultra thick,blue,<->] (-1.2,-2.2) -- (1.35,-4.65);
\node[blue] at (-0.1,-3.7) {{\large {\bf {$S^{(0)}$}}}};
%parameter R
\draw[dashed] (-0.7,-1.7) -- (-1.5,-1.7);
\draw[dashed] (3,-3) -- (-1.5,-3);
\draw[ultra thick,blue,<->] (-1.5,-2.95) -- (-1.5,-1.75);
\node[blue,rotate=90] at (-1.75,-2.3) {{\large {\bf{$R^{(0)}-3S^{(0)}$}}}};
\end{tikzpicture}}}
\caption{\sl Web diagram after a transformation $\mathcal{F}$ of \figref{Fig:41web0}. The blue parameters are the same as defined in eq.~(\ref{IndepKaehler410}).}
\label{Fig:41webA}
${}$\\[-4cm]
\end{wrapfigure} 

Here we have introduced the variables
\begin{align}
&v'_1=v_1+h_1+h_2\,,&&m'_1=m_1+h_1+h_2\,,\nonumber\\
&v'_2=v_2+h_2+h_3\,,&&m'_2=m_2+h_2+h_3\,,\nonumber\\
&v'_3=v_3+h_3+h_4\,,&&m'_3=m_3+h_3+h_4\,,\nonumber\\
&v'_4=v_4+h_4+h_1\,,&&m'_4=m_4+h_4+h_1\,.
\end{align}
The parameters $(\widehat{a}_{1,2,3,4}^{(0)},S^{(0)},R^{(0)})$, shown in blue in \figref{Fig:41webA}, are the same as those appearing in \figref{Fig:41webA}, such that the flop transformation alone does not lead to a nontrivial symmetry transformation. However, starting from the web diagram \figref{Fig:41webA}, we can present it in the form of \figref{Fig:41web2}.
\begin{figure}[htbp]
\begin{center}
\scalebox{0.62}{\parbox{21cm}{\begin{tikzpicture}[scale = 1.50]
\draw[ultra thick] (-1,-1) -- (0,0) -- (0,1) -- (1,2) -- (1,3) -- (2,4) -- (2,5) -- (3,6) -- (3,7) -- (4,8);
%horizontals
\draw[ultra thick] (0,0) -- (1,0);
\draw[ultra thick] (1,2) -- (2,2);
\draw[ultra thick] (2,4) -- (3,4);
\draw[ultra thick] (3,6) -- (4,6);
\draw[ultra thick] (0,1) -- (-1,1);
\draw[ultra thick] (1,3) -- (0,3);
\draw[ultra thick] (2,5) -- (1,5);
\draw[ultra thick] (3,7) -- (2,7);
%ends
\node at (-1.15,-1.15) {\large {$\mathbf{3}$}};
\node at (4.15,8.15) {\large {$\mathbf{3}$}};
\node at (1.15,0) {\large {\bf $\mathbf a$}};
\node at (2.15,2) {\large {\bf $\mathbf b$}};
\node at (3.15,4) {\large {\bf $\mathbf c$}};
\node at (4.15,6) {\large {\bf $\mathbf d$}};
\node at (-1.15,1) {\large {\bf $\mathbf d$}};
\node at (-0.15,3) {\large {\bf $\mathbf a$}};
\node at (0.85,5) {\large {\bf $\mathbf b$}};
\node at (1.85,7) {\large {\bf $\mathbf c$}};
%stamp
\node at (1.5,-2) {\Large {\bf $\mathbf (a)$}};
%diagonals
\node at (-0.4,-0.8) {\large  {\bf $m'_3$}};
\node at (0.6,1.2) {\large  {\bf $m'_4$}};
\node at (1.6,3.2) {\large  {\bf $m'_1$}};
\node at (2.6,5.2) {\large  {\bf $m'_2$}};
\node at (3.6,7.2) {\large  {\bf $m'_3$}};
%horizontals
\node at (0.5,-0.25) {\large  {\bf $v'_4$}};
\node at (1.5,1.75) {\large  {\bf $v'_1$}};
\node at (2.5,3.75) {\large  {\bf $v'_2$}};
\node at (3.5,5.75) {\large  {\bf $v'_3$}};
\node at (-0.5,0.75) {\large  {\bf $v'_3$}};
\node at (0.5,2.75) {\large  {\bf $v'_4$}};
\node at (1.5,4.75) {\large  {\bf $v'_1$}};
\node at (2.5,6.75) {\large  {\bf $v'_2$}};
%verticals
\node at (0.4,0.5) {\large  {\bf $-h_4$}};
\node at (1.4,2.5) {\large  {\bf $-h_1$}};
\node at (2.4,4.5) {\large  {\bf $-h_2$}};
\node at (3.4,6.5) {\large  {\bf $-h_3$}};
%%%%%%%%%%%%%%%%%%%%%%%
%%%%%%%%%%%%%%%%%%%%%%%
\begin{scope}[xshift=7.5cm,yshift=5cm]
\draw[ultra thick] (-1,0) -- (0,0) -- (0,-1) -- (1,-1) -- (1,-2) -- (2,-2) -- (2,-3) -- (3,-3) -- (3,-4) -- (4,-4);
%diagonals
\draw[ultra thick] (0,0) -- (0.7,0.7);
\draw[ultra thick] (1,-1) -- (1.7,-0.3);
\draw[ultra thick] (2,-2) -- (2.7,-1.3);
\draw[ultra thick] (3,-3) -- (3.7,-2.3);
\draw[ultra thick] (0,-1) -- (-0.7,-1.7);
\draw[ultra thick] (1,-2) -- (0.3,-2.7);
\draw[ultra thick] (2,-3) -- (1.3,-3.7);
\draw[ultra thick] (3,-4) -- (2.3,-4.7);
%ends
\node at (-1.2,0) {\large {\bf $\mathbf I$}};
\node at (4.2,-4) {\large {\bf $\mathbf I$}};
\node at (0.85,0.85) {\large {$\mathbf 1$}};
\node at (1.85,-0.15) {\large {$\mathbf 2$}};
\node at (2.85,-1.15) {\large {$\mathbf 3$}};
\node at (3.85,-2.15) {\large {$\mathbf 4$}};
\node at (-0.85,-1.5) {\large {$\mathbf 3$}};
\node at (0.15,-2.75) {\large {$\mathbf 4$}};
\node at (1.15,-3.85) {\large {$\mathbf 1$}};
\node at (2.15,-4.85) {\large {$\mathbf 2$}};
%lables hotizontal
\node at (-0.5,0.25) {\large  {\bf $-h_4$}};
\node at (0.5,-0.775) {\large  {\bf $-h_1$}};
\node at (1.5,-1.775) {\large  {\bf $-h_2$}};
\node at (2.5,-2.775) {\large  {\bf $-h_3$}};
%lables vertical
\node at (-0.25,-0.5) {\large  {\bf $m'_4$}};
\node at (0.75,-1.5) {\large  {\bf $m'_1$}};
\node at (1.75,-2.5) {\large  {\bf $m'_2$}};
\node at (2.75,-3.5) {\large  {\bf $m'_3$}};
%lables diagonal
\node at (0.6,0.3) {\large  {\bf $v'_3$}};
\node at (1.6,-0.7) {\large  {\bf $v'_4$}};
\node at (2.6,-1.7) {\large  {\bf $v'_1$}};
\node at (3.6,-2.7) {\large  {\bf $v'_2$}};
\node at (-0.25,-1.55) {\large  {\bf $v'_1$}};
\node at (0.8,-2.55) {\large  {\bf $v'_2$}};
\node at (1.7,-3.6) {\large  {\bf $v'_3$}};
\node at (2.7,-4.6) {\large  {\bf $v'_4$}};
%hexagons
\node[red] at (-0.2,0.7) {\large  {\bf $S^{(2)}_4$}};
\node[red] at (0.75,-0.3) {\large  {\bf $S^{(2)}_1$}};
\node[red] at (1.75,-1.3) {\large  {\bf $S^{(2)}_2$}};
\node[red] at (2.75,-2.3) {\large  {\bf $S^{(2)}_3$}};
\node[red] at (3.75,-3.3) {\large  {\bf $S^{(2)}_4$}};
\node[red] at (-0.75,-0.8) {\large  {\bf $S^{(2)}_2$}};
\node[red] at (0.25,-1.7) {\large  {\bf $S^{(2)}_3$}};
\node[red] at (1.2,-2.3) {\large  {\bf $S^{(2)}_4$}};
\node[red] at (2.2,-3.7) {\large  {\bf $S^{(2)}_1$}};
\node[red] at (3.2,-4.35) {\large  {\bf $S^{(2)}_2$}};
%roots
\draw[ultra thick,blue,<->] (1.05,0.95) -- (1.95,0.05);
\node[blue,rotate=315] at (1.75,0.65) {{\large {\bf {$\widehat{a}^{(2)}_3$}}}};
\draw[ultra thick,blue,<->] (2.05,-0.05) -- (2.95,-0.95);
\node[blue,rotate=315] at (2.75,-0.35) {{\large {\bf {$\widehat{a}^{(2)}_4$}}}};
\draw[ultra thick,blue,<->] (3.05,-1.05) -- (3.95,-1.95);
\node[blue,rotate=315] at (3.75,-1.35) {{\large {\bf {$\widehat{a}^{(2)}_1$}}}};
\draw[ultra thick,blue,<->] (4.05,-2.05) -- (4.95,-2.95);
\node[blue,rotate=315] at (4.75,-2.35) {{\large {\bf {$\widehat{a}^{(2)}_2$}}}};
%definition S
\draw[dashed] (2,-2) -- (0.35,-3.65);
\draw[dashed] (-0.7,-1.7) -- (-1.2,-2.2);
\draw[ultra thick,blue,<->] (-1.2,-2.2) -- (0.35,-3.65);
\node[blue] at (-0.6,-3.2) {{\large {\bf {$S^{(2)}$}}}};
%parameter R
\draw[dashed] (-0.7,-1.7) -- (-1.5,-1.7);
\draw[dashed] (1,-2) -- (-1.5,-2);
\draw[ultra thick,blue,<-] (-1.5,-2.05) -- (-1.5,-2.55);
\draw[ultra thick,blue,->] (-1.5,-1.15) -- (-1.5,-1.65);
\node[blue,rotate=90] at (-1.75,-1.7) {{\large {\bf{$R^{(2)}-2S^{(2)}$}}}};
\end{scope}
%stamp
\node at (9,-2) {\Large {\bf $\mathbf (b)$}};
\end{tikzpicture}}}
\caption{\sl (a) Web diagram \figref{Fig:41webA} after cutting the lines $m'_{1,2,3,4}$ and re-gluing along the lines $v'_{1,2,3,4}$. (b) presentation of the web diagram after cutting along the line $-h_4$ and gluing $m'_3$.  }
\label{Fig:41web2}

\end{center}
\end{figure}
The parametrisation in terms of the variables $(\widehat{a}^{(2)}_{1,2,3,4},S^{(2)},R^{(2)})$ used in \figref{Fig:41web2}(b) can be related to  $(\widehat{a}^{(0)}_{1,2,3,4},S^{(0)},R^{(0)})$ in \figref{Fig:41web0} through the transformation
\begin{align}
&\left(\begin{array}{c}\widehat{a}^{(0)}_1\\ \widehat{a}^{(0)}_2  \\ \widehat{a}^{(0)}_3 \\ \widehat{a}^{(0)}_4  \\ S^{(0)} \\ R^{(0)}\end{array}\right)=G_2\cdot \left(\begin{array}{c}\widehat{a}^{(2)}_1\\ \widehat{a}^{(2)}_2  \\ \widehat{a}^{(2)}_3  \\ \widehat{a}^{(2)}_4  \\ S^{(2)} \\ R^{(2)}\end{array}\right)\,,&&\text{where} &&G_2=\left(
\begin{array}{cccccc}
 1 & 0 & 0 & 0 & -2 & 1 \\
 0 & 1 & 0 & 0 & -2 & 1 \\
 0 & 0 & 1 & 0 & -2 & 1 \\
 0 & 0 & 0 & 1 & -2 & 1 \\
 1 & 1 & 1 & 1 & -7 & 3 \\
 4 & 4 & 4 & 4 & -24 & 9 \\
\end{array}
\right)\,.\label{Gen41G2}
\end{align}
The matrix $G_2$ has $\text{det }G_2=1$ but does not have finite order.\footnote{Indeed, by complete induction one can show that
\begin{align}
&G_2^n=1\!\!1_{6\times 6}+ n\,\left(
\begin{array}{cccccc}
 n-1 & n-1 & n-1 & n-1 & 2-4 n & n \\
 n-1 & n-1 & n-1 & n-1 & 2-4 n & n \\
 n-1 & n-1 & n-1 & n-1 & 2-4 n & n \\
 n-1 & n-1 & n-1 & n-1 & 2-4 n & n \\
 2 n-1 & 2 n-1 & 2 n-1 & 2 n-1 & -8 n & 2 n+1 \\
 4 n & 4 n & 4 n & 4 n & -8 (2 n+1) & 4 (n+1) \\
\end{array}
\right)\,,&&\text{for} &&n\in\mathbb{N}\,.
\end{align}
which only resembles the identity matrix for $n=0$.} This implies that the matrices $G_1$ and $G_2$ freely generate the group $\text{Dih}_\infty$
\begin{align}
\mathbb{G}(4)=\langle \{G_1,G_2\cdot G_1\}\rangle\cong \text{Dih}_\infty\,.\label{DefDihedral4}
\end{align}

%%%%%%%%%%%%%%%%%%%%%%%%%%%%%%%%%%%
%%%%%%%%%%%%%%%%%%%%%%%%%%%%%%%%%%%
\subsection{A Remark on Infinite Order}

\begin{wrapfigure}{l}{0.49\textwidth}
\scalebox{0.74}{\parbox{8.7cm}{\begin{tikzpicture}[scale = 1.50]
\draw[ultra thick] (-1,0) -- (0,0) -- (0,-1) -- (1,-1) -- (1,-2) -- (2,-2) -- (2,-3) -- (3,-3) -- (3,-4) -- (4,-4);
%diagonals
\draw[ultra thick] (0,0) -- (0.7,0.7);
\draw[ultra thick] (1,-1) -- (1.7,-0.3);
\draw[ultra thick] (2,-2) -- (2.7,-1.3);
\draw[ultra thick] (3,-3) -- (3.7,-2.3);
\draw[ultra thick] (0,-1) -- (-0.7,-1.7);
\draw[ultra thick] (1,-2) -- (0.3,-2.7);
\draw[ultra thick] (2,-3) -- (1.3,-3.7);
\draw[ultra thick] (3,-4) -- (2.3,-4.7);
%ends
\node at (-1.2,0) {\large {\bf $\mathbf I$}};
\node at (4.2,-4) {\large {\bf $\mathbf I$}};
\node at (0.85,0.85) {\large {$\mathbf 1$}};
\node at (1.85,-0.15) {\large {$\mathbf 2$}};
\node at (2.85,-1.15) {\large {$\mathbf 3$}};
\node at (3.85,-2.15) {\large {$\mathbf 4$}};
\node at (-0.85,-1.5) {\large {$\mathbf 3$}};
\node at (0.15,-2.75) {\large {$\mathbf 4$}};
\node at (1.15,-3.85) {\large {$\mathbf 1$}};
\node at (2.15,-4.85) {\large {$\mathbf 2$}};
%lables hotizontal
\node at (-0.5,0.25) {\large  {\bf $-h_4$}};
\node at (0.5,-0.775) {\large  {\bf $-h_1$}};
\node at (1.5,-1.775) {\large  {\bf $-h_2$}};
\node at (2.5,-2.775) {\large  {\bf $-h_3$}};
%lables vertical
\node at (-0.25,-0.5) {\large  {\bf $m'_4$}};
\node at (0.75,-1.5) {\large  {\bf $m'_1$}};
\node at (1.75,-2.5) {\large  {\bf $m'_2$}};
\node at (2.75,-3.5) {\large  {\bf $m'_3$}};
%lables diagonal
\node at (0.6,0.3) {\large  {\bf $v'_3$}};
\node at (1.6,-0.7) {\large  {\bf $v'_4$}};
\node at (2.6,-1.7) {\large  {\bf $v'_1$}};
\node at (3.6,-2.7) {\large  {\bf $v'_2$}};
\node at (-0.25,-1.55) {\large  {\bf $v'_1$}};
\node at (0.8,-2.55) {\large  {\bf $v'_2$}};
\node at (1.7,-3.6) {\large  {\bf $v'_3$}};
\node at (2.7,-4.6) {\large  {\bf $v'_4$}};
%hexagons
\node[red] at (-0.2,0.7) {\large  {\bf $S^{(3)}_4$}};
\node[red] at (0.75,-0.3) {\large  {\bf $S^{(3)}_1$}};
\node[red] at (1.75,-1.3) {\large  {\bf $S^{(3)}_2$}};
\node[red] at (2.75,-2.3) {\large  {\bf $S^{(3)}_3$}};
\node[red] at (3.75,-3.3) {\large  {\bf $S^{(3)}_4$}};
\node[red] at (-0.75,-0.8) {\large  {\bf $S^{(3)}_2$}};
\node[red] at (0.25,-1.7) {\large  {\bf $S^{(3)}_3$}};
\node[red] at (1.2,-2.3) {\large  {\bf $S^{(3)}_4$}};
\node[red] at (2.2,-3.7) {\large  {\bf $S^{(3)}_1$}};
\node[red] at (3.2,-4.35) {\large  {\bf $S^{(3)}_2$}};
%roots
\draw[ultra thick,blue,<->] (1.05,0.95) -- (1.95,0.05);
\node[blue,rotate=315] at (1.75,0.65) {{\large {\bf {$\widehat{a}^{(0)}_1$}}}};
\draw[ultra thick,blue,<->] (2.05,-0.05) -- (2.95,-0.95);
\node[blue,rotate=315] at (2.75,-0.35) {{\large {\bf {$\widehat{a}^{(0)}_2$}}}};
\draw[ultra thick,blue,<->] (3.05,-1.05) -- (3.95,-1.95);
\node[blue,rotate=315] at (3.75,-1.35) {{\large {\bf {$\widehat{a}^{(0)}_3$}}}};
\draw[ultra thick,blue,<->] (4.05,-2.05) -- (4.95,-2.95);
\node[blue,rotate=315] at (4.75,-2.35) {{\large {\bf {$\widehat{a}^{(0)}_4$}}}};
%definition S
\draw[dashed] (2,-2) -- (0.35,-3.65);
\draw[dashed] (-0.7,-1.7) -- (-1.2,-2.2);
\draw[ultra thick,blue,<->] (-1.2,-2.2) -- (0.35,-3.65);
\node[blue] at (-0.6,-3.2) {{\large {\bf {$S^{(0)}$}}}};
%parameter R
\draw[dashed] (-0.7,-1.7) -- (-1.5,-1.7);
\draw[dashed] (1,-2) -- (-1.5,-2);
\draw[ultra thick,blue,<-] (-1.5,-2.05) -- (-1.5,-2.55);
\draw[ultra thick,blue,->] (-1.5,-1.15) -- (-1.5,-1.65);
\node[blue,rotate=90] at (-1.75,-1.7) {{\large {\bf{$R^{(0)}-2S^{(0)}$}}}};
\end{tikzpicture}}}
\caption{\sl Web diagram after two transformations $\mathcal{F}$ of \figref{Fig:41web0}. The blue parameters are the same as defined in eq.~(\ref{IndepKaehler410}).}
\label{Fig:41webB}
%${}$\\[-1cm]
\end{wrapfigure} 

We have seen in the previous section that the symmetry transformation $G_2$ is of infinite order, which is markedly different than what we have seen in the previous examples. While we will present explicit checks that $G_2$ is indeed a symmetry of the free energy in the next subsection, we first want to provide an intuitive explanation of what makes the case $(N,1)=(4,1)$ different than all preceding ones. Indeed, we will provide some indication that the extended moduli space of $X_{4,1}$ contains many more regions that are represented by (a priori) very different looking web diagrams. While this will not prove that $G_2$ is of infinite order (as we have already done in the previous section by purely algebraic means), it will indicate the novel aspect of $X_{4,1}$ (in comparison to the previous examples).    

Returning to \figref{Fig:41web2}(b), the latter is a web diagram of the form $X_{4,1}^{(\delta=2)}$. Another way of obtaining such a diagram is to perform two transformations of the form $\mathcal{F}$ on \figref{Fig:41web0}, as is shown in \figref{Fig:41webB}, with the new parameters
\begin{align}
&h'_1=-h_1+v'_1+v'_4=h_1 + h_2 + h_4 + v_1 + v_4\,,&&h'_2=-h_2+v'_1+v'_2=h_1 + h_2 + h_3 + v_1 + v_2\,,\nonumber\\
&h'_3=-h_3+v'_2+v'_3=h_2 + h_3 + h_4 + v_2 + v_3\,,&&h'_4=-h_4+v'_3+v'_4=h_1 + h_3 + h_4 + v_3 + v_4\,,\nonumber
\end{align}
 as well as
 \begin{align}
 &m''_1=m'_1+v'_4+v'_2=2 h_1 + 2 h_2 + h_3 + h_4 + m_1 + v_2 + v_4\,,\nonumber\\
 &m''_2=m'_2+v'_1+v'_3=h_1 + 2 h_2 + 2 h_3 + h_4 + m_2 + v_1 + v_3\,,\nonumber\\
 &m''_3=m'_3+v'_2+v'_4=h_1 + h_2 + 2 h_3 + 2 h_4 + m_3 + v_2 + v_4\,,\nonumber\\
 &m''_4=m'_4+v'_4+v'_1=2 h_1 + h_2 + h_3 + 2 h_4 + m_4 + v_1 + v_3\,.\nonumber
 \end{align}
Notice that even upon imposing the consistency conditions (\ref{Consistency41}), the parametrisation of the web diagram \figref{Fig:41webB} is different than the one of the web diagram \figref{Fig:41web2}(b).\footnote{This can be seen by choosing the solution $v_1=v_2=v_3=v_4=v$ and $m_1=m_2=m_3=m_4=m$.} Thus, there is a duality transformation that transforms the web $X_{4,1}^{(2)}\longmapsto X_{4,1}^{(2)}$, however, with a non-trivial duality map $\mathcal{D}$ acting on the areas of all curves involved. The duality $\mathcal{D}$ can be repeatedly applied to $X_{4,1}^{(2)}$ in \figref{Fig:41web2}(b), thus producing an infinite number of diagrams of the type $X_{4,1}^{(2)}$, each one with an a priori different parametrisation of individual curves.
 
Moreover, since the blue parameters $(\widehat{a}^{(0)}_{1,2,3,4},S^{(0)},R^{(0)})$ in \figref{Fig:41webB} are the same as in \figref{Fig:41web0}, the duality map $\mathcal{D}$ from the perspective of the independent K\"ahler parameters precisely corresponds to the symmetry transformation $G_2$. Therefore, the transition from \figref{Fig:41webB} to \figref{Fig:41web2}(b) gives (a new) geometric representation of $G_2$ at the level of web diagrams, which readily allows to also compute arbitrary powers of $G_2$. 

Finally, notice that the above discussion does not generalise to the cases $N=2,3$ (but can be extended to $N>4$). Indeed, web diagrams with shifts $\delta\geq2$ for $N=2,3$ can readily be related (possibly through simple cutting and re-gluing operations) to web diagrams with $\delta\in \{0,1\}$, which only gave rise to symmetry transformations of finite order.\footnote{Notice for example that the only web diagrams in \figref{Fig:Overview21Webs} and \figref{Fig:Overview31Webs} that give rise to non-trivial symmetry transformations have either $\delta=0$ or $\delta=1$. Thus, in these cases, there is in fact no non-trivial equivalent of \figref{Fig:41web2}(b).} In other words, in the cases $N=2,3$, the equivalents of the diagrams \figref{Fig:41web2} and \figref{Fig:41webB} are of the type $\delta\leq1$, which we have seen to provide only transformations of finite order.

%%%%%%%%%%%%%%%%%%%%%%%%%%%%%%%%%%%
%%%%%%%%%%%%%%%%%%%%%%%%%%%%%%%%%%%
\subsection{Invariance of the Non-perturbative Free Energy}\label{Sect:Ord4Checks}
As non-trivial check for the fact that $G_1$ and $G_2$ are indeed symmetries of $\mathcal{Z}_{4,1}$, we consider the non-perturbative free energy associated with the latter. For simplicity, we restrict ourselves to the leading term in $\epsilon_{1,2}$. To this end, we define
\begin{align}
&\lim_{\epsilon_{1,2}\to 0} \epsilon_1\,\epsilon_2\,\mathcal{F}_{4,1}(\widehat{a}_{1,2,3,4},S,R;\epsilon_1,\epsilon_2)=\sum_{n,i_a=0}^\infty \sum_{k\in\mathbb{Z}}f^{\text{NS}}_{i_1,i_2,i_3,i_4,k,n}\,\widehat{Q}_1^{i_1}\,\widehat{Q}_2^{i_2}\,\widehat{Q}_3^{i_3}\,\widehat{Q}_3^{i_4}\,Q_S^{k}\,Q_{R}^n\,,
\end{align}
where $f^{\text{NS}}_{i_1,i_2,i_3,i_4,k,n}\in\mathbb{Z}$ and $\widehat{Q}_i=e^{-\widehat{a}_i}$ (for $i=1,2,3,4$), $Q_S=e^{-S}$ and $Q_R=e^{-R}$. In the same manner as explained in section~\ref{Sect:GeneralStrategy}, the symmetry transformations $G_1$ and $G_2$ act in the following manner on the coefficients $f^{\text{NS}}_{i_1,i_2,i_3,i_4,k,n}$
\begin{align}
&f^{\text{NS}}_{i_1,i_2,i_3,i_4,k,n}=f^{\text{NS}}_{i'_1,i'_2,i'_3,i'_4,k',n'}&&\text{for} && (i'_1,i'_2,i'_3,i'_4,k', n')^T=G^T_\ell\cdot (i_1,i_2,i_3,i_4,k, n)^T&&\forall \ell=1,2\,.\label{Rels41CheckExpand}
%\left(\begin{array}{c}i'_1 \\ i'_2 \\ i'_3 \\ k' \\ n'\end{array}\right)=G^T_\ell\cdot \left(\begin{array}{c}i_1 \\ i_2 \\ i_3 \\ k \\ n\end{array}\right)&&\forall \ell=1,2,3,4,5\,.
\end{align}
We can explicitly check the relations (\ref{Rels41CheckExpand}) by computing the relevant expansions of the free energies. However, since the matrix $G_1$ in (\ref{Gen41G1}) contains very large numbers, the relations are easier to check for the matrices $G_1\cdot G_2$ and $G_2$ with
\begin{align}
G_1\cdot G_2=\left(\begin{array}{ccccccc}1 & 0 & 0 & 0 & 0 & 0 \\ 0 & 1 & 0 & 0 & 0 & 0 \\ 0 & 0 & 1 & 0 & 0 & 0 \\ 0 & 0 & 0 & 1 & 0 & 0 \\ 1 & 1 & 1 & 1 & -1 & 0 \\ 4 & 4 & 4 & 4 & -8 & 1  \end{array}\right)\,.
\end{align}
In Table~\ref{Tab:SymTrafo41G12} and Table~\ref{Tab:SymTrafo41G2} we tabulate examples of coefficients $f^{\text{NS}}_{i_1,i_2,i_3,i_4,k,n}$ with $i_1+i_2+i_3+i_4\leq 6$ for $n=1$ and $n=2$ that are related by $G_{1}\cdot G_2$ and $G_2$ respectively.

\begin{table}[hbp]
\begin{center}
\begin{tabular}{|c|c|c|}\hline
&&\\[-10pt]
$(i_1,i_2,i_3,i_4,k,n)$ & $(i'_1,i'_2,i'_3,i'_4,k',n')$ & $f^{\text{NS}}_{i_1,i_2,i_3,i_4,k,n}$\\[4pt]\hline\hline
&&\\[-13pt]
$(0,0,1,0,-2,1) $ & $(2,2,2,3,-6,1)$ & $2$\\\hline
&&\\[-13pt]
$(0,0,1,0,-1,1) $ & $(3,3,4,3,-7,1)$ & $-8$\\\hline
&&\\[-13pt]
$(0,0,1,1,-3,1) $ & $(1,1,2,2,-5,1)$ & $-1$\\\hline
&&\\[-13pt]
$(0,0,1,2,-2,1) $ & $(2,2,3,4,-6,1)$ & $18$\\\hline
&&\\[-13pt]
$(0,0,1,2,-1,1) $ & $(3,3,4,5,-7,1)$ & $-45$\\\hline
&&\\[-13pt]
$(0,0,1,3,-3,1) $ & $(1,1,2,4,-5,1)$ & $-5$\\\hline
&&\\[-13pt]
$(0,0,1,3,-2,1) $ & $(2,2,3,5,-6,1)$ & $30$\\\hline
&&\\[-13pt]
$(0,0,1,4,-3,1) $ & $(1,1,2,5,-5,1)$ & $-7$\\\hline
&&\\[-13pt]
$(0,0,1,4,-2,1) $ & $(2,2,3,6,-6,1)$ & $42$\\\hline
&&\\[-13pt]
$(0,0,1,5,-3,1) $ & $(1,1,2,6,-5,1)$ & $-9$\\\hline
&&\\[-13pt]
$(0,0,1,5,-2,1) $ & $(2,2,3,7,-6,1)$ & $54$\\\hline
&&\\[-13pt]
$(0,0,0,6,-2,1) $ & $(2,2,2,8,-6,1)$ & $12$\\\hline
\end{tabular}
\end{center}
\caption{\sl Action of $G_1\cdot G_2$: $(i'_1,i'_2,i'_3,i'_4,k',n')^T=(G_1\cdot G_2)^T\cdot (i_1,i_2,i_3,i_4,k,n)^T$.}
\label{Tab:SymTrafo41G12}
\end{table}

\begin{table}[hbp]
\begin{center}
\begin{tabular}{|c|c|c|}\hline
&&\\[-10pt]
$(i_1,i_2,i_3,i_4,k,n)$ & $(i'_1,i'_2,i'_3,i'_4,k',n')$ & $f^{\text{NS}}_{i_1,i_2,i_3,i_4,k,n}$\\[4pt]\hline\hline
&&\\[-13pt]
$(0,0,1,1,-3,1) $ & $(1,1,2,2,-7,2)$ & $-1$\\\hline
&&\\[-13pt]
$(0,1,2,2,-4,1)$ & $(0,1,1,2,-6,2)$ & 2\\\hline
&&\\[-13pt]
$(1,1,1,2,-4,1)$ & $(1,1,1,2,-6,2)$ & 4\\\hline
&&\\[-13pt]
$(1,1,2,3,-5,1)$ & $(0,0,1,2,-3,1)$ & -3\\\hline
&&\\[-13pt]
$(1,1,2,4,-5,1)$ & $(0,0,1,3,-5,2)$ & -5\\\hline
&&\\[-13pt]
$(1,1,3,3,-5,1)$ & $(0,0,2,2,-5,2)$ & -4\\\hline
\end{tabular}
\end{center}
\caption{\sl Action of $G_2$: $(i''_1,i''_2,i''_3,i''_4,k'',n')^T=G_2^T\cdot (i_1,i_2,i_3,i_4,k,n)^T$.}
\label{Tab:SymTrafo41G2}
\end{table}
%%%%%%%%%%%%%%%%%%%%%%%%%%%%%%%%%%%%%%%%%%%%%%%%%
%%%%%%%%%%%%%%%%%%%%%%%%%%%%%%%%%%%%%%%%%%%%%%%%%
\subsection{Modularity at a Particular Point of the Moduli Space}
Similarly to the cases $N=2,3$, we can analyse how the group $\mathbb{G}(4)$ is related to $Sp(4,\mathbb{Z})$ at the particular region in the moduli space, which is characterised by $\widehat{a}_1^{(0)}=\widehat{a}_2^{(0)}=\widehat{a}_3^{(0)}=\widehat{a}_4^{(0)}=\widehat{a}$, which implies $h_1=h_2=h_3=h_4=h$ (while the consistency conditions (\ref{ConsistencyWeb21}) impose $v_1=v_2=v_3=v_4=v$ and $m_1=m_2=m_3=m_4=m$). We can introduce the period matrix
\begin{align}
&\Omega=\left(\begin{array}{cc}\tau & v \\ v & \rho\end{array}\right)\,,&&\text{with} &&\begin{array}{l}\tau=m+v\,,\\ \rho=h+m\,.\end{array}\label{Def4PeriodMatrix}
\end{align} 
Using the parametrisation (\ref{DefDihedral4}) of $\mathbb{G}(4)$, it is sufficient to analyse the relation of the generators $G_1$ and $G'_2=G_2\cdot G_1$ to $Sp(4,\mathbb{Z})$. The restriction of these generators to the subspace $(\widehat{a},S,R)$ can be written in the form
\begin{align}
&G_1^{(\text{red})}=\left(\begin{array}{ccc}1 & -2 & 1 \\ 0 & -1 & 1 \\ 0 & 0 & 1\end{array}\right)\,,&&\text{and} &&G_2^{\prime,(\text{red})}=\left(\begin{array}{ccc}1 & 0 & 0 \\ 4 & -1 & 0 \\ 16 & -8 & 1\end{array}\right)\,,
\end{align}
Rewriting them furthermore to act as elements of $Sp(4,\mathbb{Z})$ in the form of (\ref{ActionSp4Period}) on the period matrix $\Omega$ in (\ref{Def4PeriodMatrix}), they take the form
\begin{align}
&\widetilde{G}_1^{(\text{red,Sp})}=HKL^6KL^6HKHL^6KL^6KH\,,&&\text{and} &&\widetilde{G}_2^{\prime,(\text{red,Sp})}=HKL^6KL^6KL^6KH\,,
\end{align}
where $K$, $L$ and $H$ are defined in appendix~\ref{App:Sp4}. As in the cases of $N=2,3$, this implies that the restriction of $\mathbb{G}(3)$ to the particular region of the K\"ahler moduli space $(\widehat{a},S,R)$ is a subgroup of $Sp(4,\mathbb{Z})$. However, unlike the case $N=1$, we cannot conclude that the freely generated group $\langle \widetilde{G}_1^{(\text{red,Sp})},\widetilde{G}_2^{\prime,(\text{red,Sp})}, S_\rho,T_\rho,S_\tau,T_\tau \rangle$ is isomorphic to $Sp(4,\mathbb{Z})$.
%%%%%%%%%%%%%%%%%%%%%%%%%%%%%%%%%%%
%%%%%%%%%%%%%%%%%%%%%%%%%%%%%%%%%%%
%%%%%%%%%%%%%%%%%%%%%%%%%%%%%%%%%%%
%%%%%%%%%%%%%%%%%%%%%%%%%%%%%%%%%%%%%%%%%%%%%%%%
\section{General Case $(N,1)$}\label{Sect:GeneralCase}
\subsection{Symmetry Transformations of Generic Webs}
We can summarise all previous examples by introducing the following matrices
\begin{align}
&\mathcal{G}_2(N)=\left(\begin{array}{ccccc} & & & 0 & 0 \\ & 1\!\!1_{N\times N} & & \vdots & \vdots \\ & & & 0 & 0 \\ 1 & \cdots & 1 & -1 & 0 \\ N & \cdots & N & -2N & 1 \end{array}\right)\,,\label{DefG2General}
\end{align}
as well as
\begin{align}
\mathcal{G}_\infty(N)=\left(\begin{array}{ccccc} & & & -2 & 1 \\ & 1\!\!1_{N\times N} & & \vdots & \vdots \\ & & & -2 & 1 \\ 1 & \cdots & 1 & -2N+1 & N-1 \\ N & \cdots & N & -2N(N-1) & (N-1)^2 \end{array}\right)\,.\label{DefGinfGeneral}
\end{align}
The matrices $\mathcal{G}_2(N)$ and $\mathcal{G}_\infty(N)$ for the examples previously studied are given explicitly as\begin{center}
\begin{tabular}{c|c|c|c}
$N$ & $\mathcal{G}_2(N)$ & $\mathcal{G}_\infty(N)$ & {\bf defined in}\\\hline
$1$ & $b$ & $G_1$ & eq.~(\ref{Def11ab}) and eq.~(\ref{Def11G1})\\
$2$ & $G_2$ & $G_3$ & eq.~(\ref{Gen21G1}) and eq.~(\ref{Gen21G3})\\
$3$ & $G_3$ & $G_3\cdot G_2$ &  eq.~(\ref{Gen31G2}) and eq.~(\ref{Gen31G3})\\ 
$4$ & $G_1\cdot G_2$ & $G_2$ &  eq.~(\ref{Gen41G1}) and eq.~(\ref{Gen41G2})\\ 
\end{tabular}
\end{center} 
where the equation numbers refer to the definitions of the matrices in the individual cases. The matrices $\mathcal{G}_2$ and $\mathcal{G}_\infty(N)$ furnish two symmetry relations for a web diagram of the type $(N,1)$. To see this, in the following we shall check explicitly the combinations of $\mathcal{G}_\infty(N)\cdot \mathcal{G}_2(N)$ and $\mathcal{G}_\infty(N)$, which at the level of the web diagrams are generated by the same transformations we already discussed in the example of $(N,1)=(4,1)$ and which can be generalised for generic $N$:
%%%%%%%%%%%%%%%%%%%
%%%%%%%%%%
\subsubsection*{\bf 1) rearrangement:} 
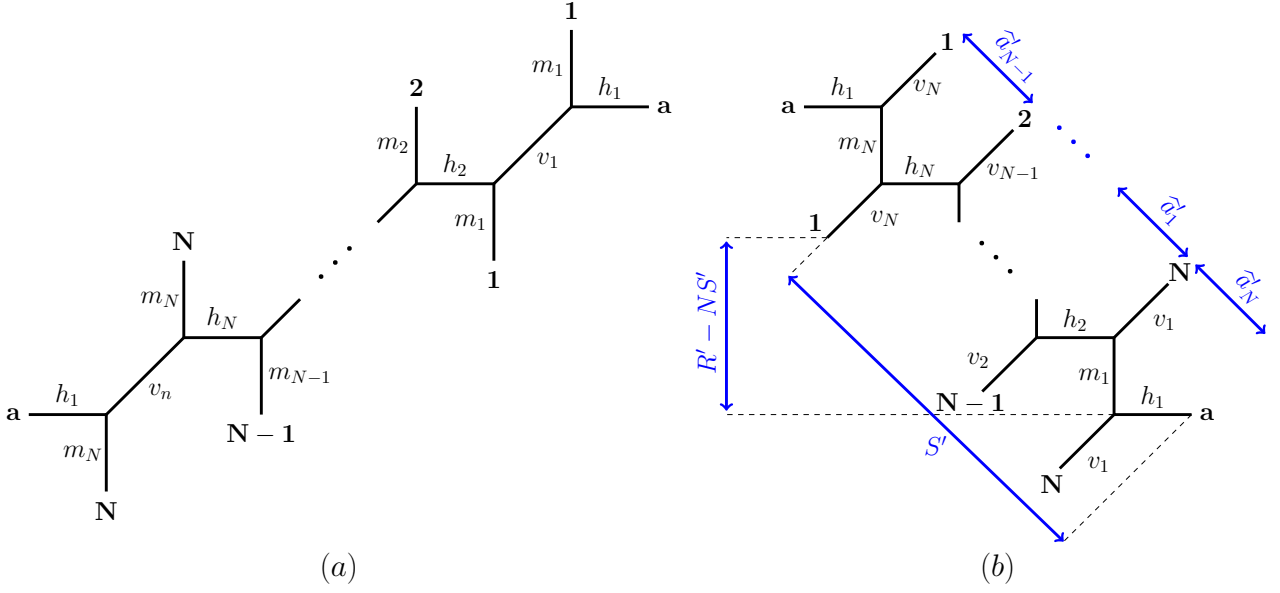
\begin{figure}
\begin{center}
\scalebox{0.68}{\parbox{25.5cm}{\begin{tikzpicture}[scale = 1.50]
\draw[ultra thick] (-1,0) -- (0,0) -- (1,1) -- (2,1) -- (2.5,1.5);
\node[rotate=45] at (3,2) {\Huge $\cdots$};
\draw[ultra thick] (3.5,2.5) -- (4,3) -- (5,3) -- (6,4) -- (7,4);
%verticals
\draw[ultra thick] (1,1) -- (1,2);
\draw[ultra thick] (0,0) -- (0,-1);
\draw[ultra thick] (2,1) -- (2,0);
\draw[ultra thick] (4,3) -- (4,4);
\draw[ultra thick] (6,4) -- (6,5);
\draw[ultra thick] (5,3) -- (5,2);
%ends
\node at (-1.2,0) {\large {\bf $\mathbf a$}};
\node at (7.2,4) {\large {\bf $\mathbf a$}};
\node at (0,-1.25) {\large {$\mathbf{N}$}};
\node at (2,-0.25) {\large {$\mathbf{N-1}$}};
\node at (5,1.75) {\large {$\mathbf 1$}};
\node at (1,2.25) {\large {$\mathbf{N}$}};
\node at (4,4.25) {\large {$\mathbf{2}$}};
\node at (6,5.25) {\large {$\mathbf 1$}};
%lables hotizontal
\node at (-0.5,0.25) {\large  {\bf $h_1$}};
\node at (1.5,1.25) {\large  {\bf $h_N$}};
\node at (4.5,3.25) {\large  {\bf $h_2$}};
\node at (6.5,4.25) {\large  {\bf $h_1$}};
%lables diagonal
\node at (0.7,0.3) {\large  {\bf $v_n$}};
\node at (5.7,3.3) {\large  {\bf $v_1$}};
%labels vertical
\node at (0.7,1.5) {\large  {\bf $m_N$}};
\node at (3.7,3.5) {\large  {\bf $m_2$}};
\node at (5.7,4.5) {\large  {\bf $m_1$}};
\node at (-0.3,-0.5) {\large  {\bf $m_N$}};
\node at (2.5,0.5) {\large  {\bf $m_{N-1}$}};
\node at (4.7,2.5) {\large  {\bf $m_1$}};
%stamp
\node at (3,-2) {\Large {\bf $\mathbf (a)$}};
%%%%%%%%%%%%%%%%%%%%%%%
%%%%%%%%%%%%%%%%%%%%%%%%%%%%%%%%%%%%%%%%
%%%%%%%%%%%%%%%%%%%%%%%%%%%%%%%%%%%%%%%%
%%%%%%%%%%%%%%%%%%%%%%%
\begin{scope}[xshift=10cm,yshift=4cm]
\draw[ultra thick] (-1,0) -- (0,0) -- (0,-1) -- (1,-1) -- (1,-1.5);
\node[rotate=315] at (1.5,-2) {\Huge $\cdots$};
\draw[ultra thick] (2,-2.5) -- (2,-3) -- (3,-3) -- (3,-4) -- (4,-4);
%diagonals
\draw[ultra thick] (0,0) -- (0.7,0.7);
\draw[ultra thick] (1,-1) -- (1.7,-0.3);
\draw[ultra thick] (3,-3) -- (3.7,-2.3);
\draw[ultra thick] (0,-1) -- (-0.7,-1.7);
\draw[ultra thick] (2,-3) -- (1.3,-3.7);
\draw[ultra thick] (3,-4) -- (2.3,-4.7);
%ends
\node at (-1.2,0) {\large {\bf $\mathbf a$}};
\node at (4.2,-4) {\large {\bf $\mathbf a$}};
\node at (0.85,0.85) {\large {$\mathbf 1$}};
\node at (1.85,-0.15) {\large {$\mathbf 2$}};
\node at (3.85,-2.15) {\large {$\mathbf{N}$}};
\node at (-0.85,-1.5) {\large {$\mathbf{1}$}};
\node at (1.15,-3.85) {\large {$\mathbf{N-1}$}};
\node at (2.2,-4.9) {\large {$\mathbf{N}$}};
%lables hotizontal
\node at (-0.5,0.25) {\large  {\bf $h_1$}};
\node at (0.5,-0.775) {\large  {\bf $h_N$}};
\node at (2.5,-2.775) {\large  {\bf $h_2$}};
\node at (3.5,-3.775) {\large  {\bf $h_1$}};
%lables vertical
\node at (-0.3,-0.5) {\large  {\bf $m_N$}};
\node at (2.75,-3.5) {\large  {\bf $m_1$}};
%lables diagonal
\node at (0.6,0.25) {\large  {\bf $v_N$}};
\node at (1.7,-0.85) {\large  {\bf $v_{N-1}$}};
\node at (3.65,-2.8) {\large  {\bf $v_1$}};
\node at (0.05,-1.45) {\large  {\bf $v_N$}};
\node at (1.25,-3.3) {\large  {\bf $v_{2}$}};
\node at (2.8,-4.65) {\large  {\bf $v_1$}};
%roots
\draw[ultra thick,blue,<->] (1.05,0.95) -- (1.95,0.05);
\node[blue,rotate=315] at (1.75,0.65) {{\large {\bf {$\widehat{a}'_{N-1}$}}}};
%
%\draw[ultra thick,blue,<->] (3.05,0.95) -- (3.95,0.05);
%\node[blue,rotate=315] at (3.75,0.65) {{\large {\bf {$\widehat{a}'_1$}}}};
%
\node[blue,rotate=315] at (2.5,-0.5) {\Huge $\cdots$};
\draw[ultra thick,blue,<->] (3.05,-1.05) -- (3.95,-1.95);
\node[blue,rotate=315] at (3.75,-1.35) {{\large {\bf {$\widehat{a}'_1$}}}};
\draw[ultra thick,blue,<->] (4.05,-2.05) -- (4.95,-2.95);
\node[blue,rotate=315] at (4.75,-2.35) {{\large {\bf {$\widehat{a}'_{N}$}}}};
%definition S
\begin{scope}[xshift=-4cm,yshift=4cm]
\draw[dashed] (8,-8) -- (6.35,-9.65);
\draw[dashed] (3.3,-5.7) -- (2.8,-6.2);
\draw[ultra thick,blue,<->] (2.8,-6.2) -- (6.35,-9.65);
\node[blue] at (4.7,-8.4) {{\large {\bf {$S'$}}}};
\end{scope}
%parameter R
\begin{scope}[xshift=-3cm,yshift=3cm]
\draw[dashed] (2.3,-4.7) -- (1,-4.7);
\draw[dashed] (6,-7) -- (1,-7);
\draw[ultra thick,blue,<->] (1,-6.95) -- (1,-4.75);
\node[blue,rotate=90] at (0.75,-5.8) {{\large {\bf{$R'-NS'$}}}};
\end{scope}
\end{scope}
%stamp
\node at (11.5,-2) {\Large {\bf $\mathbf (b)$}};
\end{tikzpicture}}}
\caption{\sl Alternative presentations of the web diagram of $X_{N,1}^{(\delta=0)}$ from \figref{Fig:N1web0} for $\delta=0$.}
\label{Fig:N1web1}
\end{center}
\end{figure}

We first verify that $\mathcal{G}_\infty(N)\cdot \mathcal{G}_2(N)$ is a symmetry. To this end, we start from the configuration shown in \figref{Fig:N1web0} for $\delta=0$, which (after mirroring and performing an $SL(2,\mathbb{Z})$ transformation) can be presented as in \figref{Fig:N1web1}(a). The latter in turn can alternatively be presented in the form. \figref{Fig:N1web1}(b).
The matrix $\mathcal{G}_\infty(N)\cdot \mathcal{G}_2(N)$ (defined in (\ref{DefG2General}) and (\ref{DefGinfGeneral}) respectively) relates the parameters in the web diagram \figref{Fig:N1web0} to those in \figref{Fig:N1web1}(b) in the following way
\begin{align}
\left(\widehat{a}_1\,, \ldots \,, \widehat{a}_N \,, S \,, R\right)^T=\mathcal{G}_2(N)\cdot \mathcal{G}_\infty(N)\cdot \left(\widehat{a}'_1\,, \ldots \,, \widehat{a}'_N \,, S' \,, R'\right)^T\,,\label{RelGenMat2}
\end{align}
where 
\begin{align}
\mathcal{G}_\infty(N)\cdot\mathcal{G}_2(N)=\left(\begin{array}{ccccc} & & & -2N+2 & 1 \\ & \mathbb{A}_{N\times N} & & \vdots & \vdots \\ & & & -2N+2 & 1 \\ N^2-3N+2 & \cdots & N^2-3N+2 & -2N^2+4N-1 & N-1 \\ N(N-2)^2 & \cdots & N(N-2)^2 & -2N(2-3N+N^2) & (N-1)^2 \end{array}\right)\,,
\end{align}
with $\mathbb{A}_{N\times N}=(N-2)\left(\begin{array}{ccc}1 & \cdots & 1 \\ \vdots & & \vdots\\ 1 & \cdots & 1\end{array}\right)+1\!\!1_{N\times N}$. Upon using the following solution of the consistency conditions in (\ref{ConsistGeneral})
\begin{align}
&v_1=v_2=\ldots=v_N=v\,,&&\text{and} &&m_1=m_2=\ldots=m_N=m\,,\label{SimpleSolConsistency}
\end{align} 
which implies from \figref{Fig:N1web0} and \figref{Fig:N1web1}(b) (for $i=1,\ldots,N$)
\begin{align}
&\widehat{a}_i=h_{i+1}+v\,,&&\widehat{a}'_i=h_{i+1}+m\nonumber\\
&S=\sum_{k=1}^Nh_k+(N-1)v\,,&&S'=\rho'-m=\sum_{k=1}^N\widehat{a}'_k=\sum_{k=1}^Nh_k+(N-1)m\,,\nonumber\\
&R-NS=m-(N-1)v\,,&&R'-NS'=v-(N-1)m\,,\label{IdentificationOrder2}
\end{align}
we have indeed  (with $\rho'=\sum_{k=1}^N\widehat{a}'_k=\sum_{k=1}^Nh_k+Nm$)
{\allowdisplaybreaks
\begin{align}
\widehat{a}_i&=\widehat{a}'_i+(N-2)\rho'-(2N-2)S'+R'=h_{i+1}+v\,,\nonumber\\
S&=(N^2-3N+2)\rho'-(2N^2-4N+1)S'+(N-1)R'=\sum_{k=1}^N h_k+(N-1)v\,,\nonumber\\
R&=N(N-2)^2\rho'-2N(2-3N+N^2)S'+(N-1)^2R'=N\sum_{k=1}^N h_k+m+(N-1)^2v\,,\nonumber
\end{align}}
which proves (\ref{RelGenMat2}).
%%%%%%%%%%%%%%%%%%%%%%%%%%%%
%%%%%%%%%%%%%%%%%%%%%%%%%%%%
\subsubsection*{\bf 2) transformation $\mathcal{F}$:} 
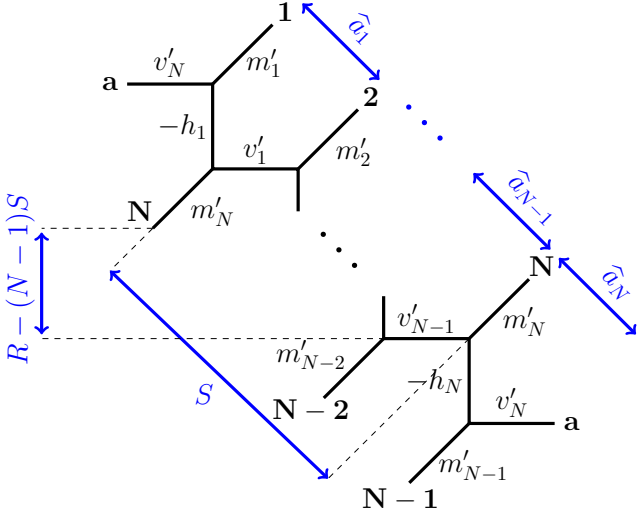
\begin{wrapfigure}{l}{0.5\textwidth}
\scalebox{0.75}{\parbox{11.3cm}{\begin{tikzpicture}[scale = 1.50]
\draw[ultra thick] (-1,0) -- (0,0) -- (0,-1) -- (1,-1) -- (1,-1.5);
\node[rotate=315] at (1.5,-2) {\Huge $\cdots$};
\draw[ultra thick] (2,-2.5) -- (2,-3) -- (3,-3) -- (3,-4) -- (4,-4);
%diagonals
\draw[ultra thick] (0,0) -- (0.7,0.7);
\draw[ultra thick] (1,-1) -- (1.7,-0.3);
\draw[ultra thick] (3,-3) -- (3.7,-2.3);
\draw[ultra thick] (0,-1) -- (-0.7,-1.7);
\draw[ultra thick] (2,-3) -- (1.3,-3.7);
\draw[ultra thick] (3,-4) -- (2.3,-4.7);
%ends
\node at (-1.2,0) {\large {\bf $\mathbf a$}};
\node at (4.2,-4) {\large {\bf $\mathbf a$}};
\node at (0.85,0.85) {\large {$\mathbf 1$}};
\node at (1.85,-0.15) {\large {$\mathbf 2$}};
\node at (3.85,-2.15) {\large {$\mathbf{N}$}};
\node at (-0.85,-1.5) {\large {$\mathbf{N}$}};
\node at (1.15,-3.85) {\large {$\mathbf{N-2}$}};
\node at (2.2,-4.9) {\large {$\mathbf{N-1}$}};
%lables hotizontal
\node at (-0.5,0.25) {\large  {\bf $v'_N$}};
\node at (0.5,-0.775) {\large  {\bf $v'_1$}};
\node at (2.5,-2.775) {\large  {\bf $v'_{N-1}$}};
\node at (3.5,-3.75) {\large  {\bf $v'_N$}};
%lables vertical
\node at (-0.35,-0.5) {\large  {\bf $-h_1$}};
\node at (2.6,-3.5) {\large  {\bf $-h_N$}};
%lables diagonal
\node at (0.6,0.25) {\large  {\bf $m'_1$}};
\node at (1.65,-0.8) {\large  {\bf $m'_2$}};
\node at (3.65,-2.8) {\large  {\bf $m'_N$}};
\node at (0,-1.5) {\large  {\bf $m'_N$}};
\node at (1.15,-3.2) {\large  {\bf $m'_{N-2}$}};
\node at (3.05,-4.5) {\large  {\bf $m'_{N-1}$}};
%roots
\draw[ultra thick,blue,<->] (1.05,0.95) -- (1.95,0.05);
\node[blue,rotate=315] at (1.75,0.65) {{\large {\bf {$\widehat{a}_{1}$}}}};
%
%\draw[ultra thick,blue,<->] (3.05,0.95) -- (3.95,0.05);
%\node[blue,rotate=315] at (3.75,0.65) {{\large {\bf {$\widehat{a}'_1$}}}};
%
\node[blue,rotate=315] at (2.5,-0.5) {\Huge $\cdots$};
\draw[ultra thick,blue,<->] (3.05,-1.05) -- (3.95,-1.95);
\node[blue,rotate=315] at (3.75,-1.35) {{\large {\bf {$\widehat{a}_{N-1}$}}}};
\draw[ultra thick,blue,<->] (4.05,-2.05) -- (4.95,-2.95);
\node[blue,rotate=315] at (4.75,-2.35) {{\large {\bf {$\widehat{a}_{N}$}}}};
%definition S
\begin{scope}[xshift=-4cm,yshift=4cm]
\draw[dashed] (7,-7) -- (5.35,-8.65);
\draw[dashed] (3.3,-5.7) -- (2.8,-6.2);
\draw[ultra thick,blue,<->] (2.8,-6.2) -- (5.35,-8.65);
\node[blue] at (3.9,-7.65) {{\large {\bf {$S$}}}};
\end{scope}
%parameter R
\begin{scope}[xshift=-3cm,yshift=3cm]
\draw[dashed] (2.3,-4.7) -- (1,-4.7);
\draw[dashed] (6,-6) -- (1,-6);
\draw[ultra thick,blue,<->] (1,-5.95) -- (1,-4.75);
\node[blue,rotate=90] at (0.75,-5.3) {{\large {\bf{$R-(N-1)S$}}}};
\end{scope}
\end{tikzpicture}}}
\caption{\sl Web diagram of $X_{N,1}^{(1)}$.}
\label{Fig:N1web0flop}
${}$\\[-3cm]
\end{wrapfigure}

In a similar fashion we can show that $\mathcal{G}_\infty(N)$ is a symmetry transformation. To this end, we first consider a transformation of the type $\mathcal{F}$ acting on the web diagram \figref{Fig:N1web0} for $\delta=0$ which results in the web digram shown in \figref{Fig:N1web0flop}, representing $X_{N,1}^{(\delta=1)}$. The blue parameters in \figref{Fig:N1web0flop}

\noindent
are the same as in \figref{Fig:N1web0}, while we also have introduced
\begin{align}
&v'_1=v_1+h_1+h_2\,,&&m'_1=m_1+h_1+h_2\,,\nonumber\\
&v'_2=v_2+h_2+h_3\,,&&m'_2=m_2+h_2+h_3\,,\nonumber\\
&\vdots &&\vdots\nonumber\\
&v'_N=v_N+h_N+h_1\,,&&m'_N=m_N+h_N+h_1\,.\nonumber
\end{align}
Cutting the diagram \figref{Fig:N1web0flop} along the lines $v'_{1,\ldots,N-1}$ and re-gluing it along the lines $m'_{1,\ldots,N}$ we obtain the web diagram shown in \figref{Fig:N1geninf}(a). Cutting the latter diagram furthermore along the line $-h_N$ it can also be represented in the form \figref{Fig:N1geninf}(b), which corresponds to a staircase diagram with shift $\delta=N-2$. The set of independent K\"ahler parameters $(\widehat{a}''_{1,\ldots,N},S'',R'')$ can be related to $(\widehat{a}_{1,\ldots,N},S,R)$ in the following manner
\begin{figure}[ht]
\begin{center}
\scalebox{0.7}{\parbox{21cm}{\begin{tikzpicture}[scale = 1.50]
\draw[ultra thick] (-1,-1) -- (0,0) -- (0,1) -- (1,2) -- (1,3) -- (1.5,3.5);
\node[rotate=45] at (2,4) {\Huge $\cdots$};
\draw[ultra thick] (2.5,4.5) -- (3,5) -- (3,6) -- (4,7) -- (4,8) -- (5,9);
%horizontals
\draw[ultra thick] (0,0) -- (1,0);
\draw[ultra thick] (1,2) -- (2,2);
\draw[ultra thick] (3,5) -- (4,5);
\draw[ultra thick] (4,7) -- (5,7);
\draw[ultra thick] (0,1) -- (-1,1);
\draw[ultra thick] (1,3) -- (0,3);
\draw[ultra thick] (3,6) -- (2,6);
\draw[ultra thick] (4,8) -- (3,8);
%ends
\node at (-1.15,-1.15) {\large {$\mathbf{a}$}};
\node at (5.15,9.15) {\large {$\mathbf{a}$}};
\node at (1.15,0) {\large {\bf $\mathbf 2$}};
\node at (2.15,2) {\large {\bf $\mathbf 3$}};
\node at (4.2,5) {\large {\bf $\mathbf N$}};
\node at (5.2,7) {\large {\bf $\mathbf 1$}};
\node at (-1.15,1) {\large {\bf $\mathbf 1$}};
\node at (-0.15,3) {\large {\bf $\mathbf 2$}};
\node at (1.5,6) {\large {\bf $\mathbf{N-1}$}};
\node at (2.8,8) {\large {\bf $\mathbf N$}};
%stamp
\node at (2,-2) {\Large {\bf $\mathbf (a)$}};
%diagonals
\node at (-0.2,-0.8) {\large  {\bf $m'_{N-1}$}};
\node at (0.6,1.2) {\large  {\bf $m'_N$}};
\node at (1.6,3.2) {\large  {\bf $m'_1$}};
\node at (3.8,6.2) {\large  {\bf $m'_{N-2}$}};
\node at (4.8,8.2) {\large  {\bf $m'_{N-1}$}};
%horizontals
\node at (0.5,-0.25) {\large  {\bf $v'_N$}};
\node at (1.5,1.75) {\large  {\bf $v'_1$}};
\node at (3.5,4.75) {\large  {\bf $v'_{N-2}$}};
\node at (4.5,6.75) {\large  {\bf $v'_{N-1}$}};
\node at (-0.5,0.75) {\large  {\bf $v'_{N-1}$}};
\node at (0.5,2.75) {\large  {\bf $v'_N$}};
\node at (2.5,5.75) {\large  {\bf $v'_{N-3}$}};
\node at (3.5,7.75) {\large  {\bf $v'_{N-2}$}};
%verticals
\node at (0.4,0.5) {\large  {\bf $-h_N$}};
\node at (1.4,2.5) {\large  {\bf $-h_1$}};
\node at (3.6,5.5) {\large  {\bf $-h_{N-2}$}};
\node at (4.6,7.5) {\large  {\bf $-h_{N-1}$}};
%%%%%%%%%%%%%%%%%%%%%%%
%%%%%%%%%%%%%%%%%%%%%%%
\begin{scope}[xshift=8cm,yshift=5.5cm]
\draw[ultra thick] (-1,0) -- (0,0) -- (0,-1) -- (1,-1) -- (1,-1.5);
\node[rotate=315] at (1.5,-2) {\Huge $\cdots$};
\draw[ultra thick] (2,-2.5) -- (2,-3) -- (3,-3) -- (3,-4) -- (4,-4);
%diagonals
\draw[ultra thick] (0,0) -- (0.7,0.7);
\draw[ultra thick] (1,-1) -- (1.7,-0.3);
\draw[ultra thick] (3,-3) -- (3.7,-2.3);
\draw[ultra thick] (0,-1) -- (-0.7,-1.7);
\draw[ultra thick] (2,-3) -- (1.3,-3.7);
\draw[ultra thick] (3,-4) -- (2.3,-4.7);
%ends
\node at (-1.2,0) {\large {\bf $\mathbf a$}};
\node at (4.2,-4) {\large {\bf $\mathbf a$}};
\node at (0.85,0.85) {\large {$\mathbf 1$}};
\node at (1.85,-0.15) {\large {$\mathbf 2$}};
\node at (3.85,-2.15) {\large {$\mathbf{N}$}};
\node at (-0.85,-1.5) {\large {$\mathbf{3}$}};
\node at (1.5,-3.8) {\large {$\mathbf{1}$}};
\node at (2.2,-4.9) {\large {$\mathbf{2}$}};
%lables hotizontal
\node at (-0.5,0.25) {\large  {\bf $-h_N$}};
\node at (0.5,-0.775) {\large  {\bf $-h_1$}};
\node at (2.5,-2.775) {\large  {\bf $-h_{N-1}$}};
\node at (3.5,-3.775) {\large  {\bf $-h_N$}};
%lables vertical
\node at (-0.3,-0.5) {\large  {\bf $m'_N$}};
\node at (2.45,-3.5) {\large  {\bf $m'_{N-1}$}};
%lables diagonal
\node at (0.75,0.25) {\large  {\bf $v'_{N-1}$}};
\node at (1.6,-0.85) {\large  {\bf $v'_{N}$}};
\node at (3.7,-2.8) {\large  {\bf $v'_{N-2}$}};
\node at (-0.1,-1.45) {\large  {\bf $v'_1$}};
\node at (1.15,-3.2) {\large  {\bf $v'_{N-1}$}};
\node at (2.85,-4.5) {\large  {\bf $v'_N$}};
%roots
\draw[ultra thick,blue,<->] (1.05,0.95) -- (1.95,0.05);
\node[blue,rotate=315] at (1.75,0.65) {{\large {\bf {$\widehat{a}''_{N-1}$}}}};
%
%\draw[ultra thick,blue,<->] (3.05,0.95) -- (3.95,0.05);
%\node[blue,rotate=315] at (3.75,0.65) {{\large {\bf {$\widehat{a}'_1$}}}};
%
\node[blue,rotate=315] at (2.5,-0.5) {\Huge $\cdots$};
\draw[ultra thick,blue,<->] (3.05,-1.05) -- (3.95,-1.95);
\node[blue,rotate=315] at (3.75,-1.35) {{\large {\bf {$\widehat{a}''_{N-3}$}}}};
\draw[ultra thick,blue,<->] (4.05,-2.05) -- (4.95,-2.95);
\node[blue,rotate=315] at (4.75,-2.35) {{\large {\bf {$\widehat{a}''_{N-2}$}}}};
%definition S
\begin{scope}[xshift=-4cm,yshift=4cm]
\draw[dashed] (8,-8) -- (6.35,-9.65);
\draw[dashed] (5.3,-7.7) -- (4.8,-8.2);
\draw[ultra thick,blue,<->] (4.8,-8.2) -- (6.35,-9.65);
\node[blue] at (5.5,-9.2) {{\large {\bf {$S''$}}}};
\end{scope}
%parameter R
\begin{scope}[xshift=-3cm,yshift=3cm]
\draw[dashed] (4.3,-6.7) -- (1,-6.7);
\draw[dashed] (6,-7) -- (1,-7);
\draw[ultra thick,blue,<-] (1,-7.05) -- (1,-7.75);
\draw[ultra thick,blue,<-] (1,-6.65) -- (1,-5.95);
\node[blue,rotate=90] at (0.75,-6.8) {{\large {\bf{$R''-2S''$}}}};
\end{scope}
\end{scope}
%stamp
\node at (9.5,-2) {\Large {\bf $\mathbf (b)$}};
\end{tikzpicture}}}
\caption{\sl (a) Alternative presentations of the web diagram in \figref{Fig:N1web0flop}. (b) Another presentation in the form of a shifted web diagram with $\delta=N-2$.}
\label{Fig:N1geninf}
\end{center}
\end{figure}
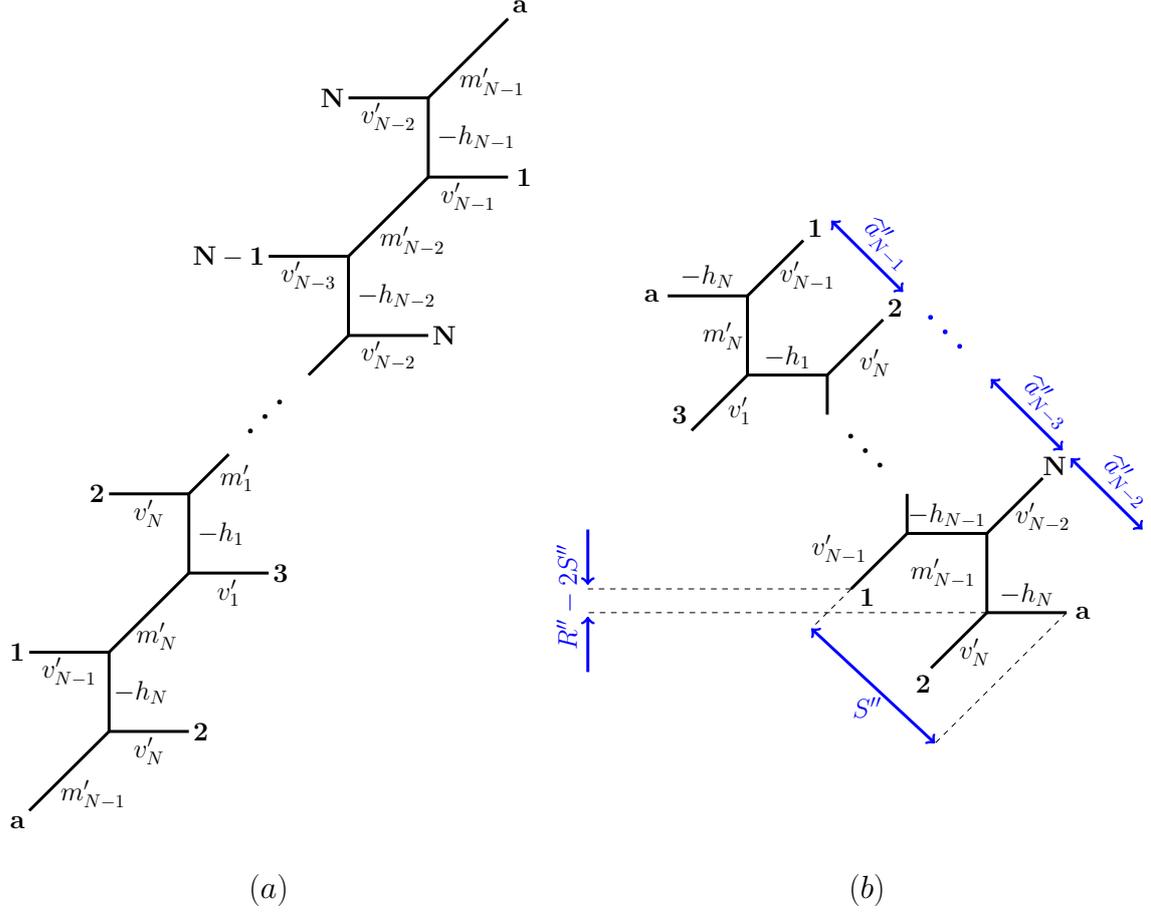
\begin{align}
\left(\widehat{a}_1\,, \ldots \,, \widehat{a}_N \,, S \,, R\right)^T=\mathcal{G}_\infty(N)\cdot \left(\widehat{a}''_1\,, \ldots \,, \widehat{a}''_N \,, S'' \,, R''\right)^T\,,\label{RelGenMatinf}
\end{align}
To show this, we use (\ref{DefGinfGeneral}) and (\ref{SimpleSolConsistency}) along with 
\begin{align}
&\widehat{a}''_i=m'_{i+1}-h_{i+2}=m+h_{i+1}\,,&&S''=m\,,&&R''-2S''=v-m
\end{align}
to compute (with $\rho''=\sum_{k=1}^N(m'_k-h_k)=Nm+\sum_{k=1}^Nh_k$)
{\allowdisplaybreaks
\begin{align}
&\widehat{a}_i=\widehat{a}''_i-2S''+R''=m+h_{i+1}+v-m=h_{i+1}+v\,,\nonumber\\
&S=\rho''-(2N-1)S''+(N-1)R''=\sum_{k=1}^Nh_k+(N-1)v\,,\nonumber\\
&R=N\rho''-2N(N-1)S''+(N-1)^2R''=N\sum_{k=1}^N h_k+m+(N-1)^2 v\,,
\end{align}}
which matches (\ref{IdentificationOrder2}) and therefore shows that $\mathcal{G}_\infty(N)$ is a symmetry transformation.
%%%%%%%%%%%%%%%%%%%%%%%%%
%%%%%%%%%%%%%%%%%%%%%%%%%
\subsection{Generators of the Dihedral Group}
After having shown that the transformations $\mathcal{G}_\infty(N)\cdot \mathcal{G}_2(N)$ and $\mathcal{G}_\infty(N)$ (and thus also $\mathcal{G}_2(N)$) are symmetry transformations of the partition function $\mathcal{Z}_{N,1}$, we shall now discuss the group structure they are generating. The matrix $\mathcal{G}_2(N)$ has order 2 (\emph{i.e.} $\mathcal{G}_2(N).\mathcal{G}_2(N)=1\!\!1_{(N+2)\times (N+2)}$), while $\mathcal{G}_\infty(N)$ has the following order
\begin{align}
\text{ord}\,\mathcal{G}_\infty(N)=\left\{\begin{array}{lcl} 3 & \text{if} & N=1\,, \\ 2 & \text{if} & N=2\,, \\ 3 & \text{if} & N=3\,, \\ \infty & \text{if} & N\geq 4\,. \end{array}\right.\label{OrderGenMatInf}
\end{align}
Here infinite order means $\nexists\, m\in\mathbb{N}$ such that $(\mathcal{G}_\infty(N))^m=1\!\!1_{(N+2)\times (N+2)}$. While we have shown all cases $N\leq 4$ explicitly in previous sections, for $N> 4$ it is sufficient to realise that 
\begin{align}
\vec{v}_N=\bigg(\underbrace{1,\ldots,1}_{N\text{ times}},\frac{N+\sqrt{N(N-4)}}{2},\frac{N}{2}(N-2+\sqrt{N(N-4)})\bigg)^T\,,
\end{align} 
is an eigenvector of $\mathcal{G}_\infty(N)$ for the eigenvalue\footnote{The remaining eigenvalues are $+1$ (with degeneracy $N$) and $\lambda_N^{-1}$.}
\begin{align}
\lambda_N=\frac{1}{2} \left((N-2)^2-2+\sqrt{N(N-4)} (N-2)\right)\in\mathbb{R}\,.
\end{align}
Since $\lambda_N>1$ for $N\geq 5 $ (and $\mathcal{G}_\infty(N)$ is diagonaliseable for $N\geq 5$) it follows that $\mathcal{G}_\infty(N)$ is not of finite order in these cases. Thus, upon introducing the matrix
\begin{align}
\mathcal{G}'_2(N)=\mathcal{G}_2(N)\cdot \mathcal{G}_\infty(N)=\left(\begin{array}{ccccc} & & & -2 & 1 \\ & 1\!\!1_{N\times N} & & \vdots & \vdots \\ & & & -2 & 1 \\ 0 & \cdots & 0 & -1 & 1 \\ 0 & \cdots & 0 & 0 & 1 \end{array}\right)\,,
\end{align}
which is of order $2$ (\emph{i.e.} $\mathcal{G}'_2(N).\mathcal{G}'_2(N)=1\!\!1_{N+2\times N+2}$), we find that $\mathcal{G}_2(N)$ and $\mathcal{G}'_2(N)$ freely generate a dihedral group 
\begin{align}
&\mathbb{G}(N)=\left\langle \{\mathcal{G}_2(N)\,,\mathcal{G}'_2(N)\}%\big|(\mathcal{G}_2(N))^2=1\!\!1_{N+2\times N+2}=(\mathcal{G}'_2(N))^2
\right\rangle\cong\left\{ \begin{array}{lcl}\text{Dih}_3 & \text{if} & N=1\,, \\ \text{Dih}_2 & \text{if} & N=2\,, \\\text{Dih}_3 & \text{if} & N=3\,, \\ \text{Dih}_\infty & \text{if} & N\geq 4\,.\end{array}\right.\,.\label{DefGNgeneric}
\end{align}
For $N\geq 4$, eq.~(\ref{OrderGenMatInf}) shows that $\nexists\,\,n\in\mathbb{N} $ with $(\mathcal{G}_2(N)\cdot\mathcal{G}'_2(N))^n=1\!\!1_{(N+2)\times (N+2)}$ (which also implies $\nexists\,\,n\in\mathbb{N} $ with $(\mathcal{G}'_2(N)\cdot\mathcal{G}_2(N))^n=1\!\!1_{(N+2)\times (N+2)}$). Furthermore, since $(\mathcal{G}_2(N))^2=1\!\!1_{(N+2)\times (N+2)}=(\mathcal{G}'_2(N))^2$, this also implies $\nexists\,\,n\in\mathbb{N} $ with $\mathcal{G}'_2(N)\cdot (\mathcal{G}_2(N)\cdot\mathcal{G}'_2(N))^n=1\!\!1_{(N+2)\times (N+2)}$ or $(\mathcal{G}_2(N)\cdot\mathcal{G}'_2(N))^n\cdot \mathcal{G}_2(N)=1\!\!1_{(N+2)\times (N+2)}$.\footnote{For example, the former relation is equivalent to $(\mathcal{G}_2(N)\cdot\mathcal{G}'_2(N))^n=\mathcal{G}'_2(N)$. Squaring this relation (due to the fact that $\mathcal{G}'_2(N)$ is of order 2) would be equivalent to $(\mathcal{G}_2(N)\cdot\mathcal{G}'_2(N))^{2n}=1\!\!1_{(N+2)\times (N+2)}$, which does not agree with (\ref{OrderGenMatInf}).} This means that there are no non-trivial (braid) relations between $\mathcal{G}_2(N)$ and $\mathcal{G}'_2(N)$, which indeed shows that the group $\mathbb{G}(N)\cong\text{Dih}_\infty$ for $N\geq 4$.

Notice that $\mathcal{G}_2(N)$ is a lower diagonal matrix, while $\mathcal{G}'_2(N)$ is an upper diagonal $(N+2)\times (N+2)$ matrix. Furthermore, the partition function is invariant under the action of the group $\widetilde{S}_N$,\footnote{$\widetilde{S}_N$ is a subgroup of $S_N$ and for $N\geq 3$ is isomorphic to $\text{Dih}_N$. For $N=2$ we have $\widetilde{S}_2\cong S_2$.} which is generated by matrices of the form
\begin{align}
R(M)=\left(\begin{array}{ccccc} & & & 0 & 0 \\ & M & & \vdots & \vdots  \\ & & & 0 & 0 \\ 0 & 0 & 0 & 1 & 0 \\ 0 & 0 & 0 & 0 & 1\end{array}\right)\,,
\end{align}
where $M$ is an $N\times N$ matrix that acts by permuting the $\widehat{a}_{1,\ldots,N}$. One can check that matrices of the form $R(M)$ commute with both $\mathcal{G}_2(N)$ and $\mathcal{G}'_2(N)$, such that we have the following symmetry group of the partition function $\widetilde{G}(N)\cong \mathbb{G}(N)\times \widetilde{S}_N$.
%%%%%%%%%%%%%%%%%%%%%%%%%%%%%%%%%%%%%%%%%%%%%%%%%
%%%%%%%%%%%%%%%%%%%%%%%%%%%%%%%%%%%%%%%%%%%%%%%%%
\subsection{Modularity at a Particular Point of the Moduli Space}
Using the general parametrisation of the group $\mathbb{G}(N)$ in (\ref{DefGNgeneric}), we once again ask the question how the latter is related to $Sp(4,\mathbb{Z})$ at the particular region in the moduli space, which is characterised by $\widehat{a}_{1,\ldots,N}^{(0)}=\widehat{a}_4^{(0)}=\widehat{a}$, which implies $h_{1,\ldots,N}=h$ (while the consistency conditions already impose $v_{1,\ldots,N}=v$ and $m_{1,\ldots,N}=m$). We can similarly introduce the period matrix
\begin{align}
&\Omega=\left(\begin{array}{cc}\tau & v \\ v & \rho\end{array}\right)\,,&&\text{with} &&\begin{array}{l}\tau=m+v\,,\\ \rho=h+m\,.\end{array}\label{DefNPeriodMatrix}
\end{align} 
Using the parametrisation (\ref{DefGNgeneric}) of $\mathbb{G}(4)$, it is sufficient to analyse the relation of the generators $\mathcal{G}_2(N)$ and $\mathcal{G}'_2(N)$ to $Sp(4,\mathbb{Z})$. The restriction of these generators to the subspace $(\widehat{a},S,R)$ can be written in the form
\begin{align}
&\mathcal{G}_2^{(\text{red})}(N)=\left(\begin{array}{ccc}1 & -2 & 1 \\ 0 & -1 & 1 \\ 0 & 0 & 1\end{array}\right)\,,&&\text{and} &&\mathcal{G}_2^{\prime,(\text{red})}(N)=\left(\begin{array}{ccc}1 & 0 & 0 \\ N & -1 & 0 \\ N^2 & -2N & 1\end{array}\right)\,,
\end{align}
or on the space $(\tau,\rho,v)$
\begin{align}
&\widetilde{\mathcal{G}}_2^{(\text{red})}(N)=D_N^{-1}\cdot \mathcal{G}_2^{(\text{red})}(N)\cdot D_N=\left(
\begin{array}{ccc}
 (N-1)^2 & (N-2)^2 N^2 & -2 N \left(N^2-3 N+2\right) \\
 1 & (N-1)^2 & 2(1- N) \\
 N-1 & N \left(N^2-3 N+2\right) & -2 N^2+4 N-1 \\
\end{array}
\right)\,, \nonumber\\
&\widetilde{\mathcal{G}}_2^{\prime,(\text{red})}(N)=D_N^{-1}\cdot \mathcal{G}_2^{\prime,(\text{red})}(N)\cdot D_N=\left(\begin{array}{ccc}1 & 4 & -4 \\ 0 & 1 & 0 \\ 0 & 2 & -1\end{array}\right)\,,\hspace{0.5cm}\hspace{0.5cm}\text{with} \hspace{0.5cm} D_N=\left(\begin{array}{ccc} 0 & 1 & 0 \\ 0 & N & -1 \\ 1 & N^2 & -2N\end{array}\right)\,.\nonumber
\end{align}
Rewriting these generators furthermore to act as elements of $Sp(4,\mathbb{Z})$ in the form of (\ref{ActionSp4Period}) on the period matrix $\Omega$ in (\ref{DefNPeriodMatrix}), they take the form
{\allowdisplaybreaks
\begin{align}
&\widetilde{\mathcal{G}}_2^{(\text{red,Sp})}(N)=(HKL^6H)^{N-2}K(HL^6 KH)^{N-2}=\left(\begin{array}{cccc}N-1 & 1-(N-1)^2 & 0 & 0 \\ 1 & 1-N & 0 & 0 \\ 0 & 0 & N-1 & 1 \\ 0 & 0 & 1-(N-1)^2 & 1-N\end{array}\right)\,,\nonumber\\
&\widetilde{\mathcal{G}}_2^{\prime,(\text{red,Sp})}(N)=HK(L^6K)^{N-1}H=\left(\begin{array}{cccc}-1 & N & 0 & 0 \\ 0 & 1 & 0 & 0 \\ 0 & 0 & -1 & 0 \\ 0 & 0 & N & 1\end{array}\right)\,,
\end{align}}
where $K$, $L$ and $H$ are defined in appendix~\ref{App:Sp4}. For $N\in\mathbb{N}$, the restriction of $\mathbb{G}(N)$ to the particular region of the K\"ahler moduli space $(\widehat{a},S,R)$ is a subgroup of $Sp(4,\mathbb{Z})$. However, for $N>1$, we cannot conclude that the freely generated group $\langle \widetilde{\mathcal{G}}_2^{(\text{red,Sp})}(N),\widetilde{\mathcal{G}}_2^{\prime,(\text{red})}(N), S_\rho,T_\rho,S_\tau,T_\tau \rangle$ is isomorphic to $Sp(4,\mathbb{Z})$.
%%%%%%%%%%%%%%%%%%%%%%%%%%%%%%%%%%%
%%%%%%%%%%%%%%%%%%%%%%%%%%%%%%%%%%%
%%%%%%%%%%%%%%%%%%%%%%%%%%%%%%%%%%%
%%%%%%%%%%%%%%%%%%%%%%%%%%%%%%%%%%%
\section{Conclusions}\label{Sect:Conclusions}
In this paper, we studied the consequences of the web of dualities among certain supersymmetric quiver gauge theories on $\mathbb{R}^5\times S^1$ which are engineered by a class of toric Calabi-Yau threefolds $X_{N,M}$. These dualities have been established in \cite{Hohenegger:2016yuv,Bastian:2017ing,Bastian:2017ary,Bastian:2018dfu}, here, however, rather than focusing on the different physical theories, we have analysed their consequences from the perspective of the partition function $\mathcal{Z}_{N,M}$. For the sake of simplicity, our analysis has been limited to the case $M=1$. We found that the partition function $\mathcal{Z}_{N,1}$ associated to the geometries $X_{N,1}$ is invariant under the group $\widetilde{\mathbb{G}}(N) \cong \mathbb{G}(N) \times \widetilde{S}_N$ which acts on the vector space spanned by a maximal set of independent K\"ahler parameters. Here $\widetilde{S}_N\subset S_N$ has an intuitive interpretation as a subgroup of the largest gauge group that can be engineered by the given geometry, which is $U(N)$ in this case. The group $\mathbb{G}(N)$, was shown to depend on $N$ as derived in (\ref{DefGNgeneric}) and was found by exploiting the fact that $X_{N,1}$ can be related to various other geometries (that are part of the same extended K\"ahler moduli space) trough flop- and symmetry transformations. These geometries are characterised by giving rise to the same topological string partition function (\emph{i.e.} the same $\mathcal{Z}_{N,1}$), but they are described by web diagrams which have their K\"ahler parameters related trough a non-trivial duality map to the ones of the initial geometry. By studying a collection of these `self-duality' maps we showed that they form the group $\widetilde{\mathbb{G}}(N)$. 

A notable feature is the appearance of the infinite dihedral group for $N\geq 4$. By using the matrix representations of the generating elements, we have explicitly shown in section \ref{Sect:GeneralCase} that for the cases $N \geq 4$, the group $\mathbb{G}(N)$ is generated by two matrices of order $2$, which have no non-trivial braid relations (implying the existence of a group element of infinite order). An intuitive understanding of the appearance of the infinite order generator can be gained by looking at the behaviour under the series of flop transformations $\mathcal{F}$, reviewed in appendix \ref{App:FlopTrafo}. They can be used to relate web diagrams that look identical but have a non-trivial mapping between their K\"ahler parameters. By iterating this procedure, it is thus possible to generate an infinite series of inequivalent web diagrams, thus giving an intuitive argument for the appearance of an infinite order group. For the cases with $N\leq 3$ there is no such iterative procedure for producing non-trivially related geometries, due to the simpler nature of the diagram.

Furthermore, we showed that $\mathbb{G}(N)$ combines non-trivially with other known symmetry groups of the partition function. For the case $N=1$, we showed explicitly that $\mathbb{G}(1) \cong \text{Dih}_3$ together with the modular group $SL(2,\mathbb{Z})$ freely generate $Sp(4,\mathbb{Z})$, which is known to be the automorphism group associated to the mirror curve of $X_{1,1}$ \cite{Hollowood:2003cv,Kanazawa:2016tnt}. For $N>1$, we showed that in a particular region of the K\"ahler moduli space, $\mathbb{G}(N)$ corresponds to a subgroup of $Sp(4,\mathbb{Z})$. Similarly, the group $\widetilde{\mathbb{G}}(N)$ mixes non-trivially with the T-duality (as specifically proposed in \cite{Hohenegger:2015btj}) that relates the IIa and IIb Little String Theories that are engineered by $X_{N,1}$. In both cases, it would be interesting to extend this analysis and to characterise the full (non-perturbative) U-duality group of the LSTs. We leave this point for future work.

From the perspective of the various gauge theories engineered by $X_{N,1}$, the symmetry group $\widetilde{\mathbb{G}}(N)$ also has important consequences: acting in the form of eq.~(\ref{SymTrafoFreeEnergy}), it identifies the multiplicities of certain single particle BPS states in the free energy. This symmetry acts a priori non-perturbatively, since in particular an element $G\in\mathbb{G}(N)$ mixes all K\"ahler parameters of $X_{N,1}$ (which from the perspective of the BPS states of the gauge theory, correspond to various fugacities in the free energy) in an arbitrary fashion. It is also important to remember, that in general there are several different gauge theories that are engineered by $X_{N,1}$: as argued in \cite{Bastian:2018dfu}, the latter engineers circular quiver gauge theories with $M'$ nodes of type $U(N')$ for any $(N',M')$ with $N'M'=N$ and $\text{gcd}(N',M')=1$. All these theories are dual to one another, in the sense that they have the same partition function $\mathcal{Z}_{N,1}$ and thus also share the symmetry $\widetilde{\mathbb{G}}(N)$. The main difference is that the latter acts very differently from the perspective of the BPS spectrum: indeed, these theories differ in how the physical parameters (like coupling constants or Coulomb branch parameters) are expressed in terms of the K\"ahler parameters of $X_{N,1}$. The action of $\widetilde{\mathbb{G}}(N)$ on the latter thus leads to different (physical) symmetries from the perspective of the various gauge theories (in particular their BPS states).

Another important aspect concerns the relation of the symmetry group $\widetilde{\mathbb{G}}(N)$ with other symmetries that have previously been observed in the literature: 
\begin{itemize}
\item In \cite{Hohenegger:2016eqy} it was found that (at a particular region in the K\"ahler moduli space of $X_{N,1}$), the free energy $\mathcal{F}_{N,1}$ in the NS-limit is fully captured by $\mathcal{F}_{1,1}$.
\item In \cite{Ahmed:2017hfr} it was argued that in the NS limit a particular part of $\mathcal{Z}_{N,M}$ (called the reduced partition function) can be be written as the partition function of a symmetric orbifold CFT, giving rise to numerous Hecke like relations between various terms in the corresponding free energies.
\item In \cite{Bastian:2017jje} it was demonstrated at a large number of examples that (in the unrefined limit) for a particular choice of some of the K\"ahler parameters, the partition function $\mathcal{Z}_{N,M}$ can be written as the sum over the weights of a single integrable representation of the affine Lie algebra $\widehat{\mathfrak{a}}_{N-1}$ associated with the gauge group $U(N)$. 
\end{itemize}
It is important that in all these cases, it was necessary to choose particular values for (some of) the K\"ahler moduli and/or the regularisation parameters $\epsilon_{1,2}$, in one way or another. The elements of the group $\widetilde{\mathbb{G}}(N)$ we found in the current work, are more general in the sense that they are symmetries of $\mathcal{Z}_{N,1}$ (or the corresponding free energy $\mathcal{F}_{N,1}$) at a generic point in the K\"ahler moduli space of $X_{N,1}$ and for generic values of $\epsilon_{1,2}$.\footnote{Indeed, the group $\widetilde{\mathbb{G}}(N)$ is based on dualities among web diagrams, which themselves are blind to $\epsilon_{1,2}$. Furthermore, while we considered the NS-limit (combined with the unrefined limit) in Sections~\ref{Sect:Order3Checks} and \ref{Sect:Ord4Checks}, the latter was only a convenience to keep the computational complexity at bay when performing certain checks of the symmetry transformations. The latter, however, hold in full generality.} In the future, it will be interesting to analyse, how $\widetilde{\mathbb{G}}(N)$ combines with the additional symmetries mentioned above in the respective regions of the moduli space.

At a generic point in the moduli space, it would be interesting to analsye how $\widetilde{\mathbb{G}}(N)$ combines with other symmetries of the partition function (such as the modular groups $SL(2,\mathbb{Z})_\tau$ and $SL(2,\mathbb{Z})_\rho$) to form an even larger symmetry group. As the symmetries discussed in this work impose severe constraints on the structure of $\mathcal{Z}_{N,1}$, it would be interesting to investigate how much perturbative information (from the perspective of either one of the gauge theories engineered by $X_{N,1}$) on the spectrum is required to recover the whole non-perturbative partition function. Questions of this type have recently been considered, \emph{e.g.} in \cite{Kim:2018gak}, where the authors showed that the partition function can be reconstructed by using information from the 2d world-sheet theories of the little string in combination with T-duality.

Another interesting implication of the symmetries discussed in this work concerns the consequences at the level of the gauge theories themselves. For example, in \cite{Mitev:2014jza}, the authors used the well known fiber-base duality of (a limit of) $X_{N,1}$ in order to argue for an enhancement of the global symmetry group of a certain class of five-dimensional theories at their superconformal fixed point. They showed explicitly the appearance of characters of the enhanced global symmetry group when expanding the Nekrasov partition function in a specific set of Coulomb branch parameters that are invariant under fiber-base duality. While the theories we analysed here are six-dimensional and also do not have a superconformal fixed point (rather their UV completions are LSTs), one might hope to gain information about some enhanced global symmetry. We leave some of these points for future work.
%%%%%%%%%%%%%%%%%%%%%%%%%%%%%%%%%%%
%%%%%%%%%%%%%%%%%%%%%%%%%%%%%%%%%%%
%%%%%%%%%%%%%%%%%%%%%%%%%%%%%%%%%%%
\section*{Acknowledgements}
We would like to thank Amer~Iqbal and Soo-Jong~Rey for collaboration on related topics and many useful exchanges. Furthermore, SH would like to thank Fabrizio~Nieri for useful discussions and Pietro~Longhi for many interesting exchanges and a careful reading of a preliminary version of the manuscript.
%%%%%%%%%%%%%%%%%%%%%%%%%%%%%%%%%%%
\appendix
\section{Duality Transformation $\mathcal{F}$}\label{App:FlopTrafo}
Since it is frequently used in the main body of this paper, in this appendix we review a particular duality transformation (called $\mathcal{F}$) that was first proposed in \cite{Hohenegger:2016yuv} (see also \cite{Bastian:2017ing}) and which acts on a shifted web diagram as shown in \figref{Fig:N1web0} by changing $\delta\to \delta+1$. We specifically recall the duality map.

Starting from the web diagram in \figref{Fig:N1web0} with shift $\delta\in\{0,\ldots,N-1\}$, the duality trans-

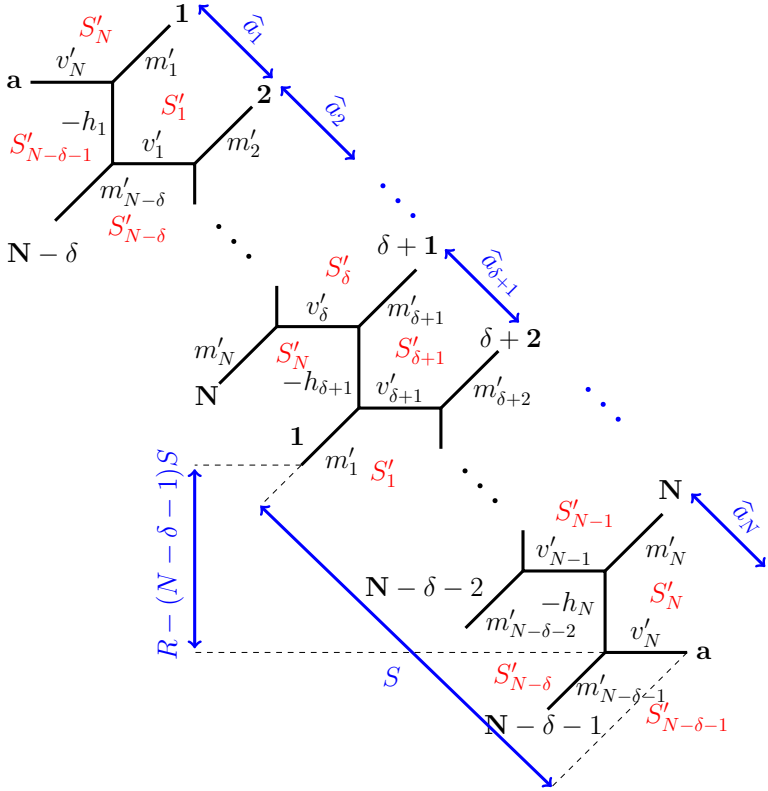
\begin{wrapfigure}{l}{0.62\textwidth}
\begin{center}
\vspace{-0.7cm}
\scalebox{0.72}{\parbox{15cm}{\begin{tikzpicture}[scale = 1.50]
\draw[ultra thick] (-1,0) -- (0,0) -- (0,-1) -- (1,-1) -- (1,-1.5);
\node[rotate=315] at (1.5,-2) {\Huge $\cdots$};
\draw[ultra thick] (2,-2.5) -- (2,-3) -- (3,-3) -- (3,-4) -- (4,-4) -- (4,-4.5);
\node[rotate=315] at (4.5,-5) {\Huge $\cdots$};
\draw[ultra thick] (5,-5.5) -- (5,-6) -- (6,-6) -- (6,-7) -- (7,-7);
%diagonals
\draw[ultra thick] (0,0) -- (0.7,0.7);
\draw[ultra thick] (1,-1) -- (1.7,-0.3);
\draw[ultra thick] (3,-3) -- (3.7,-2.3);
\draw[ultra thick] (0,-1) -- (-0.7,-1.7);
\draw[ultra thick] (2,-3) -- (1.3,-3.7);
\draw[ultra thick] (3,-4) -- (2.3,-4.7);
\draw[ultra thick] (4,-4) -- (4.7,-3.3);
\draw[ultra thick] (5,-6) -- (4.3,-6.7);
\draw[ultra thick] (6,-6) -- (6.7,-5.3);
\draw[ultra thick] (6,-7) -- (5.3,-7.7);
%ends
\node at (-1.2,0) {\large {\bf $\mathbf a$}};
\node at (7.2,-7) {\large {\bf $\mathbf a$}};
\node at (0.85,0.85) {\large {$\mathbf 1$}};
\node at (1.85,-0.15) {\large {$\mathbf 2$}};
\node at (3.6,-2) {\large {$\mathbf{\delta+1}$}};
\node at (4.85,-3.15) {\large {$\mathbf{\delta+2}$}};
\node at (6.8,-5) {\large {$\mathbf{N}$}};
\node at (-0.85,-2.1) {\large {$\mathbf{N-\delta}$}};
\node at (1.15,-3.85) {\large {$\mathbf N$}};
\node at (2.25,-4.35) {\large {$\mathbf 1$}};
\node at (3.8,-6.2) {\large {$\mathbf N-\delta-2$}};
\node at (5.25,-7.9) {\large {$\mathbf N-\delta-1$}};
%lables hotizontal
\node at (-0.5,0.25) {\large  {\bf $v'_N$}};
\node at (0.5,-0.76) {\large  {\bf $v'_1$}};
\node at (2.5,-2.76) {\large  {\bf $v'_{\delta}$}};
\node at (3.5,-3.76) {\large  {\bf $v'_{\delta+1}$}};
\node at (5.5,-5.76) {\large  {\bf $v'_{N-1}$}};
\node at (6.5,-6.76) {\large  {\bf $v'_N$}};
%lables vertical
\node at (-0.35,-0.5) {\large  {\bf $-h_1$}};
\node at (2.5,-3.7) {\large  {\bf $-h_{\delta+1}$}};
\node at (5.55,-6.4) {\large  {\bf $-h_N$}};
%lables diagonal
\node at (0.6,0.25) {\large  {\bf $m'_1$}};
\node at (1.6,-0.75) {\large  {\bf $m'_2$}};
\node at (3.7,-2.8) {\large  {\bf $m'_{\delta+1}$}};
\node at (4.75,-3.8) {\large  {\bf $m'_{\delta+2}$}};
\node at (6.75,-5.75) {\large  {\bf $m'_N$}};
\node at (0.25,-1.35) {\large  {\bf $m'_{N-\delta}$}};
\node at (1.25,-3.3) {\large  {\bf $m'_{N}$}};
\node at (2.8,-4.65) {\large  {\bf $m'_1$}};
\node at (5.1,-6.65) {\large  {\bf $m'_{N-\delta-2}$}};
\node at (6.2,-7.45) {\large  {\bf $m'_{N-\delta-1}$}};
%hexagons
\node[red] at (-0.2,0.65) {\large  {\bf $S'_N$}};
\node[red] at (0.75,-0.3) {\large  {\bf $S'_1$}};
%\node[red] at (1.75,-1.3) {\large  {\bf $S_2$}};
\node[red] at (2.75,-2.3) {\large  {\bf $S'_{\delta}$}};
\node[red] at (3.75,-3.3) {\large  {\bf $S'_{\delta+1}$}};
\node[red] at (5.75,-5.3) {\large  {\bf $S'_{N-1}$}};
\node[red] at (6.75,-6.3) {\large  {\bf $S'_N$}};
\node[red] at (-0.75,-0.8) {\large  {\bf $S'_{N-\delta-1}$}};
\node[red] at (0.3,-1.8) {\large  {\bf $S'_{N-\delta}$}};
%\node[red] at (1.25,-2.75) {\large  {\bf $S_2$}};
\node[red] at (2.2,-3.35) {\large  {\bf $S'_{N}$}};
\node[red] at (3.3,-4.8) {\large  {\bf $S'_1$}};
\node[red] at (5,-7.3) {\large  {\bf $S'_{N-\delta}$}};
\node[red] at (7,-7.8) {\large  {\bf $S'_{N-\delta-1}$}};
%roots
\draw[ultra thick,blue,<->] (1.05,0.95) -- (1.95,0.05);
\node[blue,rotate=315] at (1.75,0.65) {{\large {\bf {$\widehat{a}_1$}}}};
\draw[ultra thick,blue,<->] (2.05,-0.05) -- (2.95,-0.95);
\node[blue,rotate=315] at (2.75,-0.35) {{\large {\bf {$\widehat{a}_2$}}}};
\node[blue,rotate=315] at (3.5,-1.5) {\Huge $\cdots$};
%\draw[ultra thick,blue,<->] (3.05,-1.05) -- (3.95,-1.95);
%\node[blue,rotate=315] at (3.75,-1.35) {{\large {\bf {$\widehat{a}_3$}}}};
%
\draw[ultra thick,blue,<->] (4.05,-2.05) -- (4.95,-2.95);
\node[blue,rotate=315] at (4.75,-2.35) {{\large {\bf {$\widehat{a}_{\delta+1}$}}}};
\node[blue,rotate=315] at (6,-4) {\Huge $\cdots$};
\draw[ultra thick,blue,<->] (7.05,-5.05) -- (7.95,-5.95);
\node[blue,rotate=315] at (7.75,-5.35) {{\large {\bf {$\widehat{a}_N$}}}};
%definition S
\draw[dashed] (7,-7) -- (5.35,-8.65);
\draw[dashed] (2.3,-4.7) -- (1.8,-5.2);
\draw[ultra thick,blue,<->] (1.8,-5.2) -- (5.35,-8.65);
\node[blue] at (3.4,-7.3) {{\large {\bf {$S$}}}};
%parameter R
\draw[dashed] (2.3,-4.7) -- (1,-4.7);
\draw[dashed] (6,-7) -- (1,-7);
\draw[ultra thick,blue,<->] (1,-6.95) -- (1,-4.75);
\node[blue,rotate=90] at (0.7,-5.8) {{\large {\bf{$R-(N-\delta-1)S$}}}};
\end{tikzpicture}}}
\caption{\sl Web diagram after $\mathcal{F}$ acting on $X_{N,1}^{(\delta)}$.}
\label{Fig:N1webTrafoF}
\end{center}
${}$\\[-2cm]
\end{wrapfigure} 

\noindent
formation $\mathcal{F}$ is comprised of flop transformations on the curves with areas $\{h_1,\ldots,h_N\}$, along with $SL(2,\mathbb{Z})$ transformations and cutting and re-gluing of the web diagram. The resulting web diagram can again be presented in the form of a shifted 'staircase' diagram with shift $\delta+1$, as shown in \figref{Fig:N1webTrafoF}.

It is important to notice that the independent K\"ahler parameters $(\widehat{a}_{1,\ldots,N},S,R)$ (shown in blue in \figref{Fig:N1webTrafoF}) are in fact the same parameters as in \figref{Fig:N1web0}, which in \cite{Bastian:2017ing} were indeed shown to be invariant under the duality transformation. Similarly, these parameters are a solution of the consistency conditions imposed by the hexagons $S'_{1,\ldots,N}$, the latter being equivalent to the conditions (\ref{ConsistGeneral}) stemming from the hexagons $S_{1,\ldots,N}$ in the web diagram in \figref{Fig:N1web0}. While the basis of the K\"ahler parameters $(\widehat{a}_{1,\ldots,N},S,R)$ is invariant under $\mathcal{F}$, the individual curves $(h_{1,\ldots,N},v_{1,\ldots,N},m_{1,\ldots,N})$ are not invariant under the transformation $\mathcal{F}$. Indeed, with respect to \figref{Fig:N1webTrafoF} we have the following duality map
{\allowdisplaybreaks
\begin{align}
&v'_1=v_1+h_1+h_2\,,&&m'_1=m_1+h_1+h_{\delta+2}\,,\nonumber\\
&v'_2=v_2+h_2+h_3\,,&&m'_2=m_2+h_2+h_{\delta+3}\,,\nonumber\\
&\hspace{0.6cm}\vdots &&\hspace{0.75cm}\vdots\nonumber\\
&v'_{N-\delta}=v_{N-\delta}+h_{N-\delta}+h_{N-\delta+1}\,,&&m'_{N-\delta}=m_{N-\delta}+h_1+h_{N-\delta}\nonumber\\
&\hspace{0.6cm}\vdots &&\hspace{0.75cm}\vdots\nonumber\\
&v'_N=v_N+h_1+h_N\,,&&m'_N=m_N+h_N+h_{\delta+1}\,,
\end{align}}
where $h_{i+N}=h_i$ for $i=1,\ldots,N$.
%%%%%%%%%%%%%%%%%%%%%%%%%%%%%%%%%%%
\section{Presentation of $Sp(4,\mathbb{Z})$ and Modularity}\label{App:Sp4}
In \cite{Bender} a presentation of $Sp(4,\mathbb{Z})$ in terms of 2 generators (satisfying 8 defining relations) has been given. The latter are of order 2 and 12 respectively
\begin{align}
&K=\left(\begin{array}{cccc}1 & 0 & 0 & 0 \\ 1 & -1 & 0 & 0 \\ 0 & 0 & 1 & 1 \\ 0 & 0 & 0 & -1\end{array}\right)\,,&&\text{and} &&L=\left(\begin{array}{cccc} 0 & 0 & -1 & 0 \\ 0 & 0 & 0 & -1 \\ 1 & 0 & 1 & 0 \\ 0 & 1 & 0 & 0 \end{array}\right)\,,
\end{align}
which satisfy
\begin{align}
&K^2=L^{12}=1\!\!1_{4\times 4}\,, \hspace{0.5cm}(KL^7 KL^5 K)L=L (KL^5KL^7K)\,, \hspace{0.5cm}(L^2 K L^4)H=H(L^2 KL^4)\,,\nonumber\\
&(L^3 KL^3)H=H(L^3KL^3)\,, \hspace{0.4cm}(L^2 H)^2=(HL^2)^2\,, \hspace{0.4cm}L(L^6 H)^2=(L^6 H)^2 L\,, \hspace{0.4cm}(KL^5)^5=(L^6H)^2\,,\nonumber
\end{align}
where $H=KL^5KL^7K$. We also mention that another presentation \cite{Behr} (in terms of 6 generators and 18 defining relations) is given by $X_{1,2,3,4,5,6}$, which can be expressed in terms of $L$ and $K$ as follows
\begin{align}
&X_1=L^5 K L\,,&&X_2=L^9 H L^{10} H\,,&&X_3=L^8 K L^{10}\,,\nonumber\\
&X_4=H L^9 H L^{10}\,,&&X_5=HL^6\,,&&X_6=L^9H L^6H\,.\label{DefXGeneratorsBehr}
\end{align}
Furthermore, the group $Sp(4,\mathbb{Z})$ acts in a very natural form on the period matrix $\Omega=\left(\begin{array}{cc}\tau & v \\ v & \rho\end{array}\right)$ of a genus 2 Riemann surface
\begin{align}
&\left(\begin{array}{cc} A & B \\ C & D\end{array}\right):\hspace{1cm}\Omega\longmapsto (A\Omega+B)(C\Omega+D)^{-1}\,.\label{ActionSp4Period}
\end{align}
Here $A,B,C,D$ are $2\times 2$ matrices that satisfy
\begin{align}
&A^TD-C^TB=1\!\!1_{2\times 2}=DA^T-CB^T\,,&&A^T C=C^T A\,,&&B^T D=D^T B\,.
\end{align}
For convenience, we provide the action of some of the generators on the period matrix $\Omega$
{\allowdisplaybreaks
\begin{align}
&K:\hspace{0.25cm} \Omega\to \left(
\begin{array}{cc}
 \tau  & \tau -v \\
 \tau -v & -2 v+\rho +\tau  \\
\end{array}
\right)\,,
&&L^3:\hspace{0.25cm}\Omega\to \left(
\begin{array}{cc}
 \tau -\frac{v^2}{\rho } & \frac{v}{\rho } \\
 \frac{v}{\rho } & -\frac{1}{\rho } \\
\end{array}
\right)\,,\nonumber\\[4pt]
%%%
&L^6:\hspace{0.25cm}\Omega\to \left(
\begin{array}{cc}
 \tau  & -v \\
 -v & \rho  \\
\end{array}
\right)\,,
&&L^9:\hspace{0.25cm}\Omega\to \left(
\begin{array}{cc}
 \tau -\frac{v^2}{\rho } & -\frac{v}{\rho } \\
 -\frac{v}{\rho } & -\frac{1}{\rho } \\
\end{array}
\right)\,,\nonumber\\[4pt]
%%%
&H:\hspace{0.25cm}\Omega\to \left(
\begin{array}{cc}
 \rho  & v \\
 v & \tau  \\
\end{array}
\right)\,,
&&L^2 K L^4:\hspace{0.25cm}\Omega\to \left(
\begin{array}{cc}
 \tau  & v-1 \\
 v-1 & \rho  \\
\end{array}
\right)\,,\nonumber\\[4pt]
%%%
&L^9HL^{10}H:\hspace{0.25cm}\Omega\to \left(
\begin{array}{cc}
 \tau  & v \\
 v & \rho+1  \\
\end{array}
\right)\,,
&&HL^9HL^{10}:\hspace{0.25cm}\Omega\to \left(
\begin{array}{cc}
 \tau+1  & v \\
 v & \rho  \\
\end{array}
\right)\,.
\end{align}}
%%%%%%%%%%%%%%%%%%%%%%%%%%%%%%%%%%%
%%%%%%%%%%%%%%%%%%%%%%%%%%%%%%%%%%%
%%%%%%%%%%%%%%%%%%%%%%%%%%%%%%%%%%%

\end{document}